\documentclass[12pt]{article}

\usepackage{amsmath,amssymb,amsthm,amsfonts,amssymb}
\usepackage[mathscr]{eucal}
\usepackage{graphicx}
\usepackage{grffile}

\graphicspath{ {figures_new/} }

\usepackage{color}
\usepackage{caption}
\usepackage[numbers,super,comma,sort&compress]{natbib}

\usepackage{cancel}

\usepackage{tabulary}
\usepackage{setspace}
\usepackage{multirow}
\newcolumntype{C}{>{\centering\let\newline\\\arraybackslash\hspace{0pt}}m{0.8cm}}
\newcolumntype{D}[1]{>{\centering\let\newline\\\arraybackslash\hspace{0pt}}m{#1}}

\usepackage{hyperref}
\hypersetup{
 bookmarks=true,		
 unicode=false,			
 pdftoolbar=true,		
 pdfmenubar=true,		
 pdffitwindow=false,		
 pdfstartview={FitH},		
 pdftitle={Tensor product methods and entanglement optimization for ab initio quantum chemistry},    
 pdfauthor={Szil{\'a}rd Szalay, Max Pfeffer, Valentin Murg, Gergely Barcza, Frank Verstraete, Reinhold Schneider, \"Ors Legeza},	
 pdfsubject={tutorial review},	
 pdfcreator={pdflatex},		
 pdfproducer={vim},		
 pdfkeywords={quantum} {renormalization} {entanglement} {ab inition quantum chemistry} {DMRG} {MPS} {TTNS}, 
 pdfnewwindow=true,		
 colorlinks=true,		
 linktoc=page,			
 linkcolor=blue,		
 citecolor=blue,		
 filecolor=blue,		
 urlcolor=blue			
}

\newcommand{\be}{\begin{enumerate}}
\newcommand{\ee}{\end{enumerate}}
\newcommand{\beq}{\begin{eqnarray*}}
\newcommand{\eeq}{\end{eqnarray*}}
\newcommand{\beqnr}{\begin{eqnarray}}
\newcommand{\eeqnr}{\end{eqnarray}}

\newcommand{\dd}{\mathrm{d}}

\newtheoremstyle{remark}{.5\topsep}{.5\topsep}{\normalfont}{0pt}{\itshape}{.}{5pt plus 1pt minus 1pt}{#1 #2}

\theoremstyle{plain}
\newtheorem{theorem}{Theorem}[section]

\theoremstyle{definition}

\theoremstyle{remark}

\newtheorem{criterion}[theorem]{Criterion}

\makeatletter
\def\th@plain{%
  \thm@notefont{}
  \itshape 
}
\def\th@definition{%
  \thm@notefont{}
  \itshape 
}
\makeatother

\DeclareMathOperator*{\argmin}{argmin}
\DeclareMathOperator{\Span}{Span}
\DeclareMathOperator{\Lin}{Lin}
\DeclareMathOperator{\rank}{rank}
\DeclareMathOperator{\cond}{cond}

\newcommand{\expect}[1]{\langle #1 \rangle}

 \newcommand{\down}{ $\downarrow$ }
 \newcommand{\up}{$\uparrow$ }
 \newcommand{\zero}{$-$ }
 \newcommand{\double}{$\uparrow\downarrow$ }

\newcommand{\Varepsilon}{\mathscr{E}}

\title{Tensor product methods and entanglement optimization for ab initio quantum chemistry}

\author{Szil{\'a}rd Szalay\thanks{Strongly correlated systems ``Lend{\"u}let'' research group, Wigner Research Centre for Physics, Konkoly-Thege Mikl{\'o}s {\'u}t 29-33, 1121 Budapest, Hungary} \and
        Max Pfeffer\thanks{Fakult{\"a}t II - Mathematik und Naturwissenschaften, Institut f{\"u}r Mathematik, Technische Universit{\"a}t Berlin, Strasse des 17. Juni 136, Berlin, Germany} \and
        Valentin Murg\thanks{Fakult{\"a}t f{\"u}r Physik, Universit{\"a}t Wien, Boltzmanngasse 3, A-1090 Vienna, Austria} \and
        Gergely Barcza\footnotemark[1] \and
        Frank Verstraete\footnotemark[3] \and
        Reinhold Schneider\footnotemark[2] \and
        {\"O}rs Legeza\footnotemark[1]
}

\begin{document}

\maketitle

\begin{abstract}

The treatment of high-dimensional problems such as the Schr\"odinger
equation can be approached by concepts of tensor product approximation.
We present general techniques that can be used for the treatment of
high-dimensional optimization tasks and time-dependent equations, and
connect them to concepts already used in many-body quantum physics.
Based on achievements from the past decade, 
entanglement-based methods, -- developed from different
perspectives for different purposes in distinct communities already matured
to provide a variety of tools -- can be combined to attack highly
challenging problems in quantum chemistry.
The aim of the present paper is to give a pedagogical introduction to
the theoretical background of this novel field and demonstrate the
underlying benefits through numerical applications on a
text book example.
Among the various optimization tasks we will discuss only those which are
connected to a controlled manipulation of the entanglement which is
in fact the key ingredient of the methods considered in the paper.
The selected topics will be covered according
to a series of lectures given
on the topic \emph{``New wavefunction methods and
entanglement optimizations in quantum chemistry''}
at the
Workshop on Theoretical Chemistry, 18 - 21 February 2014, Mariapfarr, Austria.
%
%

\end{abstract}

\tableofcontents


\section{Introduction}
\label{sec:intro}

For the approximation of the wave function of the electronic structure of an atomic or mo\-le\-cu\-lar system, 
any method chosen will have to compromise between the demanded 
accuracy on the one hand and the high computational complexity of the task on the other. 
While \emph{Density Functional Theory} (DFT)\cite{Ciarlet-2003} 
and \emph{Coupled Cluster} (CC) or \emph{Quantum Monte Carlo} methods\cite{Szabo-1982,Helgaker-2000,Rohwedder-2013b}
are in this sense standard methods 
for the quantitative study of large weakly correlated systems,
there has been no method-of-choice solution for finding a sufficiently accurate, 
data-sparse representation of the exact many-body wave function 
if many electrons are \emph{strongly correlated,} as, for instance, 
in open-shell systems as transition metal 
complexes\cite{Cramer-2006,Reiher-2007,Marti-2008a,Reiher-2009,Kurashige-2009,Podewitz-2011,Barcza-2011,Boguslawski-2012a,Boguslawski-2012b,Sharma-2012a,Nakatani-2014,Harris-2014,Wouters-2014b}.

Due to the many-electron interactions present, strongly correlated 
problems cannot be sufficiently described by small perturbations 
of a single Slater determinant. For the treatment of other many-particle 
systems, e.g., spin systems, alternative representations have been 
proposed, resulting in the development of so-called 
\emph{Matrix Product States} (MPS)\cite{Ostlund-1995,Verstraete-2004b,Schollwock-2005,Schollwock-2011}. 
The MPS method represents the wavefunction of a system of $d$ components or \emph{``sites''} 
(corresponding, e.g., to \emph{molecular orbitals}) by forming products of $d$ matrices, 
each belonging to one component of the system. The computational 
complexity of the task is now governed by the size of these matrices,
related to the eigenvalue spectrum of the corresponding subsystem 
density matrices\cite{Peschel-1999b} 
characterizing in a formal way the so-called \emph{entanglement} among the different 
components\cite{Vidal-2003a,Legeza-2003a,Legeza-2003b,
Plenio-2005,Eisert-2010,Hastings-2007,Amico-2008,VanAcoleyen-2013}. 
MPS consists in a 
linear arrangement of the components,
while more recently 
the approach has been generalized to so-called 
\emph{Tensor Network States} (TNS)\cite{Verstraete-2004a,Murg-2007,Murg-2009,Verstraete-2008,Vidal-2008,Changlani-2009,Marti-2010d,Marti-2011,Evenbly-2011,Shi-2006,Ferris-2013,Legeza-2013,Legeza-2014,Orus-2014},
allowing a more flexible 
connection of the components of the respective system. Identical, but 
independent approaches were devised in numerical mathematics under the 
term of tensor product approximation, where low-rank factorization 
of matrices is generalized to higher order tensors 
\cite{Hackbusch-2012,Hackbusch-2014b,Hackbusch-2014a}. 
In quantum chemistry, the MPS%
\cite{White-1999,Chan-2002a,Legeza-2003a,Chan-2003,Legeza-2003c,Mitrushenkov-2003a,Legeza-2003b,Chan-2004,Chan-2004b,Legeza-2004b,Moritz-2005a,Chan-2005,Moritz-2005b,Moritz-2006,Hachmann-2006,Rissler-2006,Moritz-2007,Dorando-2007,Hachmann-2007,Marti-2008a,Zgid-2008a,Zgid-2008b,Zgid-2008c,Ghosh-2008,Chan-2008b,Yanai-2009,Dorando-2009,Kurashige-2009,Yanai-2010,Neuscamman-2010,Marti-2010b,Luo-2010,Mizukami-2010}\\\cite{Marti-2010d,Marti-2011,Barcza-2011,Boguslawski-2011,Kurashige-2011,Mitrushchenkov-2012,Sharma-2012a,Wouters-2012,Boguslawski-2012a,Yanai-2012,Boguslawski-2012b,Mizukami-2013,Boguslawski-2013,Kurashige-2013,Ma-2013,Saitow-2013,Liu-2013,Tecmer-2014,Nakatani-2014,Knecht-2014,Wouters-2014a,Harris-2014,Mottet-2014,Lan-2014,Sharma-2014,Kurashige-2014a,Wouters-2014b,Fertitta-2014,Keller-2014,Duperrouzel-2014}\\
and TNS 
\cite{Marti-2010d,Murg-2010a,Nakatani-2013,Murg-2014} 
representation can be used 
to approximate the full-CI wave function\cite{Legeza-2008,Chan-2008,Chan-2009,Marti-2010c,Marti-2011,Chan-2011,Chan-2012,Legeza-2013,Legeza-2014,Kurashige-2014b,Wouters-2014c,Wouters-2014e}.
By this new concept of data-sparse representation, an accurate 
representation of the electronic structure will then be possible in 
polynomial time if the exact wave function can be approximated to a 
sufficient extent by moderately entangled TNS representations. 
The underlying \emph{Molecular Orbital} (MO) basis can be optimized by well known techniques 
from multi-configurational methods\cite{Helgaker-2000} as,
e.g., \emph{Multi Configuration Self Consistent Field} (MCSCF) method,
which constitutes a tensor approximation method as well at the level of first quantization.

Entanglement-based methods,
-- developed from different perspectives for different purposes in distinct communities,
already matured to provide a variety of tools -- 
can be combined to attack highly challenging problems in quantum 
chemistry\cite{Workshop-Leiden-2004,Workshop-CECAM-2010,Workshop-Beijing-2010,Workshop-EBAQC-2012,Workshop-CECAM-2013}.
A very promising direction is, especially, to develop and implement an 
efficient \emph{Quantum Chemistry algorithm based on Tree Tensor Network States} (QC-TTNS), 
in particular enabling the treatment of problems in quantum chemistry that 
are intractable by standard techniques as DFT or CC\cite{Murg-2010a,Legeza-2013,Nakatani-2013,Murg-2014}.  

The aim of the present paper is to give a pedagogical introduction to
the theoretical background of this novel field and demonstrate the 
underlying benefits through numerical applications on a 
text book example.
We give a technical introduction to low rank 
tensor factorization and do not intend to present a detailed review of 
the field. Only some selected topics will be covered according 
to lectures given 
on the topic \textit{``New wave function methods and 
entanglement optimizations in quantum chemistry''}
at the
\href{http://tagung-theoretische-chemie.uni-graz.at/en/past-workshops/workshop-2014/}{Workshop on Theoretical Chemistry}, 18 - 21 February 2014, Mariapfarr, Austria%
\cite{Workshop-Mariapfarr-2014}.
In accordance with this, the organization of the present paper is as follows.
In Secs.~\ref{sec:qchem} and \ref{sec:tensor} a very detailed description  
of the theory follows so that those interested readers who just 
entered in the field could follow recent developments.
A brief summary in order to highlight the most important concepts used
in numerics is presented
in Sec.~\ref{sec:num} together with numerical applications  
by outlining ideas and existing 
algorithmic structures that have been used to arrive at an 
efficient implementation. 
At this stage among the various optimization tasks only those will
be analyzed which are connected directly to the manipulation 
of entanglement, which is in fact the key ingredient of the methods presented
in the paper.
To unify notations, in what follows, 
we mainly use terms and notations as common in physics and chemistry.

\subsection{Tensor product methods in quantum chemistry}
\label{sec:intro.mps}

Multi-particle Schr\"odinger-type equations constitute 
an important example of problems posed on high-dimensional tensor spaces. 
Numerical approximation of solutions of these problems 
suffers from the \emph{curse of dimensionality}, 
i.e., the computational complexity scales exponentially with the dimension 
of the space.
Circumventing this problem is a challenging topic in modern numerical 
analysis with a variety of applications, covering aside from the 
electronic and nuclear Schr\"odinger equation
e.g., the Fokker-Planck equation and the chemical master equation\cite{Lubich-2008}.
Considerable progress in the treatment of such problems has been made 
by concepts of tensor product approximation\cite{Hackbusch-2010,Hackbusch-2012,Hackbusch-2014b}.

In the year~1992, S.~R.~White introduced a very powerful numerical method,
the Density-Matrix Renormalisation Group (DMRG)\cite{White-1992a,White-1992b,White-1993}.
It allows us to determine the physical properties of
low-dimensional correlated systems
such as quantum spin chains or chains
of interacting itinerant electrons to unprecedented
accuracy\cite{Peschel-1999,Dukelsky-2004,Hallberg-2004,Schollwock-2005,Noack-2005,Hallberg-2006}.

Further success of the DMRG method in quantum physics motivated its application 
to \emph{Quantum Chemical} problems (QC-DMRG)\cite{White-1999,Mitrushenkov-2001,Chan-2002a,Legeza-2003a,Moritz-2005a}.
In the treatment of problems where large active spaces are mandatory to obtain 
reliable and accurate results, it has proven capable of going well beyond the 
limits of present day quantum chemistry methods and even reach the 
full-CI limit\cite{Legeza-2008,Chan-2008,Chan-2009,Marti-2010c,Marti-2011,Chan-2011,Chan-2012,Legeza-2013,Legeza-2014,Kurashige-2014b,Wouters-2014c,Wouters-2014d}.

In the past decade, the method has gone through major algorithmic 
developments by various groups\cite{Schollwock-2005,Legeza-2008,Chan-2008,Marti-2010b,Marti-2010b,Kurashige-2014b,Wouters-2014e}.
For example, the two-body reduced density matrices calculated with 
the DMRG method\cite{Chan-2002a,Legeza-2003a,Zgid-2008b} can be used in the standard orbital 
optimization procedure\cite{Helgaker-2000}.
Resulting methods are the \emph{DMRG Complete Active Space Self Consistent Field} (DMRG-CASSCF) 
or \emph{DMRG Self Consistent Field} (DMRG-SCF)\cite{Zgid-2008b,Zgid-2008c,Ghosh-2008,Yanai-2009}. 
Another direction is the post-DMRG treatment of dynamic correlation.
DMRG as it can be considered as a \emph{CAS Configuration Interaction} (CAS-CI) technique can recover static correlation,
and, depending on the size of the active space, one can afford 
also some portion of the dynamic correlation. Quite recently, 
various advanced methods accounting for dynamic correlation
on top of the DMRG framework have been 
developed\cite{Kurashige-2011,Saitow-2013,Liu-2013,White-2002,Yanai-2010,Neuscamman-2010,Yanai-2012}. 
The first implementation of the relativistic quantum
chemical \emph{two- and four-component} density matrix renormalization group
algorithm (2c- and 4c-DMRG) has also been presented\cite{Knecht-2014}.

The DMRG method can be used to calculate  
ground as well as excited states. This can be achieved either 
by targeting several of them in a state
average fashion\cite{Legeza-2003a,Legeza-2003c,Moritz-2005b,Dorando-2007,Ghosh-2008,Liu-2013,Wouters-2014a,Sharma-2014,Murg-2014}
or alternatively based on the MPS tangent 
vectors\cite{Nakatani-2014,Pirvu-2012,Haegeman-2012,Wouters-2013,Haegeman-2013}.  
Since the DMRG method is very flexible it can be used even in such situations 
when the wave function character changes dramatically\cite{Legeza-2003c,Moritz-2005b,Boguslawski-2013,Wouters-2014a,Murg-2014}.
Additionally, the ansatz is size consistent by construction and
symmetries as particle number,  
spin projection\cite{White-1999}, 
spin reflection symmetries\cite{Legeza-1997},
Abelian point group symmetries\cite{Legeza-2003b,Legeza-2003c,Chan-2004b}
and even non-Abelian symmetries can be factored out explicitly\cite{Sierra-1997,McCulloch-2000,McCulloch-2001,McCulloch-2002a,McCulloch-2007,Pittel-2006,Rotureau-2006,Toth-2008,Legeza-2008b,Thakur-2008,Zgid-2008a,Singh-2010a,Singh-2010b,Singh-2012,Sharma-2012a,Weichselbaum-2012,Wouters-2012,Wouters-2014a,Weichselbaum-2012}. 
Quite recently,  MPS and further  tensor  product  approximations have been 
applied in \emph{post Hartree-Fock} (post-HF) methods 
to the decomposition of the two electron integrals,
the AO-MO \emph{(Atomic Orbital-Molecular Orbital)} transformation,
and the \emph{M{\o}ller-Plesset perturbation theory} (MP2) energy expression\cite{Benedikt-2011}.

In the MPS like methods the computational complexity of the task is governed 
by the size of the matrices used to approximate the wavefunction, which can, 
however, be controlled based on various truncation criteria to achieve a 
priory set error margin\cite{Legeza-2003a,Holtz-2012b}.
In a system with identical sites, this feature is directly connected to the scaling of entanglement 
when subsystems include larger and larger portion of the total system,
also called as area law\cite{Hastings-2007,Plenio-2005,Eisert-2010,VanAcoleyen-2013}.
The situation is more complicated in quantum chemical applications 
since the ranks 
of the matrices also depend strongly on the ordering of the 
matrices\cite{Legeza-2003a,Legeza-2003b,Legeza-2003c,Barcza-2011,Boguslawski-2012b},
thus different orderings lead to better or worse results if the ranks 
are kept fixed\cite{Chan-2002a,Mitrushenkov-2003a,Moritz-2005a,Rissler-2006,
Kurashige-2009,Yanai-2010,Mizukami-2013,Ma-2013,Wouters-2014a}. 
Another main aspect that effects the performance of the method is the 
optimization of the basis\cite{Legeza-2006b,Zgid-2008c,Ghosh-2008,Yanai-2009,Murg-2010a,Ma-2013} and initialization of the 
network\cite{Mitrushenkov-2001,Chan-2002a,Legeza-2003b,Moritz-2006,Barcza-2011,Wouters-2014a}. 
Even though the significant efforts dedicated to the various 
optimization tasks, it remains an open question to determine the minimum 
of computational effort to obtain results with a given accuracy threshold.   

Shortly after DMRG was introduced, it was found that DMRG may also be 
phrased in terms of MPS\cite{Ostlund-1995}, 
first formulated for special spin systems as the \emph{Affleck-Kennedy-Lieb-Tasaki} (AKLT) model\cite{Affleck-1987}.
More recently, the \emph{Higher Order Singular Value Decomposition} (HOSVD)\cite{Vidal-2003b,Verstraete-2004b,Verstraete-2008} have made 
MPS the basis of variational frameworks and revealed a profound connection 
to quantum information theory\cite{Vidal-2003a,Legeza-2003b,Schollwock-2011}. 
In this context, it became apparent 
that MPS is only one of a more general set of formats: while 
MPS corresponds to an arrangement of orbitals in a linear topology, 
quantum states may more generally be arranged as 
more complex topologies, leading to tensor network states 
TNS\cite{Murg-2010a,Marti-2010d,Nakatani-2013,Murg-2014}. 
For applications to smaller systems, prototypical tensor-network state 
approaches 
to quantum chemistry have already been developed, including the so called 
\emph{Complete Graph Tensor Network State} (CGTNS) approach\cite{Marti-2010d}, 
and the \emph{Tree Tensor Network State} (TTNS) approach\cite{Murg-2010a,Nakatani-2013,Murg-2014}. 
The QC-TTNS combines a number of favorable features that suggest
 it might represent a novel, flexible approach in quantum chemistry: the more 
general concept of data-sparsity inherent in the TNS representation allows for 
the efficient representation of a much bigger class of wave functions than 
accessible by state-of-the-art methods. 
The desired accuracy may be adjusted, so that the ansatz in principle permeates the whole full-CI space.

These developments 
foster the hope that with their help some of the major questions 
in quantum chemistry and condensed matter physics may be solved. 
The concept of MPS and tree structured tensor network states has been rediscovered independently in numerical mathematics
for tensor product approximation\cite{Oseledets-2009,Hackbusch-2010}.

\subsection{Entanglement and quantum information entropy in quantum chemistry} 
\label{sec:intro.entanglement}

%
%
In quantum systems, correlations having no counterpart in classical physics 
arise. Pure states showing these strange kinds of correlations are 
called entangled ones\cite{Preskill,Hastings-2007,Plenio-2005,Amico-2008,Horodecki-2009,Szalay-2013,Eisert-2010, VanAcoleyen-2013},
and the existence of these states has so deep and important 
consequences\cite{Einstein-1935,Bell-1967,Clauser-1969}
that Schr{\"o}\-din\-ger has identified entanglement to be the characteristic 
trait of quantum mechanics\cite{Schrodinger-1935a,Schrodinger-1935b}.
The QC-DMRG and QC-TTNS algorithms approximate a composite system 
with strong interactions between many pairs of orbitals,
and it turned out that the results of quantum information 
theory\cite{Nielsen-2000,Wilde-2013}
can be used to understand the criteria of their convergence.
\cite{Vidal-2003a,Legeza-2003b,Verstraete-2004b,Schollwock-2005,Schollwock-2011}. 
 
Recently, quantum information theory has also appeared in quantum chemistry 
giving a fresh impetus to the development of methods in electronic structure
theory\cite{Ziesche-1995,Nagy-1996,Nalewajski-2000,Legeza-2003b,Legeza-2004b,Rissler-2006}\\\cite{Juhasz-2006,Luzanov-2007,Alcoba-2010,Barcza-2011,Boguslawski-2012a,Boguslawski-2012b,Boguslawski-2013,Kurashige-2013,Barcza-2014,Mottet-2014,Knecht-2014,Fertitta-2014,Duperrouzel-2014,Boguslawski-2014}.
The amount of contribution of an orbital to the total correlation can be characterized,
for example, by the \emph{single-orbital entropy}\cite{Legeza-2003b},
and the the sum of all single-orbital entropies gives the amount of \emph{total correlation} encoded in the
wave function\cite{Legeza-2004b,Legeza-2006b}.
This quantity can be used to monitor changes in
entanglement as system parameters are adjusted, for example, changing bond
length or other geometrical properties\cite{Murg-2014,Fertitta-2014,Duperrouzel-2014}.
A useful quantity to numerically characterize the correlations (classical and quantum together) 
between pairs of orbitals is the \emph{mutual information}\cite{Legeza-2006a,Rissler-2006,Barcza-2011}
and it together with the orbital entropy 
provides chemical information about the system, especially about bond
formation and nature of static and dynamic correlation\cite{Barcza-2011,Boguslawski-2012b,Boguslawski-2013,Duperrouzel-2014}.

The two-orbital mutual information also yields a weighted graph of the
overall two-orbital correlation of both classical and quantum origin 
reflecting the entanglement topology of the molecules.
Therefore, this quantity can also be used to carry out optimization 
tasks based on  
the entanglement between the different components -- itself determining 
the complexity of the computation -- since it depends strongly on the chosen 
network topology and is in principle unknown for a given system.
To promote the efficiency of tensor product methods various entanglement-based 
approaches are used to determine, for example, the appropriate ordering, 
network topology, optimal basis, and efficient network initialization. 
These important and non-trivial tasks will 
be considered in Sec.~\ref{sec:num}.

\subsection{Tensor decomposition methods in mathematics}
\label{sec:intro.tns-mathematics}

Approximation of quantities in high dimensional spaces is a hard problem 
with a variety of applications, 
and the development of generic methods that circumvent the 
enormous complexity of this task have recently, independent of the 
developments in the study of quantum systems, gained significant interest in 
numerical mathematics\cite{Hackbusch-2012}. 
A recent analysis shows that, beyond the matrix case (corresponding to tensors of order 2), 
almost all tensor problems, 
even that of finding the best rank-1 approximation of a given tensor, 
are in general NP hard\cite{Hillar-2011}. 
Although this shows that tensor product approximation in 
principle is an extremely difficult task, a variety of generic concepts 
for the approximation of solutions of certain problem classes have 
recently been proposed\cite{Hackbusch-2014a,Hackbusch-2014b}, 
some of which\cite{Hackbusch-2012,Grasedyck-2013,Lubich-2008,Rohwedder-2013,Dolgov-2012,Lubich-2012,Uschmajew-2013}
bear a surprising similarity to methods used to treat problems in quantum physics\cite{Haegeman-2011,Haegeman-2014}.

The classical \emph{Tucker format} attains sparsity via a subspace approximation. 
Multi-configurational methods like MCSCF or CASSCF are in fact a Tucker approximation in the framework of antisymmetry.
Its unfavorable scaling has recently been circumvented by 
a multilevel or hierarchical subspace approximation framework named \emph{Hierarchical Tucker format}\cite{Hackbusch-2010,Hackbusch-2012}, 
interestingly corresponding to the TTNS. 
A similar format called \emph{Tensor Trains,} developed independently\cite{Oseledets-2010,Oseledets-2011,Savostyanov-2014}, 
is a formal version of the MPS with open boundary conditions.
Investigation of the theoretical properties 
of TNS and MPS in a generic context have shown that they inherit desirable 
properties of matrix factorization. E.g., closedness of the set of tensors of 
fixed block size\cite{Hackbusch-2012} implies the existence of minimizers 
in these 
sets for convex optimization problems. Also, these sets possess a manifold 
structure that helps to remove redundancy in the parametrization by the 
introduction of so-called gauge conditions\cite{Holtz-2012a}. They can be 
used to set up variational frameworks for the treatment of optimization 
problems\cite{Legeza-2013} and of time-dependent problems 
\cite{Lubich-2012,Dolgov-2012,Haegeman-2014,Lubich-2014}, 
bearing again a close connection to approaches in the quantum physics 
community\cite{Haegeman-2011}. In this general context,
the robustness and quasi-best approximation of the HOSVD,
studied in the mathematics community\cite{Grasedyck-2010,Hackbusch-2012,Hackbusch-2014b},
and of the (one site) DMRG\cite{White-2005} 
as simple and very efficient numerical methods for the treatment of optimization problems
are now well-understood\cite{Hackbusch-2012,Hackbusch-2014a}. 
These fundamental properties establish MPS and TTNS as advantageous concepts in 
tensor product approximation. It is important to note that all these properties are no longer valid in general
if the tensor networks contains closed loops, as in case of 
the projected entangled pair states (PEPS)\cite{Verstraete-2004a} and the
multiscale entanglement renormalization ansatz (MERA)\cite{Vidal-2007}.  
It is still widely unexplored under which conditions 
the favorable 
properties of tree structured TNS can be extended to general TNS.  
In mathematics the 
phrases 
\emph{hierarchical tensor representation} or \emph{Hierarchical Tucker format} as well as Tensor Trains instead of MPS  
are used since there the focus is not only on quantum mechanical systems, but rather on universal tools to handle high-dimensional approximations.   
Many of the recent developments in mathematics parallel others in 
quantum computations on a more formal, generic level, often faced with similar 
experiences and similar problems.

\section{Quantum chemistry}
\label{sec:qchem}

\subsection{The electronic Schr\"odinger equation}
\label{sec:qchem.eSchrodinger}

A quantum mechanical system of $N$ non-relativistic electrons 
is completely described by a \emph{state-function} $\Psi$
depending on $3N$ spatial variables $\mathbf{r}_a \in \mathbb{R}^3$, $a=1, \ldots ,N$,  
together with $N$ discrete spin variables $s_a \in \{\pm \frac{1}{2}\}$, $a=1, \ldots, N$,
\begin{align}
 \Psi : \mathbb{R}^{3N}\otimes \Bigl\{\pm \frac{1}{2}\Bigr\}^N 
\cong \Bigl(\mathbb{R}^{3} \otimes \Bigl\{\pm\frac{1}{2}\Bigr\}\Bigr)^N \quad\longrightarrow&\quad \mathbb{C} \nonumber\\
 ( \mathbf{r}_1,s_1; \ldots ; \mathbf{r}_N, s_N ) \quad\longmapsto&\quad 
 \Psi ( \mathbf{r}_1,s_1; \ldots ; \mathbf{r}_N, s_N ).
\end{align}
The function $\Psi $ belongs to the Hilbert space $L_2\bigl((\mathbb{R}^3 \times \{\pm \frac{1}{2}\})^N\bigr)$
having the standard inner product
\begin{equation}\label{eq:innerp}
 \langle \Psi , \Phi \rangle= \sum_{s_i = \pm \frac{1}{2}} \int_{\mathbb{R}^{3N}} 
\overline{\Psi(\mathbf{r}_1 ,s_1 ; \ldots ; \mathbf{r}_N , s_N )}
          \Phi(\mathbf{r}_1 ,s_1 ; \ldots ; \mathbf{r}_N , s_N )
\dd\mathbf{r}_1 \ldots \dd\mathbf{r}_N,
\end{equation} 
and the norm $\Vert\Psi\Vert = \sqrt{\langle\Psi ,\Psi \rangle}$.
The \emph{Pauli antisymmetry principle} states that the wave function of \emph{fermions,} 
in particular electrons, must be antisymmetric with respect to the permutation of variables, i.e., 
for $a\neq b$
\begin{equation}
\label{eq:antiprinc}
 \Psi ( \ldots ; \mathbf{r}_a, s_a ; \ldots ; \mathbf{r}_b , s_b ; \ldots ) =
-\Psi ( \ldots ; \mathbf{r}_b, s_b ; \ldots ; \mathbf{r}_a , s_a ; \ldots ).
\end{equation}
Such wave-functions are the elements of the antisymmetric tensor subspace
$\bigwedge_{i=1}^N L_2 \bigl( \mathbb{R}^3 \times \{\pm\frac{1}{2}\}\bigl)$.
The \emph{Pauli exclusion principle} immediately follows:
$\Psi$ must vanish for the points of 
$(\mathbb{R}^3 \times \{\pm \frac{1}{2}\})^N$
which have the coordinates $\mathbf{r}_a=\mathbf{r}_b$ and $s_a=s_b$ for some $a\neq b$ fermions~\cite{Ciarlet-2003,Reed-1978}.

In quantum mechanics, we are usually interested in wave-functions
having definite energies. 
This is expressed by the \emph{stationary Schr\"odinger equation},
\begin{equation}\label{eq:Schrodinger}
H \Psi= E \Psi,
\end{equation}
i.e., the wave function is an eigenfunction of a differential operator,
namely the \emph{Hamilton operator} $H$,
and the eigenvalue $E\in\mathbb{R}$ is the energy of the
corresponding state $\Psi$.
One of the most important quantities is the \emph{ground state energy} $E_0$,
which is the lowest eigenvalue.
The well known \emph{Born-Oppenheimer-approximation} considers 
a nonrelativistic quantum mechanical system of $N$ electrons 
in an exterior field generated by the $K$ nuclei.
In this case $H$ is as follows
\begin{subequations}
\begin{gather}
 \label{eq:Hamilton}
 H = H_{\text{kin}} + H_{\text{pot}},  \qquad  H_{\text{pot}}= H_{\text{ext}} + H_{\text{int}},\\
 \label{eq:HamiltonParts}
 H_{\text{kin}} = \sum_{a=1}^N - \frac{1}{2}\Delta_a,\qquad
 H_{\text{ext}} = - \sum_{a=1}^N\sum_{c=1}^K \frac{Z_c}{|\mathbf{R}_c - \mathbf{r}_a|},\qquad
 H_{\text{int}} = \frac12 \sum_{\substack{a,b=1\\ b\not=a}}^N\frac{1}{|\mathbf{r}_b-\mathbf{r}_a|}.
\end{gather}
\end{subequations}

Since the Hamilton operator is a linear second order differential operator,
the analysis for the electronic Schr\"odinger equation has been already established to a certain extent. 
We would like to briefly summarize some basic results
and refer to the literature\cite{Reed-1978,Ciarlet-2003}.
The \emph{Sobolev spaces} $H^m := H^m \bigl( (\mathbb{R}^3 \times \{\pm\frac{1}{2}\})^N \bigr)$,
$m \in \mathbb{N}_0$ are defined as the spaces of functions 
for which all derivatives up to order $m$ are in
$H^0:= L_2 \bigl( (\mathbb{R}^3 \times \{\pm \frac{1}{2}\})^N \bigr)$.
Consequently, the operator $H$ maps the Sobolev space $H^1 
$ continuously into its dual space $H^{-1}$, i.e.,
$ H: H^1 \to H^{-1}$ boundedly\cite{Ciarlet-2003,Reed-1978,Yserentant-2004}.
The potential operator $H_{\text{pot}}$ maps the Sobolev spaces $ H^1 
$ continuously into $ H^0 
$, i.e., $H_{\text{pot}}: H^1 \to H^0 = L_2$ boundedly\cite{Reed-1978,Yserentant-2004}.
The electronic Schr\"odinger operator admits a rather complicated
spectrum. We are interested mainly in  the ground state energy $E_0$. 
If $ \sum_{c=1}^K Z_c \geq N$, in particular for electrical
neutral systems, it is known\cite{Reed-1978,Yserentant-2010} that $E_0$ 
is an eigenvalue of finite multiplicity of the operator $ H : H^2
\to H^0$ below the essential spectrum  $\sigma_\text{ess}(H)$ of $H$,
i.e.,  $-\infty < E_0 < \inf \sigma_\text{ess} (H) $.
Summing up, the energy space for the electronic Schr\"odinger equation is
\begin{equation}\label{eq:V-space}
 \mathcal{V}_N = H^1 \Bigl( \Bigl(\mathbb{R}^3 \times \Bigl\{\pm\frac{1}{2}\Bigl\}\Bigl)^N \Bigl) 
\cap \bigwedge_{i=1}^N L_2 \Bigl( \mathbb{R}^3 \times \Bigl\{\pm\frac{1}{2}\Bigl\}\Bigl).
\end{equation}
This situation will be  considered in the sequel.

For the sake of simplicity, we will also always 
assume that $E_0$ is a simple eigenvalue, i.e., of multiplicity one.
In the case we deal with here, 
i.e., the stationary electronic Schr\"odinger equation in non-relativistic and Born-Openheimer setting,
we can assume without the loss of generality that the wave function is real valued.
(This does not hold for linear response theory or time-dependent problems,
as well as for the relativistic regime,
 where complex phases play an important role.)
According to the well known mini-max principle\cite{Reed-1978},
the ground state energy and the corresponding wave function 
satisfies the \emph{Rayleigh-Ritz variational principle}\cite{Yserentant-2004,Reed-1978}, 
i.e., the lowest eigenvalue is the minimum of the Rayleigh quotient 
$\frac{\langle \Psi , H\Psi \rangle}{\langle \Psi , \Psi \rangle}$,
or equivalently,
\begin{subequations}\label{eq:var}
\begin{align}
\label{eq:var.E0}
E_0    =    \min&\bigl\{ \langle \Psi , H\Psi \rangle : \langle \Psi , \Psi \rangle = 1, \Psi\in \mathcal{V}_N \bigr\},\\
\label{eq:var.Psi0}
\Psi_0 = \argmin&\bigl\{ \langle \Psi , H\Psi \rangle : \langle \Psi , \Psi \rangle = 1, \Psi\in \mathcal{V}_N \bigr\}.
\end{align}
\end{subequations}
Since the Hamilton operator maps $ H : \mathcal{V}_N \to(\mathcal{V}_N)^*$ boundedly,
we will put the eigenvalue problem into the following weak formulation\cite{Yserentant-2004}, 
to find the normalized $ \Psi_0 \in \mathcal{V}_N$, 
satisfying
\begin{equation}\label{eq:weak-Schroedinger}
    \langle \Phi , ( H - E_0 ) \Psi_0 \rangle = 0, \qquad \langle \Psi_0 , \Psi_0 \rangle =1,\qquad
\forall \Phi \in  \mathcal{V}_N.
\end{equation}
We will consider the above framework~\cite{Ciarlet-2003,Yserentant-2010} throughout the present paper.

\subsection{Full configuration interaction approach and the Ritz-Galerkin approximation}
\label{sec:qchem.FullCIRG}

A convenient way to approximate the wave function is to use an anti-symmetric 
tensor product of basis functions depending only on single particle variables $(\mathbf{r}_a, s_a)$,
which can be realized by determinants. 
To this end, let us consider a finite subset of an orthonormal set of basis functions 
$\varphi_i:(\mathbf{r}, s ) \mapsto \varphi_i (\mathbf{r},s )$ in
$H^1 ( \mathbb{R}^3 \times \{\pm\frac{1}{2}\} )$, that is,
\begin{align*} 
 B^d := \bigl\{\varphi_i : i= 1 , \ldots , d \bigr\} \quad&\subseteq\quad
 B   := \bigl\{\varphi_i : i \in \mathbb{N}\bigr\} \quad\subseteq\quad
 H^1\Bigl(\mathbb{R}^3 \times \Bigl\{\pm \frac 12\Bigr\}\Bigr),\\
\mathcal{V}^d :=\Span  B^d \quad&\subseteq\quad 
\mathcal{V}   :=\Span B = H^1 \Bigl( \mathbb{R}^3 \times \Bigl\{\pm\frac{1}{2}\Bigr\} \Bigr),
\end{align*}
where
\begin{equation}\label{eq:orbitalog}
  \langle \varphi_i , \varphi_j \rangle
:= \sum_{s= \pm \frac{1}{2}} \int_{\mathbb{R}^3} 
\overline{\varphi_i (\mathbf{r},s)} \varphi_j ( \mathbf{r},s) \dd\mathbf{r}
= \delta_{i,j}. 
\end{equation}
(For simplicity of notation, we will  use  the same  brackets
$\langle \cdot \,, \cdot \rangle $ for designating inner products in Hilbert
spaces, independent of the underlying Hilbert space.)
In quantum chemistry these functions are called \emph{spin orbitals},
because they depend on the spin variable $s =\pm \frac{1}{2}$ and
the spatial variable $ \mathbf{r} \in \mathbb{R}^3$. 
In the sequel,
we will  first confine ourselves to spin orbital formulations.
How we go from spin orbitals to \emph{spatial orbitals} will be explained later.

We build \emph{Slater determinants} of an $N$-electron system, 
by selecting $N$ different indices, 
for example $i_a$ for $a=1,\ldots,N$,
out of the set $\{1,\ldots,d\}$.
By this we have chosen $N$ ortho-normal spin orbitals
$\varphi_{i_a}$, $a= 1 , \ldots ,N$, 
to define the \emph{Slater determinant}~\cite{Szabo-1982,Ciarlet-2003}
\begin{equation}
\label{eq:Slater} 
\begin{split}
\Phi_{[i_1 , \ldots , i_N ]} ( \mathbf{r}_1, s_1 ; \ldots ; \mathbf{r}_N , s_{N} ) 
&= \frac{1}{\sqrt{N!}} \det\bigl( \varphi_{i_a}(\mathbf{r}_b,s_b)\bigr)_{a , b =1 }^N \\ 
&= \frac{1}{\sqrt{N!}} \sum_{\sigma\in S_N} P(\sigma) 
\bigl(\varphi_{i_{\sigma(1)} } \otimes \dots \otimes \varphi_{i_{\sigma(N)}}\bigr)(\mathbf{r}_1 ,s_1 ; \ldots ; \mathbf{r}_N , s_N ),
\end{split}
\end{equation}
where the summation goes for all $\sigma$ permutations of $N$ elements,
and $P(\sigma)$ is the parity of the permutation.
To fix the sign  of the determinant, we suppose e.g. that  
$ i_a < i_{a+1}$ for $a = 1, \ldots ,N-1$; i.e., the indices are ordered increasingly.
Therefore the Slater determinants are uniquely defined by 
referring to the orbital functions $\varphi_{i_a}$, 
respectively indices $i_a  \in \{ 1, \ldots , d\}$, which are contained in the determinant.

It is easy to check that the Slater determinants constructed in this way
by the ortonormalized spin-orbitals $\varphi_i\in\mathcal{V}^d$
are also orthonormalized.
We define the  \emph{Full Configuration Interaction (FCI) space for an $N$-electron system}~\cite{Szabo-1982,Helgaker-2000}
as the finite dimensional space $ \mathcal{V}^d_N$ spanned by the Slater-determinants
\begin{align*}
 B^d_N := \bigl\{\Phi_{[i_1,\ldots,i_N]} :  1 \leq i_a < i_{a+1} \leq d  \bigr\} \quad&\subseteq\quad
 B_N   := \bigl\{\Phi_{[i_1,\ldots,i_N]} :  1 \leq i_a < i_{a+1}  \bigr\} \quad\subseteq\quad
 \mathcal{V}_N,\\
\mathcal{V}^d_N :=\Span  B^d_N \quad&\subseteq\quad 
\mathcal{V}_N   :=\Span B_N.
\end{align*}
The dimension of $\mathcal{V}^d_N$ is
\begin{equation}\label{eq:dimVdN}
\dim \mathcal{V}^d_N = \binom{d}{N}
= \frac{d!}{ N! ( d - N)!} \sim \mathcal{O} ( d^{N}) .
\end{equation}

To obtain an approximate solution of the electronic Schr\"odinger equation,
one may apply the \emph{Ritz-Galerkin method} using the finite dimensional subspace
$\mathcal{V}^d_N \subset \mathcal{V}_N$. 
I.e., consider the solution of the finite dimensional eigenvalue problem
\begin{equation} \label{eq:SchrodingerVdN} 
  \Pi^d_N H \Psi = E \Psi, \quad
  \Psi \in \mathcal{V}_N^d, \quad
\end{equation}
where $\Pi^d_N : \mathcal{V}_N \to \mathcal{V}_N^d$ is $L_2$-orthogonal projection,
or, equivalently,
\begin{equation} \label{eq:weak-SchrodingerVdN}
  \langle \Phi , (H-E) \Psi \rangle = 0,\quad \Psi \in \mathcal{V}_N^d,\quad
  \text{for all} \; \Phi \in \mathcal{V}_N^d.
\end{equation}
So the approximate ground state energy is
\begin{subequations}\label{eq:varVdN}
\begin{equation}\label{eq:varVdN.E0}
E_{0,d} := \min \bigl\{ \langle \Psi , H\Psi \rangle : \langle \Psi , \Psi \rangle = 1, \Psi\in \mathcal{V}^d_N \bigr\},
\end{equation}
and the \emph{full CI} ground state wavefunction $\Psi_{0,d} \in \mathcal{V}^d_N$
is the solution of the Galerkin scheme
\begin{equation}\label{eq:varVdN.Psi0}
\Psi_{0,d} := \argmin\bigl\{ \langle \Psi , H\Psi \rangle : \langle \Psi , \Psi \rangle = 1, \Psi\in \mathcal{V}^d_N \bigr\}.
\end{equation}
\end{subequations}
From the above definitions it becomes obvious that the approximate
ground state energy $E_{0,d}$ obtained by the Ritz-Galerkin method provides an
upper bound for the exact energy value $E_0 \leq E_{0,d}$, given in (\ref{eq:var.E0}).
The convergence theory  of the Galerkin scheme for the numerical
solution of eigenvalue problems is well established\cite{Yserentant-2004,Ciarlet-2003}. 
Roughly speaking, the Galerkin method provides an approximate
eigenfunction $\Psi_{0,d}$ which approximates the exact eigenfunction
quasi-optimally. Moreover, the eigenvalue converges quadratically
compared to the convergence of the eigenfunction. 
Since  $\dim \mathcal{V}_{d}^N \sim \mathcal{O}(d^N)  \geq \mathcal{O}(2^N)$, 
the full CI approach scales exponentially with respect to $N$.
Therefore, for molecules this approach is practically not feasible,
except for a very small number of electrons.

\subsection{Fock spaces} 
\label{sec:qchem.Fock}

We embed the full CI space $\mathcal{V}^d_N$ of $N$-electrons 
into a larger space $\mathcal{F}^d$, called \emph{(discrete) Fock space},
where we do not care about the number of electrons,
\begin{equation}\label{eq:Fock}
\mathcal{F}^d := \bigoplus_{M=0}^d \mathcal{V}_M^d
=\{ \Psi= \Psi_{(0)}\oplus\Psi_{(1)}\oplus\ldots\oplus\Psi_{(d)} : \Psi_{(M)}\in\mathcal{V}_M^d\}.
\end{equation}
Its dimension is
\begin{equation}\label{eq:dimFd}
 \dim \mathcal{F}^d = \sum_{M=0}^d\binom{d}{M} = 2^d.
\end{equation}
The Fock space is again a Hilbert space with the inner product
inherited from $\mathcal{V}_M^d$
\begin{equation}\label{eq:innerpFd}
\langle \bigoplus_{M=0}^d\Phi_{(M)} , \bigoplus_{M'=0}^d\Psi_{(M')} \rangle 
= \sum_{M=0}^d \langle \Phi_{(M)} , \Psi_{(M)} \rangle.
\end{equation}
The full Fock space can be obtained by taking the limit for $d\to \infty$.
Since we consider only finite dimensional approximation,
we are not intended here to understand in what sense this limit might be defined or not.

We have the Hamiltonian (\ref{eq:Hamilton})
acting on the full-CI space of N electrons 
as $\Pi_N^d H:\mathcal{V}_N^d\to \mathcal{V}_N^d$.
Since now we allow different numbers of electrons,
we denote this explicitly as $H_N$,
then the Hamiltonian acting on the whole Fock space reads
\begin{equation}\label{eq:HamiltonFd}
H^d:=\bigoplus_{M=0}^d\Pi_M^d H_M\ : \ \mathcal{F}^d \to \mathcal{F}^d.
\end{equation}

It is convenient to define the \emph{creation operator} $a_i^\dagger: \mathcal{F}^d\to \mathcal{F}^d$,
which is given on Slater determinants as
$a_i^\dagger\Phi_{[i_1, \ldots , i_N ]} := \Phi_{[i,i_1, \ldots , i_N ]}$.
This connects the subspaces with different numbers of particles in the Fock space,
$a_i^\dagger:\mathcal{V}_N^d \to \mathcal{V}_{N+1}^d$.
The result of this operator acting on a Slater determinant in $B^d_N$ is a Slater determinant again,
and, up to a $\pm$ sign, it is contained in $B^d_N$:
\begin{subequations}\label{eq:opFd}
\begin{equation}\label{eq:opFd.adag}
 a_i^\dagger  \Phi_{[i_1 , \ldots , i_N ]} = 
\Phi_{[i, i_1, \ldots , i_N ]}   = (-1)^{k_i}\Phi_{[i_1, \ldots,i , \ldots, i_N ]}, 
\end{equation}
where the indices in $[i_1, \ldots,i , \ldots i_N ]$ are ordered increasingly, 
and $k_i=|\{i_b|i_b<i\}|$.
From the definition, it immediately follows that $a_i^\dagger\Phi_{[i_1, \ldots , i_N ]}=0$
if $i\in[i_1, \ldots , i_N ]$, which is the manifestation of the exclusion principle.
One can then obtain the adjoint $a_i:=(a_i^\dagger)^\dagger$ of the creation operator, 
which is called \emph{annihilation operator}
\begin{equation}\label{eq:opFd.a}
 a_i \Phi_{[i_1 , \ldots,i_b,\ldots , i_N ]}  
  = (-1)^{k_i}\Phi_{[i_1, \ldots, i_N ]}, \qquad\text{if $i_b=i$ for some $b=1,\ldots,N$, othervise $0$}.
\end{equation}
\end{subequations}
It is straightforward to check that these operators obey the \emph{fermionic anticommutation relations:}
\begin{equation}\label{eq:fermAnticommFd}
\{a_i^\dagger,a_j^\dagger\} = 0, \qquad
\{a_i,a_j\}= 0, \qquad
\{a_i,a_j^\dagger\} = \delta_{i,j},
\end{equation}
(with the \emph{anticommutator} $\{A,B\}=AB+BA$),
which is the manifestation of the Pauli antisymmetry principle (\ref{eq:antiprinc}).
One can check that the operator $n_i:=a_i^\dagger a_i$ 
leaves invariant all the Slater determinants for which $i\in[i_1,\ldots,i_N]$,
while it annihilates all the Slater determinants for which $i\notin[i_1,\ldots,i_N]$.
One can conclude then that the operator
$P=\sum_{i=1}^da_i^\dagger a_i$
acts on $\mathcal{V}_M^d$ as $M$ times the identity,
that is
\begin{equation} \label{eq:PFd}
P=\sum_{i=1}^da_i^\dagger a_i  = \bigoplus_{M=0}^d M I_{\mathcal{V}_N^d}
\end{equation}
on the whole Fock space $\mathcal{F}^d$,
and it is called \emph{particle number operator}, since
\begin{equation*}
 P (0\oplus\dots\oplus\Psi_{(M)}\oplus\dots\oplus0) 
=M\;0\oplus\dots\oplus\Psi_{(M)}\oplus\dots\oplus0.
\end{equation*}

\subsection{Occupation numbers and second quantization} 
\label{sec:qchem.2quant}

Instead of the above notations,
it is usual to introduce the very convenient binary labeling, or \emph{occupation number labeling},
for the Slater determinants $\Phi_{[i_1,\ldots,i_N]}$.
Let $(\mu_1,\ldots,\mu_d)$ be a binary string, i.e., $\mu_i\in\{0,1\}$,
depending on the presence or absence of $\varphi_i$ in the Slater determinant $\Phi_{[i_1,\ldots,i_N]}$:
For all $i=1,\ldots,d$, if $i\in[i_1,\ldots,i_N]$ then $\mu_i=1$ 
(then we say that spin-orbital $\varphi_i$ is \emph{occupied} in the Slater determinant),
else  $\mu_i=0$
(\emph{unoccupied}).
So $\mu_i \in \{ 0,1\}$ has the meaning of an \emph{occupation number},
and we use the notation
\begin{equation}\label{eq:occup}
\Phi_{(\mu_1, \ldots, \mu_d)} :=  \Phi_{[i_1, \ldots , i_N ]},\qquad
\mu_i \in \{0,1\}, i=1,\dots,d.
\end{equation}
Furthermore, in an $N$ particle Slater determinant,
$\mu_i =1$ appears exactly $N$ times.
With this, the Fock space becomes
\begin{equation}\label{eq:Fockoccup}
\mathcal{F}^d = \bigoplus_{M=0}^d \mathcal{V}_{M}^{d} =  
\Bigl \{ \Psi :  \Psi = \sum_{\mu_1,\dots,\mu_d} u_{\mu_1, \ldots, \mu_d} \Phi_{(\mu_1, \ldots, \mu_d)} , 
u_{\mu_1, \ldots, \mu_d}\in\mathbb{C}, \mu_i \in \{0,1\}, i=1,\dots,d \Bigr\}. 
\end{equation}
The effect of the creation and annihilation operators (\ref{eq:opFd})
can also be formulated using the occupation numbers in a more expressive way than before:
\begin{subequations}\label{eq:opFdocc}
\begin{align}
\label{eq:opFdocc.adag}
a_i^\dagger  \Phi_{(\mu_1,\ldots,\mu_i,\dots,\mu_d )} = 
(-1)^{k_i} \Phi_{(\mu_1,\ldots,\mu_i+1,\dots,\mu_d )}, \quad\text{if $\mu_i=0$, othervise 0,} \\
\label{eq:opFdocc.a}
a_i          \Phi_{(\mu_1,\ldots,\mu_i,\dots,\mu_d )} = 
(-1)^{k_i} \Phi_{(\mu_1,\ldots,\mu_i-1,\dots,\mu_d )}, \quad\text{if $\mu_i=1$, othervise 0,}
\end{align}
\end{subequations}
where $k_i=\sum_{j=1}^{i-1}\mu_j$.
On the other hand, 
$a_i^\dagger a_i\Phi_{(\mu_1,\ldots,\mu_i,\dots,\mu_d )} = \mu_i \Phi_{(\mu_1,\ldots,\mu_i,\dots,\mu_d )}$,
and with the definition (\ref{eq:PFd}) we have
\begin{equation}\label{eq:PFdocc}
P\Phi_{(\mu_1,\ldots,\mu_d)} = \Bigl(\sum_{i=1}^d\mu_i\Bigr)\Phi_{(\mu_1,\dots,\mu_d)}.
\end{equation}

This binary labelling 
gives us the opportunity of using 
another, more convenient, representation of the Fock space, 
which is called \emph{second quantization.}
To this end,
we consider the Hilbert space
for the representation of the events of the occupation of the orbitals.
This space has a two dimensional tensor factor $\Lambda_i\cong\mathbb{C}^2$ for each orbital,
containing two ortogonal states representing the unoccupied and occupied states of the orbital.
So, let us define the tensor product space
\begin{equation}\label{eq:Hd} 
\Lambda^{(d)} := \bigotimes_{i=1}^d \Lambda_i
\cong\bigotimes_{i=1}^d \mathbb{C}^2, 
\end{equation}
and the canonical basis 
$\{ |\phi_0\rangle \cong \mathbf{e}_0, |\phi_1\rangle \cong \mathbf{e}_1 \} $ 
of the vector space $\Lambda_i\cong\mathbb{C}^2 $,
where $ (\mathbf{e}_0)_\mu = \delta_{\mu,0} $, $(\mathbf{e}_1)_\mu = \delta_{\mu,1} $.
(For the elements of only these spaces we use the so called \emph{ket}-notation $|\dots\rangle$,
which is very common in physics for denoting elements of Hilbert spaces.)
So we write any $|U\rangle\in\mathbb{C}^2$ as 
$| U  \rangle  = \sum_{\mu =0}^1 U(\mu)|\phi_\mu\rangle.$
The dimension of this space is
\begin{equation}\label{eq:dimHd}
\dim\Lambda^{(d)} = 2^d.
\end{equation}
We also have the canonical inner product in $\mathbb{C}^2$:
\begin{equation}\label{eq:innerC2}
\langle U | V \rangle = \sum_{\mu=0}^1 \overline{U(\mu)}V(\mu),
\end{equation}
for which the canonical basis is orthogonal,
$ \langle \phi_\mu | \phi_\nu  \rangle = \delta_{\mu,\nu} $, for $ \mu,\nu = 0,1$.
Using the canonical basis $\{| \phi_{\mu_i}^{\{i\}} \rangle\}$  of $\Lambda_i$, the set 
$\{| \phi_{\mu_1}^{\{1\}} \rangle \otimes \cdots \otimes |\phi_{\mu_d}^{\{d\}} \rangle:
\mu_i\in\{0,1\}, i=1,\dots,d\}$
gives the canonical basis in $\Lambda^{(d)}$, 
and any $| U \rangle \in \Lambda^{(d)}$ can be represented by 
\begin{equation}\label{eq:elemHd}
| U \rangle  =  \sum_{\mu_1 , \ldots , \mu_d} U (\mu_1 , \ldots , \mu_d ) 
| \phi_{\mu_1}^{\{1\}} \rangle \otimes \cdots \otimes |\phi_{\mu_d}^{\{d\}} \rangle.
\end{equation}
If there is no ambiguity about the underlying basis,
we consider the canonical tensor product basis above, 
and we can identify $ | U  \rangle \in \Lambda^{(d)}$ with $U \in \bigoplus_{i=1}^d\mathbb{C}^2$, 
where $U$ are simply $d$-variate functions (see section \ref{sec:tensor.param}) 
\begin{equation}\label{eq:elemHd2}
(\mu_1,\ldots,\mu_d)\quad \longmapsto\quad U(\mu_1,\ldots,\mu_d)  \in \mathbb{C},
\end{equation}
depending on the discrete variables $\mu_i= 0,1$, for $i=1,\ldots,d$,  
called also indices in the sequel.
Due to the orthogonality of the canonical basis, 
the canonical inner product of  $\mathbb{C}^2$ induces
\begin{equation}\label{eq:innerHd}
\langle U|V \rangle := \sum_{\mu_1 , \ldots , \mu_d} 
\overline{U (\mu_1, \ldots , \mu_d )} V (\mu_1, \ldots , \mu_d )  \ 
\end{equation}
and the norm $ \| U \| = \sqrt{\langle U , U \rangle}$ in $\Lambda^{(d)}$.

Now, let us introduce the \emph{isomorphism} $\iota : \mathcal{F}^d \to  \Lambda^{(d)}$, defined
by its action on the basis functions, i.e., the Slater determinants $\Phi_{(\mu_1,\ldots,\mu_d)}$
of occupation numbers $(\mu_1,\ldots,\mu_d)$ simply as
\begin{equation}\label{eq:iota}
\iota  ( \Phi_{(\mu_1,\ldots,\mu_d)} ) := | \phi_{\mu_1}^{\{1\}} \rangle \otimes \cdots \otimes |\phi_{\mu_d}^{\{d\}} \rangle,
\qquad \mu_i \in \{ 0,1\},\quad i=1,\dots,d. 
\end{equation}
(This is an elementary tensor product, in physics called \emph{tensor product state}.)
It is easy to check that this is indeed an isomorphism,
and compatible with the inner products of the two spaces,
so we conclude that the discrete Fock space $\mathcal{F}^d $ is isomorphic  to 
$\Lambda^{(d)} \cong \otimes_{i=1}^d \mathbb{C}^2$, 
and
\begin{subequations}
\begin{equation}\label{eq:iotaU}
|\Psi\rangle:= \iota(\Psi) = \iota  \Bigl( \sum_{\mu_1 , \ldots , \mu_d} u_{\mu_1 , \ldots , \mu_d}\Phi_{(\mu_1,\ldots,\mu_d)} \Bigr) 
= \sum_\mu U(\mu_1,\ldots,\mu_d)
| \phi_{\mu_1}^{\{1\}} \rangle \otimes \cdots \otimes |\phi_{\mu_d}^{\{d\}} \rangle
\end{equation}
leads to 
\begin{equation}\label{eq:Uu}
U(\mu_1,\ldots,\mu_d)= u_{\mu_1,\ldots,\mu_d}.
\end{equation}
\end{subequations}
On the other hand, we have used the convention above that for a function $\Psi\in\mathcal{F}^d$,
its image by $\iota$ is written as the \emph{ket} $|\Psi\rangle$.
The Full-CI space for $N$ electrons $\mathcal{V}_{N}^{d}$ is a subspace of the Fock space $\mathcal{F}^d$,
and its image in $\Lambda^{(d)}$ is denoted as $\Lambda_{\text{FCI}}:=\iota(\mathcal{V}_{N}^{d})\subset\Lambda^{(d)}$.
This is the $N$-electron subspace of $\Lambda^{(d)}$, having dimension $\dim\Lambda_{\text{FCI}} = \binom{d}{N}$.

Through this isomorphism, we can obtain the creation and annihilation operators (\ref{eq:opFdocc}) 
acting on $\Lambda^{(d)}$ as follows~\cite{Reed-1978}
\begin{subequations}\label{eq:opHd}
\begin{align}
\label{eq:opHd.adag}
\mathbf{a}_i^\dagger &:= \iota \circ a_i^\dagger   \circ \iota^{-1} = 
\mathbf{s} \otimes \ldots \otimes \mathbf{s} \otimes \mathbf{a}^\dagger \otimes \mathbf{I} \otimes \ldots \otimes \mathbf{I} \ : \ 
\Lambda^{(d)} \to \Lambda^{(d)},\\ 
\label{eq:opHd.a}
\mathbf{a}_i &:=         \iota \circ a_i           \circ \iota^{-1} = 
\mathbf{s} \otimes \ldots \otimes \mathbf{s} \otimes \mathbf{a}         \otimes \mathbf{I} \otimes \ldots \otimes \mathbf{I} \ : \  
\Lambda^{(d)} \to \Lambda^{(d)}, 
\end{align}
\end{subequations}
(where $a^\dagger$ and $a$ appeare in the $i$-th position)
with the operators
\begin{equation}\label{eq:matrices}
\mathbf{a} := \begin{pmatrix}
 0 &  1   \\
 0 &  0   
\end{pmatrix},\quad
\mathbf{a}^\dagger = \begin{pmatrix}
 0 &  0   \\
 1 &  0   
\end{pmatrix},\quad
\mathbf{s} := \begin{pmatrix}
 1 &  0   \\
 0 & -1   
\end{pmatrix},\quad
\mathbf{I} := \begin{pmatrix}
 1 &  0   \\
 0 &  1   
\end{pmatrix},
\end{equation}
acting on $\Lambda_i$.
The operators $\mathbf{a}_i$ and $\mathbf{a}_i^\dagger$ again obey the 
fermionic anticommutation relation (\ref{eq:fermAnticommFd}).
Let us highlight that the $2\times2$-matrix $\mathbf{s}$ is required to provide the correct phase factor,
i.e., sign of the Slater determinant. 
The particle number operator acting on $\Lambda^{(d)}$ is
\begin{equation}\label{eq:PHd}
\mathbf{P} := \iota\circ P \circ\iota^{-1} = \sum_{i=1}^{d}  \mathbf{a}_i^\dagger  \mathbf{a}_i,
\end{equation}
which is, since $\mathbf{s}^2=I$, the sum of matrices
\begin{equation}
\mathbf{n}_i = \mathbf{a}_i^\dagger  \mathbf{a}_i
= \mathbf{I} \otimes \ldots \otimes \mathbf{I} \otimes \mathbf{a}^\dagger \mathbf{a} \otimes \mathbf{I} \otimes \ldots \otimes \mathbf{I} \ : \  
\Lambda^{(d)} \to \Lambda^{(d)}, 
\end{equation}
with the matrix
\begin{equation} 
\mathbf{n} = \mathbf{a}^\dagger \mathbf{a} := \begin{pmatrix}
 0 &  0   \\
 0 &  1   
\end{pmatrix}
\end{equation}
in the $i$-th position,
representing the occupation number of the given orbital.

Let us remark that since the isomorphism $\iota$ is defined through a given Slater determinant basis,
the above representation of a wave function is basis dependent, i.e.,
it depends on the choice of the one-particle basis set $B^d$.

\subsection{Example: Hartree-Fock determinant and change of one-particle basis}
\label{sec:qchem.xmpleHFchbasis}

The Hartree Fock determinant in terms of canonical molecular orbital functions is given
by $\Psi_{\text{HF}} = \Phi_{[1, \ldots , N]} = \Phi_{(1, \ldots, 1,0 , \ldots,0)}$, 
i.e., $i_b=b$ for $b=1,\dots,N$, the \emph{first} $N$ spin-orbitals are occupied.
In $\Lambda^{(d)}$ this is represented 
by the tensor for which $U(\mu_1,\dots,\mu_d) = 1$
if and only if $(\mu_1,\dots,\mu_d) = (1, \ldots, 1,0 , \ldots ,0)$, or 
\begin{equation}
\label{eq:PsiHF}
|\Psi_{\text{HF}}\rangle =
| \phi_1^{\{1\}}   \rangle \otimes\cdots\otimes | \phi_1^{\{N\}} \rangle  \otimes 
| \phi_0^{\{N+1\}} \rangle \otimes\cdots\otimes | \phi_0^{\{d\}} \rangle.
\end{equation}

If we move to another basis, respectively basis set, 
say 
\begin{equation}\label{eq:basistraf}
           B^d =\bigl\{\varphi_i: i=1,\ldots,d\bigl\} \qquad\longmapsto\qquad 
\widetilde{B^d}=\Bigl\{\widetilde{\varphi}_i = \sum_{j=1}^d U_{i,j} \varphi_j: i=1,\ldots,d\Bigl\},
\end{equation}
with the unitary $d\times d$ matrix $\mathbf{U}=\big(U_{i,j}\big)$,
the representation 
of the old Slater determinants
 $\Phi_{(\mu_1,\dots,\mu_d)} =  \sum_{\nu_1,\dots,\nu_d} \widetilde{v}_{\nu_1,\dots,\nu_d} \widetilde{\Phi}_{(\nu_1,\dots,\nu_d)}$ 
in terms of the new Slater determinants $\widetilde{\Phi}_{(\nu_1,\dots,\nu_d)}$  
is no longer of rank one. 
It is a short exercise to show a representation of the form 
$\Psi=\sum_{\mu_1,\dots,\mu_d} u_{\mu_1,\dots,\mu_d} \Phi_{(\mu_1,\dots,\mu_d)}$ is transformed into
\begin{equation*}
\begin{split}
| \widetilde{\Psi} \rangle  
 &= e^{ \sum_{i=N+1}^d  \sum_{j=1}^N t_{i,j}  \mathbf{a}_i^{\dagger} \mathbf{a}_j  } | \Psi \rangle  \\
 &= \Pi_{i=N+1}^d\Pi_{j=1}^N  ( \mathbf{I}^d + t_{i,j} \mathbf{a}_i^{\dagger} \mathbf{a}_j )  \; | \Psi \rangle  \\
 &= ( \mathbf{I}^d + t_{N+1,1} \mathbf{a}_{N+1}^{\dagger} \mathbf{a}_1  )  \cdot   \ldots  \cdot  
 ( \mathbf{I}^d  + t_{d,N} \mathbf{a}_d^{\dagger} \mathbf{a}_N ) | \Psi  \rangle
\end{split}
\end{equation*}
up to a normalization constant. 
This transformation serves as the transformation of the basis sets.
Let us remark that
in Coupled Cluster theory 
the above expression is known as the \emph{Coupled Cluster single excited states}, 
and 
the above transformation is used for the definition of the \emph{Br\"uckner orbital}\cite{Helgaker-2000}.
Also each factor is at most of rank two, 
so in the worst case the rank could increased by a factor of two 
with each matrix multiplication. 
The single Slater determinant expressed by another basis set 
is a linear combination of many Slater determinants with respect to the new basis. 
Indeed it can happen that in the new bases 
it is represented by the maximally entangled state tensor $ | \widetilde{\Psi} \rangle$.

\subsection{Ritz-Galerkin approximation in second quantization}
\label{sec:qchem.RG2quant}

Now we are able to formulate the full-CI Schrodinger equation (\ref{eq:SchrodingerVdN})
in the second-quantized form.
By the isomorphism $\iota$, 
the Hamiltonian (\ref{eq:HamiltonFd}) acting on $\mathcal{F}^d$ 
is defined on $\Lambda^{(d)}$ as follows
\begin{equation}\label{eq:HamiltonHddef}
\mathbf{H} = \iota \circ H^d \circ \iota^{-1}\ : \ \Lambda^{(d)} \to \Lambda^{(d)}.
\end{equation}
Using the ansatz $\Psi = \sum_{\nu_1,\dots,\nu_d} u_{\nu_1,\dots,\nu_d} \Phi_{(\nu_1,\dots,\nu_d)} \in \mathcal{V}^d_N$ for the eigenvector, 
with the help of $\iota$,
we obtain the discrete eigenvalue problem from (\ref{eq:weak-SchrodingerVdN})
in the $N$-electron subspace of $\Lambda^{(d)}$
\begin{equation*}
\begin{split}
\langle \Phi_{(\mu_1,\dots,\mu_d)} , (H - E) \Psi \rangle  
&= \sum_{\nu_1,\dots,\nu_d} \langle \Phi_{(\mu_1,\dots,\mu_d)} , (H - E) \Phi_{(\nu_1,\dots,\nu_d)} \rangle u_{\nu_1,\dots,\nu_d} \\
&= \big(  (\mathbf{H} - E \mathbf{I} )  | \Psi \rangle  \big)_{\mu_1,\dots,\mu_d}
 =  0 \quad \forall \mu_i\in \{0,1\},\quad i=1,\dots,d. 
\end{split}
\end{equation*}
If we allow $|\Psi\rangle\in \Lambda^{(d)}$, not only in its $N$-electron subspace,
we do not get only the eigenvalues $E$  for a discretized  $N$ electron system,
but also the eigenvalues for other numbers of electrons between $1$ and $d$. 
To fix this kind of problem, we take into account the particle number operator (\ref{eq:PHd}),
then $\Psi = \sum_{\mu_1,\dots,\mu_d} u_{\mu_1,\dots,\mu_d} \Phi_{(\mu_1,\dots,\mu_d)} \in \mathcal{V}^d_N $ holds if and only if
\begin{equation}\label{eq:Nconstraint}
 \mathbf{P} |\Psi\rangle = N |\Psi \rangle,
\end{equation}
i.e., $|\Psi \rangle$ is an eigenvector of $\mathbf{P}$ with the eigenvalue $N$.

From the well known Slater-Condon rules\cite{Szabo-1982} one concludes the precise representation of 
the Hamiltonian in  the discrete space $\Lambda^{(d)}$, which reads as 
\begin{equation}\label{eq:HamiltonHd}
 \mathbf{H} = \sum_{i,j=1}^d  T_{ij}  \mathbf{a}_i^\dagger\mathbf{a}_j + 
 \sum_{i,j,k,l=1}^d  V_{ijkl}  \mathbf{a}_i^\dagger\mathbf{a}_j^\dagger \mathbf{a}_k  \mathbf{a}_l.
\end{equation}
Here the coefficients
\begin{subequations}
\begin{equation} \label{eq:singleint}
T_{ij}
= \sum_{s = \pm \frac{1}{2} } \int_{\mathbb{R}^3}\overline{\varphi_i (\mathbf{r}, s)}
\Bigl( -\frac12\Delta-\sum_{c=1}^K\frac {Z_c}{| \mathbf{r} - \mathbf{R}_c|} \Bigr)
\varphi_j (\mathbf{r} , s ) \dd \mathbf{r}
\end{equation}
are the well known \emph{single electron integrals,} and
\begin{equation} \label{eq:doubleint}
 V_{ijkl} =  
\sum_{s,s' = \pm \frac{1}{2}} \int_{\mathbb{R}^3}
\overline{\varphi_i(\mathbf{r},s) \varphi_j(\mathbf{r}',s')}
\frac1{|\mathbf{r}'-\mathbf{r}|}
\varphi_k(\mathbf{r},s)\mathbf{\varphi}_l(\mathbf{r}',s')
 \dd \mathbf{r} \dd \mathbf{r}' 
\end{equation}
\end{subequations}
are the \emph{two electron integrals,}
both are coming from the parts of the original Hamiltonian (\ref{eq:HamiltonParts}).
With this, the discrete (Full CI) Schr{\"o}dinger equation for the approximation of the ground state
can be cast into the binary variational form
of finding $|\Psi\rangle \in \Lambda^{(d)}$ such that
\begin{subequations}\label{eq:varHd}
\begin{align}
\label{eq:varHd.E0}
E_{0,d} =  \min    &\bigl\{ \langle \Psi| \mathbf{H} |\Psi \rangle : 
\langle \Psi|\Psi \rangle =1,\, \mathbf{P}|\Psi\rangle = N |\Psi\rangle,\, |\Psi\rangle \in \Lambda^{(d)} \bigr\},\\
\label{eq:varHd.U0}
|\Psi_{0,d}\rangle =  \argmin &\bigl\{ \langle \Psi| \mathbf{H} |\Psi \rangle : 
\langle \Psi|\Psi \rangle =1,\, \mathbf{P}|\Psi\rangle = N |\Psi\rangle,\, |\Psi\rangle \in \Lambda^{(d)} \bigr\}.
\end{align}
\end{subequations}

Let us remark that this representation depends on the basis orbital functions $B^d$.
For a change of basis, as is given in (\ref{eq:basistraf}),
the creation and annihilation operators transform as
\begin{equation}\label{eq:optraf}
 \widetilde{\mathbf{a}}_i^\dagger = \sum_{j=1}^d U_{i,j}    \mathbf{a}_j^{\dagger},\qquad
 \widetilde{\mathbf{a}}_i = \sum_{j=1}^d \overline{U_{i,j}} \mathbf{a}_j.
\end{equation}
With respect to the new basis set $\widetilde{B^d}$,
we can build another Slater determinant basis 
$\{ \widetilde{\Phi}_{(\nu_1,\dots,\nu_d)} : \nu_i \in \{0,1\},i=1,\dots,d\}$
of the discrete Fock space $\mathcal{F}^d$.
With respect to this new Slater determinant basis, 
the operators $\widetilde{\mathbf{a}}_i$ has the canonical form
(\ref{eq:opHd.a})
\begin{equation}
 \widetilde{\mathbf{a}}_i  = \mathbf{s} \otimes \cdots \otimes \mathbf{s} \otimes \mathbf{a} \otimes \mathbf{I} \otimes \cdots \otimes \mathbf{I}. 
\end{equation}
The one and two electron integrals transform easily, e.g.,
\begin{equation}\label{eq:htraf}
  \tilde{T}_{ij} = \sum_{kl}(U^\dagger)_{ik} T_{kl} U_{lj}.
\end{equation}

\subsection{Spatial orbitals}
\label{sec:qchem.spatial}

For the sake of simplicity of representation,
throughout this section we have dealt with spin orbital basis $\varphi_i$ for $i=1,\dots,d$,
and the above setting $\Lambda^{(d)} = \bigotimes_{i=1}^d \Lambda_i \cong \bigotimes_{i=1}^d \mathbb{C}^2$. 
In the QC-DMRG, it has been experienced that 
it is favorable to use spin functions explicitly, 
and deal only with \emph{spatial orbitals}.
More precisely, we take a set
$\{\kappa_i\in H^1(\mathbb{R}^3):i=1,\dots,d\}$
of orthonormalized functions, depending on the space-variable $\mathbf{r}\in\mathbb{R}^3$ only,
and define the basis of $2d$ elements
\begin{subequations}
\begin{align}
 \varphi_{2i}  (\mathbf{r},s) &= \kappa_i(\mathbf{r})\chi_+(s),& \qquad 
\chi_+(s) &= 1\quad\text{if}\; s = + \frac12,\qquad
\chi_+(s)  = 0\quad\text{if}\; s = - \frac12,\\
 \varphi_{2i+1}(\mathbf{r},s) &= \kappa_i(\mathbf{r})\chi_-(s),& \qquad
\chi_-(s) &= 0\quad\text{if}\; s = + \frac12,\qquad
\chi_-(s)  = 1\quad\text{if}\; s = - \frac12,
\end{align}
\end{subequations}
which are orthonormalized in $H^1(\mathbb{R}^3\times\{\pm\frac12\})$.
Now, repeating the previous construction leads to the $2d$ spaces 
$W_{2i-1}\cong\mathbb{C}^2$ and $W_{2i}\cong\mathbb{C}^2$.
Let us cast the tensor product of two adjacent spaces into one 
$\Lambda_i:=W_{2i-1}\otimes W_{2i}\cong\mathbb{C}^4$,
and with this, 
\begin{equation}
\Lambda^{(d)} = \bigotimes_{i=1}^d \Lambda_i \cong \bigotimes_{i=1}^d \mathbb{C}^4,
\end{equation}
having the dimension $\dim\Lambda^{(d)} = 4^d$.
The $N$-electron subspace $\Lambda_{\text{FCI}}$ is then of dimension $\dim\Lambda_{\text{FCI}} = \binom{2d}{N}$.
In the case when the $N_\downarrow$ and $N_\uparrow$ numbers of electrons of spins $-1/2$ and $+1/2$ are conserved,
only a subspace of this is needed, which is called then the Full-CI space
that is of dimension $\dim\Lambda_{\text{FCI}} = \binom{d}{N_\downarrow}\binom{d}{N_\uparrow}<\binom{2d}{N}$.

Using the matrices (\ref{eq:matrices}) we define 
\begin{subequations}\label{eq:Cs}
\begin{gather}
\label{eq:Cs.cs}
 \mathbf{c}_s := \mathbf{a} \otimes \mathbf{I} \quad \text{if}\;  s = +  \frac {1}{2},\qquad
 \mathbf{c}_s := \mathbf{s} \otimes \mathbf{a} \quad \text{if}\;  s = -  \frac {1}{2}, \\
\label{eq:Cs.ph}
 \mathbf{z} := \mathbf{s} \otimes \mathbf{s},\qquad 
 \mathbb{I} = \mathbf{I}\otimes \mathbf{I},\\
\label{eq:Cs.cis}
  \mathbf{c}_{i,s} := \mathbf{z}\otimes \cdots \otimes \mathbf{z} \otimes \mathbf{c}_s \otimes 
 \mathbb{I} \otimes \cdots \otimes \mathbb{I},\\
\label{eq:Cs.nis}
\mathbf{n}_s := \mathbf{c}_s^\dagger \mathbf{c}_s,\qquad
\mathbf{n}_{i,s} := \mathbf{c}_{i,s}^\dagger\mathbf{c}_{i,s} = 
 \mathbb{I} \otimes \cdots \otimes \mathbb{I} \otimes
 \mathbf{c}_s^\dagger \mathbf{c}_s \otimes 
 \mathbb{I} \otimes \cdots \otimes \mathbb{I}.
\end{gather}
\end{subequations}
With these, the Hamilton operator reads as
\begin{equation}\label{eq:HamiltonC}
\mathbf{H} = \sum_{i,j=1}^d  \sum_{s_i,s_j=\pm\frac12}
 T_{ij} \mathbf{c}_{i,s_i}^\dagger \mathbf{c}_{j,s_j} 
+ \sum_{i,j,k,l=1}^d \sum_{s_i,s_j,s_k,s_l=\pm\frac12} V_{ijkl}
  \mathbf{c}_{i,s_i}^\dagger \mathbf{c}_{j,s_j}^\dagger \mathbf{c}_{k,s_k} \mathbf{c}_{l,s_l}. 
\end{equation}
Let us remark that in the nonrelativistic quantum chemistry,  
the one- and two electron integrals do not depend on the spin variables $s_i$.
This is the reason for using the spatial orbital formulation.
(However, this does not hold when relativistic effects, e.g. spin-orbit coupling are taken into account,
as is used in a recent development in DMRG\cite{Knecht-2014}.)

\section{Tensor product approximation}
\label{sec:tensor}

\subsection{Tensor product parametrization}
\label{sec:tensor.param}

We generalize (\ref{eq:iotaU}) by considering vector spaces with arbitrary dimension $\dim \Lambda_i = q_i$:
\begin{equation}
\label{eq:tens}
| \Psi \rangle = \sum_{\alpha_1,\ldots,\alpha_d} U(\alpha_1,\ldots,\alpha_d) \ | \phi_{\alpha_1}^{\{1\}} \rangle \otimes \cdots \otimes | \phi_{\alpha_d}^{\{d\}} \rangle.
\end{equation} Thus, the tensor 
\begin{equation}
\label{eq:tensspace}
| \Psi \rangle \in \Lambda^{(d)} 
\end{equation}
is equivalent to the multi-indexed array 
\begin{align}
\label{eq:multiarr}
& U \in \mathbb C^{q_1 \times \ldots \times q_d}, \\
& U(\alpha_1,\ldots,\alpha_d), \ \alpha_i \in \{ 1, \ldots , q_i \}  \ \ i=1, \ldots , d.
\end{align}
The norm and inner product of these two tensor spaces are defined analogously to Sec.~\ref{sec:qchem.2quant}.

Computation with tensors suffer from the {\em curse of dimensionality} \cite{Hackbusch-2014b}, 
since the storage of the complete array grows exponentially with the order $d$.

We seek to reduce computational costs by parametrizing the tensors in some data-sparse representation.
For this purpose, we adhere to the {\em separation of variables}, a classical approach which traces back to Bernoulli and
Fourier among others. 
In principle, we want to represent or approximate tensors as multi-variate functions by a sum of 
products of univariate functions.
This concept is well established for tensors of order $d=2$ where it leads to fundamental results known as the 
{\em singular value decomposition} (SVD) or Schmidt decomposition, proper orthogonal decomposition, the Karhunen-Loeve transform 
and so on. In the discrete case discussed here, i.e. in matrix theory, this is known as {\em low rank approximation}. 
However, the generalization of the concept of ranks to higher order tensors is not as straightforward as one may expect \cite{Hackbusch-2012}.
There are many possible and a priori equally justifiable tensor decompositions that all yield different
definitions of a tensor rank.

The {\em canonical tensor representation} separates the variables
\begin{equation}
\label{eq:canonical}
 U(\alpha_1,\ldots,\alpha_d) = \sum_{i = 1}^R u_i^1(\alpha_1) \cdots u_i^d(\alpha_d).
\end{equation}
The canonical tensor rank of $U$ is the smallest $R$ such that this representation is exact. This is then called the 
{\em canonical decomposition} of the tensor \cite{Hackbusch-2012}.

However, while this is a beautiful and rather simplistic tensor representation, it has several severe drawbacks.
First of all, finding the canonical rank and thus also its decomposition is $NP$-hard \cite{Kolda-2009}. Additionally, the set of 
tensors with rank smaller or equal $R$ is not closed, i.e. it is possible to find a sequence of rank-$R$-tensors that
converges to a tensor with rank greater than $R$, see the {\em border rank problem} \cite{Landsberg-2012b, Hackbusch-2012}. While the former property 
obviously poses problems in computation, the latter can be very undesirable as well when it comes to optimization algorithms. 
Altogether, the canonical format has not only led to deep and difficult mathematical problems \cite{Landsberg-2012a, Landsberg-2012b}, but also computational experience has often been 
disappointing, by observing slow convergence, low accuracy and the like. It is not clear how to circumvent these problems while still retaining its outstanding complexity scaling.
In recent years, the canonical format has therefore been put into question, albeit not completely disqualified, and we are looking for alternatives with favorable properties.
 
We parametrize a tensor in a very  general form  to define a {\em tensor representation} via 
\begin{equation} 
\label{eq:multi-linear}
 U(\alpha_1,\ldots,\alpha_d) = \sum_{\mathbf m}^{\mathbf r} \prod_{i}^K U_i(\alpha_{i,1},\ldots,\alpha_{i,y_i} ; m_{i,1},\ldots,m_{i,z_i}),
\end{equation}
where $\mathbf m$ denotes the multi-index 
\begin{equation}
\label{eq:multiind}
\mathbf m = \bigcup_{i = 1}^K \{ m_{i,1},\ldots,m_{i,z_i} \}.
\end{equation}
Since the ordering of the indices is irrelevant in this context, we maintain the slight abuse of notation and interpret multi-indices as sets of natural numbers.

This tensor representation is parametrized by $K$ {\em component tensors}  $U_1,\ldots,U_K$.
The $\alpha_i$ are called {\em physical} indices and the $m_j$ are called {\em virtual}.
A component $U_i$ is called virtual, if it does not have any physical indeces, i.e. $y_i = 0$.
Otherwise it is called physical. Summation over a common virtual index $m_j$ is called the {\em contraction} over $m_j$.

We can demand a number of further properties that allow for simpler treatment of the tensor. 
First of all, it is conventional to only deal with  multi-linear representations: 
\begin{criterion}
\label{crit:lin}
 For each $j \in \{1,\ldots,d \}$ there exists {\em exactly} one $i \in \{1,\ldots,K \}$ such that 
 $\alpha_j \in \{ \alpha_{i,1},\ldots,\alpha_{i,y_i} \}$.
\end{criterion}
This means that no two components can depend on the same physical index. The present 
 multi-linear parametrization provides simple representation of the derivatives, an indispensable tool for local optimization \cite{Espig-2011, Espig-2012}, 
and alternating directional search methods \cite{Holtz-2012b}.

It is our  central aim to reduce complexity of the tensor and we therefore need to choose the representation carefully. Firstly, the number of components 
should not be exceedingly high, as this makes the representation more complicated. But more importantly, 
we try to minimize the dimensions $\mathbf r$ of
the multi-index $\mathbf m$ over all possible representations (\ref{eq:multi-linear}). 
If these dimensions $r_i$ are minimal for the given parametrization, the tuple 
$\mathbf r$ is called the {\em rank}, or better {\em multi-linear rank} of the
representation and the representation is called a {\em decomposition}.
However, as mentioned for the canonical format above, this notion of rank leads to extreme difficulties even for the simplest forms.

\subsection{Tensor networks}
\label{sec:tensor.networks}

For a proper definition of multi-linear ranks, we consider subclasses of tensor representations and introduce a 
further restriction that each $m_i \in \mathbf m$ appears exactly twice:
\begin{criterion}
\label{crit:noncan}
 For each virtual index $m_i \in \mathbf m$ there exist {\em exactly} two component tensors $U_{i_1}, U_{i_2}$ with $m_i$ as an index.
\end{criterion}

Any parametrization satisfying criterion \ref{crit:lin} and \ref{crit:noncan} 
can be expressed as a simple undirected weighted graph with half-edges, 
and we obtain what is called a {\em tensor network} or {\em tensor network states} in quantum physics. 
The component tensors give
the vertices of the graph, the contractions are represented by the edges between the vertices and the physical indices yield half edges. Therefore, we get a graph
$TNS(U) := (V,E,H)$, 
\begin{equation}
\label{eq:tensgraph}
V = \{  U_i : i = 1,\ldots,K \}, \ E = \mathbf m, \ H = \{ \alpha_1, \ldots, \alpha_d \}.
\end{equation}

\begin{figure}
\centering
\includegraphics{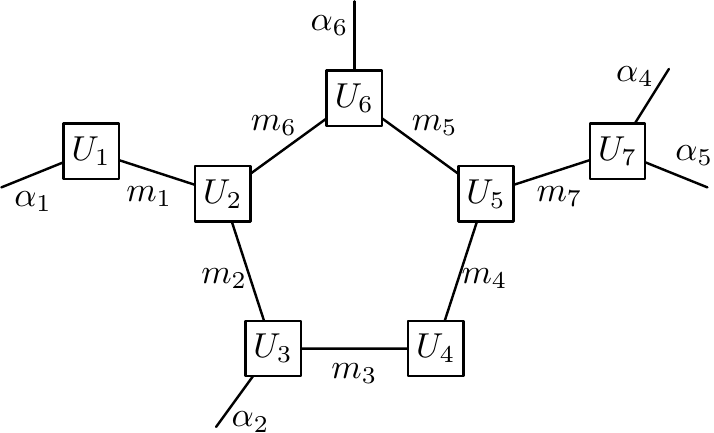}
\caption{A general tensor network representation of a tensor of order 5.}
\label{fig:networks-general}
\end{figure}

Because of criterion \ref{crit:lin}, each half-edge has exactly one incident vertex, and because of \ref{crit:noncan}, each edge has exactly two incident vertices \cite{Espig-2011}. 
Thus, this is well-defined. 
The weight of the half-edge $\alpha_i$ is given by its dimension $q_i$ and the weight of the edge $m_j$ is given by its dimension $r_j$. 
In accordance with the tensor decompositions, we call the vector $\mathbf r$ the {\em rank} of the tensor network if it is minimal. 
$q_1 \cdots q_d$ is naturally the dimension of the tensor network, as shown in Fig.~\ref{fig:networks-general}.

Since a contraction over an index with dimension 1 is trivial, we can choose to either omit this index or even to introduce extra indices. In general, we require the 
tensor network graph to be connected and if it is not we invent an arbitrary index of dimension 1 to make it so. Apart from that, any index of dimension 1 that is
not necessary for connectedness will usually be omitted.

\begin{figure}
\centering
\includegraphics{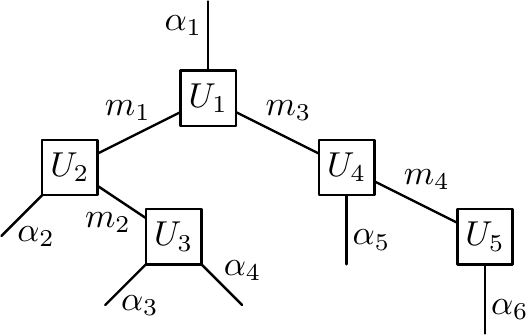}
\caption{An arbitrary example of a tensor tree.}
\label{fig:networks-tree}
\end{figure}

Although heavily used in physics this general concept still suffers from some instabilities. 
Recently it has been shown that tensor networks which contain closed loops are not necessarily Zariski closed \cite{Landsberg-2012a}, i.e. they do not form algebraic varieties without further restrictions.
This is closely related  to the border rank problem for the canonical format. 
While we will not go into these details here, we highlight that all these difficulties can be avoided, if we restrict ourselves to
tensors fulfilling the following criterion \cite{Landsberg-2012a}:
\begin{criterion}
\label{crit:nocirc}
 The tensor network $TNS(U)$ is cycle-free.
\end{criterion}
Since we have the trivial connectedness mentioned above, any tensor network that fulfills criterion \ref{crit:nocirc} is a tree. It is thus called a {\em Tree Tensor Network} or, in accordance
with nomenclature from Quantum Physics, {\em Tree Tensor Network States (TTNS)}. See Fig.~\ref{fig:networks-tree} for an arbitrary example.

While general tensor network representations, like the canonical format, might still be very useful and shall not be disqualified, we presently only consider the special case of non-circular graphs
that prevents these fundamental difficulties.

\subsection{Subspace optimization and the Tucker format}
\label{sec:tensor.tucker}

Tensor trees can be introduced from a different perspective in order to illustrate that they share favorable properties with the matrix case.
In our discrete setting, the tensor space and the tensor product space are equivalent
\begin{equation}
\label{eq:spaceeq}
\mathbb C^{q_1 \times \ldots \times q_d} \cong \bigotimes_{i = 1}^d \mathbb C^{q_i} \cong \bigotimes_{i = 1}^d \Lambda_i
\end{equation}
via the trivial formula (\ref{eq:tens}).
$ | \phi_{\alpha_i}^{\{i\}} \rangle $ are the standard euclidean basis vectors of $ \Lambda_i $ for each $i$.

In this context, we define the {\em subspace optimization} as the best approximation of $U$ \cite{Hackbusch-2012, Hackbusch-2014a}
\begin{equation}
\label{eq:besttucker}
  \argmin \{ \| U - U_{\epsilon} \| : U_{\epsilon}  \in \bigotimes_{i=1}^d \Xi_i , \ \dim \Xi_i \leq r_i \},
\end{equation}
where we optimize over the tensor product of all univariate subspaces $ \Xi_i \subset \Lambda_i $ of dimension at most $r_i$. 
If we can recover the tensor $U$ exactly, i.e. $\| U - U_{\epsilon} \| = 0$, we call $U_{\epsilon}$ 
the {\em subspace representation} of $U$. In accordance with the above, a subspace representation is called a decomposition if the dimensions $r_i$ are the ranks, i.e. they are the smallest 
numbers such that the tensor can still be recovered exactly.

This immediately motivates the {\em Tucker decomposition format} of a tensor. 
For each $i = 1,\ldots,d$, we aim at finding an optimal basis set $\{ | \xi_{m_i}^{\{i\}} \rangle : m_i = 1, \ldots, r_i \}$ of a subspace  $\Xi_i \subseteq \mathbb C^{q_i}$ where $r_i \leq q_i$. (\ref{eq:tens}) can thus be restated as
\begin{equation}
\label{eq:tucker}
 | \Psi \rangle = \sum_{m_1 = 1}^{r_1} \ldots \sum_{m_d = 1}^{r_d} C(m_1,\ldots,m_d) \ | \xi_{m_1}^{\{1\}} \rangle \otimes \cdots \otimes | \xi_{m_d}^{\{d\}} \rangle.
\end{equation}
$C \in \mathbb C^{r_1 \times \ldots \times r_d}$ is a reduced core tensor, that is hopefully much smaller than the original coefficient tensor, due to the optimal
choice of basis.

For exact recovery, obtaining the basis vectors in the discrete setting is relatively straightforward. 
It  can be achieved  by applying a singular value decomposition (SVD) in every mode - thus called Higher Order SVD (HOSVD) - of 
the tensor: For the $i$-th mode, we compute the SVD of the {\em $i$-mode matricization} 
\begin{equation}
\label{eq:modematric}
[ U ]_{\alpha_1,\ldots,\cancel{\alpha_i},\ldots,\alpha_d}^{\alpha_i} \in \mathbb C^{(q_1 \cdots \cancel{q_i} \cdots q_d) \times q_i}
\end{equation} 
and obtain the basis vectors $| \xi_{1}^{\{i\}} \rangle,  \ldots, | \xi_{r_i}^{\{i\}} \rangle $, which span the optimal subspace of $\Lambda_i$ \cite{Lathauwer-2000, Hackbusch-2012, Lubich-2008}.

In many applications, 
we want to approximate the tensor with lower rank $\mathbf{\tilde{r}} \leq \mathbf r$. In the matrix case $d = 2$, this can be done by
truncating the above SVD and omitting the basis vectors that belong to the smallest $r - \tilde r$ singular values. The discovery that this already yields the optimal
result is mostly accredited to Eckard and Young in mathematics, while most physics articles recognize the fact that it had been proven by Schmidt long before
for the more complicated case of integral operators \cite{Schmidt-1907}.

Unfortunately, this result cannot be generalized to tensors with $d > 2$. 
It has been shown in \cite{Hillar-2011} that even finding the best rank one, i.e. $ \mathbf{r} = (1, 1 , \ldots , 1)$, can be $NP$-hard if $d>2$. 
Nevertheless, truncating the HOSVD in every mode only yields a {\em quasi-optimal approximation}
 with respect to the $l_2$-norm \cite{Lathauwer-2000}. 
However, in many cases, this is satisfactory. 

\begin{figure}
\centering
\includegraphics{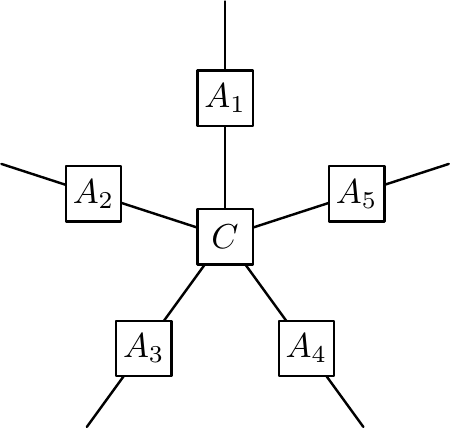}
\caption{A Tucker tensor of order 5.}
\label{fig:networks-tree_Tucker}
\end{figure}

The Tucker format is a subspace decomposition as the tensor is expressed in the basis of a subspace of the tensor space. At the same time, it yields a tensor tree, i.e. its representation fulfills criterion \ref{crit:lin},
\ref{crit:noncan} and \ref{crit:nocirc}. $C \in \mathbb C^{r_1 \times \ldots \times r_d}$ is the only virtual component and 
\begin{equation}
\label{eq:tuckcomps}
A_i(\alpha_i,m_i) = \langle \phi_{\alpha_i}^{\{i\}} | \xi_{m_i}^{\{i\}} \rangle
\end{equation}
yields the $d$ physical components, see Fig.~\ref{fig:networks-tree_Tucker}.

The HOSVD gives us a constructive algorithm that computes the Tucker decomposition, 
i.e. a representation of the form (\ref{eq:tucker}) with minimal rank $\mathbf r$, in polynomial time.
Additionally, the set of tensors with Tucker rank at most $\mathbf r$ is known to be Zariski-closed \cite{Landsberg-2012a}. Therefore, it is closed
and we overcome the border rank problem.
In terms of storage complexity however, this format is far from being optimal. It now scales exponentially in $r_i$, i.e for $r := \max \{r_i \}$ the scaling is in $\mathcal{O} (dq r + r^d)$.  
Especially for small $q_i$, where we do not have $r_i \ll q_i$, we cannot hope for much reduction of complexity.
In particular, for $q_i =2$ we do not gain any nontrivial progress.

\subsection{Matricization and tensor multiplication}
\label{sec:tensor.matricization}

To a certain extent, these representation allow to apply matrix analysis techniques to tensors. We therefore generalize the aforementioned matricization. Let $t \subseteq \{1,\ldots,d\}$ be a collection of physical dimensions
and $t^c := \{ 1,\ldots,d \} \setminus t$ its complement. Then 
\begin{equation}
\label{eq:matricization}
[U]_{\alpha_t}^{\alpha_{t^c}} \in \mathbb C^{q_t} \otimes \mathbb C^{q_{t^c}}
\end{equation} 
is the matricization with $q_t = \{ q_i \in \{ q_1,\ldots,q_d \} : i \in t \}$ as row dimensions 
and $q_{t^c} = \{ q_i \in \{ q_1,\ldots,q_d \} \setminus q_t \}$ as column dimensions.
A special case is the $i$-th matricization 
\begin{equation}
\label{eq:imat}
[U]_{\alpha_1,\ldots,\alpha_i}^{\alpha_{i+1},\ldots,\alpha_d} \in \mathbb C^{q_1 \cdots q_i} \otimes \mathbb C^{q_{i+1} \cdots q_d}
\end{equation} 
utilized further down that is casting the first $i$-variables into 
the row index, and the remaining $d-i$ in the column index.

This Einstein-like notation allows us to introduce a {\em tensor multiplication}. Let $U \in \mathbb C^{q_{1,1} \times \ldots \times q_{1,d_1}}$ and $V \in \mathbb C^{q_{2,1} \times \ldots \times q_{2,d_2}}$. Then if for to matricizations $t_1 \in \{ 1,\ldots,d_1 \}, t_2 \in \{ 1, \ldots, d_2 \}$ it holds $q_{1,t_1^c} = q_{2,t_2} =: q_{t_1,t_2}$ we get
\begin{equation}
\label{eq:tensmult}
 [U]_{\alpha_{1,t_1}}^{\alpha_{t_1,t_2}} [V]_{\alpha_{t_1,t_2}}^{\alpha_{2,t_2^c}} = \sum_{\alpha_{t_1,t_2}} U(\alpha_{1,t_1},\alpha_{t_1,t_2}) V(\alpha_{t_1,t_2},\alpha_{2,t_2^c}).
\end{equation}
This is exactly the matrix multiplication of the matricizations and it is the contraction over the indeces $\alpha_{t_1,t_2}$. In the case where no dual space is involved, i.e. no contraction is performed, we obtain the tensor product
\begin{equation}
\label{eq:nomult}
[U]_{\alpha_{1,1},\ldots,\alpha_{1,d_1}} [V]_{\alpha_{2,1},\ldots,\alpha_{2,d_2}} = U \otimes V.
\end{equation}

Note that in the complex case described here, the matricization should only be seen as the reordering and grouping of indeces, instead of introducing a concept of duality as done in some literature \cite{Landsberg-2012b}. This is due to the fact that it is impossible to take the complex conjugate only in a few indeces of $U$, which would be required for this concept \cite{Szalay-2013}. Thus, the reader should note that switching the ordering of the indeces gives only the transpose and not the hermitian of the original matricization:
\begin{equation}
\label{eq:noherm}
[U]_{\alpha_t}^{\alpha_{t^c}} = \bigl( [U]_{\alpha_{t^c}}^{\alpha_t} \bigr)^{\mathsf T} = \overline{ \bigl( [U]_{\alpha_{t^c}}^{\alpha_t} \bigr)^\dagger }.
\end{equation}

Finally, we want to simplify the notation for the unambiguous case where we multiply over {\em all} common indeces. This will be denoted with a circle, since it can be seen as a composition of two linear operators:
\begin{equation}
\label{eq:circ}
U \circ V := [U]_{\alpha_{1,t_1}}^{\alpha_{t_1,t_2}} [V]_{\alpha_{t_1,t_2}}^{\alpha_{2,t_2^c}} , \ q_{1,i} \neq q_{2,j} \; \forall \ i \in t_1, j \in t_2^c.
\end{equation}

\subsection{Matrix product states or the tensor train format}
\label{sec:tensor.tt}

\begin{figure}
\centering
\includegraphics{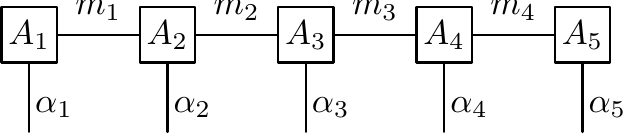}
\caption{A tensor of order 5 in TT representation.}
\label{fig:networks-train}
\end{figure}

Another example of a tensor network is the {\em Tensor Train (TT)} decomposition of a tensor. The tensor $U$ is 
given element-wise as
\begin{equation}
\label{eq:TTelem}
 U(\alpha_1,\ldots,\alpha_d) = \sum_{m_1 = 1}^{r_1} \ldots \sum_{m_{d-1} = 1}^{r_{d-1}} A_1(\alpha_1,m_1) A_2(m_1,\alpha_2,m_2) \cdots A_d(m_{d-1},\alpha_d).
\end{equation}
We get $d$ component tensors of order 2 or 3. Their graph has the structure of a chain or train, hence the name. Figure (\ref{fig:networks-train}) gives an example of a TT tensor. 

The TT format maintains the positive characteristics of the Tucker format and overcomes most of the disadvantages of the canonical format.
However, the complexity now scales only quadratically in the ranks, or with $\mathcal O (qdr^2)$, for $r = \max\{r_i\}$. While the Tensor Train decomposition is not the only format that has this advantage, it is one of the 
most widely used ones and it will also be the standard format in this paper.

This format has been introduced to the mathematical realm by Oseledets et al. \cite{Oseledets-2011}. While it was developed independently, it can be seen as a special case of the {\em Hierarchical Tucker (HT)} decomposition. 
However, we will restrict ourselves to the TT format and deal with the HT format only briefly further down. As stated above, nearly everything of the following can be generalized to a general tensor tree format without effort, 
but notation becomes more complex.

In physics the Tensor Train decomposition has been known as Matrix Product States (MPS) since the late nineties and many results can be taken directly from there. 
The name Matrix Product States is justified if we fix the physical indices. This yields a chain of matrix products:
\begin{equation}
\label{eq:TTMPS}
 U(\alpha_1,\ldots,\alpha_d) = \mathbf A_1(\alpha_1) \mathbf A_2(\alpha_2) \cdots \mathbf A_{d-1}(\alpha_{d-1}) \mathbf A_d(\alpha_d)
\end{equation}
with $[\mathbf A_i(\alpha_i)]_{m_{i-1},m_i} := A_i(m_{i-1},\alpha_i,m_i) \in \mathbb C^{r_{i-1} \times r_i}$.

Let it be noted that an important modification of the Tensor Train format follows if we introduce a contraction of rank greater than 1 between the first and last component, also called {\em periodic boundary conditions},
\begin{equation}
\label{eq:TTuMPS}
 U(\alpha_1,\ldots,\alpha_d) = \sum_{m_1 = 1}^{r_1} \ldots \sum_{m_d = 1}^{r_d} A_1(m_d,\alpha_1,m_1) A_2(m_1,\alpha_2,m_2) \cdots A_d(m_{d-1},\alpha_d,m_d).
\end{equation}
These {\em uniform Matrix Product States (uMPS)} are especially significant in physics. Verstraete et al. deal with uMPS that are also translation invariant, 
i.e. all components are equal $A_1 = \ldots = A_d$ \cite{Perez-Garcia-2007}. The graph of this decomposition is circular and therefore does not suffice criterion \ref{crit:nocirc}. 
As mentioned above, this poses a number of problems \cite{Landsberg-2012b} that are - in a nutshell - similar to those of the canonical format.
For this reason, we will only deal with regular, non-circular Matrix Product States from now on.

The TT format can be considered as a multi-layered subspace representation. This is achieved in a hierarchical way \cite{Hackbusch-2012}. In $\Lambda_1$ we consider the subspace $\Xi_1$ 
given by the basis set
$\{ | \xi_{m_1}^{\{1\}} \rangle : m_1 = 1,\ldots,r_1 \}$, where 
\begin{equation}
\label{eq:firstbase}
| \xi_{m_1}^{\{1\}} \rangle := \sum_{\alpha_1=1}^{q_1} A_1(\alpha_1,m_1) \ | \phi_{\alpha_1}^{\{1\}} \rangle.
\end{equation} 
We proceed with a subspace of the partial tensor product space
$\Xi_{\{ 1,2 \}} \subset \Lambda_1 \otimes \Lambda_2$ 
of  dimension $r_{\{1,2\}} \leq q_1  q_2$. 
Indeed $\Xi_{\{ 1,2\}}$ is defined through a new basis set $\{ | \xi_{m_{\{1,2\}}}^{\{1,2\}} \rangle : 1,\ldots,r_{\{1,2\}} \}$
where the new basis vectors are given in the form 
\begin{equation}
\label{eq:secondbase}
 | \xi_{m_{\{1,2\}}}^{\{1,2\}} \rangle  = \sum_{\alpha_1=1}^{q_1} \sum_{\alpha_2=1}^{q_2} U_{\{1,2\}} (\alpha_1,\alpha_2,m_{\{1,2\}}) \ | \phi_{\alpha_1}^{\{1\}} \rangle \otimes | \phi_{\alpha_2}^{\{2\}} \rangle. 
\end{equation}
We observe that $\Xi_{\{1,2\}} \subset \Xi_1 \otimes \Lambda_2$ with
\begin{equation}
\label{eq:seconddiff}
 | \xi_{m_{\{1,2\}}}^{\{1,2\}} \rangle = \sum_{m_1=1}^{r_1} \sum_{\alpha_2=1}^{q_2} A_2 (m_1,\alpha_2,m_{\{1,2\}}) \ | \xi_{m_1}^{\{1\}} \rangle \otimes | \phi_{\alpha_2}^{\{2\}} \rangle
\end{equation}
and thus
\begin{equation}
\label{eq:seccomp}
 U_{\{1,2\}} (\alpha_1,\alpha_2,m_{\{1,2\}}) = \sum_{m_1=1}^{r_1} A_1(\alpha_1,m_1) A_2 (m_1,\alpha_2,m_{\{1,2\}}).
\end{equation}
For this reason, when dealing with TT tensors, we simplify the notation and often set $\{1,2\} \simeq 2$, and in general $\{1,2,\ldots,i\} \simeq i$.
 
The tensor is recursively defined by the component tensors $A_i $,  
\begin{equation}
\label{eq:tensrecursive}
\begin{split}
 | \xi_{m_3}^{\{3\}} \rangle &= \sum_{m_2, \alpha_3} A_3 (m_2,\alpha_3,m_3) \ | \xi_{m_2}^{\{2\}} \rangle \otimes | \phi_{\alpha_3}^{\{3\}} \rangle \\
  &= \sum_{m_1,m_2,\alpha_2,\alpha_3} A_2(m_1,\alpha_2,m_2) A_3(m_2,\alpha_3,m_3) \ | \xi_{m_1}^{\{1\}} \rangle \otimes | \phi_{\alpha_2}^{\{2\}} \rangle \otimes | \phi_{\alpha_3}^{\{3\}} \rangle \\
  &= \sum_{m_1,m_2,\alpha_1,\alpha_2,\alpha_3} A_1(\alpha_1,m_1) A_2(m_1,\alpha_2,m_2) A_3(m_2,\alpha_3,m_3) \ | \phi_{\alpha_1}^{\{1\}} \rangle \otimes | \phi_{\alpha_2}^{\{2\}} \rangle \otimes | \phi_{\alpha_3}^{\{3\}} \rangle,
\end{split}
\end{equation}   
and so forth, by taking $\Xi_{\{1,\ldots,i+1\}} \subset \Xi_{\{1,\ldots,i\}} \otimes \Lambda_{i+1}$. 

We may also proceed differently, e.g. $\Xi_{\{ 1,2,3,4,\ldots \} } \subset  \Xi_{\{ 1,2\} } \otimes \Xi_{\{ 3,4\}} \otimes \ldots$. 
Especially, it can be advantageous to start from the right hand side, i.e. taking  $\Lambda_i \otimes \Xi_{\{i+1,\ldots,d\}} $ etc.,
obtaining basis vectors 
\begin{equation}
\label{eq:rightbase}
| \zeta_{m_{i-1}}^{\{i\}} \rangle \in \Xi_{\{ i, \ldots d\}}.
\end{equation}

\begin{figure}
\centering
\includegraphics{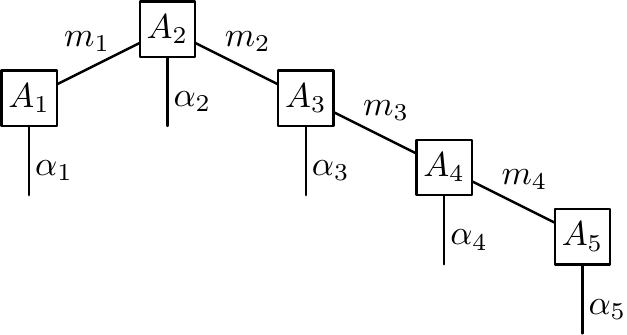}
\caption{Hierarchical picture of a Tensor Train with $A_2$ as the root.}
\label{fig:networks-train_hier}
\end{figure}

Let us fix some $i \in \{ 1, \ldots, d\} $ and call it the {\em root}. This gives a hierarchical picture (see Fig.~\ref{fig:networks-train_hier} ).

We consider the spaces $ \mathcal L_i :=  \Xi_{ \{ 1, \ldots , i-1 \}} $ and $\mathcal R_j : = \Xi_{ \{ i+1, \ldots , d \}} $. 
Their dimensions are given by 
\begin{equation}
\label{eq:dimspaces}
\dim \mathcal L_i = r_{i-1}, \dim \mathcal R_i = r_i
\end{equation} 
and hence, the full tensor $| \Psi \rangle$ is contained in the $ r_{i-1} q_i r_{i}$-dimensional subspace \cite{Hackbusch-2014a,Grasedyck-2011}
\begin{align}
\label{eq:subspace}
  | \Psi \rangle & \in \mathcal L_i \otimes \Lambda_i \otimes \mathcal R_i \cong \mathbb{C}^{r_{i-1} \times q_i \times r_i}  \ \\
  | \Psi \rangle & = \sum_{m_{i-1},m_i,\alpha_i} A_i(m_{i-1},\alpha_i,m_i) \ | \xi_{m_{i-1}}^{\{i-1\}} \rangle \otimes | \phi_{\alpha_i}^{\{i\}} \rangle \otimes | \zeta_{m_i}^{\{i+1\}} \rangle   
\end{align}

A canonical but not necessary choice is that the basis vectors $| \xi_{1}^{\{i-1\}} \rangle, \ldots, | \xi_{r_{i-1}}^{\{i-1\}} \rangle$ and $| \zeta_{1}^{\{i+1\}} \rangle, \ldots, | \zeta_{r_i}^{\{i+1\}} \rangle$
are orthogonal and normalized. 

We will  see in the following that this hierarchical or multi-layered subspace approximation constitutes the mechanism behind the 
{\em renormalization group} formalism in the one-site DMRG (density matrix renormalization group).  

An obvious observation\cite{Holtz-2012a} following from the above will be that the minimal dimension $r_i$ is the rank of the $i$-th matricization (\ref{eq:imat}):

\begin{theorem}[Separation Theorem]
\label{thm:separation}
 For any tensor $U \in \mathbb C^{q_1 \times \cdots \times q_d}$, there exists a minimal TT (MPS) 
 representation, thus called TT decomposition $TT(U)$, 
 such that for any $i = 1, \ldots , d-1 $ the dimensions $r_i$ of the contractions $m_i = 1 , \ldots , r_i$
 are minimal and given by 
\begin{equation}
\label{eq:septhm}
 r_i = \rank ([ U ]_{\alpha_1, \ldots , \alpha_i}^{\alpha_{i+1},\ldots, \alpha_d}).
\end{equation}
\end{theorem}

We can change the hierarchy, e.g. by choosing the next component $A_{i+1}$ as the root. In most applications, it will then become necessary to shift the orthogonalization such that
$\{ | \xi_{m_i}^{\{i\}} \rangle : m_i = 1,\ldots,r_{i} \}$ and $\{ | \zeta_{m_{i+1}}^{\{i+2\}} \rangle : m_{i+1} = 1,\ldots,r_{i+1} \}$ are orthonormal. This can be done by applying
the singular value decomposition to the matricization of the $i$-th component
\begin{equation}
\label{eq:compsvd}
 [A_i]_{m_{i-1},\alpha_i}^{m_{i+1}} = [\tilde A_i]_{m_{i-1},\alpha_i}^{m_{i+1}} \mathbf \Sigma_i \mathbf Y_i^\dagger
\end{equation}
and shifting $\mathbf \Sigma_i, \mathbf Y_i^\dagger \in \mathbb C^{r_i \times r_i}$ to the next component
\begin{equation}
\label{eq:svdshift}
 [\tilde A_{i+1}]_{m_i}^{\alpha_{i+1},m_{i+2}} = \mathbf \Sigma_i \mathbf Y_i^\dagger [A_{i+1}]_{m_i}^{\alpha_{i+1},m_{i+2}}.
\end{equation}
For $| \Psi \rangle$ we obtain
\begin{equation}
\label{eq:orthshift}
\begin{split}
 | \Psi \rangle &= \sum_{m_{i-1},m_i,\alpha_i} A_i(m_{i-1},\alpha_i,m_i) \ | \xi_{m_{i-1}}^{\{i-1\}} \rangle \otimes | \phi_{\alpha_i}^{\{i\}} \rangle \otimes | \zeta_{m_i}^{\{i+1\}} \rangle \\
 &= \sum_{\substack{m_{i-1},m_i,m_{i+1} \\ \alpha_i,\alpha_{i+1}}} A_i(m_{i-1},\alpha_i,m_i) A_{i+1}(m_i,\alpha_{i+1},m_{i+1}) \ | f\xi_{m_{i-1}}^{\{i-1\}} \rangle \otimes | \phi_{\alpha_i}^{\{i\}} \rangle \otimes | \phi_{\alpha_{i+1}}^{\{i+1\}} \rangle \otimes| \zeta_{m_{i+1}}^{\{i+2\}} \rangle \\
 &= \sum_{\substack{m_{i-1},m_i,m_{i+1} \\ \alpha_i,\alpha_{i+1}}} \tilde A_i(m_{i-1},\alpha_i,m_i)  \tilde A_{i+1}(m_i,\alpha_{i+1},m_{i+1}) \ | \xi_{m_{i-1}}^{\{i-1\}} \rangle \otimes | \phi_{\alpha_i}^{\{i\}} \rangle \otimes | \phi_{\alpha_{i+1}}^{\{i+1\}} \rangle \otimes| \zeta_{m_{i+1}}^{\{i+2\}} \rangle \\
 &= \sum_{m_i,m_{i+1},\alpha_{i+1}} \tilde A_{i+1}(m_i,\alpha_{i+1},m_{i+1}) \ | \xi_{m_i}^{\{i\}} \rangle \otimes | \phi_{\alpha_{i+1}}^{\{i+1\}} \rangle \otimes| \zeta_{m_{i+1}}^{\{i+2\}} \rangle.
\end{split}
\end{equation}

\begin{figure}
\centering
\includegraphics{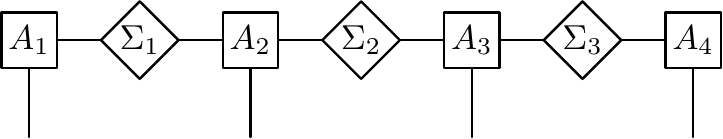}
\caption{A TT tensor of order 4 in standard representation.}
\label{fig:networks-train_Vidal}
\end{figure}

Alternatively one may  use QR factorization for the orthogonalization, but it often advantageous  
to keep the small diagonal matrix $\mathbf \Sigma_i \in \mathbb C^{r_i \times r_i}$ containing the singular values in between to adjacent component tensors.
In fact this provides a {\em standard representation} or {\em HSVD representation} of $U$, see Fig.~\ref{fig:networks-train_Vidal}
\begin{equation} 
\label{eq:TT-SVD} 
 U = A_1 \circ \mathbf \Sigma_1 \circ A_2 \circ \mathbf \Sigma_2 \circ \cdots \circ \mathbf \Sigma_{d-1} \circ A_d.
\end{equation}

\begin{figure}
\centering
\includegraphics{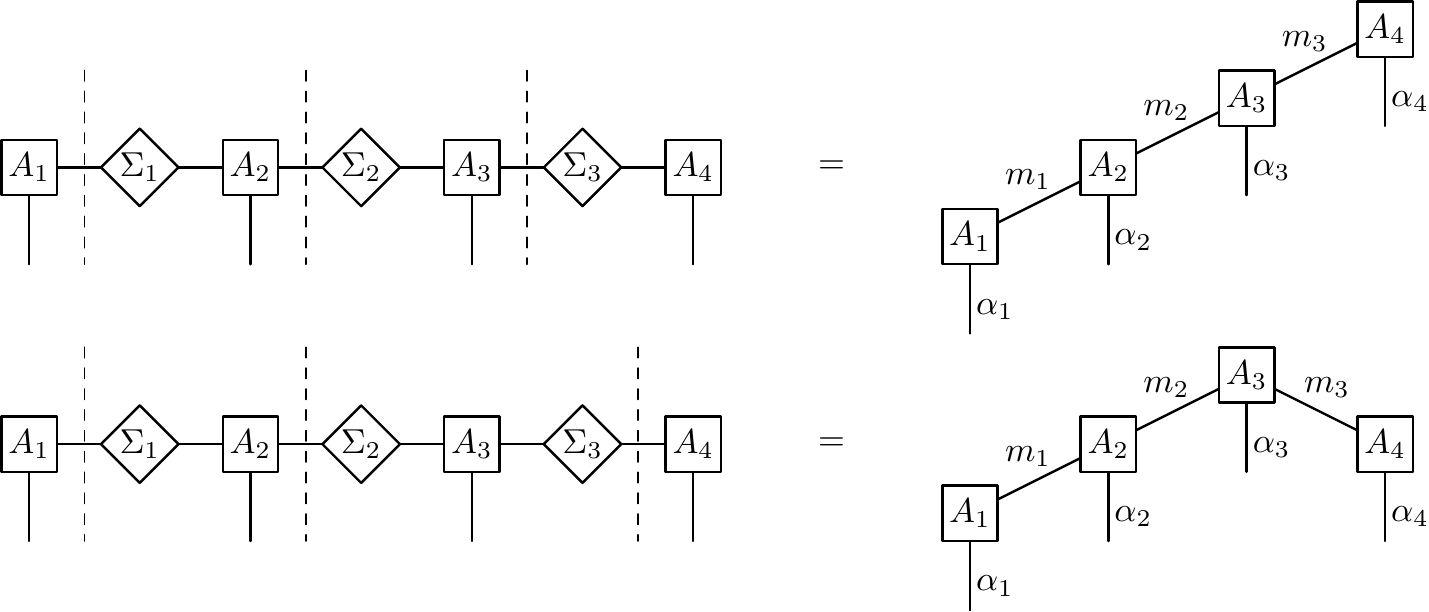}
\caption{A shift of orthogonality in the standard representation.}
\label{fig:networks-train_shift}
\end{figure}

This representation has been developed independently by different authors \cite{Vidal-2003b,Oseledets-2009,Grasedyck-2010}. 
In physics, it is accredited to Vidal and is hence also known as the {\em Vidal representation}. Very beneficial is the criterion that
\begin{equation}
\label{eq:standben1}
 A_1^\dagger A_1 = \mathbf I_{r_1}, \ A_d A_d^\dagger = \mathbf I_{r_{d-1}}
\end{equation}
and for all $1 < i < d$
\begin{align}
\label{eq:standben2}
 {[ A_i \circ \mathbf \Sigma_i ]}_{m_{i-1}}^{\alpha_i,m_i} \bigl( {[ A_i \circ \mathbf \Sigma_i ]}_{m_{i-1}}^{\alpha_i,m_i} \bigr)^\dagger &= \mathbf I_{r_{i-1}}, \\
 \bigl( {[ \mathbf \Sigma_{i-1} \circ A_i ]}_{m_{i-1},\alpha_i}^{m_i} \bigr)^\dagger {[ \mathbf \Sigma_{i-1} \circ A_i ]}_{m_{i-1},\alpha_i}^{m_i} &= \mathbf I_{r_i}.
\end{align}
This means, we can shift the root, and thus the orthogonality, by simply shifting the {\em density matrices} $\mathbf \Sigma_i$, see Fig.~\ref{fig:networks-train_shift}.

This representation can be computed by applying a sequence of singular value decomposition and storing the singular values. The procedure is called {\em Hierarchical SVD (HSVD)}. It recovers the tensor exactly.
However, as mentioned for the Tucker format and the HOSVD, the HSVD can be used for approximation by thresholding the singular values.
For density matrices $\mathbf \Sigma = \mathrm{diag}(\sigma_1,\ldots,\sigma_r)$ we define two thresholding operators
\begin{align}
\label{eq:densthresh}
 H_{\tilde{r}}(\mathbf \Sigma) &= \underset{1 \leq i \leq r}{\mathrm{diag}(\sigma_i)} \ ,\ \tilde{r} \leq r, \\
 H_\epsilon(\mathbf \Sigma) &= \underset{\sigma_i \geq \epsilon}{\mathrm{diag}(\sigma_i)} \ , \ \epsilon > 0
\end{align}
and for TT tensors
\begin{align}
\label{eq:tensthresh}
 H_{\tilde{\mathbf r}}(U) &= A_1 \circ H_{\tilde{r}_1}(\mathbf \Sigma_1) \circ A_2 \circ H_{\tilde{r}_2}(\mathbf \Sigma_2) \circ \cdots \circ H_{\tilde{r}_{d-1}}(\mathbf \Sigma_{d-1}) \circ A_d, \\
 H_{\mathbf \epsilon}(U) &= A_1 \circ H_{\epsilon_1}(\mathbf \Sigma_1) \circ A_2 \circ H_{\epsilon_2}(\mathbf \Sigma_2) \circ \cdots \circ H_{\epsilon_{d-1}}(\mathbf \Sigma_{d-1}) \circ A_d.
\end{align}

Again, this will not yield the best approximation of the tensor, as it does in the matrix case.
As with Tucker tensors, we maintain a so called {\em quasi optimality}:

\begin{theorem}[Quasi Optimality]\cite{Oseledets-2009,Grasedyck-2010,Hackbusch-2012,Hackbusch-2014b}
\label{thm:quasioptthm}
The truncation of the HSVD can be estimated by
\begin{equation}
\label{eq:quasiopt}
 \| U - H_{\tilde{\mathbf r}}(U) \| \leq \sqrt{d-1} \inf_{V \in \mathcal M_{\leq \tilde{\mathbf r}}} \| U - V \|,
\end{equation}
where $\mathcal M_{\leq \tilde{\mathbf r}}$ is the space of all tensors with TT rank not exceeding $\tilde{\mathbf r}$.
\end{theorem}

As most other results, the separation theorem and the quasi optimality can be readily generalized to all tree tensor networks. It is also possible to formulate a standard representation for other trees.
In contrast to the parametrization (\ref{eq:multi-linear}), the subspace representation provides further essential information about minimal representability and approximability. 
It justifies the use of the notion of {\em entanglement}
for the tensor $U$ or an appropriate low rank approximation of it.
Entanglement here means the quantum correlation between
the subsystem consisting of the first $i$ orbitals
and the subsystem consisting of the remaining orbitals,
and it can be characterized by, e.g.,
the quantum Hartley entropy $\log r_i$,
see in section \ref{sec:num.ent.toSVD}). 
Without further conditions, these quantities are not well defined for tensor representations that do not have a tree structure. However, Verstraete developed an {\em injectivity condition} that aims at overcoming that problem
for uniform MPS with periodic boundary conditions \cite{Haegeman-2014}.

\subsection{Dimension trees and the hierarchical tensor decomposition}
\label{sec:tensor.ht}

We briefly discuss the {\em Hierarchical Tucker} (HT) representation that has been introduced by Hackbusch and Kuhn \cite{Hackbusch-2010} in 2009 and has since received a lot of attention.
This is also due to the fact that it is a reasonable generalization of the TT format.

The HT representation is defined by a {\em dimension tree}, usually a binary tree, where the leafs $U_{\{1\}},\ldots,U_{\{d\}} = A_1,\ldots,A_d$ constitute the physical components and the inner vertices $U_t$ are virtual.
Hackbusch gives the following comprehensive notation in \cite{Hackbusch-2012}:
The vertices of the tree tensor network $TTNS(U) = (V,E,H)$ are labeled
\begin{enumerate}
\renewcommand{\labelenumi}{\roman{enumi})}
\label{enum:htlist}
 \item $t_r = \{ 1,\ldots,d \}$ for the root,
 \item $t \in L := \bigl\{ \{1\},\ldots,\{d\} \bigr\}$ for the leafs and 
 \item $t \in V \setminus L$ for inner vertices, which have sons $t_1,\ldots,t_p$ that are an ordered partition of $t$, i.e.
 $$\bigcup_i^p t_i = t \text{ and } \mu < \nu \; \forall \ \mu \in t_i, \ \nu \in t_j, \ i < j.$$
\end{enumerate}

\begin{figure}
\centering
\includegraphics{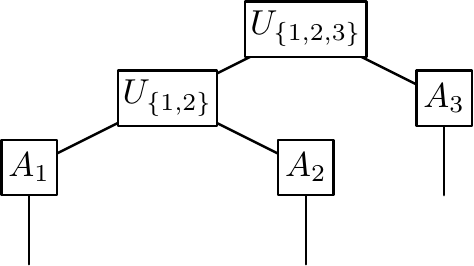}
\caption{A tensor of order 3 in HT format.}
\label{fig:networks-tree_HT}
\end{figure}

For an inner vertex $ t \subset V \setminus L $, with sons $ t_1 , \ldots , t_p $ (usually $p=2$), there is a subspace 
$ \Xi_t $ defined by its basis set $\{ | \xi_{m_t}^{\{t\}} \rangle : m_t = 1,\ldots,r_t \}$\cite{Falco-2014} given by 
\begin{equation}
\label{eq:htbase}
 | \xi_{m_t}^{\{t\}} \rangle = \sum_{m_1=1}^{r_{t_1}} \cdots \sum_{m_p=1}^{r_{t_p}} U_t (m_1,\ldots,m_p,m_t) \ | \xi_{m_1}^{\{t_1\}} \rangle\otimes \cdots \otimes | \xi_{m_p}^{\{t_p\}} \rangle.
\end{equation}
The root  $ t_r = \{ 1, \ldots , d \}$, with sons $ t_1 , \ldots , t_p $, is to reconstruct the tensor 
\begin{equation}
\label{eq:reconstht}
 U =  \sum_{m_1=1}^{r_{t_1}} \cdots \sum_{m_p=1}^{r_{t_p}} U_{t_r}  ( m_1, \ldots , m_p) \ | \xi_{m_1}^{\{t_1\}} \rangle\otimes \cdots \otimes | \xi_{m_p}^{\{t_p\}} \rangle.
\end{equation}
Therefore the tensor $U$ is defined completely by the component tensors $U_t$, using the above representations 
recursively, see Fig.~\ref{fig:networks-tree_HT}. There are at most $ \mathcal{O} (d)$ vertices and consequently the complexity is $ \mathcal{O} (  qdr +  d r^{p+1}   )  $. For $p=2$ we obtain 
$ \mathcal{O} (  qdr +  d r^{3}   ) $ \cite{Hackbusch-2012,Hackbusch-2014a}.

As with the Tucker and the TT format, obtaining the HT format can be done by applying the singular value decomposition successively, in a hierarchical fashion. 
Again, we maintain a well-defined rank through a separation theorem, a quasi optimality of a truncated HSVD and so on. 

In fact, the Tensor Train decomposition can be seen as a special case of the Hierarchical Tucker decomposition, where we use an unbalanced tree and omit the optimal subspaces in the leafs. 
However, in some cases, the binary tree structure can be advantageous \cite{Wang-2003}. Additionally, the leafs $A_1,\ldots,A_d$ form exactly the optimal subspaces already observed in the Tucker decomposition. 
We refer to the literature cited above. 

Arguably, this could make the HT format superior to the TT format. However, the notation becomes very messy and all notable theoretical results are valid for any tree tensor network.
Hence, we refrain from dealing with the Hierarchical format and proceed with the Tensor Train format, keeping the similarities in mind.

\subsection{Fixed rank manifolds and varieties}
\label{sec:tensor.manifolds}

For many applications, we consider the set of tensors of fixed TT rank $$\mathcal M_{\mathbf r} := \{ U \in \mathbb C^{n_1 \times \ldots \times n_d} : \mathbf r_{TT} = \mathbf r \}.$$
This set is no longer a linear space nor is it convex. In order to parametrize this space, we introduce the component set 
$\mathcal C = \{ \underline U = ( A_1,\ldots,A_d ) : A_i \in \mathbb C_*^{r_{i-1} \times q_i \times r_i} \}$ and the map
\begin{align}
\label{eq:manparam}
 \tau : \mathcal C &\rightarrow \mathcal M_{\mathbf r} \subseteq \mathbb C^{q_1 \times \ldots \times q_d}, \\
 (A_1,\ldots,A_d) &\mapsto \tau(A_1,\ldots,A_d) := U.
\end{align}
For each $i$, $\mathbb C_*^{r_{i-1} \times q_i \times r_i}$ is the space of all elements with full multilinear rank:
\begin{equation}
\label{eq:fullspace}
\mathbb C_*^{r_{i-1} \times q_i \times r_i} := \{ A_i \in \mathbb C^{r_{i-1} \times q_i \times r_i} : \rank([A_i]_{m_{i-1}}^{\alpha_i,m_i}) = r_{i-1}, \ \rank([A_i]_{m_{i-1},\alpha_i}^{m_i}) = r_i \}
\end{equation}
Let it be noted that this space is a smooth manifold \cite{Absil-2008}.

The map $\tau$ is clearly surjective onto $\mathcal M_{\mathbf r}$, but it is not injective: For any non-singular matrix $\mathbf X \in GL(r_i) \subseteq \mathbb C^{r_i \times r_i}$
we have 
\begin{equation}
\label{eq:redundancy}
A_1 \circ \cdots \circ A_d = A_1 \circ \cdots \circ A_i \circ \mathbf X \circ \mathbf X^{-1} \circ A_{i+1} \circ \cdots \circ A_d.
\end{equation}

Any parametrization of the form (\ref{eq:multi-linear}) shares this kind of non-uniqueness. But for tree tensor networks, this problem can be overcome:
Let $\mathcal G$ be the Lie group 
\begin{equation}
\label{eq:liegroup}
\mathcal G = \{ g = (\mathbf X_1,\ldots,\mathbf X_{d-1}) : \mathbf X_i \in \mathrm{GL}(r_i) \} \cong \bigotimes_{i=1}^{d-1} \mathrm{GL}(r_i).
\end{equation}
We define the group action of $g \in \mathcal G$ on the components $\underline U$ as
\begin{equation}
\label{eq:groupaction}
g \cdot \underline U := (A_1 \circ \mathbf X_1, \mathbf X_1^{-1} \circ A_2 \circ \mathbf X_2,\ldots,\mathbf X_{d-1}^{-1} \circ A_d).
\end{equation}
This action is smooth and it acts freely and properly on $\mathcal C$, see \cite{Uschmajew-2013}. 
The orbits are the equivalence classes, given by 
\begin{equation}
\label{eq:orbits}
[\underline U] = \mathcal G \cdot \underline U = \{ g \cdot \underline U : g \in \mathcal G \}.
\end{equation}
Thus, we obtain the quotient space 
\begin{equation}
\label{eq:quotientspace}
\mathcal C / \mathcal G = \{[\underline U] = \mathcal G  \cdot \underline U: \underline U \in \mathcal C\}
\end{equation}
with the quotient map 
\begin{equation}
\label{eq:quotientmap}
\pi : \mathcal C \rightarrow \mathcal C / \mathcal G, \ \underline U \mapsto [\underline U].
\end{equation}
This yields a bijection 
\begin{equation}
\label{eq:tensbijection}
\hat{\tau} : \mathcal C / \mathcal G \rightarrow \mathcal M_{\mathbf r},
\end{equation} 
where $\tau = \hat{\tau} \circ \pi$. As a result, we get that $\mathcal M_{\mathbf r}$ is a smooth quotient manifold \cite{Lee-2003}.

This manifold can be globally embedded into the tensor space $\mathcal M_{\mathbf r} \subset \mathbb C^{q_1 \times \ldots \times q_d}$ and we call it the {\em TT manifold} \cite{Holtz-2012a,Uschmajew-2013,Haegeman-2013,Haegeman-2011b,Arnold-2014}.
Thus, it is possible to define the tangent space $\mathrm T_U \mathcal M_{\mathbf r}$, which is a linear subset of $\mathbb C^{q_1 \times \ldots \times q_d}$. It is isomorphic to the {\em horizontal space} 
\begin{equation}
\label{eq:horizontalspace}
\mathrm H_{\underline U} \mathcal C = \{ (W_1, \ldots, W_d) \in \mathcal C : \bigl( [W_i]_{k_{i-1},x_i}^{k_i} \bigr)^\dagger [A_i]_{k_{i-1},x_i}^{k_i} = \mathbf 0 \; \forall \ i = 1,\ldots,d-1 \}
\end{equation}
via
\begin{equation}
\label{eq:tangentiso}
\bigl( D \tau (\underline U) \bigr) (W_1,\ldots,W_d) = \sum_{i=1}^d A_1 \circ \cdots \circ W_i \circ \cdots \circ A_d.
\end{equation}
We remark that different definitions of the horizontal space are possible and that the choice of the gauge conditions above is not unique. It also depends on the choice of the root. In the above case, the root is set to be the last component $A_d$. The only requirement for a horizontal space is that it forms the tangent space of $\mathcal C$ via the direct sum
\begin{equation}
\label{eq:tanghoriz}
\mathrm T_{\underline U} \mathcal C = \mathrm V_{\underline U} \mathcal C \oplus \mathrm H_{\underline U} \mathcal C,
\end{equation}
where $\mathrm V_{\underline U} \mathcal C$ is the {\em vertical space} tangential to the orbits.

The manifold $\mathcal M_{\mathbf r}$ is an open set. However, in finite dimensions, its closure is given by
\begin{equation}
\label{eq:manifoldclos}
\overline{\mathcal M_{\mathbf r}} = \mathcal M_{\leq \mathbf r}.
\end{equation}
This is based on the observation that the matrix rank is an upper semi-continuous function \cite{Falco-2014,Hackbusch-2012}. The singular points are exactly those where the actual rank is not maximal.

As mentioned above, the set $\mathcal M_{\leq \mathbf r}$ is Zariski-closed and thus forms an {\em algebraic variety}, i.e. it is the set of common zeros of polynomials. This is easy to see:
Indeed, we know from the separation theorem, that  $\mathcal{M}_{\leq \mathbf r}$ is the intersection of all tensors where
the corresponding matricizations $[U]_{\alpha_1,\ldots,\alpha_i}^{\alpha_{i+1},\ldots,\alpha_d}$ have at most rank $ r_i $. The sets of matrices
with rank at most $r_i$ are known to be algebraic varieties \cite{Schneider-2014}, each some zero-set of polynomials \cite{Landsberg-2012b}. Then, trivially, the intersection is the zero-set of the union of all such polynomials.
Again, this property generalizes to all tensor trees.

\subsection{Dirac-Frenkel variational principle or dynamical low rank approximation}
\label{sec:tensor.dirac-frenkel}

Solving problems in the large tensor space is often too expensive due to the curse of dimensionality. We therefore restrict ourselves to tensors of fixed rank, i.e. to the space $\mathcal M_{\leq \mathbf r}$.
In general, the appropriate ranks are unknown. Thus, we start with an initial guess and increase the ranks when necessary. There are some greedy techniques available that serve this purpose \cite{Falco-2014}.

For the approximation with fixed rank, we consider the smooth manifold $\mathcal M_{\mathbf r}$, as this facilitates the theoretical framework.
Let 
\begin{equation}
\label{eq:minfunct}
J(U) \rightarrow \min, \ J \in C^1(\mathbb C^{q_1,\ldots,q_d}, \mathbb R)
\end{equation}
be a minimization problem on the tensor space, for example the minimization of the energy functional (\ref{eq:varHd.E0}) in quantum chemistry. 

For the restriction of $J$ to $\mathcal M_{\mathbf r}$ we obtain the necessary condition 
\begin{equation}
\label{eq:gradzero}
U = \argmin_{V \in \mathcal M_{\mathbf r}}  J (V) \Rightarrow \langle \nabla J(V) , \delta U \rangle = 0 \; \forall \ \delta U \in \mathrm T_U \mathcal M_{\mathbf r},
\end{equation}
i.e. if $U$ minimizes $J$ on $\mathcal M_{\mathbf r}$, then the gradient of $J$ must be orthogonal to the tangent space at $U$.
Equivalently, if we denote the orthogonal projection onto $\mathrm T_U \mathcal M_{\mathbf r}$ with $P_{\mathrm T_U}$, we get
\begin{equation}
\label{eq:orthproj}
P_{\mathrm T_U} \nabla J(U) = 0.
\end{equation}

This variational approach can be generalized to the dynamical problem
\begin{align}
\label{eq:timedep}
 \frac{d}{dt} U &= f ( U ), \\
 U(0) &= U_0 \in \mathcal M_{\mathbf r}.
\end{align}
This is a differential equation on $M_{\mathbf r}$ if and only if 
\begin{equation}
\label{eq:intangentsp}
f(U) \in \mathrm T_U \mathcal M_{\mathbf r}, \; \forall U \in \mathcal M_{\mathbf r}.
\end{equation}

Thus (\ref{eq:timedep}) can be solved approximately by projecting $f(U)$ on the tangent space $\mathrm T_U \mathcal M_{\mathbf r}$, 
\begin{equation}
\label{eq:projf}
F(U) := P_{\mathrm T_U} f(U)
\end{equation} 
and solving the projected differential equation 
\begin{equation}
\label{eq:projequation}
\frac{d}{dt} U = F ( U ).
\end{equation}
In accordance with the above, we obtain
\begin{equation}
\label{eq:dirac-fren}
 \frac{d}{dt} U - F(U) = 0 \Leftrightarrow \langle \frac{d}{dt} U - f(U) , \delta U \rangle = 0 \; \forall \ \delta U \in \mathrm T_U \mathcal M_{\mathbf r}.
\end{equation}
In the context of time-dependent quantum chemistry, this is well-known as the {\em Dirac-Frenkel variational principle} \cite{Meyer-1990, Beck-1999, Hairer-2003,Lubich-2008,Lubich-2010,Lubich-2012}.

\begin{figure}
\centering
\includegraphics{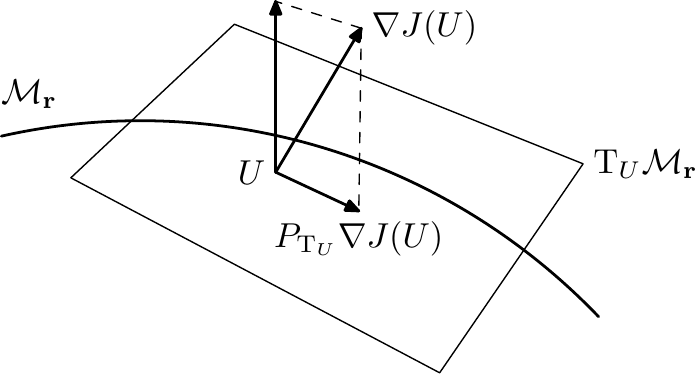}
\caption{An illustration of the gradient flow on manifold $\mathcal M_{\mathbf r}$.}
\label{fig:tangent_DiracFrenkel}
\end{figure}

Replacing $f(U)$ with $-\nabla J (U)$ in (\ref{eq:timedep}) gives the gradient flow of $J$. Then (\ref{eq:dirac-fren}) becomes 
\begin{equation}
\label{eq:gradientflow}
\langle \frac{d}{dt} U + \nabla J(U), \delta U \rangle = 0 \; \forall \ \delta U \in \mathrm T_U \mathcal M_{\mathbf r}
\end{equation}
and a solution can be computed with the aforementioned methods \cite{DaSilva-2014,Lubich-2012,Lubich-2013,Haegeman-2014,Hackbusch-2014a,Arnold-2014}, see Fig.~\ref{fig:tangent_DiracFrenkel}.

\subsection{The alternating least squares algorithm}
\label{sec:tensor.ALS}

Consider the functional
\begin{align}
\label{eq:smallfunc}
j : \mathcal C &\rightarrow \mathbb R \\
( A_1, \ldots, A_d) &\mapsto j(A_1, \ldots, A_d) := J(\tau(A_1, \ldots, A_d)).
\end{align}
For $i \in \{1,\ldots,d\}$ we fix $A_1,\ldots,A_{i-1}$ and $A_{i+1},\ldots,A_d$ and solve the 
subproblem 
\begin{equation}
\label{eq:subproblem}
A_i^+ := \argmin_{V_i \in \mathbb C^{r_{i-1} \times q_i \times r_i}} j(A_1,\ldots,V_i,\ldots,A_d).
\end{equation}

This is done in a successive manner and with alternating directions, which - for the best least squares fit $J(U) = \| U - B \|$ - justifies the 
name {\em Alternating Least Squares} (ALS) algorithm. The well-known Gau\ss-Seidel iteration is based on this strategy.

The TT format allows for a special formulation of this algorithm, sometimes dubbed the {\em Alternating
Linear Scheme} to maintain the abbreviation. In this case, we can give a closed form for each subproblem 
and they can be solved using standard tools from linear algebra and numerical optimization.

In every step, one has to solve a small problem in order to achieve the minimum. Note that we allow $V_i \in \mathbb C^{r_{i-1} \times q_i \times r_i}$, i.e. the ranks can decrease in each step. This automatically restricts $J$ to the variety $\mathcal M_{\leq \mathbf r}$ since the components can have full rank or less, but obviously not more than that.

The small subproblems will be of the same kind as the original problem, i.e. linear equations will be turned into small linear equations and eigenvalue problems give rise to relatively small
(generalized) eigenvalue problems. In physics this supports the renormalization picture, where an original large systems is reduced to a small system
with the same ground state energy and possibly further physical quantities.

As we have observed before, this simple approach should be realized with some care. Since the representation is redundant, we can generally not minimize over the full parameter space $\mathbb C^{r_{i-1} \times q_i \times r_i}$
but rather some non-linear quotient space and it becomes necessary to introduce gauge conditions like above. However, this can be avoided if we choose to minimize only the root of the tensor as there is no redundancy in this part.
After the minimization, it would then be crucial to restructure the hierarchy of the tensor and consider the next component as the root. This can be done by shifting the orthogonality as explained in (\ref{eq:orthshift}).
The extension to general hierarchical trees is straightforward.

\begin{figure}
\centering
\setlength{\unitlength}{40pt}
\begin{picture}(7.5,2.5)
\put(0.0,0.0){\includegraphics{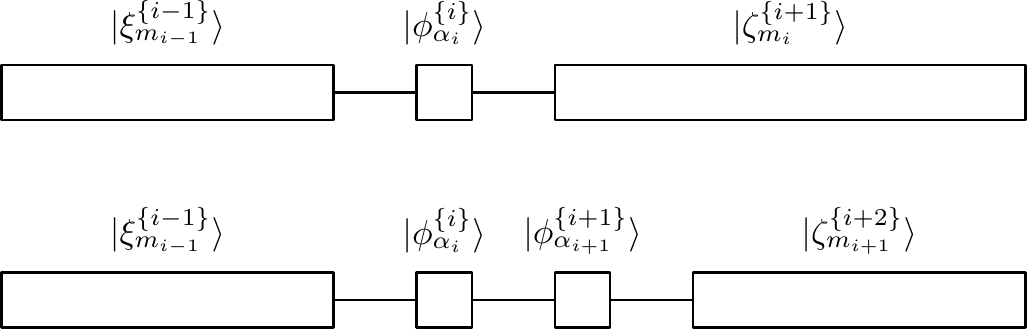}}
\put(0.0,2.4){\makebox(0,0)[r]{\strut{}(b)}}
\put(0.0,0.9){\makebox(0,0)[r]{\strut{}(a)}}
\end{picture}
\caption{Reduced basis representation for (a) the ALS algorithm, and (b) the two-site DMRG.}
\label{fig:networks-DMRGsd}
\end{figure}

Conforming with the earlier notation (\ref{eq:subspace}), each subproblem becomes a problem over a small subset that constitutes a subspace 
\begin{equation}
\label{eq:subbasis}
\mathcal L_i \otimes \Lambda_i \otimes \mathcal R_i \subseteq \Lambda^{(d)} = \bigotimes_{i=1}^d \Lambda_i.
\end{equation}
We define the orthogonal projector onto this space 
\begin{equation}
\label{eq:projbase}
P_i : \Lambda^{(d)} \rightarrow \mathcal L_i \otimes \Lambda_i \otimes \mathcal R_i.
\end{equation}
If we choose orthogonal bases for $| \xi_{1}^{\{i-1\}} \rangle,\ldots,| \xi_{r_{i-1}}^{\{i-1\}} \rangle$ and $| \zeta_{1}^{\{i+1\}} \rangle,\ldots,| \zeta_{r_i}^{\{i+1\}} \rangle$, we obtain
\begin{equation}
\label{eq:projequiv}
P_i \simeq E_i E_i^\dagger,
\end{equation}
where 
\begin{align}
\label{eq:insertionop}
 E_i : \mathbb C^{r_{i-1} \times q_i \times r_i} &\rightarrow \mathbb C^{q_1 \times \ldots \times q_d} \\
 V_i &\mapsto E_i V_i = A_1 \circ \cdots \circ A_{i-1} \circ V_i \circ A_{i+1} \circ \cdots \circ A_d
\end{align}
is the {\em insertion operator} also use elsewhere \cite{Holtz-2012b}.
This can easily be seen, as for $V \in \mathbb C^{q_1 \times \ldots \times q_d}$ it holds
\begin{align}
\label{eq:spaceprojequiv}
P_i | \Psi \rangle &= \sum_{m_{i-1},m_i,\alpha_i} \tilde{V}_i (m_{i-1},\alpha_i,m_i) \ | \xi_{m_{i-1}}^{\{i-1\}} \rangle \otimes | \phi_{\alpha_i}^{\{i\}} \rangle \otimes \zeta_{m_i}^{\{i+1\}} \rangle \\
&= \sum_{\alpha_1,\ldots,\alpha_d} E_i \tilde{V}_i (\alpha_1,\ldots,\alpha_d) \ | \phi_{\alpha_1}^{\{1\}} \rangle \otimes \ldots \otimes | \phi_{\alpha_d}^{\{d\}} \rangle
\end{align}
and
\begin{equation}
\label{eq:insertopadj}
E_i^\dagger V = \tilde{V}_i.
\end{equation}
Note that $E_i$ is a bijection onto its image, and since it is also orthogonal, its hermitian is well-defined as its inverse. See Fig.~\ref{fig:networks-DMRGsd}(a) for an illustration of the reduced basis.

To formulate the procedure explicitly, we consider a linear system, i.e. a functional
\begin{equation}
\label{eq:linsys}
J(U) = \frac{1}{2} \langle \mathbf X U, U \rangle - \langle B, U \rangle,
\end{equation}
where $\mathbf X \in \Lin(\mathbb C^{q_1 \times \ldots \times q_d}, \mathbb C^{q_1 \times \ldots \times q_d})$ is a linear operator. This operator can be stored and viewed in a canonical-like format, i.e. as a sum of rank-one tensor products
\begin{equation}
\label{eq:canonop}
\mathbf X = \sum_k \mathbf X_1^k \otimes \cdots \otimes \mathbf X_d^k
\end{equation}
or even in a TT-Matrix or Matrix Product Operator (MPO) format \cite{Oseledets-2011}. This is irrelevant for the purpose of notation, but it can be of computational interest.

Since we have equivalence $\Lambda^{(d)} \cong \mathbb C^{q_1 \times \ldots \times q_d}$, we also denote $\mathbf X \in \Lin(\Lambda^{(d)}, \Lambda^{(d)})$ without changing the notation. For the right side we
denote $B \simeq | \mathbf \Upsilon \rangle \in \Lambda^{(d)}$. A single subproblem can then be expressed as
\begin{equation}
\label{eq:ALSsubproblem}
\begin{split}
 A_i^+ &= \argmin_{V_i \in \mathbb C^{r_{i-1} \times q_i \times r_i}}  j(A_1,\ldots,V_i,\ldots,A_d) \\
 &= \argmin_{V_i \in \mathbb C^{r_{i-1} \times q_i \times r_i}} \bigl( \frac{1}{2} \langle \mathbf X E_i V_i , E_i V_i \rangle - \langle B , E_i V_i \rangle \bigr) \\
 &= \argmin_{V_i \in \mathbb C^{r_{i-1} \times q_i \times r_i}} \bigl( \frac{1}{2} \langle E_i^\dagger \mathbf X E_i V_i , V_i \rangle - \langle E_i^\dagger B , V_i \rangle \bigr)
\end{split}
\end{equation}

\begin{figure}
\centering
\setlength{\unitlength}{40pt}
\begin{picture}(11.4,5.4)
\put(0.0,0.0){\includegraphics{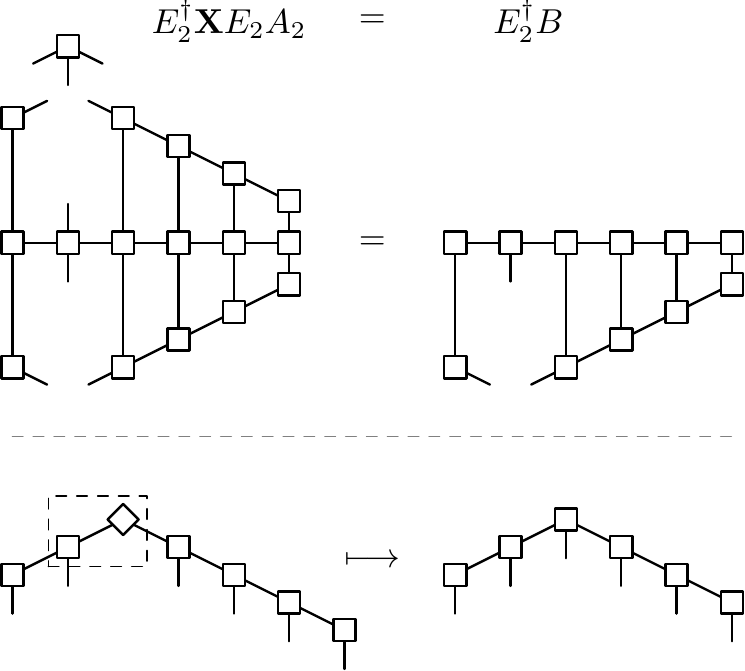}
\hspace{.25cm} \vline \hspace{.25cm}
\includegraphics{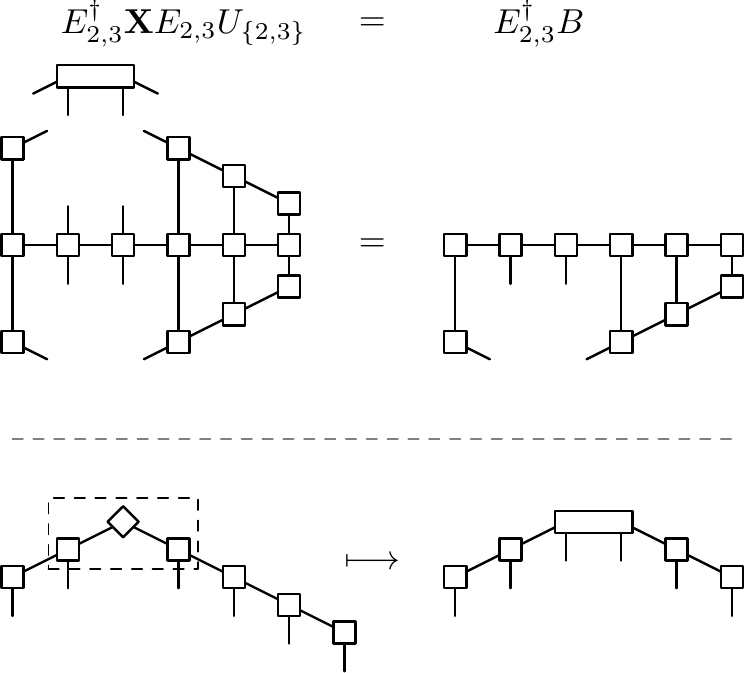}}
\put(0.0,5.2){\makebox(0,0)[l]{\strut{}(a)}} 
\put(6.0,5.2){\makebox(0,0)[l]{\strut{}(b)}}
\end{picture}
\caption{A micro-iteration of (a) the ALS algorithm, and (b) the two-site DMRG.}
\label{fig:networks-general_ALSDMRG}
\end{figure}

At stationary points $V_i$ of the functional $j \circ E_i$, there holds the first order condition
\begin{equation}
\label{eq:firstordcond}
\nabla (j \circ E_i)(V_i) = E_i^\dagger \mathbf X E_i \mathcal V_i - E_i^\dagger B = 0.
\end{equation}
As such, one micro-iteration of the ALS algorithm can be defined as
\begin{align}
\label{eq:ALSmicro}
U^+ &:= A_1 \circ \cdots \circ A_i^+ \circ \cdots \circ A_d \\
A_i^+ &= \bigl( E_i^\dagger \mathbf X E_i \bigl)^{-1} E_i^\dagger B.
\end{align}
See Fig.~\ref{fig:networks-general_ALSDMRG}(a) for an illustration.

In the subspace notation, we get
\begin{equation}
\label{eq:ALSsubspacenot}
| \Psi \rangle ^+ = \argmin_{| \Phi \rangle \in \mathcal L_i \otimes \Lambda_i \otimes \mathcal R_i} \bigl( \frac{1}{2} \langle \Phi | \mathbf X | \Phi \rangle - \langle \Upsilon , \Phi \rangle \bigr) = P_i \mathbf X^{-1} P_i | \Upsilon \rangle.
\end{equation}

For this to work, $\mathbf X$ does not necessarily have to be invertible on the whole tensor space but only on the small subspaces $\mathcal L_i \otimes \Lambda_i \otimes \mathcal R_i$. This is guaranteed if $\mathbf X$ is invertible as a whole.
Additionally, one can see that the spectrum of $\mathbf X$ on $\mathcal L_i \otimes \Lambda_i \otimes \mathcal R_i$ is a subset of the whole spectrum of $\mathbf X$ and in particular it holds $\cond_i (\mathbf X) \leq \cond (\mathbf X)$ \cite{Holtz-2012b}.

This notation suggests that the ALS is closely related to the DMRG algorithm. In fact, it is often called the {\em one-site} DMRG as it can be seen as a simple modification of the that algorithm.
In comparison, the ALS has the advantage that it optimizes the tensor on very small subspaces. On the other hand, the ranks $\mathbf r = (r_1,\ldots,r_{d-1})$ remain fixed and have to be guessed at the beginning.
In order to introduce higher ranks, one has to do this in a greedy fashion, e.g. by adding a rank-one approximation of the residual, see \cite{Uschmajew-2014}.

The classical two-site DMRG is a clever modification. Here, we minimize over the bigger subspace $\mathcal L_i \otimes \Lambda_i \otimes \Lambda_{i+1} \otimes R_{i+1}$, with the basis representation as in Fig.~\ref{fig:networks-DMRGsd}(b),
\begin{equation}
\label{eq:DMRGbasisrep}
| \Psi \rangle = \sum_{m_{i-1},m_{i+1},\alpha_i,\alpha_{i+1}} U_{\{i,i+1\}}(m_{i-1},\alpha_i,\alpha_{i+1},m_{i+1}) \ | \xi_{m_{i-1}}^{\{i-1\}} \rangle \otimes | \phi_{\alpha_i}^{\{i\}} \rangle \otimes | \phi_{\alpha_{i+1}}^{\{i+1\}} \rangle \otimes \zeta_{m_{i+1}}^{\{i+2\}} \rangle,
\end{equation}
i.e. we optimize two components at the same time, see Fig.~\ref{fig:networks-general_ALSDMRG}(b). The advantage is that a subsequent SVD after the optimization step in order to separate the two components yields a new - 
and possibly higher - rank.
To control the size of these new ranks, a further truncation is often required. Several strategies for {\em dynamical rank selection} can be implemented by considering the error in different norms \cite{Holtz-2012b,Legeza-2003a}.

\begin{table}
\centering
\begin{tabular}{|c|c|c|c|c|}
 \hline  
 & canonical  & Tucker & TT & HT \\
 \hline \hline 
 complexity & $\textcolor{blue}{\mathcal{O}(qdr)}$ & $\mathcal{O} (r^d + qdr)$ & $\textcolor{blue}{\mathcal{O}(qdr^2)}$ & $\textcolor{blue}{\mathcal{O}(qdr + dr^3)}$ \\
 \hline 
 rank & no & \textcolor{blue}{defined}  & \textcolor{blue}{defined} & \textcolor{blue}{defined} \\
 & \multicolumn{4}{c|}{ $ \max \{ \mathbf r_{Tucker} \} \leq \max \{ \mathbf r_{HT} \} \leq r_{canonical} $ }\\
 \hline 
 (weak) closedness & no & \textcolor{blue}{yes} &  \textcolor{blue}{yes} & \textcolor{blue}{yes} \\
 \hline 
 ALS (1site DMRG)  & yes - but slow & \textcolor{blue}{yes} &  \textcolor{blue}{yes} & \textcolor{blue}{yes} \\
 \hline 
 DMRG  & no  & no &  \textcolor{blue}{yes} & no \\
 \hline 
 H(O)SVD & no & \textcolor{blue}{yes} &  \textcolor{blue}{yes} & \textcolor{blue}{yes}\\
 \hline 
 embedded manifold  & no  &  \textcolor{blue}{yes} &  \textcolor{blue}{yes} &  \textcolor{blue}{yes} \\ 
 dyn. low rank approx.  & no  &  \textcolor{blue}{yes} &  \textcolor{blue}{yes} &  \textcolor{blue}{yes} \\ 
 algebraic variety  & no  &  \textcolor{blue}{yes} &  \textcolor{blue}{yes} &  \textcolor{blue}{yes} \\ 
 \hline 
 recovery & ?? &  \textcolor{blue}{yes} &  \textcolor{blue}{yes} &  \textcolor{blue}{yes}\\
 quasi best approx. & no &  \textcolor{blue}{yes} &  \textcolor{blue}{yes} &  \textcolor{blue}{yes}\\
 best approx. & no & exist & exist & exist \\
  &  & but $NP$-hard & but $NP$-hard & but $NP$-hard \\
 \hline
\end{tabular}
\caption{Comparison between the different tensor formats introduced in Sec.~\ref{sec:tensor}.}
\label{tab:compareformats}
\end{table}

General convergence theory of both the ALS and the DMRG is subject to research \cite{Rohwedder-2013}. They converge only to stationary points or at most local minima, as global convergence can not be guaranteed \cite{Holtz-2012b}.
Some convergence results have been published for a modified scheme, that proceeds in a Gau\ss-Southwell-like fashion and optimizes only the component with the largest residual \cite{Schneider-2014, Li-2014}.
There are also many open questions in dealing with physical applications like the SCF iteration for Hartree Fock models.
The prescribed approach is completely variational, which has important consequences for computing gradients, e.g. forces. An efficient implementation plays a crucial role. 
The interested reader should consult fundamental contributions in the DMRG literature \cite{White-1992b,White-1993,Legeza-2003b,Chan-2002a}.

As a summary of the section about tensor formats, we present Tab.~\ref{tab:compareformats} that compares the different decompositions, their complexity and their advantages and disadvantages in numerical computations.

\section{Numerical techniques} 
\label{sec:num}

In order to utilize efficiently the theoretical framework discussed
in the previous sections one has to carry out various optimization tasks.
Therefore, we start this section with a brief overview
and highlight important concepts using simple examples. Then various
iterative methods based on blocking procedure will be reviewed briefly
and the concept of entanglement will be studied with respect to
entanglement localization, geometrical network optimization,
choosing optimal bases and network initialization.
In this section, our focus is on the numerical method, thus
entropic measures of electronic properties of molecules will
be discussed only very briefly.

For pedagogical reasons, tutorial examples will be presented for a text book 
example, the LiF molecule.
Due to the ionic-neutral curve crossing between the two lowest 
$^1\Sigma^+$ states of LiF, this system provides a good testing ground 
to demonstrate the efficiency of the \emph{quantum chemistry version of 
the density matrix renormalization group} method (QC-DMRG) and \emph{tree 
tensor network state} (QC-TTNS) algorithm. 
Our analysis is especially useful for systems in which the wave 
function character of molecules changes as a function of geometry. 
In the LiF example, it differs greatly on two sides of an avoided crossing 
in a diatomic molecule. 
Atomic orbital (AO) basis was
adapted from the literature\cite{Bauschlicher-1988}
in order to match with previous DMRG computations\cite{Legeza-2003c}.
The AO basis set\cite{Bauschlicher-1988} is suitable to describe
the ionic and covalent LiF states as well. 
It consists of 
9s and 4p functions contracted to 4s and 2p functions on the Li atom 
and 9s, 6p and 1d functions contracted to 4s, 3p and 1d on the F atom.
For more details of the AO basis set, we refer to the original publication\cite{Bauschlicher-1988}.
The two lowest $^1\Sigma^+$ states of LiF around the
equilibrium bond length can be qualitatively described by the
1$\sigma^2$2$\sigma^2$3$\sigma^2$4$\sigma^2$1$\pi^4$ and
1$\sigma^2$2$\sigma^2$3$\sigma^2$4$\sigma^1$5$\sigma^1$1$\pi^4$
configurations\cite{Gordon-2005}. For this reason, the
 MO basis was obtained by CASSCF optimizations, with  two active electrons on
two active orbitals (4$\sigma$ and 5$\sigma$)  (CAS(2,2)).
MO's were optimized
simultaneously for both $^1\Sigma^+$ states.
$T_{ij}$ and $V_{ijkl}$ matrix elements  of Eq.~(\ref{eq:singleint}) and (\ref{eq:doubleint})
are expressed in this MO basis.
CASSCF optimizations were
carried out with the \texttt{GAMESS-US} quantum chemistry package\cite{Gordon-2005}.
Orbitals 1$\sigma$, 2$\sigma$ and 3$\sigma$ were kept frozen
in all presented configurational interaction (CI), MPS(DMRG) and TTNS
computations. Six of the valence electrons were excited to all
orbitals in the CI calculation, which we use as reference to benchmark
the QC-DMRG and QC-TTNS results. 
Therefore the active space in most of our CI, MPS(DMRG) and TTNS computations
consists of 6 electrons and 25 orbitals: CAS(6,25).
In certain cases, a smaller active space, CAS(6,12), will also be used.
Using the same MO basis obtained as a result of CASSCF optimizations
in the previous CAS(2,2) active space,
the CAS(6,12) active space is constructed by excluding the three
lowest lying occupied and 13 highest virtual orbitals
from the total 28 orbitals.
CI results were obtained by utilizing 
standard full-CI programs.
C$_{2v}$ point group symmetry constraints were assigned during this study.


\subsection{Basic terms and brief overview}
\label{sec:num.basic}

\subsubsection{The problem in the language of tensor factorization}
\label{sec:num.basic.tensorform}
Let us start this section with a very brief summary in order to highlight the most important concepts.
In the rest of the paper, a spin-orbital will be called a \emph{local tensor space} 
$\Lambda \cong \mathbb{C}^q$, with $\dim\Lambda = q$, and will be denoted by $\bullet$.
Using the fermionic occupation number basis ($q=2$), $|\mu,s\rangle$
for all spins $s\in\{\downarrow,\uparrow\}$,
with $\mu\in\{0,1\}$ occupation numbers, 
the operators (see (\ref{eq:matrices})) are defined as
\begin{equation}
\label{eq:matricesagain}
\mathbf{a}^\dagger = 
\begin{pmatrix}
 0 & 0  \\
 1 & 0  \\
\end{pmatrix},\qquad
 \mathbf{I} = 
\begin{pmatrix}
 1 & 0  \\
 0 & 1  \\
\end{pmatrix},\qquad
 \mathbf{s} = 
\begin{pmatrix}
 1 & 0  \\
 0 & -1  \\
\end{pmatrix},
\end{equation}
where $\mathbf{a}^\dagger$ creates an electron, $\mathbf{I}$ is the identity matrix and $\mathbf{s}$ stands for the phase
factor due to the antisymmetric fermionic wavefunction.
As was constructed in section \ref{sec:qchem.spatial},
it is also possible to use a $\mathbb{C}^4$ representation in which case $\bullet$ will represent 
a molecular orbital ($q=4$). 
In this representation a state can be empty, singly occupied with spin-up
or down particle, or doubly occupied,
represented by the basis states $\{|\phi_\alpha\rangle\}$ for $\alpha\in\{1,2,3,4\}$ as
$\{|\phi_1\rangle \equiv |-\rangle,
   |\phi_2\rangle \equiv |\downarrow\rangle,
   |\phi_3\rangle \equiv |\uparrow\rangle,
   |\phi_4\rangle \equiv |\uparrow\downarrow\rangle\}$.
(In this sloppy but extremely convenient notation, 
on the one hand, $|\phi_\alpha\rangle\equiv|\alpha\rangle$ is written for simplicity, usual in quantum information theory,
on the other hand, the $1,2,3,4$ index-values (useful for computers) 
are identified with the $-,\downarrow,\uparrow,\uparrow\downarrow$ labels of the states (carrying physical meaning).
Therefore we can write the same basis state in four different ways,
e.g., $|\phi_2\rangle\equiv|\phi_\downarrow\rangle\equiv|2\rangle\equiv|\downarrow\rangle$.)
The relevant orbital operators (\ref{eq:Cs.cs})-(\ref{eq:Cs.ph}) in this
basis are
\begin{subequations}
\label{eq:matricesmolorb}
\begin{align}
  \mathbf{c}^\dagger_{\uparrow} &=  \mathbf{a}^\dagger\otimes \mathbf{I} = 
\begin{pmatrix}
0  &   0 &    0  &   0\\
0  &   0 &    0  &   0\\
1  &   0 &    0  &   0\\
0  &   1 &    0  &   0
\end{pmatrix},& \qquad
  \mathbf{c}^\dagger_{\downarrow} &=  \mathbf{s}\otimes \mathbf{a}^\dagger =
\begin{pmatrix}
     0  &   0  &   0  &   0\\
     1  &   0  &   0  &   0\\
     0  &   0  &   0  &   0\\
     0  &   0  &  -1  &   0
\end{pmatrix},\\
  \mathbb{I} &= \mathbf{I} \otimes \mathbf{I} = 
\begin{pmatrix}
 1 & 0 & 0 & 0 \\
 0 & 1 & 0 & 0 \\
 0 & 0 & 1 & 0 \\
 0 & 0 & 0 & 1
\end{pmatrix},&\qquad
  \mathbf{z} &= \mathbf{s}\otimes \mathbf{s} =
\begin{pmatrix}
 1 & 0 & 0 & 0 \\
 0 & -1 & 0 & 0 \\
 0 & 0 & -1 & 0 \\
 0 & 0 & 0 & 1
\end{pmatrix}.
\end{align}
\end{subequations}

We can put together two $\mathbb{C}^4$ tensor spaces, i.e., forming a two-orbital system ($\bullet\bullet$), where
$\Lambda^{\{1,2\}}=\Lambda_1\otimes\Lambda_2$ with 
$\dim\Lambda^{\{1,2\}} = \dim\Lambda_1\dim\Lambda_2=q^2=16$. 
The basis of the $\bullet\bullet$ system is given as
 $|\phi^{\{1,2\}}_{\alpha_{\{1,2\}}}\rangle=|\phi^{\{1\}}_{\alpha_1}\rangle\otimes|\phi^{\{2\}}_{\alpha_2}\rangle$
where $\alpha_{\{1,2\}} = (\alpha_1-1)q+\alpha_2$.
\begin{table}[t]
\centering
\begin{tabular}{|c||cc||c||c||cc||cc|}
\hline
$\alpha_{\{1,2\}}$&$\alpha_1$&$\alpha_2$&
$N_{\alpha_{\{1,2\}}\uparrow}$&$N_{\alpha_{\{1,2\}}\downarrow}$&
$N_{\alpha_1\uparrow}$&$N_{\alpha_1\downarrow}$&
$N_{\alpha_2\uparrow}$&$N_{\alpha_2\downarrow}$\\
\hline
1 & $-$&$-$                    &0&0&0&0&0&0\\
2 & $-$&$\downarrow$           &0&1&0&0&0&1\\
3 & $-$&$\uparrow$             &1&0&0&0&1&0 \\
4 & $-$&$\uparrow\downarrow$   &1&1&0&0&1&1 \\
5 & $\downarrow$&$-$           &0&1&0&1&0&0 \\
6 & $\downarrow$&$\downarrow$  &0&2&0&1&0&1\\
\vdots&\vdots&\vdots           &\vdots&\vdots&\vdots&\vdots&\vdots&\vdots\\
16 & $\uparrow\downarrow$&$\uparrow\downarrow$ &2&2&1&1&1&1 \\
\hline
\end{tabular}
\caption{Basis states for a two-orbital system.
Index values of basis states are $\alpha_1,\alpha_2\in\{1,2,3,4\}$,
and we use the shorthand notation $|\phi_{\alpha_1}^{\{1\}}\rangle,|\phi_{\alpha_2}^{\{2\}}\rangle\in
\{|\phi_1\rangle \equiv |-\rangle, 
  |\phi_2\rangle \equiv |\downarrow\rangle, 
  |\phi_3\rangle \equiv |\uparrow\rangle,
  |\phi_4\rangle \equiv |\uparrow\downarrow\rangle\}$,
as usual.
For the two-site basis $\alpha_{\{1,2\}}=(\alpha_1-1)q+\alpha_2 \in\{1,2,3,4\ldots 16\}$,
and $|\phi_{\alpha_{\{1,2\}}}^{\{1,2\}}\rangle\in
\{|\phi_1^{\{1,2\}}\rangle \equiv |--\rangle, 
  |\phi_2^{\{1,2\}}\rangle \equiv |-\downarrow\rangle,
  |\phi_3^{\{1,2\}}\rangle \equiv |-\uparrow\rangle,
  |\phi_4^{\{1,2\}}\rangle \equiv |-\uparrow\downarrow\rangle,\dots,
  |\phi_{16}^{\{1,2\}}\rangle \equiv |\uparrow\downarrow\uparrow\downarrow\rangle\}$.
Particle numbers for different spins are also shown.
These are proper quantum numbers if the corresponding operators commute with the Hamiltonian.}
\end{table}
The relevant operators for the $\bullet\bullet$ system are formed as
\begin{equation}
\label{eq:olorb}
\mathbf{c}^\dagger_{1,\uparrow}   = \mathbf{c}^\dagger_\uparrow   \otimes \mathbb{I},\qquad
\mathbf{c}^\dagger_{2,\uparrow}   = \mathbf{z} \otimes   \mathbf{c}^\dagger_\uparrow,\qquad
\mathbf{c}^\dagger_{1,\downarrow} = \mathbf{c}^\dagger_\downarrow \otimes \mathbb{I},\qquad
\mathbf{c}^\dagger_{2,\downarrow} = \mathbf{z} \otimes \mathbf{c}^\dagger_\downarrow.
\end{equation}
A wavefunction (\ref{eq:iotaU}), (\ref{eq:tens}) can be expressed in a general form as
\begin{equation}
\label{eq:psi12}
|\Psi^{\{1,2\}}\rangle = \sum_{\alpha_1,\alpha_2}U^{\{1,2\}}(\alpha_1,\alpha_2)|\phi^{\{1\}}_{\alpha_1}\rangle\otimes|\phi^{\{2\}}_{\alpha_2}\rangle,
\end{equation}
where the matrix $U^{\{1,2\}}(\alpha_1,\alpha_2)$ describes the quantum mechanical probability distrubution 
of the basis of the combined system. 
Such wavefunctions can arise from
the diagonalization of the Hamiltonian $\mathbf{H}$, which is a $q^2$ by $q^2$ matrix (\ref{eq:HamiltonHd}),
using the above representation of the creation and annihilation operators.  
The full diagonalization of $\mathbf{H}$ gives the exact solution (full-CI),
and the $m^{\rm{th}}$ eigenstate of a two-orbital Hamiltonian is
\begin{equation}
\label{eq:psi12m}
|\Psi^{\{1,2\}}_m\rangle = \sum_{\alpha_1,\alpha_2}U^{\{1,2\}}(\alpha_1,\alpha_2,m)|\phi^{\{1\}}_{\alpha_1}\rangle\otimes|\phi^{\{2\}}_{\alpha_2}\rangle,
\end{equation}
where $\alpha_1,\alpha_2=1,\ldots,q$ and $m=1,\ldots,q^2$.

\subsubsection{Change of basis, truncation and iterative diagonalization} 
\label{sec:num.basic.bchangtrunc}
The representation of the problem is, however, not unique.
Using a unitary operator acting on $\Lambda^{\{1,2\}}$, $\mathbf{O}$, which leaves the eigenvalue spectrum of the Hamiltonian unchanged, 
we can carry out a \emph{change of basis} 
\begin{equation}
\label{eq:transform}
\mathbf{H} \qquad \longmapsto \qquad \mathbf{O} \mathbf{H} \mathbf{O}^\dagger.
\end{equation}
One possibility to achive this is to apply the unitary operator to all operators used
to construct the Hamiltonian \ref{eq:HamiltonHd}, i.e., 
$\mathbf{c}^\dagger_{1,\downarrow} \mapsto \mathbf{O} \mathbf{c}^\dagger_{1,\downarrow} \mathbf{O}^\dagger$, 
$\mathbf{c}^\dagger_{2,\downarrow} \mapsto \mathbf{O} \mathbf{c}^\dagger_{2,\downarrow} \mathbf{O}^\dagger$, 
$\mathbb{I} \mapsto \mathbf{O} \mathbb{I} \mathbf{O}^\dagger = \mathbb{I}$,
$\mathbf{c}^\dagger_{1,\uparrow} \mathbf{c}^{\phantom\dagger}_{2,\uparrow} \mapsto
\mathbf{O} \mathbf{c}^\dagger_{1,\uparrow} \mathbf{c}^{\phantom\dagger}_{2,\uparrow} \mathbf{O}^\dagger$, etc.
If the rows of the matrix $\mathbf{O}$ is constructed from the $|\Psi^{\{1,2\}}_m\rangle$ eigenstates (\ref{eq:psi12m}) then we arrive at the
eigenbasis representation of $\mathbf{H}$, i.e., $\mathbf{H}$ becomes diagonal and its elements are equivalent to the eigenvalues
of the original problem.

The eigenvalue spectrum of $\mathbf{H}$ determines the physical properties of the system exactly. 
It is, however, 
possible to use an approximate representation of $\mathbf{H}$, 
i.e., using a smaller basis as we select only $M<q^2$ eigenstates 
to form the $\mathbf{O}$ matrix, which becomes then rectangular.
That is, we change over to a subspace $\Xi^{\{1,2\}}$ of the original tensor space $\Lambda^{\{1,2\}}$,
(see section \ref{sec:tensor}).
This \emph{truncation}
leads to loss of information as $\mathbf{O}\mathbf{O}^\dagger\neq\mathbb{I}$, but the kept eigenstates can still provide a 
good description of the low-energy physics of the problem.  

If we are interested in the low-lying eigenstates of $\mathbf{H}$ it is not
necessary to carry out a full diagonalization, but  
systematic 
application of the Hamiltonian to a randomly chosen state provides the lowest lying eigenstate. 
An extension of such power methods, like the \textit{L\'anczos}\cite{Lanczos-1950} or \textit{Davidson}\cite{Davidson-1975} methods,
provides faster convergence rates, and excited states can also be calculated\cite{Schollwock-2005,Noack-2005,Hallberg-2006}.

\subsubsection{Unitary transformation for two molecular orbitals}
\label{sec:num.basic.toMPS}
For the sake of simplicity let us consider an example of two $S=1/2$-spins.
That is, the basis of the $\bullet\bullet$ system is formed from the 
$|\phi_1\rangle\equiv|\downarrow\rangle$, $|\phi_2\rangle\equiv|\uparrow\rangle$
vectors with $q=2$, and the eigenvectors (\ref{eq:psi12m}) can be formed as
\begin{equation}
\label{eq:psi12mO}
|\Psi_m^{\{1,2\}}\rangle = \sum_{\alpha_1,\alpha_2} O(m, 2(\alpha_1-1)+\alpha_2)|\phi^{\{1\}}_{\alpha_1}\rangle\otimes{ |\phi^{\{2\}}_{\alpha_2}\rangle},
\end{equation}
where $\alpha_1,\alpha_2\in\{1,2\}\equiv\{\downarrow,\uparrow\}$.
An example for the $\mathbf{O}$ matrix is shown in Table~\ref{tab:Omatrix}.
\begin{table}[t]
\centering
\begin{tabular}{c|cccc||cc}
$O$ & $\downarrow\downarrow$ & $\downarrow\uparrow$ & $\uparrow\downarrow$ & $\uparrow\uparrow$ & $S^z$ & S \\
\hline
$\Psi_1^{\{1,2\}}$ & 1 & 0 & 0 & 0 & -1 & 1\\
$\Psi_2^{\{1,2\}}$ & 0 & $1/\sqrt 2$ & $1/\sqrt 2$ & 0 & 0 & 1\\
$\Psi_3^{\{1,2\}}$ & 0 & $1/\sqrt 2$ & $-1/\sqrt 2$ & 0 & 0 & 0\\
$\Psi_4^{\{1,2\}}$ & 0 & 0 & 0 & 1 & 1 & 1
\end{tabular}
\caption{
An example for the unitary matrix $\mathbf{O}$, used to transfrom the Hamilton to an $S^z$ eigenbasis.
This transformation arises when the Hamiltonian commutes with the operators of
the z-component and the magnitude of the total spin.
Then the eigenvalues of these operators, $S^z$ and $S$ respectively,
are proper quantum numbers, and are listed in the last two columns of the table.} 
\label{tab:Omatrix}
\end{table}
The dimension of the O matrix is $M\times q^2$ where $M$ can take values between 1 and $q^2$ (truncation).
The $Mq^2$ elements of the matrix can also be represented by $q$ (two) $M\times q$ matrices, 
denoted with $\mathbf{B}_2(\alpha_2)$, 
i.e., for each basis of the second spin we assign a matrix. This means that we take 
columns 1 and 3 to form $\mathbf{B}_2(\downarrow)$ and
columns 2 and 4 for $\mathbf{B}_2(\uparrow)$, so, 
for the example given in Table \ref{tab:Omatrix} we have (without truncation)
\begin{equation}
\label{eq:matricesBB}
(\mathbf{B}_2(\downarrow))_{m,\alpha_1} =  
\begin{pmatrix}
  1 & 0 \\
  0 & 1/\sqrt 2 \\
  0 & -1/\sqrt 2 \\
  0 & 0 
\end{pmatrix},\qquad
(\mathbf{B}_2(\uparrow))_{m,\alpha_1} = 
\begin{pmatrix}
  0 & 0\\
  1/\sqrt 2 & 0 \\
  1/\sqrt 2 & 0 \\
  0 & 1
\end{pmatrix}.
\end{equation}
We also denote this by $(\mathbf{B}_2(\alpha_2))_{m,\alpha_1}=B_2(m,\alpha_2,\alpha_1)$.
It is easy to recognize that such $\mathbf{B}$ matrices form the basis of the 
matrix product state representation
discussed in Sec.\ref{sec:tensor}.  
In the literature, usually $\mathbf{A}_2\equiv \mathbf{B}_2^T$ is used, that is, $A_2(\alpha_1,\alpha_2,m)=B_2(m,\alpha_2,\alpha_1)$, and the wavefunction is written as 
\begin{equation}
\label{eq:psi12mA}
|\Psi_m^{\{1,2\}}\rangle = \sum_{\alpha_1,\alpha_2} A_2(\alpha_1,\alpha_2,m) 
|\phi^{\{1\}}_{\alpha_1}\rangle\otimes{ |\phi^{\{2\}}_{\alpha_2}\rangle}.
\end{equation}

\subsubsection{Symmetries} 
\label{sec:num.basic.symm}
In many systems the time evolution governed by the Hamilton operator does not change the value of a 
measurable quantity, i.e., the Hamilton operator commutes with the operator associated 
to that measurable quantity. These operators are called symmetry operators 
and can be used to cast the Hilbert space to smaller independent subspaces.
Consequently, instead of solving a large matrix eigenvalue problem, the 
eigenvalue spectrum can be determined by solving several smaller problems.
Thus, the distinct quantum numbers helps to partition the Hilbert space into 
multiple independent subspaces corresponding to a given combination 
of quantum number values.

A given symmetry operator has the same eigenvectors as the Hamiltonian, 
thus the eigenstates of the Hamiltonian can be labelled by the eigenvalues 
of the symmetry operator (\textit{quantum number} $Q$), and the Hilbert 
space can be decomposed into subspaces (\textit{sectors}) spanned by the 
eigenvectors of each quantum number value\cite{Cornwell-1997}. Introducing 
a quantum number based representation, the sparse operators 
can be decomposed to a set of smaller 
but dense matrices, furthermore the Hamiltonian operator 
becomes blockdiagonal.

For two orbitals, quantum numbers are formed from orbital 
quantum numbers as $Q_{\{\alpha_i,\alpha_j\}} = f(Q_{\alpha_i}, Q_{\alpha_j})$,
where function $f$ depends on the given symmetry. For $U(1)$ 
symmetries the $f(Q_{\alpha_i}, Q_{\alpha_j}) = Q_{\alpha_i} + Q_{\alpha_j}$ 
while for non-Abelian symmeties, such as for the conservation of total spin,
more complex algebra is involved, based on the Wigner-Eckart theorem\cite{Weyl-1927,Wigner-1939,Wigner-1931,Wigner-1959,Toth-2008}. 
For more details, see section \ref{sec:num.optim.opt-symm}.

\subsubsection{Unitary transformation for $d$ number of molecular orbitals and tensor product approximation}
\label{sec:num.basic.dorbs}
The formalism discussed above can be extended to describe a system with $d$ molecular orbitals 
denoted as $\bullet\bullet\bullet\ldots\bullet$. 
The Hilbert space is formed as 
$\Lambda^{\{1,2,\ldots,d\}}=\otimes_{i=1}^d\Lambda_i$ with
${\rm dim}\Lambda^{\{1,2,\ldots,d\}}=\prod_{i=1}^d {\rm dim}\Lambda_i=q^d$.
A wavefunction is written as
\begin{equation}
\label{eq:psid}
|\Psi^{\{1,2,\ldots,d\}}\rangle = \sum_{\alpha_1\ldots \alpha_d}U^{\{1,2,\ldots,d\}}(\alpha_1,\alpha_2,\ldots \alpha_d)
|\phi^{\{1\}}_{\alpha_1}\rangle\otimes|\phi^{\{2\}}_{\alpha_2}\rangle\otimes\ldots\otimes|\phi^{\{d\}}_{\alpha_d}\rangle,
\end{equation} 
where
 $U^{\{1,2,\ldots,d\}}$ is a tensor of order $d$.
\begin{figure}[t]
\centering
\includegraphics{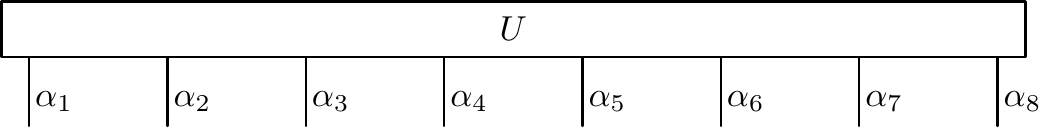}
\caption{Example $d=8$.}
\label{fig:full_tensor}
\end{figure}
Since the dimension of $U$ scales exponentially with $d$ we need approximative methods.
The major aim is to find a good approximation of $U$ in terms of products of lower order tensors with
smaller rank than the original problem.

\begin{figure}[t]
\centering
\setlength{\unitlength}{40pt}   
\begin{picture}(7.5,3)
\put(0,1.7){\includegraphics{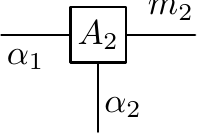}}
\put(3,1.7){\includegraphics{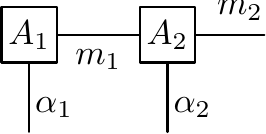}}
\put(6,1.7){\includegraphics{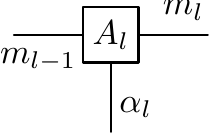}}
\put(0,0){\includegraphics{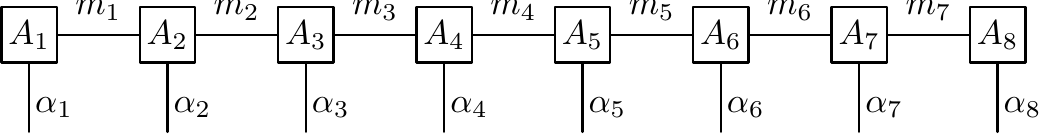}}
\put(0,3){\makebox(0,0)[r]{\strut{}(a)}}
\put(3,3){\makebox(0,0)[r]{\strut{}(b)}}
\put(6,3){\makebox(0,0)[r]{\strut{}(c)}}
\put(0,1.3){\makebox(0,0)[r]{\strut{}(d)}}
\end{picture}
\caption{(a) Graphical representation of the two-orbital composed system using 
the procedure outlined in Sec.~\ref{sec:num.basic.bchangtrunc}, 
(b) and after using idenity (\ref{eq:endidentity}). 
(c) Graphical representation of the component tensor and 
(d) the $d=8$-orbital wavefuntion as a network built from matrices.}
\label{fig:mps-components}
\end{figure}
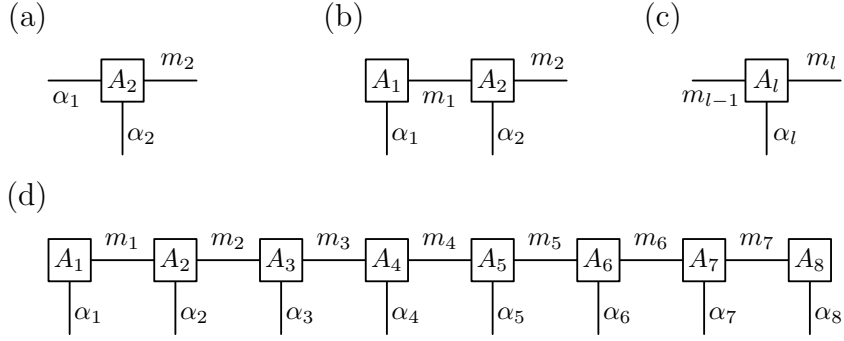

One possibility is to systematically apply the procedure outlined 
in Sec.~\ref{sec:num.basic.bchangtrunc} and
in Sec.~\ref{sec:tensor} to 
describe one, two, three, \dots $d$-orbital wavefunction.
Starting with two orbitals, $\bullet\bullet$, 
the new (truncated) basis of the composed system is written as
\begin{subequations}
\begin{equation} 
\label{eq:makingTT1}
|\xi^{\{1,2\}}_{m_2}\rangle = \sum_{\alpha_{1},\alpha_2}A_2(\alpha_{1},\alpha_2,m_2)
|\phi^{\{1\}}_{\alpha_1}\rangle\otimes{ |\phi^{\{2\}}_{\alpha_2}\rangle}, 
\end{equation}
which is shown schematically in Fig.~\ref{fig:mps-components}(a).
This can be rewritten as
\begin{equation} 
\label{eq:makingTT2}
|\xi^{\{1,2\}}_{m_2}\rangle = \sum_{m_{1},\alpha_2} A_2 (m_1,\alpha_1,m_2)
|\xi^{\{1\}}_{m_1}\rangle\otimes{ |\phi^{\{2\}}_{\alpha_2}\rangle}, 
\end{equation}
using the identity
\begin{equation}
\label{eq:endidentity}
|\xi^{\{1\}}_{m_1}\rangle = \sum_{\alpha_1}A_1(1,\alpha_1,m_1)|\phi^{\{1\}}_{\alpha_1}\rangle\quad
\text{with}\quad A_1(1,\alpha_1,m_1)=\delta_{\alpha_1,m_1},
\end{equation}
as is depicted in Fig.~\ref{fig:mps-components}(b).
We have the above form for $A_1(1,\alpha_1,m_1)$
since here the transformation and truncation comes from
the subspace approximation in $\Lambda^{\{1,2\}}$ 
common in a wide part of renormalization group methods in physics,
shown in sections \ref{sec:num.basic.bchangtrunc} and \ref{sec:num.basic.toMPS}.
On the other hand, 
when the transformations and truncation come from successive subspace optimization
starting with the space $\Lambda_1$, e.g., based on SVD,
then we have nontrivial basis change even inside $\Lambda_1$, see section \ref{sec:tensor.tt}.
For three orbitals $\bullet\bullet\bullet$,
\begin{equation}
\label{eq:makingTT3}
|\xi^{\{1,2,3\}}_{m_3}\rangle = \sum_{m_{2},\alpha_3}A_3(m_2,\alpha_3,m_3)
|\xi^{\{1,2\}}_{m_2}\rangle\otimes|\phi^{\{3\}}_{\alpha_3}\rangle. 
\end{equation}
This procedure can be extended iteratively 
using series of component tensors
\begin{equation}
\label{eq:makingTT5}
|\xi^{\{1,2,\dots,l\}}_{m_l}\rangle = \sum_{\alpha_1,\ldots \alpha_l}(\mathbf{A}_1(\alpha_1)\mathbf{A}_2(\alpha_2)\ldots \mathbf{A}_l(\alpha_l))_{1,m_l} 
|\phi^{\{1\}}_{\alpha_1}\rangle\otimes|\phi^{\{2\}}_{\alpha_2}\rangle\otimes\ldots\otimes|\phi^{\{l\}}_{\alpha_l}\rangle,
\end{equation}
where the component tensor $A_l(m_{l-1},\alpha_l,m_l)=(\mathbf{A}_l(\alpha_l))_{m_{l-1},m_l}$ is defined as 
\begin{equation} 
\label{eq:makingTT6}
|\xi^{\{1,2,\dots,l\}}_{m_l}\rangle = \sum_{m_{l-1},\alpha_l} (\mathbf{A}_l(\alpha_l))_{m_{l-1},m_l}
|\xi^{\{1,2,\dots,l-1\}}_{m_{l-1}}\rangle\otimes{ |\phi^{\{l\}}_{\alpha_l}\rangle},
\end{equation}
see in Fig.~\ref{fig:mps-components}(c).
As a result of this procedure, the $d$-orbital wavefunction is expressed as
\begin{equation}
\label{eq:makingTT7}
|\Psi\rangle = \sum_{\alpha_1,\alpha_2,\dots,\alpha_d} \mathbf{A}_1(\alpha_1) \mathbf{A}_2(\alpha_2) \cdots \mathbf{A}_d(\alpha_d)\;
|\phi^{\{1\}}_{\alpha_1}\rangle\otimes|\phi^{\{2\}}_{\alpha_2}\rangle\otimes\ldots\otimes|\phi^{\{d\}}_{\alpha_d}\rangle,
\end{equation}
\end{subequations}
i.e., 
for each molecular orbital we can assign a matrix $\mathbf{A}_l(\alpha_l)$,
coming from the basis change in $\Lambda^{(l-1)}\otimes\Lambda_{l}$,
and we form a network built from matrices
as shown in Fig.~\ref{fig:mps-components}(d).
For more detailed derivations we refer to the original papers and 
review articles\cite{Ostlund-1995,Verstraete-2004a,Vidal-2003b,Evenbly-2011,Hackbusch-2010,Schollwock-2011} and
Sec.~\ref{sec:tensor}.

Succesively repeating the construction of section \ref{sec:num.basic.symm},
the quantum numbers for the $q^d$ states of the $d$-orbital systems can be determined.
As before, the full Hilbert space is decomposed into sectors based on these quantum numbers.
If we consider only the case where 
the number of electrons with down and up spins is conserved,
the quantum number is the vector $Q=(N_\downarrow,N_\uparrow)$ with $N=N_\downarrow+N_\uparrow$, 
then the dimension of the related sector $\Lambda_{\text{FCI}}\subset\Lambda^{(d)}$ in the Hilbert space is
$\dim\Lambda_{\text{FCI}}=\binom{d}{N_\downarrow}\binom{d}{N_\uparrow}$.

\subsubsection{Tensor topology}
\label{sec:num.basic.topol}
If we render the tensor spaces corresponding to the orbitals 
in a ``one- or two-dimensional space'' (higher dimensional extension is also possible) 
we form a \emph{chain-} or \emph{lattice-topology} of the tensor product representation. 
In some cases this topology is also reflected by the physical lattice topology of the problem,
i.e., one-dimensional-like polimers can be studied very well using the one-dimensional tensor-topology.
As will be discussed below, 
one of the major aim is to find the best tensor topology for a given molecule.

\subsection{Entanglement and correlations}
\label{sec:num.ent}

In the previous subsection,
we have considered basis change based on the Hamiltonian of the system.
Another approach of basis change is based on an actual pure state of the system,
and connected to the \emph{entanglement} of that state\cite{Horodecki-2009,Szalay-2013}. 

In quantum systems, correlations having no counterpart in classical physics arise.
Pure states showing these strange kinds of correlations are called entangled\cite{Horodecki-2009,Szalay-2013},
and the existence of these states has so deep and important consequences\cite{Einstein-1935,Bell-1967,Clauser-1969}
that Schr{\"o}dinger has identified entanglement to be the characteristic trait of quantum mechanics\cite{Schrodinger-1935a,Schrodinger-1935b}.
The QC-DMRG and QC-TTNS algorithms approximate a composite 
system with long-range interactions, 
and it turned out that the results of quantum information theory\cite{Nielsen-2000,Wilde-2013}
can be used to understand the criteria of its convergence.

\subsubsection{Singular value decomposition and entanglement}
\label{sec:num.ent.toSVD}

The basic concept on which entanglement theory is built up
is the entanglement with respect to a \emph{bipartition} of the system.
In this manybody situation,
the system composed of $d$ orbitals can be treated as the sum of two subsystems (also called blocks), 
$(\mathrm{A}),(\mathrm{B})\subset\{1,2,\dots,d\}$.
(They are disjoint, and their union gives the whole system.)
The Hilbert spaces associated to them are $\Lambda^{(\mathrm{A})}$ and $\Lambda^{(\mathrm{B})}$,
so $\Lambda^{\{1,2,\dots,d\}}\cong \Lambda^{(\mathrm{A})}\otimes\Lambda^{(\mathrm{B})}$.
After choosing bases in the subsystems, 
$\{|\phi^{(\mathrm{A})}_{\alpha_{(\mathrm{A})}}\rangle\in\Lambda^{(\mathrm{A})}\}$ 
and $\{|\phi^{(\mathrm{B})}_{\alpha_{(\mathrm{B})}}\rangle\in\Lambda^{(\mathrm{B})}\}$,
the wavefunction (\ref{eq:psid}) characterizing the \emph{pure state} of the system can be written as
\begin{equation}
\label{eq:psiAB}
|\Psi\rangle =\sum_{\alpha_{(\mathrm{A})}=1}^{\dim\Lambda^{(\mathrm{A})}}\sum_{\alpha_{(\mathrm{B})}=1}^{\dim\Lambda^{(\mathrm{B})}}
U(\alpha_{(\mathrm{A})},\alpha_{(\mathrm{B})})
|\phi^{(\mathrm{A})}_{\alpha_{(\mathrm{A})}}\rangle\otimes|\phi^{(\mathrm{B})}_{\alpha_{(\mathrm{B})}}\rangle.
\end{equation}
Based on the UDV-decomposition of the matrix $U(\alpha_{(\mathrm{A})},\alpha_{(\mathrm{B})})$,
one can find a product unitary transformation
$\mathbf{O}_A\otimes \mathbf{O}_B$, which brings it to the \emph{Schmidt form}\cite{Schmidt-1907}
\begin{equation}
|\Psi \rangle = \sum_{m=1}^{r_{\rm Sch}} \sqrt{\omega_m} |\xi^{(\mathrm{A})}_m\rangle \otimes |\xi^{(\mathrm{B})}_m\rangle.
\label{eq:schmidt}
\end{equation}
Here, the vectors $|\xi^{(\mathrm{A})}_{m_{(\mathrm{A})}} \rangle$ and $|\xi^{(\mathrm{B})}_{m_{(\mathrm{B})}}\rangle$ form 
orthonormal bases, also called \emph{Schmidt bases}, in the Hilbert spaces of the two blocks,
$\langle \xi^{(\mathrm{A})}_m|\xi^{(\mathrm{A})}_{m'} \rangle = \langle \xi^{(\mathrm{B})}_m | \xi^{(\mathrm{B})}_{m'} \rangle = \delta_{m,m'}$,
moreover, the squares of the \emph{Schmidt coefficients} $\sqrt{\omega_m}$ satisfy $0\leq \omega_m \leq 1$ with the constraint $\sum_m \omega_m = 1$.
The summation goes until the \emph{Schmidt rank}, $r_{\rm Sch}\leq\min(\dim\Lambda^{(\mathrm{A})},\Lambda^{(\mathrm{B})})$.
The $\sqrt{\omega_m}$ numbers are also called the singular values of 
$U(\alpha_{(\mathrm{A})},\alpha_{(\mathrm{B})})$,
and the above form \emph{singlar value decomposition} (SVD).
If the \emph{Schmidt rank} $r_{\rm Sch}>1$,
then $|\Psi\rangle$ is entangled (inseparable) 
and we say that the two blocks are entangled\cite{Horodecki-2009}.

If we consider the two-electron subspace of a two-orbital system,
then the state
\begin{subequations}
\label{eq:psixmpl}
\begin{equation}
\label{eq:psient}
|\Psi_{\text{ent}}\rangle = \frac1{\sqrt2}\bigl(|\phi^{\{1\}}_1\rangle\otimes|\phi^{\{2\}}_2\rangle - |\phi^{\{1\}}_2\rangle\otimes|\phi^{\{2\}}_1\rangle \bigr)
\equiv \frac1{\sqrt2}\bigl( |\downarrow\rangle\otimes|\uparrow\rangle - |\uparrow\rangle \otimes|\downarrow\rangle \bigr)
\end{equation}
is an entangled state,
while
\begin{equation}
\label{eq:psisep}
|\Psi_{\text{sep}}\rangle = |\phi^{\{1\}}_1\rangle\otimes|\phi^{\{2\}}_2\rangle 
\equiv  |\downarrow\rangle\otimes|\uparrow\rangle 
\end{equation}
\end{subequations}
is separable.
Both vectors are almost in Schmidt form, 
(unitary transformation $\mathbf{O}_2$ acting as 
$|\phi^{\{2\}}_2\rangle\mapsto |\phi^{\{2\}}_1\rangle$ and
$|\phi^{\{2\}}_1\rangle\mapsto-|\phi^{\{2\}}_2\rangle$ brings the first one to a Schmidt form)
and the squared Schmidt coefficients can immediately be read: $\omega_1=\omega_2=1/2$ in the first case
and $\omega_1=1$, $\omega_2=0$ in the second.

For a system characterized by a pure state $|\Psi\rangle$, 
the \emph{state of the subsystem} $(\mathrm{A})$
is encoded in the \emph{reduced density matrix} of the subsystem,
\begin{equation}
\label{eq:rhoA}
  \rho^{(\mathrm{A})} = \mbox{Tr}_{\mathrm{B}} \, |\Psi\rangle\langle\Psi|.
\end{equation}
The subsystem of interest is usually labelled by $(\mathrm{A})$ and the other subsystem $(\mathrm{B})$,
which can also be considered as the ``environment'' of $(\mathrm{A})$.
The operation $\mbox{Tr}_{\mathrm{B}}$ means carrying out the trace over subsystem $(\mathrm{B})$,
that is, $\mbox{Tr}_{\mathrm{B}}(\mathbf{X}\otimes\mathbf{Y}) = \mathbf{X} \mbox{Tr}(\mathbf{Y})$,
leading to the form
\begin{equation}
\label{eq:rhoAcomp}
  \rho^{(\mathrm{A})} = \sum_{\alpha_{(\mathrm{A})},\alpha'_{(\mathrm{A})}}
\biggl[\sum_{\alpha_{(\mathrm{B})}}
U(\alpha_{(\mathrm{A})},\alpha_{(\mathrm{B})})
\overline{U(\alpha'_{(\mathrm{A})},\alpha_{(\mathrm{B})})}\biggr]
|\phi^{(\mathrm{A})}_{\alpha_{(\mathrm{A})}}\rangle \langle\phi^{(\mathrm{A})}_{\alpha'_{(\mathrm{A})}}|,
\end{equation}
having the matrix elements in the square bracket
\begin{equation}
\label{eq:rhoAcompaap}
  \rho^{(\mathrm{A})}(\alpha_{(\mathrm{A})},\alpha'_{(\mathrm{A})}) 
\equiv \langle\phi^{(\mathrm{A})}_{\alpha_{(\mathrm{A})}}|\rho^{(\mathrm{A})}|\phi^{(\mathrm{A})}_{\alpha'_{(\mathrm{A})}}\rangle
= \sum_{\alpha_{(\mathrm{B})}}
U(\alpha_{(\mathrm{A})},\alpha_{(\mathrm{B})})
\overline{U(\alpha'_{(\mathrm{A})},\alpha_{(\mathrm{B})})}
\end{equation}
(Similar expressions can be written for subsystem $(\mathrm{B})$.)
If we write (\ref{eq:rhoA}) using the Schmidt form (\ref{eq:schmidt}), 
we get immediately a diagonal form
\begin{equation}
\label{eq:rhoAcompSch}
  \rho^{(\mathrm{A})} = \sum_m
\omega_m
|\xi^{(\mathrm{A})}_m\rangle \langle\xi^{(\mathrm{A})}_m|
\end{equation}
in the Schmidt basis.

On the other hand, in this pure case, 
the information on the \emph{entanglement} between the $(\mathrm{A})$ and $(\mathrm{B})$ blocks of the system
is encoded in the density matrices of the blocks.
It turns out that the eigenvalue spectrum of $\rho^{(\mathrm{A})}$ is enough for the complete characterization
of the entanglement between blocks $(\mathrm{A})$ and $(\mathrm{B})$,
and, as we have seen in (\ref{eq:rhoAcompSch}), it follows from the Schmidt decomposition that
the eigenvalues of $\rho^{(\mathrm{A})}$ are exactly the squared Schmidt coefficients $\omega_m$ in (\ref{eq:schmidt}).
(The same holds for $\rho^{(\mathrm{B})}$.)
Several quantitative measures of entanglement can be extracted
from this eigenvalue spectrum\cite{Vidal-2000,Horodecki-2001}.
These are usually the different kinds of \emph{entropies} of the reduced density matrix,
characterizing its mixedness.
The most commonly used measure is the \emph{von Neumann entropy}\cite{Ohya-1993,Petz-2008}
\begin{equation}
  S^{(\mathrm{A})} \equiv S(\rho^{(\mathrm{A})})= - \mbox{Tr} \rho^{(\mathrm{A})} \ln \rho^{(\mathrm{A})},
\label{eq:vonN-entropy}
\end{equation}
others include the more general one-parameter family of \emph{R\'enyi entropies}\cite{Horodecki-1996,Vidal-2000,Horodecki-2001} for parameter lower than $1$, 
the \emph{Hartley entropy} $\ln r_\text{Sch}$
(which can be considered as the R\'enyi entropy in the limit when its parameter tends to $0$),
the \emph{Schmidt rank} $r_\text{Sch}$ itself,
the one-parameter family of \emph{Tsallis entropies}\cite{Furuichi-2007},
the \emph{concurrence-squared}, or \emph{linear entropy}
(the latter two are, up to normalization, the Tsallis entropy for parameter $2$).
On the other hand, the von Neumann entropy is the  R\'enyi or Tsallis entropy in the limit when their parameters tend to $1$.

The definitive property, based on which the entropies are proper measures of entanglement,
is the \emph{monotonity under LOCC:}
entanglement is \emph{quantum} correlation, so
any measure of entanglement must not increase under applying \emph{Local Operations} (that is, inside subsystems)
and using \emph{Classical Communication} between subsystems\cite{Bennett-2000,Vidal-2000,Horodecki-2001,Chitambar-2014,Szalay-2013}.
Here we have to give an important remark.
This locality concept is understood with respect to the notion of \emph{subsystems.}
The subsystems have very different meanings in the first- and second-quantized description of quantum systems.
In the first quantized case, the subsystems are the electrons (they can occupy different orbitals), 
their entanglement (\emph{particle-entanglement}) can not increase if we apply LOCC for them, 
for example, if we change the local basis $\varphi_i$ in $\mathcal{V}^d$ from which the Slater determinants are built up
(see section \ref{sec:qchem.FullCIRG}),
especially, changing from atomic orbitals to molecular orbitals or reverse.
In the second quantized case, the subsystems are the orbitals or sites (they can be occupied by electrons),
their entanglement (\emph{orbital-entanglement} or \emph{site-entanglement}) can not increase if we apply LOCC for them
(see section \ref{sec:qchem.2quant}),
for example, if we change the local basis $|\phi_{\mu_i}^{\{i\}}\rangle$ in $\Lambda_i$ for a local subspace approximation.
However, since the isomorphism $\iota$ in Eq.~(\ref{eq:iota}) is \emph{nonlocal},
i.e., it does not respect the tensor product structure either in $\mathcal{F}^d$ or in $\Lambda^{(d)}$,
a basis change in $\mathcal{V}^d$
altough does not change the particle-entanglement but does change the orbital-entanglement.
(C.f.~Eq.~(\ref{eq:optraf}).)
This will be utilized in section \ref{sec:num.optim.opt-basis} for reducing the overall orbital-entanglement
by changing locally the particle basis.

Once the eigenvalues $\omega^{(\mathrm{A})}_m$ of $\rho^{(\mathrm{A})}$ are known,
the von Neumann entropy (\ref{eq:vonN-entropy}) can be calculated, leading to
\begin{equation}
S^{(\mathrm{A})}=-\sum_m \omega^{(\mathrm{A})}_m\ln \omega^{(\mathrm{A})}_m.
\label{eq:vonN-entropy-omega}
\end{equation}
In the examples (\ref{eq:psixmpl}) above, one can conclude that
the entanglement measured by the von Neumann entropy
for $|\Psi_{\text{ent}}\rangle$ is $S^{(\mathrm{A})}(\Psi_{\text{ent}})=\ln2$,
while for $|\Psi_{\text{sep}}\rangle$ is $S^{(\mathrm{A})}(\Psi_{\text{sep}})=0$.
It turns out also that
$|\Psi_{\text{ent}}\rangle$ is maximally entangled
in the two-electron subspace of a two-orbital system.
(The base of the logarithm in the above expressions are often set to $2$,
in which case the von Neumann entropy is measured in the units called \emph{qubit},
the quanum analogy of the \emph{bit} in classical information theory.)

In Eq.\ (\ref{eq:vonN-entropy}), subsystem $(\mathrm{A})$ can be formed, in general,
from an arbitrary subset of the total set of orbitals.
If it is only one orbital, $(\mathrm{A})=\{i\}$, then its entropy is called \emph{orbital entropy}, $S_i$.
The number of orbitals included in $(\mathrm{A})$ 
can be tailored to obtain specific information on the
distribution of entanglement,
which can then be used to
characterize the physical nature of the
system.

\subsubsection{Block entropy}
\label{sec:num.ent.blockentr}

\begin{figure}[t]
\centering
\setlength{\unitlength}{40pt}   
\begin{picture}(11,3)
\put(0,0.5){\includegraphics{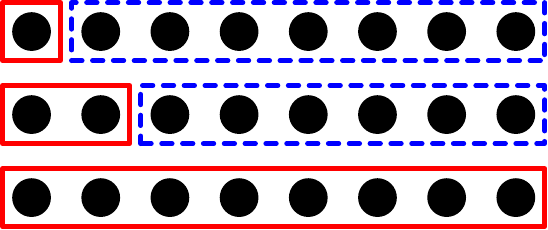}}
\put(4.4,0){\includegraphics[scale=0.66]{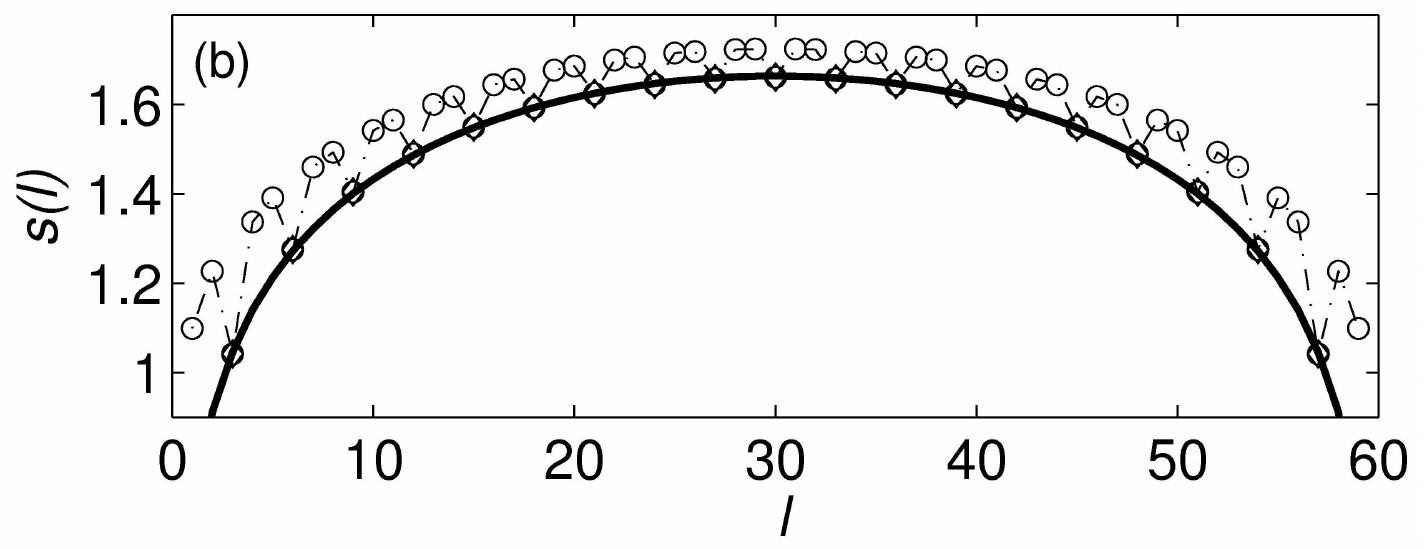}}
\put(0,2.8){\makebox(0,0)[r]{\strut{}(a)}}
\put(4.6,2.8){\makebox(0,0)[r]{\strut{}(b)}}
\end{picture}
\caption{(a) Contiguous block of orbitals to determine block entropy.
(b) Block entropy profile $S^{\{1,2,\dots,l\}}$ obtained with the DMRG method 
for a one-dimensional critical model with soft modes at $k=\pm 2\pi/3$. 
}
\label{fig:sl}
\end{figure}

\begin{figure}[t]
\centering
\setlength{\unitlength}{40pt}   
\begin{picture}(10,3.5)
\put(  0,  0){\includegraphics[scale=0.4]{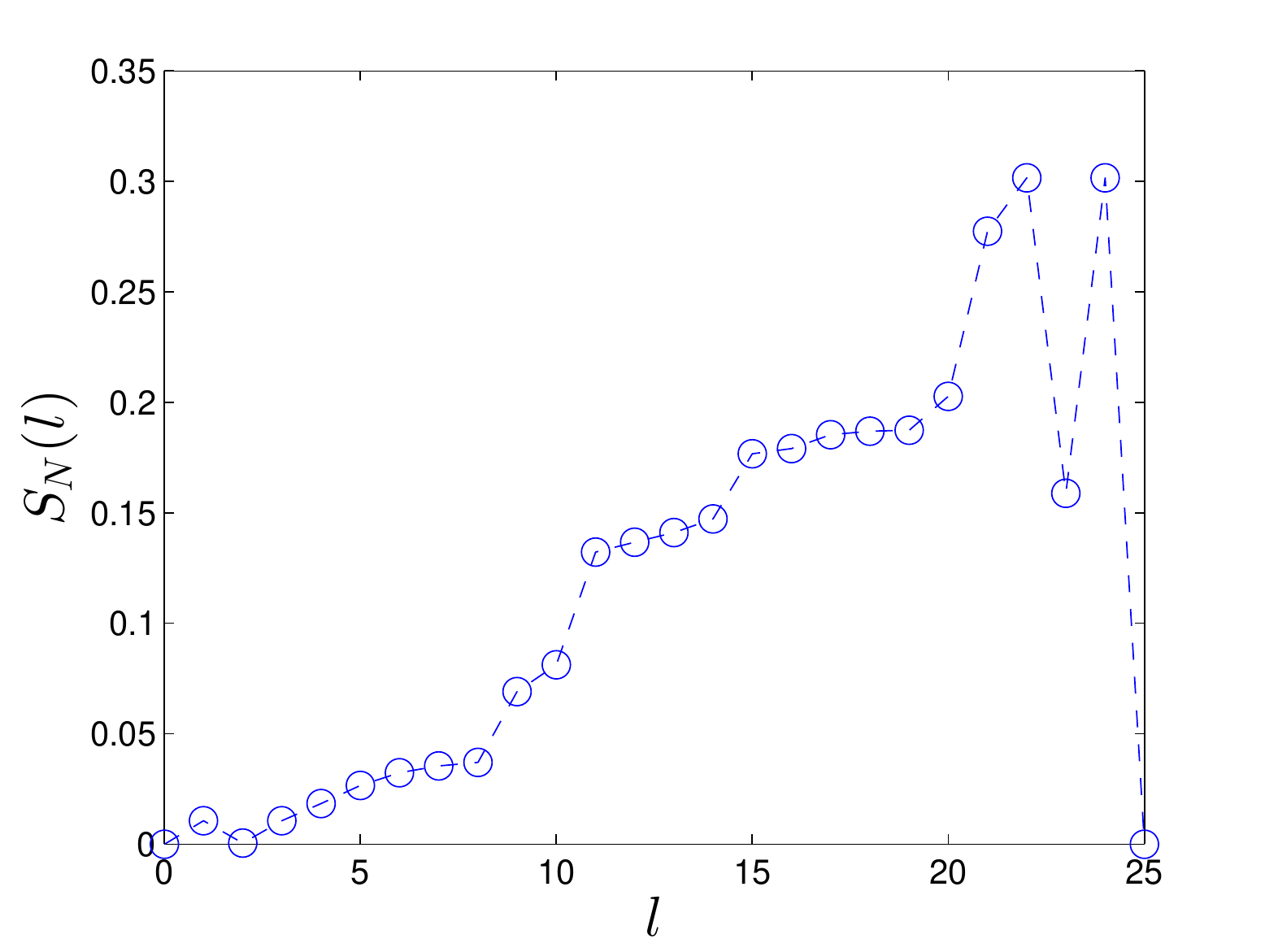}}
\put(5.5,0  ){\includegraphics[scale=0.4]{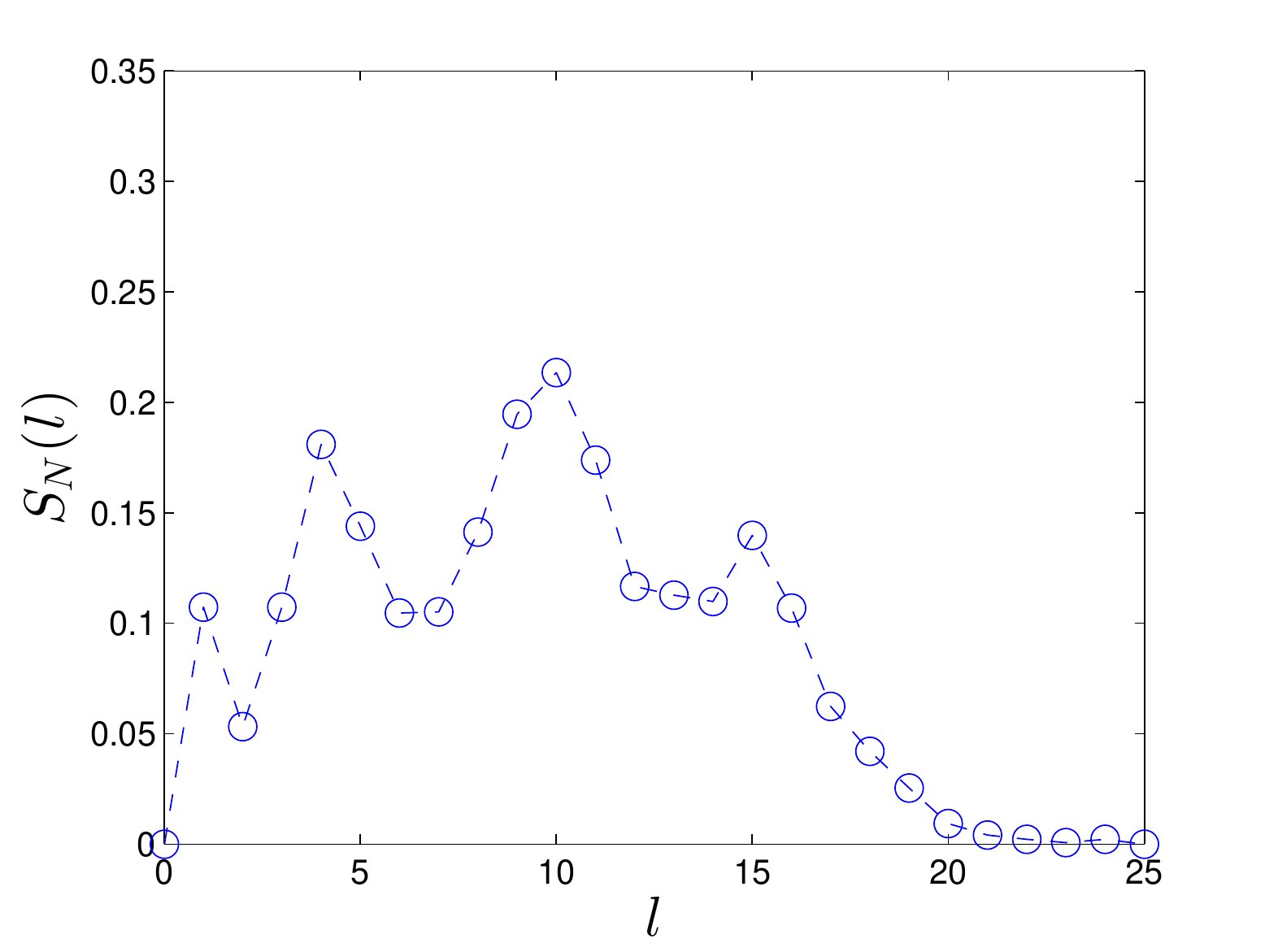}}
\put(  0,3.5){\makebox(0,0)[r]{\strut{}(a)}}
\put(5.5,3.5){\makebox(0,0)[r]{\strut{}(b)}}
\end{picture}
\caption{
Block entropy profile obtained by the DMRG method for the LiF molecule at 
bond length $d_{\text{Li-F}}=3.05$ a.u. for a non-optimized tensor topology (a) and 
for an optimized tensor topology (b).
}
\label{fig:LiF_block_entropy}
\end{figure}

The usual practice is to take one, two, or more neighboring
orbitals into a subsystem (called also block), as is shown
in Fig.~\ref{fig:sl}(a) for a one-dimensional topology used in DMRG.
The scaling behavior of the von Neumann entropy
$S^{\{1,2,\dots,l\}}$
of a contiguous block of the first $l$ orbitals with the number
of orbitals has also been used to study the quantum phases of
one-dimensional systems.
For systems with local interactions, this ``block entropy'' diverges
logarithmically with block size $l$ for critical systems, but saturates
for gapped systems\cite{Vidal-2003a,Calabrese-2004}, and
in certain cases its profiles provide further information about 
the energy spectrum\cite{Laflorence-2006,Legeza-2007}. For example,
the oscillation with a period of three as is shown in Fig.~\ref{fig:sl}
identifies soft modes with a wavevector, $k=\pm 2\pi/3$.
In contrast to this, the block entropy has more complex behavior when
non-local interactions are present\cite{Legeza-2003b,Barcza-2011}
and its profile depends strongly on the ordering of the orbitals along
the one dimensional chain as will be discussed below.
As an example, block entropy profiles obtained with the DMRG method for 
the LiF molecule at bond length $d_{\text{Li-F}}=3.05$ a.u. 
are shown in Fig.~\ref{fig:LiF_block_entropy}.
At this point it is worth to note that not only the profiles are different 
but the maximum of 
the block entropy is much smaller in the latter case. This property will
be used to optimize tensor methods as will be discussed below.

\subsubsection{One- and two-orbital entropy and mutual information}
\label{sec:num.ent.mutinf}

\begin{figure}[t]
\centering
\setlength{\unitlength}{40pt}   
\begin{picture}(10,1.25)
\put(  0,  0){\includegraphics{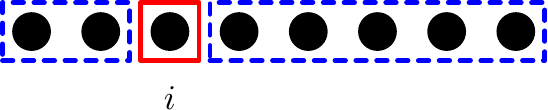}}
\put(5.5,  0){\includegraphics{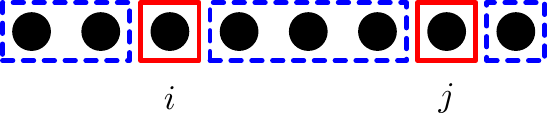}}
\put(  0,1.25){\makebox(0,0)[r]{\strut{}(a)}}
\put(5.5,1.25){\makebox(0,0)[r]{\strut{}(b)}}
\end{picture}
\caption{Partitioning of the system into single-orbital $(\mathrm{A})=\{i\}$ and double-orbital $(\mathrm{A})=\{i,j\}$ subsystems,
in order to determine single-orbital entropy $S_i$ (a) 
and two-orbital entropy $S_{ij}$ (b).}
\label{fig:s1_i}
\end{figure}

Orbitals lying closer to and further away from the Fermi surface 
possess larger and smaller orbital entropy, respectively\cite{Legeza-2003b}. 
The orbital entropy is related to the mixedness of a local state 
and it is expressed by the eigenvalues of the one-orbital reduced density matrix (as shown in  (\ref{eq:vonN-entropy-omega})) for 
a given orbital $(\mathrm{A})=\{i\}$, as shown in Fig.~\ref{fig:s1_i}(a). Namely,
\begin{equation}
S_i = - \sum_\alpha \omega_{\alpha,i} \ln \omega_{\alpha,i},
\end{equation}
where $i=1,\ldots,d$ is the orbital index,
while $\omega_{\alpha,i}$ for $\alpha = 1,\ldots,q$ stands for 
the eigenvalues of the reduced density matrix of orbital $i$.
The amount of contribution to the total correlation energy of an orbital 
can be detected by the single-orbital entropy.
Since the total system is in a pure state,
i.e., we calculate the ground state or an excited state, 
the sum of all single-orbital entropy, 
\begin{equation}
\label{eq:Itot}
I_{\rm tot}=\sum_i S_i, 
\end{equation}
gives the amount of \emph{total correlation} encoded in the 
wavefunction\cite{Legeza-2004b,Legeza-2006b}.
Since the full system is in a pure state,
this is equal to the \emph{total entanglement} encoded in the state/wavefunction.
This quantity can be used to monitor chagnes in 
entanglement as system parameters are adjusted, for example, changing bond 
length or other geometrical properties\cite{Murg-2014,Fertitta-2014,Duperrouzel-2014}. 
 
\begin{figure}[t]
\centering
\setlength{\unitlength}{40pt}   
\begin{picture}(10.7,4)
\put(  0,  0){\includegraphics[scale=0.6]{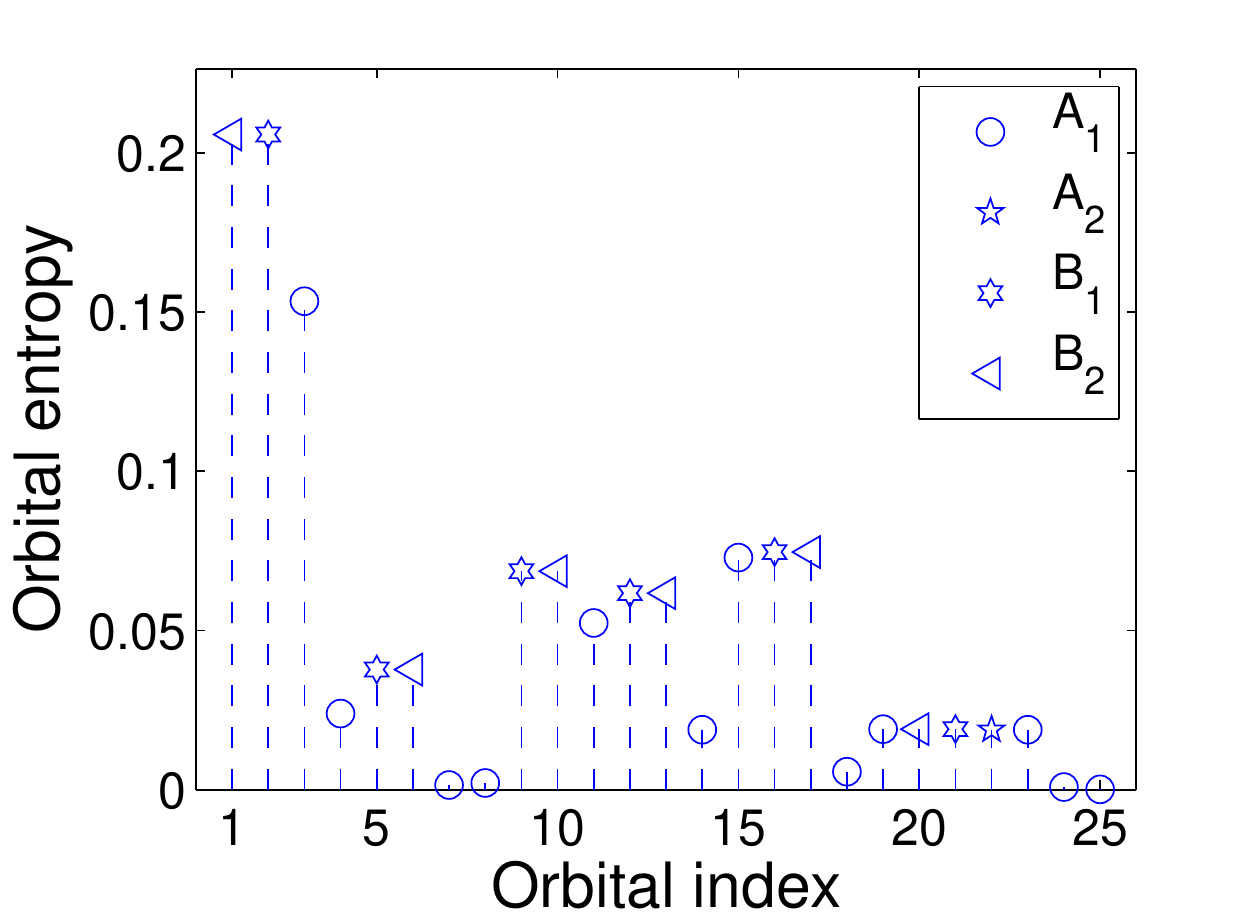}}
\put(5.5,0  ){  \includegraphics[scale=0.6]{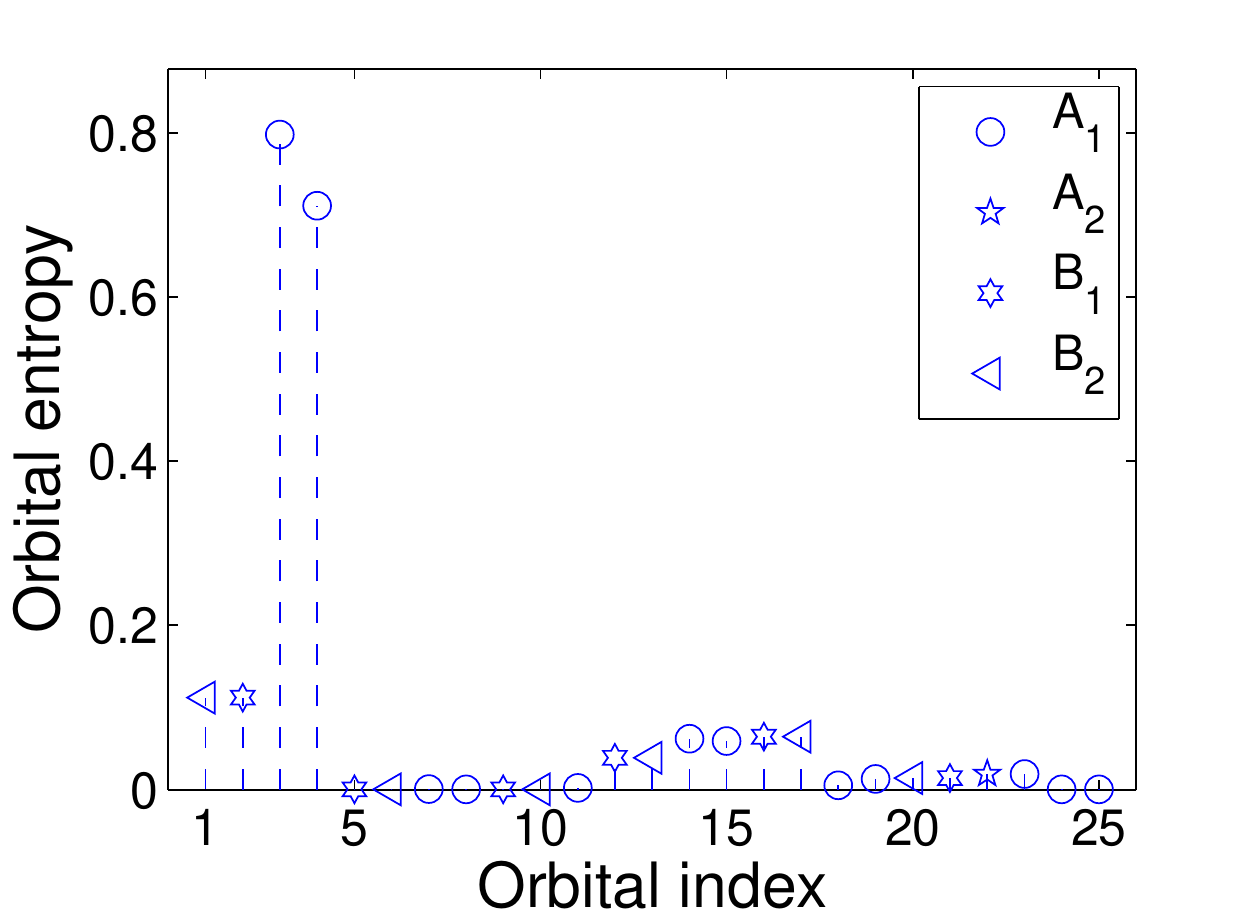}}
\put(0.3,4){\makebox(0,0)[r]{\strut{}(a)}}
\put(5.8,4){\makebox(0,0)[r]{\strut{}(b)}}
\end{picture}
\caption{One orbital entropy profile for the LiF molecule at bond length (a) $_{\text{Li-F}}=3.05$ a.u. and at (b) $d_{\text{Li-F}}=13.7$ a.u.
 Symbols label the irreducible representations of the molecular orbitals in the C$_{2v}$ point group.
}
\label{fig:s_i}
\end{figure}
\begin{figure}[t]
\centering
\setlength{\unitlength}{40pt}   
\begin{picture}(10.7,4)
\put(  0,  0){\includegraphics[scale=0.6]{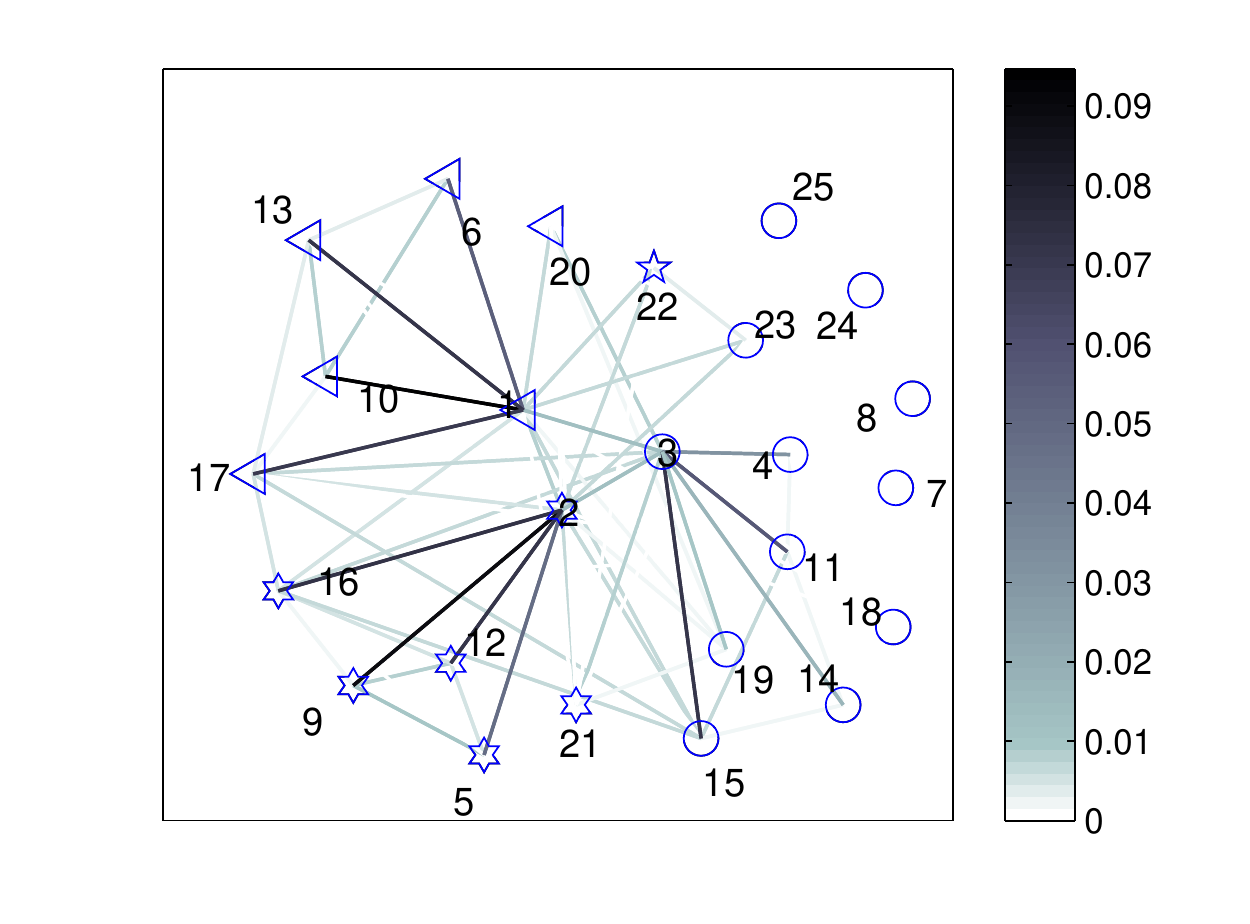}}
\put(5.5,0  ){\includegraphics[scale=0.6]{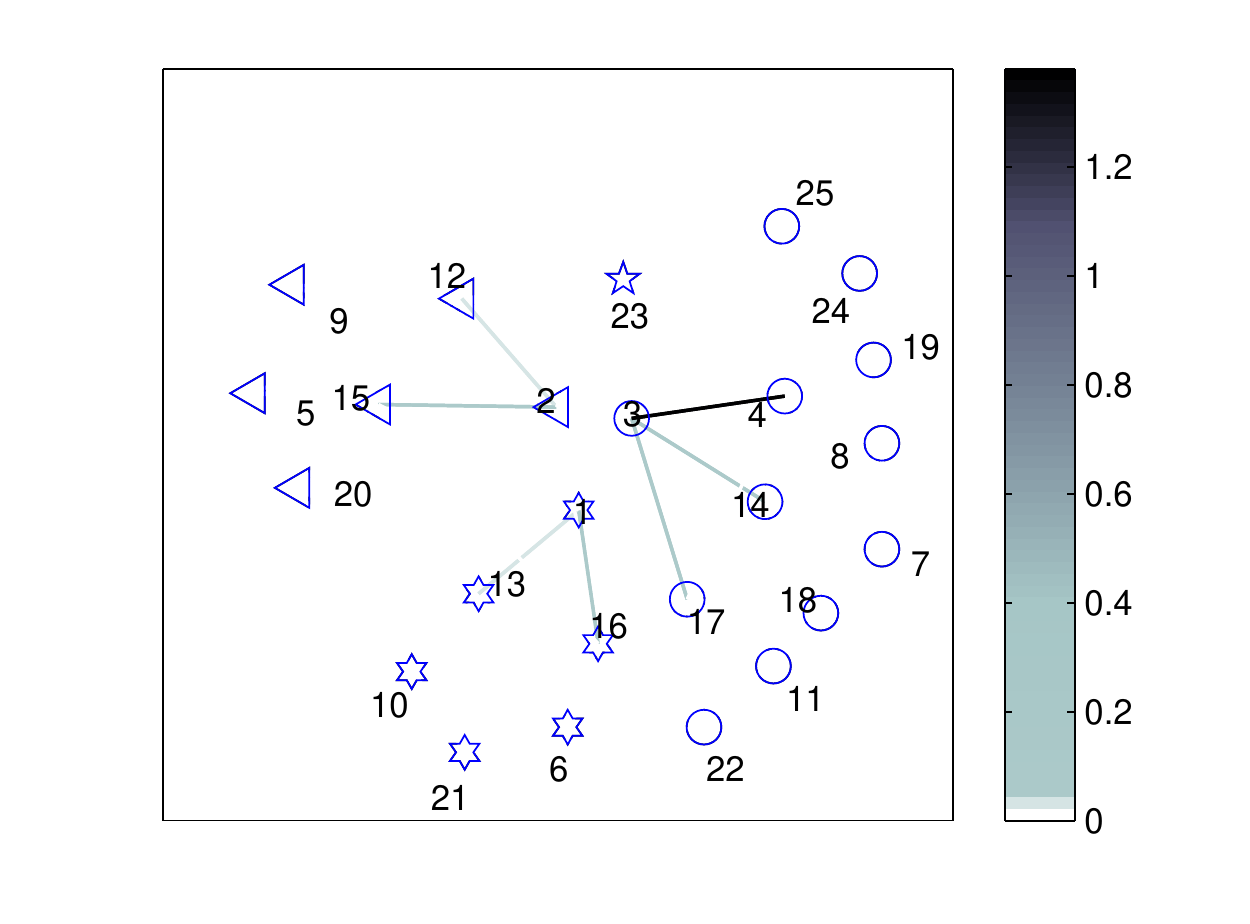}}
\put(0.3,4){\makebox(0,0)[r]{\strut{}(a)}}
\put(5.8,4){\makebox(0,0)[r]{\strut{}(b)}}
\end{picture} 
\caption{(Color online) Mutual information represented as a two-dimensional
weighted graph for the LiF molecule at bond length
(a) $d_{\text{Li-F}}=3.05$ a.u. and at (b) $d_{\text{Li-F}}=13.7$ a.u. Colors indicate different strengths of $I_{ij}$ and
the symbols label the irreducible representations of the molecular orbitals in the C$_{2v}$ point group.
}
\label{fig:I_free}
\end{figure}

A useful quantity to numerically characterize all kinds of
correlations between pairs of orbitals is the \emph{mutual information}
\begin{equation}
I_{ij} = S_i + S_j - S_{ij},
\label{eq:mut}
\end{equation}
calculated between two generally placed orbitals, $i$ and $j$
as shown in Fig.~\ref{fig:s1_i}(b).
Here $S_i$ is the von Neumann entropy, Eq.(\ref{eq:vonN-entropy}), for
a subsytem $(\mathrm{A})$ chosen to be the single orbital $i$, and $S_{ij}$ is the
entropy for $(\mathrm{A})$ chosen to consist of orbitals $i$ and $j$.
The mutual information $I_{ij}$ describes the
correlation between the
two selected subsystems, orbitals $i$ and $j$, embedded in a larger system.
$I_{ij}$ yields a weighted graph of the
overall
correlation of both classical and quantum origin among the orbitals.
The mutual information defined in this way has been introduced
previously to study correlation between
neighboring orbitals in spin and
fermionic chains with local interactions\cite{Legeza-2006a} and in quantum
chemical problems in order to optimize the network 
structure\cite{Rissler-2006,Barcza-2011}
as well as to study molecular bonding properties in
various transition metal
complexes\cite{Barcza-2011,Boguslawski-2012a,Boguslawski-2012b,Boguslawski-2013}.
Therefore, these 
quantities provide chemical information about the system, especially about bond 
formation and nature of static and dynamic correlation\cite{Barcza-2011,Boguslawski-2012b,Boguslawski-2013,Kurashige-2013,Duperrouzel-2014}.
As an example, $S_i$ and $I_{ij}$ are shown in Figs.~\ref{fig:s_i} and \ref{fig:I_free}, respectively,
for the equilibrium bond length $d_{\text{Li-F}}=3.05$ a.u. and at large separation $d_{\text{Li-F}}=13.7$ a.u..
It is clear form Fig.~\ref{fig:I_free} that some orbitals are strongly entangled
with several other orbitals while some orbitals are entangled with only a few others
and some are almost disentangled from the system.

\subsubsection{One- and two-orbital reduced density matrix and generalized correlation functions}
\label{sec:num.ent.gencorr}

It has been shown\cite{Barcza-2014,Fertitta-2014} that 
one can also analyze the sources of entanglement encoded in $I_{ij}$ 
by studying the behavior of the matrix elements of the two-orbital reduced density matrix $\rho_{ij}$. 
The $d$-orbital wave function can be written in terms of the
single-orbital $q$-dimensional basis as
\begin{equation}
\left|\Psi\right\rangle =\sum_{\alpha_1,\ldots,\alpha_d}
U(\alpha_1,\ldots\alpha_d)|\phi^{\{1\}}_{\alpha_1}\rangle\otimes\ldots\otimes|\phi^{\{d\}}_{\alpha_d}\rangle,
\label{eq:fulltensor}
\end{equation}
where the $\alpha_{j}$ labels single-orbital basis states and
the set of coefficients $U(\alpha_1,\ldots,\alpha_d)$ is viewed as a
tensor of order $d$.
The one- and two-orbital reduced density matrices 
$\rho_i=\mbox{Tr}_{1,\dots,\cancel{i},\dots,d}|\Psi\rangle\langle\Psi|$ and 
$\rho_{ij}=\mbox{Tr}_{1,\dots,\cancel{i},\dots,\cancel{j},\dots,d}|\Psi\rangle\langle\Psi|$ 
can be calculated by taking the appropriate partial traces of $|\Psi\rangle\langle\Psi|$, 
leading to the matrix elements
\begin{subequations}
\begin{align}
\label{eq:rho-i-full}
\begin{split}
&\rho_i(\alpha_i,\alpha_i^\prime)
= \langle \phi^{\{i\}}_{\alpha_i} |\varrho_i|\phi^{\{i\}}_{\alpha'_i}\rangle \\
&\qquad =\sum_{\substack{\alpha_1,\ldots,\cancel{\alpha_i},\ldots,\alpha_d}}
U(\alpha_1,\ldots,\alpha_i,\ldots,\alpha_d) 
\overline{U(\alpha_1,\ldots,\alpha^\prime_i,\ldots,\alpha_d)},
\end{split}\\
\label{eq:rho-ij-full}
\begin{split}
&\rho_{ij}(\alpha_i,\alpha_j,\alpha_i^\prime,\alpha_j^\prime)
= \langle \phi^{\{i\}}_{\alpha_i} \phi^{\{j\}}_{\alpha_j} |\varrho_{ij}|\phi^{\{i\}}_{\alpha'_i} \phi^{\{j\}}_{\alpha'_j}\rangle \\
&\qquad =\sum_{\substack{\alpha_1,\ldots,\cancel{\alpha_i}, \ldots, \\
                                                   \cancel{\alpha_j},\ldots,\alpha_d}}
U(\alpha_1,\ldots,\alpha_i,\ldots,\alpha_j,\ldots,\alpha_d) 
\overline{U(\alpha_1,\ldots,\alpha^\prime_i,\ldots,\alpha^\prime_j,\ldots,\alpha_d)}.
\end{split}
\end{align}
\end{subequations}

\begin{figure}[t]
\centering
\setlength{\unitlength}{40pt}   
\begin{picture}(7.5,4.5)
\put(  0,  2.5){\includegraphics{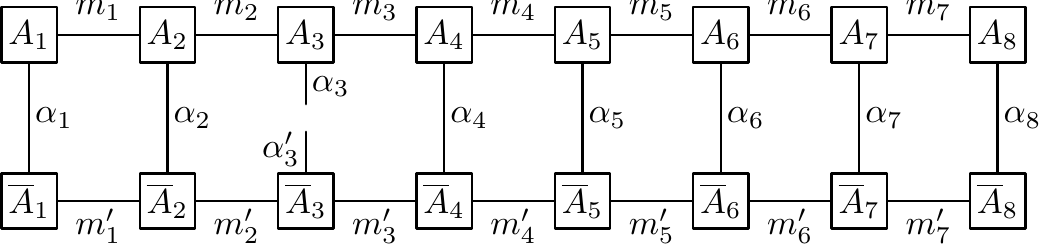}}
\put(  0,  0){\includegraphics{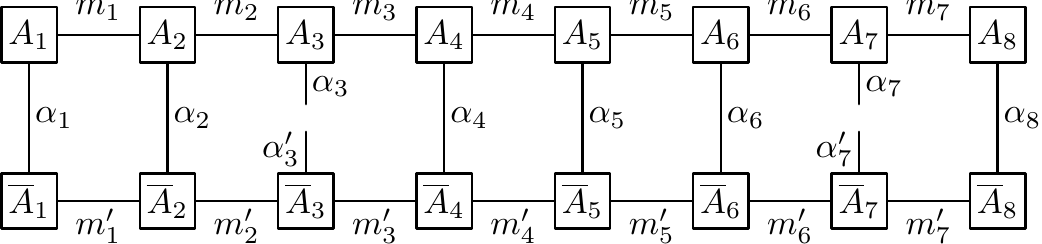}}
\put(  0,4.5){\makebox(0,0)[r]{\strut{}(a)}}
\put(  0,2){\makebox(0,0)[r]{\strut{}(b)}}
\end{picture} 
\caption{Contraction of the MPS network to calculate the 
one- (a) and two-orbital (b) reduced density matrices $\varrho_i$ and $\varrho_{ij}$ for a chain with $d=8$.}
\label{fig:mps-rho-ij}
\end{figure}

The dimension of $U$ grows exponentially with system size $d$, thus, 
such full tensor representations of the wave function,
needed for the computation of the reduced density matrices above,
are only possible for small system sizes.
Using the methods described in the previous and following sections,
the $d$th-order tensor $U$ can, in many cases, be efficiently
factorized into a product of matrices, as e.g., in (\ref{eq:makingTT7})
\begin{equation}
U(\alpha_1,\ldots,\alpha_d)=
\mathbf{A}_1(\alpha_1)\mathbf{A}_2(\alpha_2)\ldots \mathbf{A}_d(\alpha_d),
\label{eq:psi-mps}
\end{equation}
leading to an MPS representation of the wave function, where the
$\mathbf{A}_i(\alpha_i)$ are $M\times M$ matrices in general\cite{Verstraete-2004b}.
For systems with open boundary conditions,
$\mathbf{A}_1(\alpha_1)$ and $\mathbf{A}_d(\alpha_d)$ are row and column vectors, respectively.
In the MPS representation, the calculation of $\rho_i$ and $\rho_{ij}$ by means of
Eqs.~(\ref{eq:rho-i-full}) and (\ref{eq:rho-ij-full}) corresponds to the contraction of the
network over all states except those at orbital $i$ in the first case and at orbital $i$ and $j$ in the second,
as depicted in Fig.~\ref{fig:mps-rho-ij} for a chain with $d=8$ orbitals.

From a different point of view,
the matrix elements of $\varrho_i$ and $\varrho_{ij}$ in Eqs.~(\ref{eq:rho-i-full}) and (\ref{eq:rho-ij-full}) 
can be written as expectation values of projection-like operators 
acting on the corresponding orbitals
Let the \emph{transition operators} be defined as
\begin{equation}
\label{eq:transop}
{\cal T}^{(m)}=|\phi_{\alpha'}\rangle\langle\phi_\alpha|,\qquad\text{for $m=1,\dots,q^2$},
\end{equation}
which describe a possible \emph{transition} between the initial states $|\phi_\alpha\rangle$ 
and the final states $|\phi_{\alpha^\prime}\rangle$ understood for a given orbital,
with the \emph{numbering rules}
\begin{subequations}
\label{eq:numbrules}
\begin{gather}
\label{eq:numbrulesma}
\alpha-1 = ((m-1)\mod q),\qquad
\alpha'-1=\left \lfloor (m-1)/q \right \rfloor,\\
\label{eq:numbrulesam}
m-1 = (\alpha-1)q + \alpha^\prime-1.
\end{gather}
\end{subequations}
(Here $\left \lfloor x \right \rfloor$ denotes the floor function, the integral part of $x$.)
These operators can be extended to operate on the complete Hilbert space 
consisting of $d$ local Hilbert spaces labeled by $i=1, \ldots, d$ as
\begin{equation}
\label{eq:transopi}
{\cal T}_i^{(m)}= \mathbb{I} \otimes  \dots \otimes \mathbb{I} \otimes {\cal T}^{(m)}  \otimes \mathbb{I} \otimes  \dots \otimes \mathbb{I},
\end{equation}
with the operator ${\cal T}^{(m)}$ in the $i$-th position.

One can now easily check that 
the matrix elements of the one- and two-orbital reduced density matrices, 
given in (\ref{eq:rho-i-full})-(\ref{eq:rho-ij-full}),
can be expressed as the expectation values of the transition operators
for one and for two sites, respectively, as follows 
\begin{subequations}
\begin{align}
\label{eq:rhoT}
\rho_i(\alpha_i,\alpha_i^\prime) &= 
\langle {\cal T}_{i}^{(m_i)}\rangle,\\
\label{eq:rhoTT}
\rho_{ij}(\alpha_i,\alpha_j,\alpha_i^\prime,\alpha_j^\prime) &= 
\langle {\cal T}_{i}^{(m_i)}{\cal T}_{j}^{(m_j)}\rangle,
\end{align}
\end{subequations}
using the numbering rules (\ref{eq:numbrules}) for each orbitals.
That is, the matrix representation of 
the one-orbital reduced density operator $\varrho_i$
can be constructed from expectation values of operators describing 
\emph{transitions between the single-orbital basis} $|\phi^{\{i\}}_{\alpha_i}\rangle$, 
while the two-orbital reduced density operator $\varrho_{ij}$
can be constructed from expectation values of operators describing 
\emph{transitions between two-orbital basis states} $|\phi^{\{i\}}_{\alpha_i} \phi^{\{j\}}_{\alpha_j} \rangle 
\equiv|\phi^{\{i\}}_{\alpha_i}\rangle\otimes|\phi^{\{j\}}_{\alpha_j}\rangle$.
This is a generalization of the procedure introduced in the DMRG context
for spin-1/2 fermion models\cite{Rissler-2006,Boguslawski-2013}.
In the following, 
we refer to the expectation values of pairs of state-transition operators in Eq.~(\ref{eq:rhoTT})
as \emph{generalized} correlation functions
in order to distinguish them from conventional correlation functions,
i.e., those based on physically motivated self-adjoint operators such as local spin or density operators.
For (\ref{eq:rhoT}), note that 
when the individual local basis states are completely
distinguished by abelian quantum numbers, the one-orbital density
matrix is diagonal and has the form
$\rho_i(\alpha_i,\alpha_i^\prime) = \delta_{\alpha_i,\alpha_i^\prime} \langle {\cal T}_i^{(\alpha_i(q-1)+\alpha_i')} \rangle$,
providing the spectrum immediately.

A given generalized correlation function measures the 
expectation value of the resonance amplitude between the initial and 
final states within a particular environment. 
In general, $\langle {\cal T}_{i}^{(m_i)}\,{\cal T}_{j}^{(m_j)}\rangle$ contains both 
connected and disconnected contributions between subsystems $i$ and $j$. 
Therefore, it can, in general, scale to a finite
value as the distance $l=|i-j|$ is increased,
even if the physical correlation function goes to zero for large $l$.
In order to circumvent this behavior, one generally study the connected part 
of the generalized correlation functions,
$\langle {\cal T}_{i}^{(m_i)}\,{\cal T}_{j}^{(m_j)}\rangle_{\rm C}= 
 \langle {\cal T}_{i}^{(m_i)}\,{\cal T}_{j}^{(m_j)}\rangle- 
 \langle {\cal T}_{i}^{(m_i)}\rangle \langle{\cal T}_{j}^{(m_j)}\rangle,$
where the disconnected part, given by the product of the expectation values 
of the local transition operators, is subtracted out. Note that the mutual 
information (\ref{eq:mut}) is formulated in such a way that the disconnected parts of the 
generalized correlation functions do not contribute. 
These can be used to identify the relevant physical processes that lead 
to the generation of the entanglement\cite{Barcza-2014,Fertitta-2014}.

As an example, let us take the spin-1/2 fermionic model.
Here the single-electron basis states can be empty,
occupied with a single spin-down or spin-up electron,
or doubly occupied, with the corresponding basis states denoted as $|$\zero$\!\rangle$, $|\!\downarrow\rangle$,
$|\!\uparrow\rangle$, and $|\!\uparrow\downarrow\rangle$, as before.
Since the local basis is $q=4$-dimensional,
$q^2=16$ possible transition operators ${\cal T}^{(m)}$ arise, as is displayed in Table.~\ref{tab:primitiveop-f12}.
They can be written explicitly in terms of local fermion
creation $\mathbf{c}_{i,s}^\dagger$, annihilation $\mathbf{c}_{i,s}$ and number $\mathbf{n}_{i,s}$
operators (\ref{eq:Cs.cis})-(\ref{eq:Cs.nis}) as
\begin{align}
{\cal T}^{(1)}  &=(\mathbb{I}-\mathbf{n}_{\uparrow})(\mathbb{I}-\mathbf{n}_{\downarrow}),\;\;\nonumber
&{\cal T}^{(2)}  &= (\mathbb{I}-\mathbf{n}_{\uparrow}) \mathbf{c}_{\downarrow},\;\;\\
{\cal T}^{(3)}  &= \mathbf{c}_{\uparrow}(\mathbb{I}-\mathbf{n}_{\downarrow}),\;\;\nonumber
&{\cal T}^{(4)}   &= -\mathbf{c}_{\uparrow}\mathbf{c}_{\downarrow},\;\;\\
{\cal T}^{(5)}  &= (\mathbb{I}-\mathbf{n}_{\uparrow})\mathbf{c}_{\downarrow}^{\dagger}   ,\;\;\nonumber
&{\cal T}^{(6)}   &= (\mathbb{I}-\mathbf{n}_{\uparrow})\mathbf{n}_{\downarrow},\;\;\\
{\cal T}^{(7)}  &= - \mathbf{c}_{\uparrow}\mathbf{c}_{\downarrow}^{\dagger},\;\;\nonumber
&{\cal T}^{(8)}   &= \mathbf{c}_{\uparrow}\mathbf{n}_{\downarrow},\;\;\\
{\cal T}^{(9)}  &= \mathbf{c}_{\uparrow}^{\dagger}(\mathbb{I}-\mathbf{n}_{\downarrow}),\;\;\nonumber
&{\cal T}^{(10)}  &= \mathbf{c}_{\uparrow}^{\dagger}\mathbf{c}_{\downarrow},\;\;\\
{\cal T}^{(11)} &= \mathbf{n}_{\uparrow}(\mathbb{I}-\mathbf{n}_{\downarrow}),\;\;\nonumber
&{\cal T}^{(12)}  &= -\mathbf{n}_{\uparrow}\mathbf{c}_{\downarrow},\;\;\\
{\cal T}^{(13)} &= \mathbf{c}_{\uparrow}^{\dagger}\mathbf{c}_{\downarrow}^{\dagger},\;\;\nonumber
&{\cal T}^{(14)}  &= \mathbf{c}_{\uparrow}^{\dagger}\mathbf{n}_{\downarrow},\;\;\\
{\cal T}^{(15)} &= -\mathbf{n}_{\uparrow}\mathbf{c}_{\downarrow}^{\dagger},\;\;
&{\cal T}^{(16)}  &= \mathbf{n}_{\uparrow}\mathbf{n}_{\downarrow}.
\label{eq:fermion-ops-1}
\end{align}
The non-vanishing matrix elements of the two-orbital density matrix
$\rho_{ij}$ are
given in Table \ref{tab:fermion-ops-2}.
Note that the two-orbital density matrix is block-diagonal in the
particle number $N_c$ and in the $z$ component of the spin $S^z$.
The block-diagonal structure is evident, and the values of $m_i$ and $m_j$
appropriate for each matrix element are displayed.

\begin{figure}[h]
\centering
\setlength{\unitlength}{40pt}   
\begin{picture}(11.2,8.8)
\put(5.6,8.8){\makebox(0,0)[c]{\strut{}(a) Hoppings}}
\put(1.4,8.4){\makebox(0,0)[c]{\strut{}$d_{\text{Li-F}}=3.05$}}
\put(4.2,8.4){\makebox(0,0)[c]{\strut{}$d_{\text{Li-F}}=13.7$}}
\put(7.0,8.4){\makebox(0,0)[c]{\strut{}$d_{\text{Li-F}}=3.05$}}
\put(9.8,8.4){\makebox(0,0)[c]{\strut{}$d_{\text{Li-F}}=13.7$}}
\put(  0,5.8){\includegraphics[width=0.24\columnwidth]{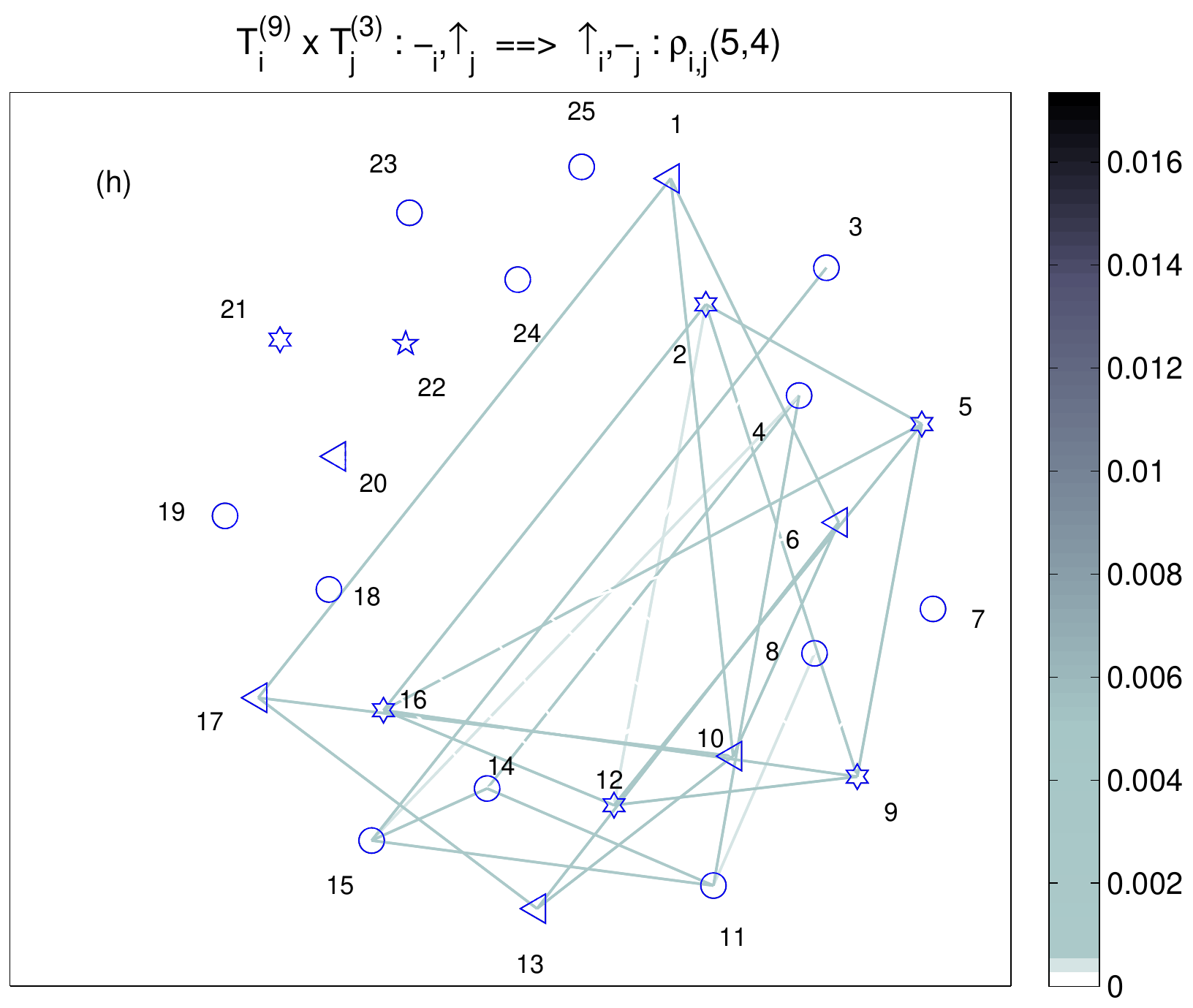}}
\put(2.8,5.8){\includegraphics[width=0.24\columnwidth]{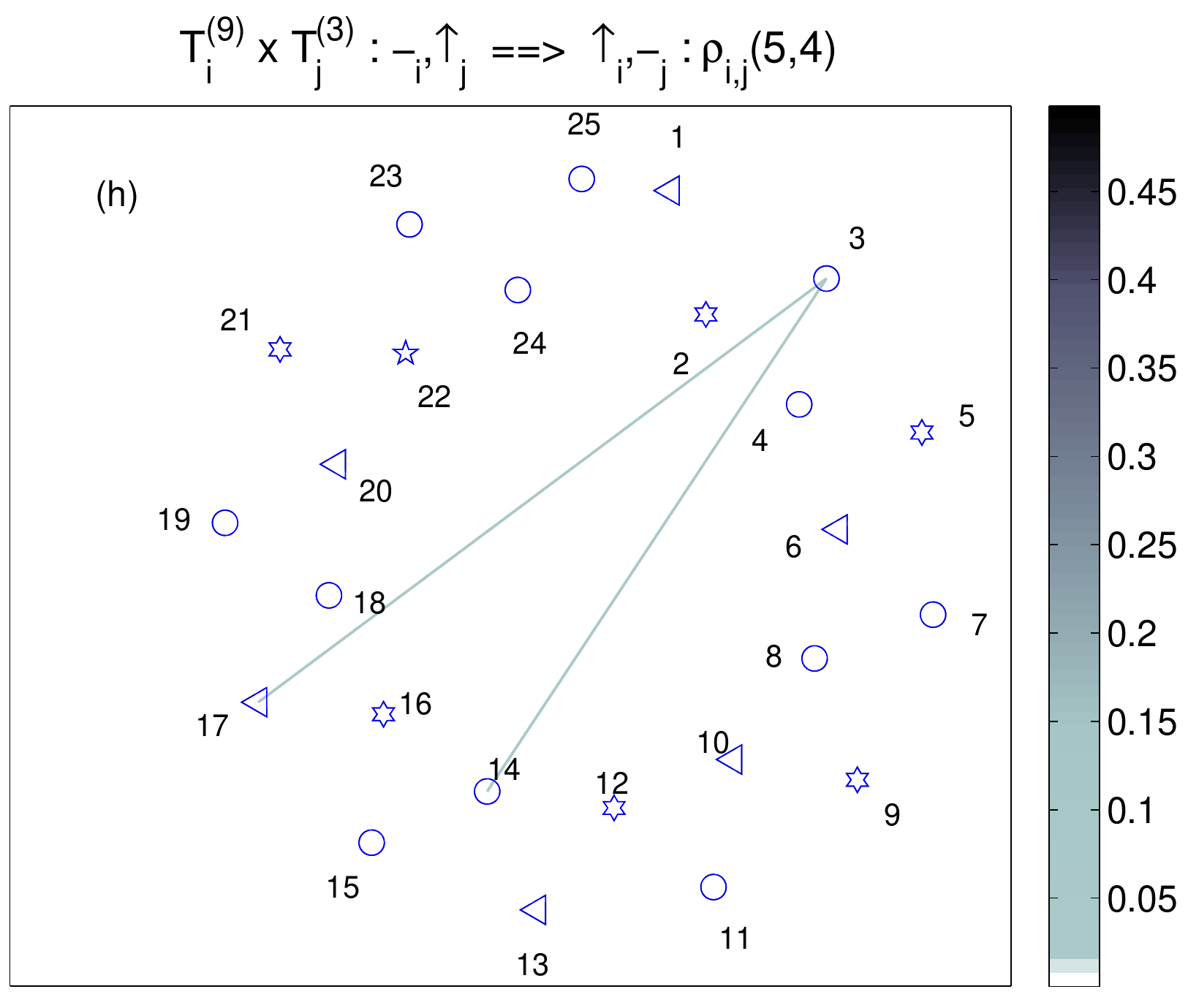}}
\put(5.6,5.8){\includegraphics[width=0.24\columnwidth]{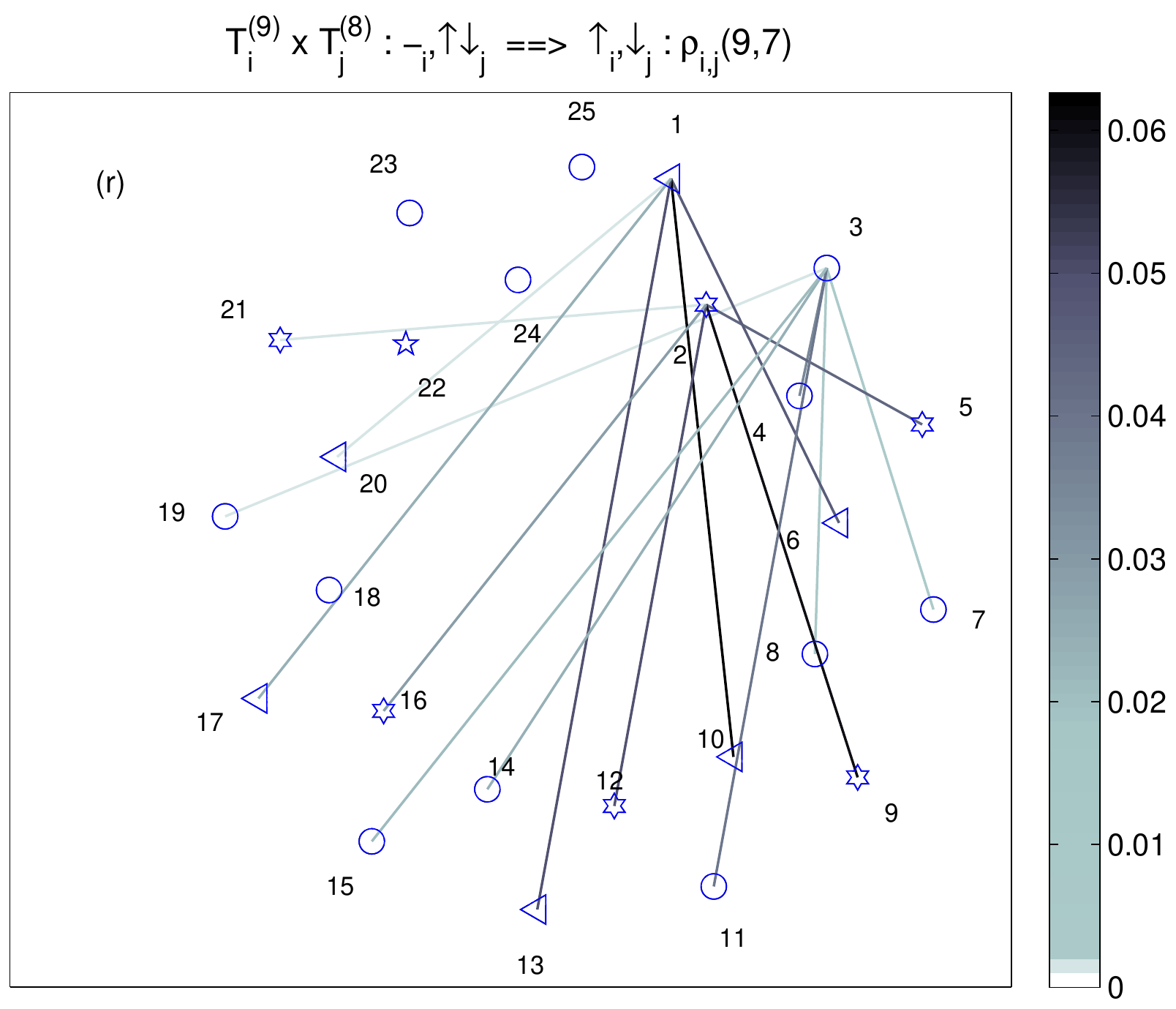}}
\put(8.4,5.8){\includegraphics[width=0.24\columnwidth]{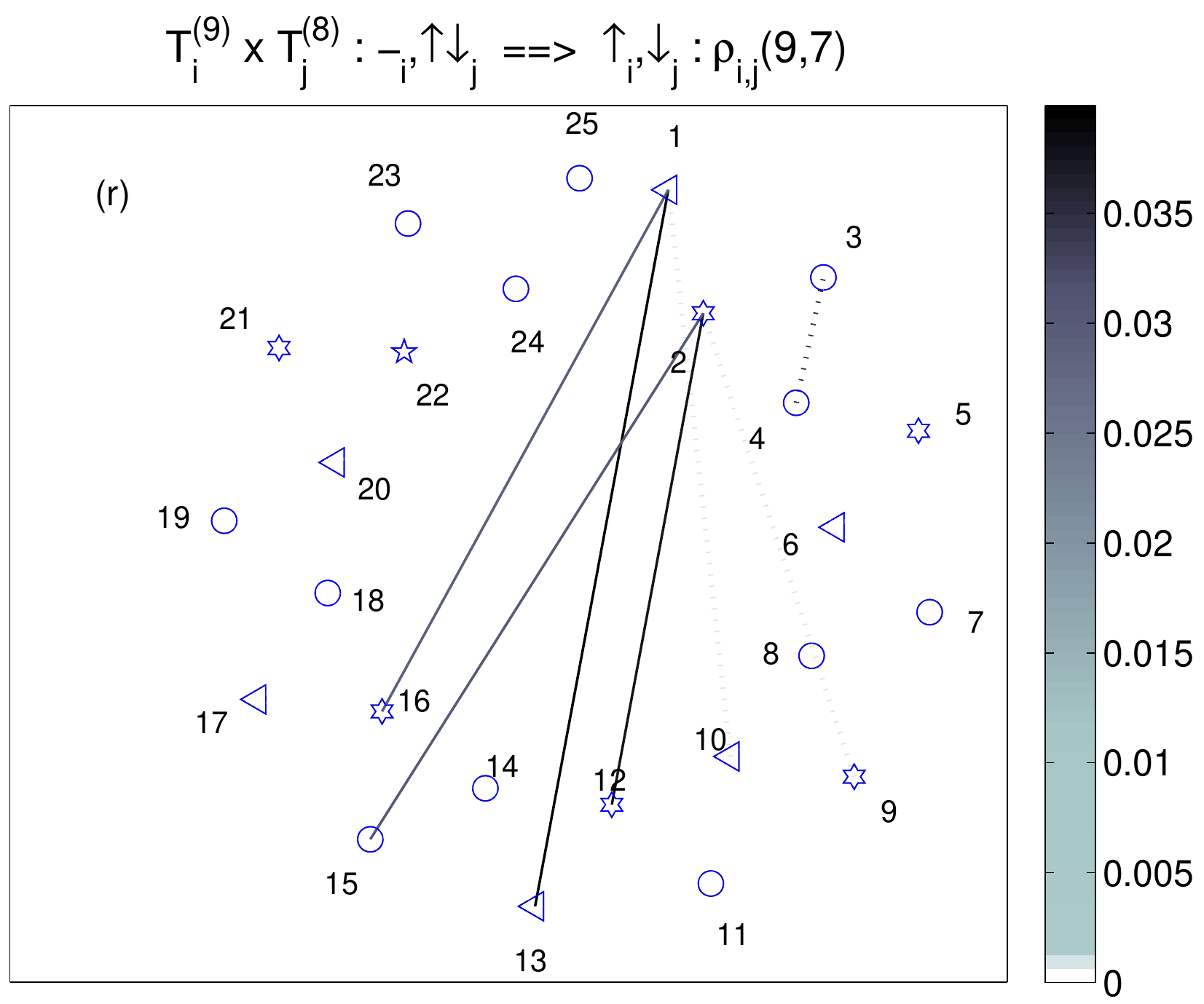}}
\put(  0,3.2){\includegraphics[width=0.24\columnwidth]{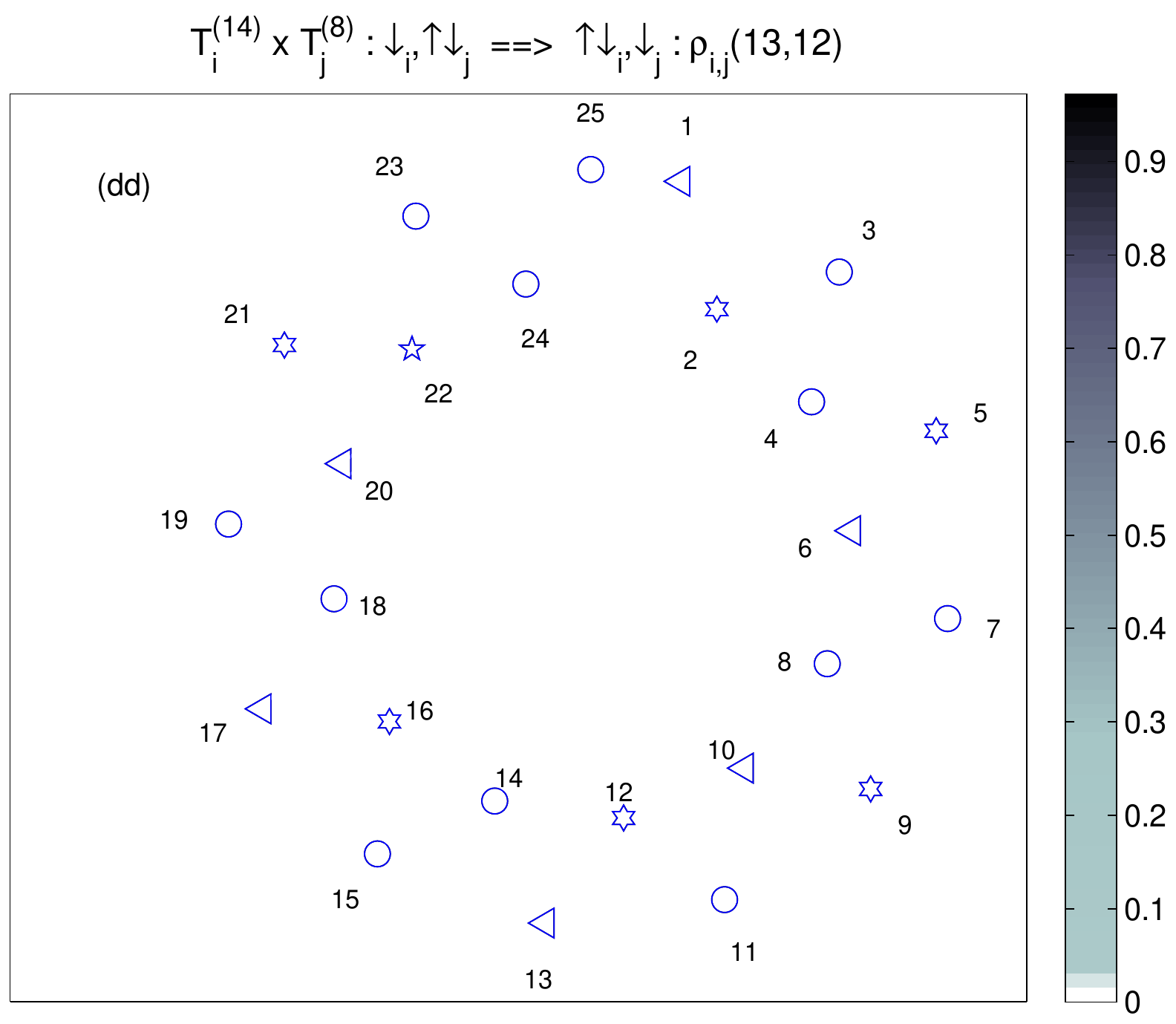}}
\put(2.8,3.2){\includegraphics[width=0.24\columnwidth]{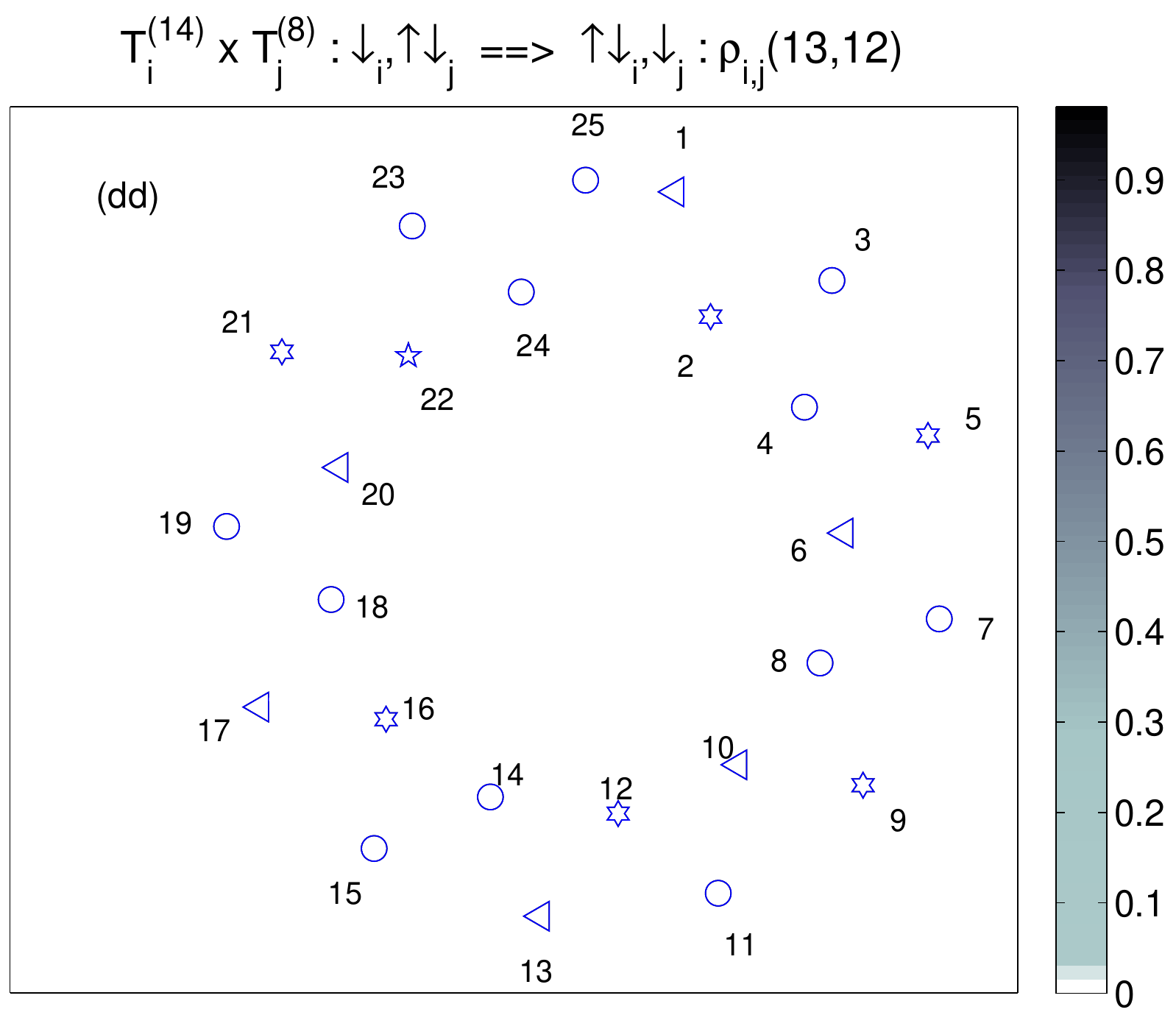}}
\put(5.6,3.2){\includegraphics[width=0.24\columnwidth]{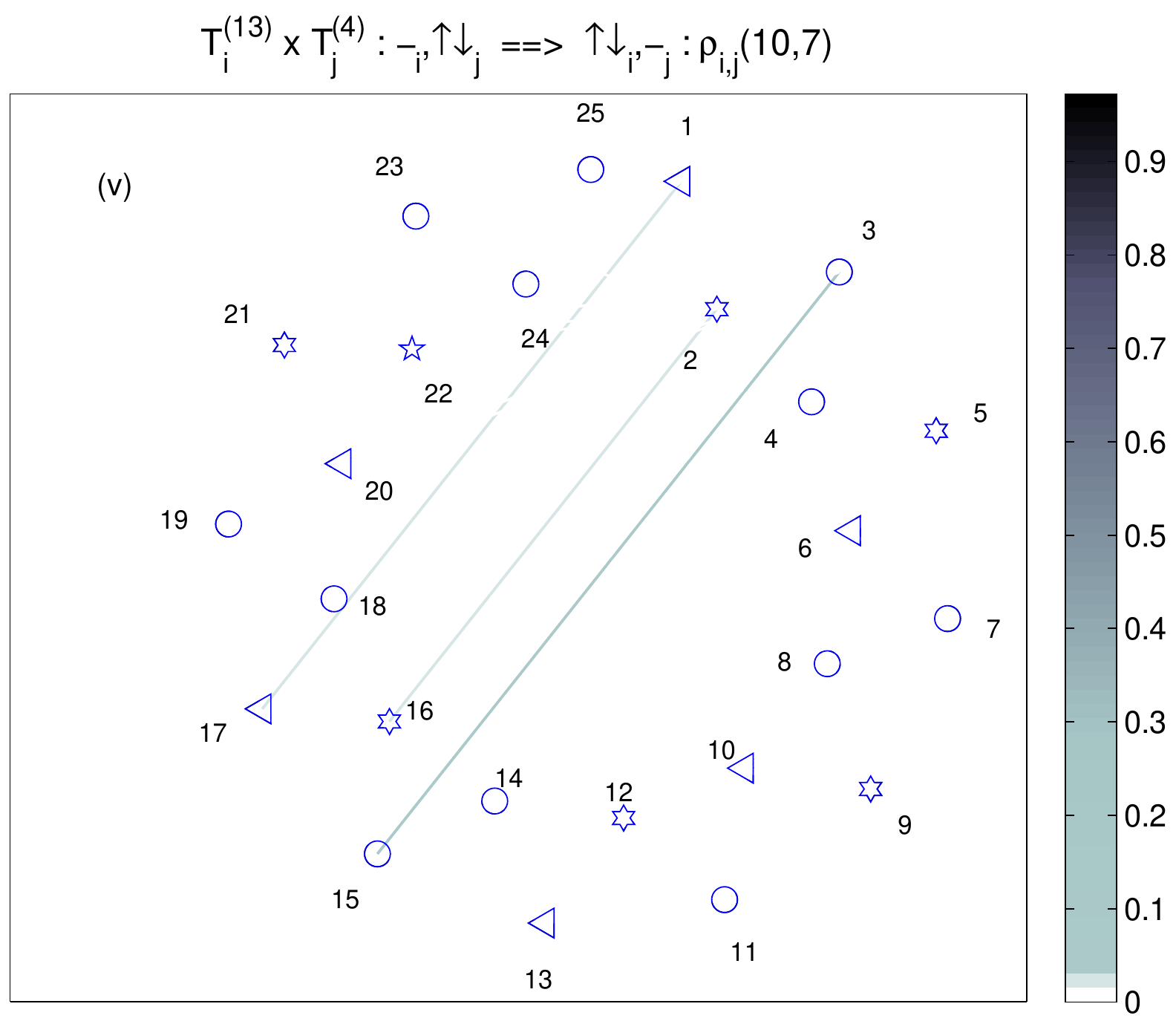}}
\put(8.4,3.2){\includegraphics[width=0.24\columnwidth]{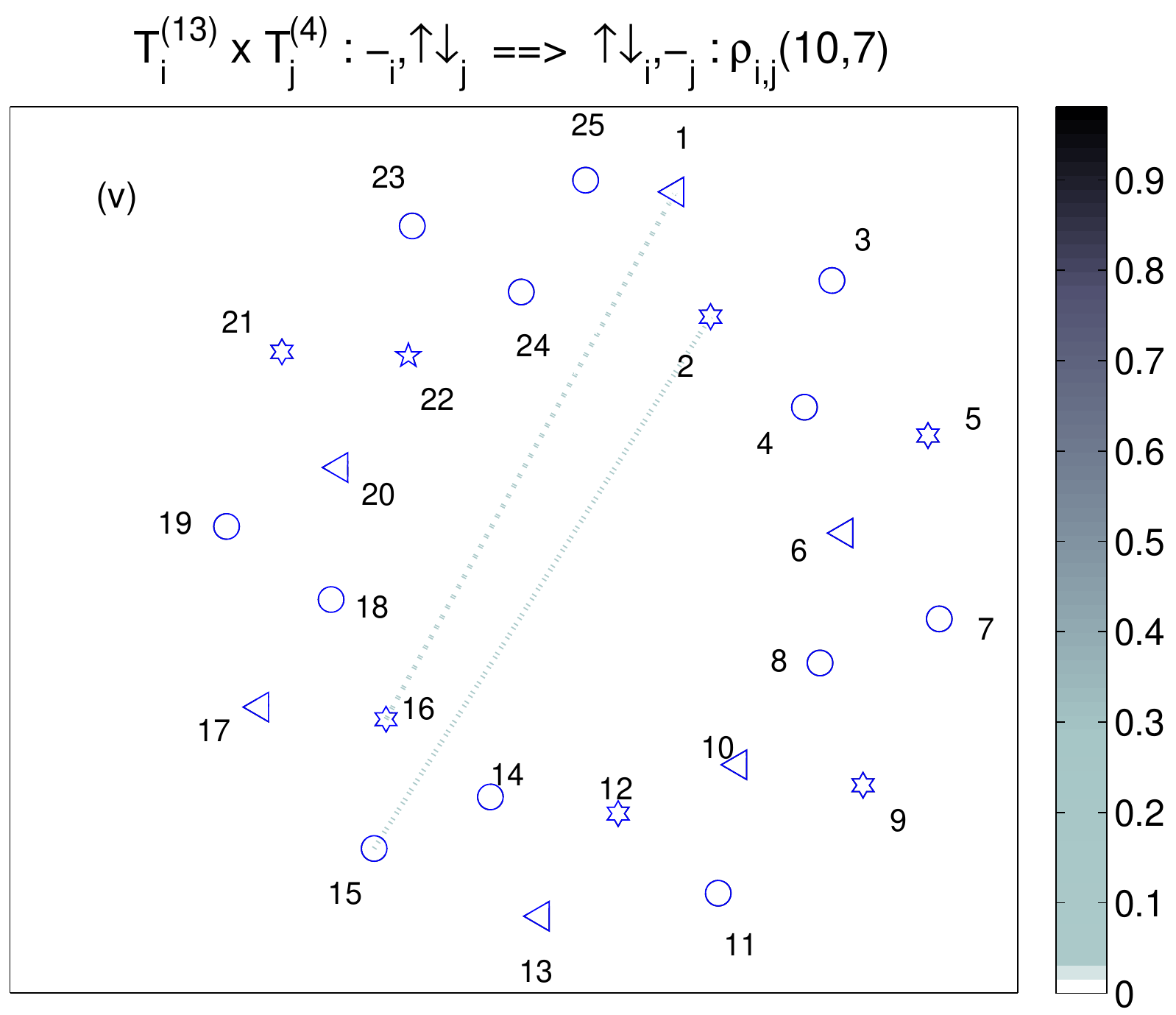}}
\put(5.6,3.0){\makebox(0,0)[c]{\strut{}(b) Flipping}}
\put(4.2,2.6){\makebox(0,0)[c]{\strut{}$d_{\text{Li-F}}=3.05$}}
\put(7.0,2.6){\makebox(0,0)[c]{\strut{}$d_{\text{Li-F}}=13.7$}}
\put(  3,  0){\includegraphics[width=0.24\columnwidth]{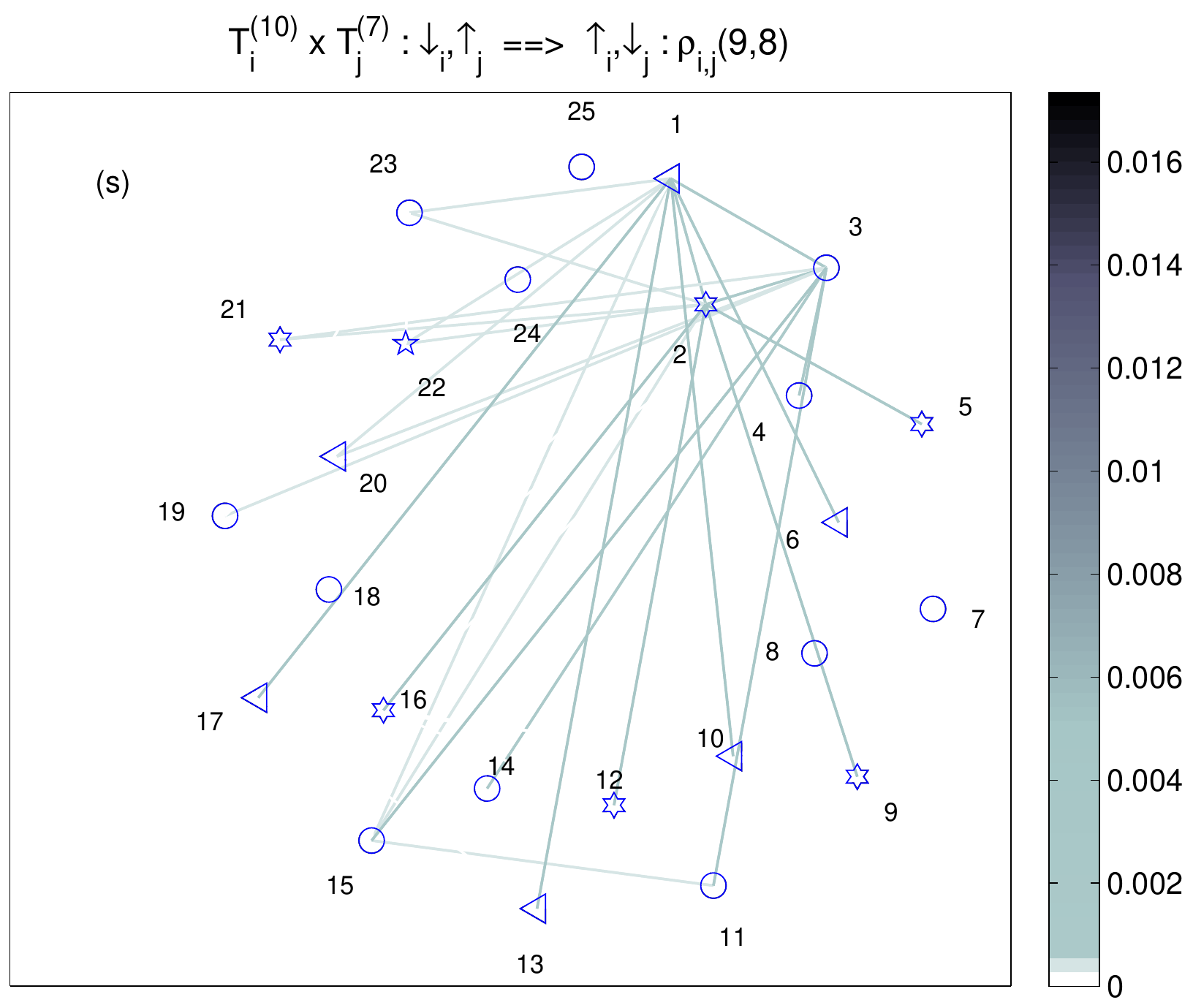}}
\put(  6,  0){\includegraphics[width=0.24\columnwidth]{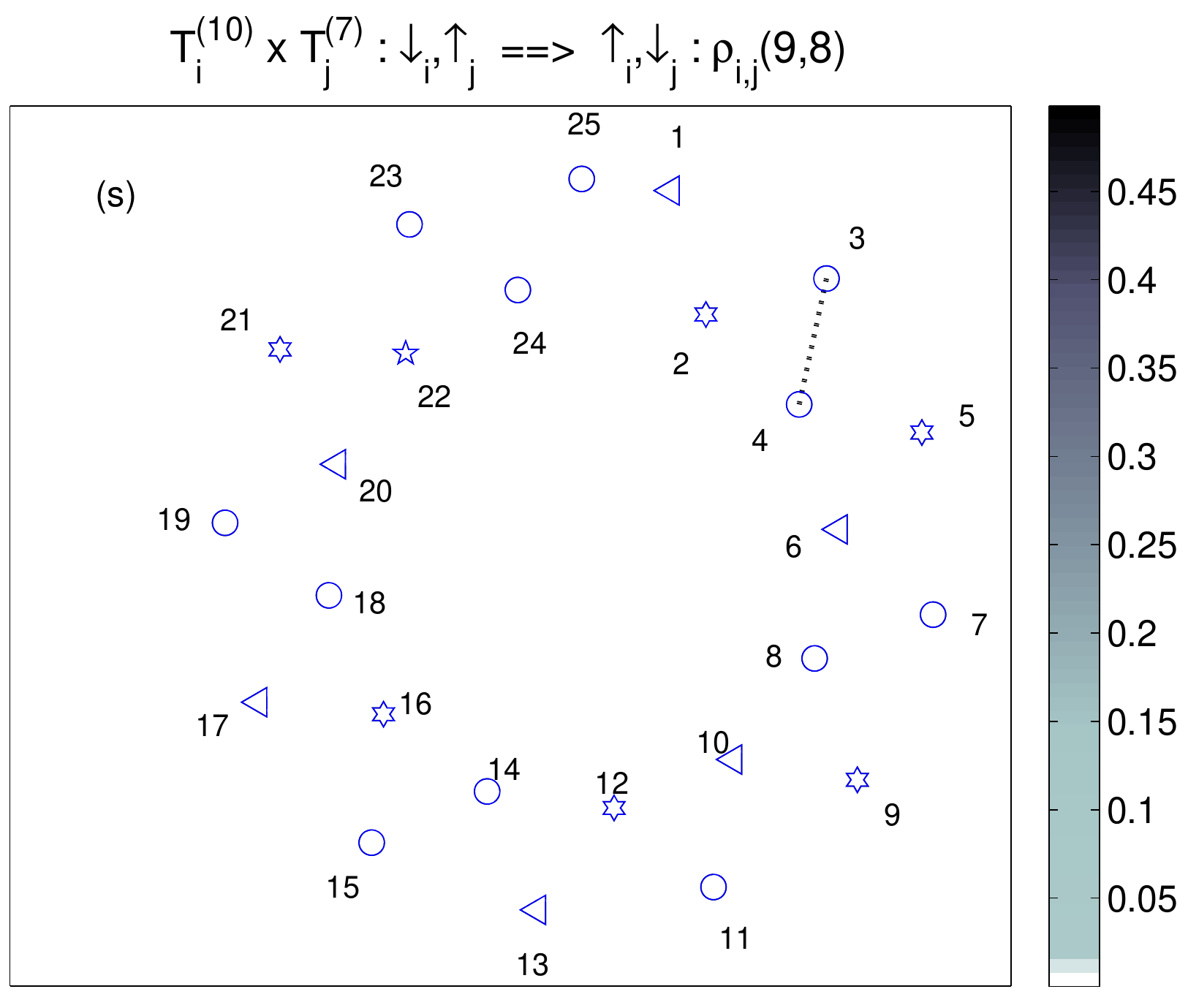}}
\end{picture}
\caption{Pictorial representation of the absolute value of the generalized correlation functions
used to construct the lower-triangular elements of the two-orbital reduced density matrix 
for LiF at $d_{\text{Li-F}}=3.05$ a.u. and at $d_{\text{Li-F}}=13.7$ a.u.. 
Strength of transition amplitues between initial ($|\phi^{\{i\}}_{\alpha_i} \phi^{\{j\}}_{\alpha_j}\rangle$) 
and final states ($|\phi^{\{i\}}_{\alpha'_i} \phi^{\{j\}}_{\alpha'_j}\rangle$) on orbital $i$ and $j$ 
are indicated with different line colors. 
Note the different scales used for colorbars in case of the various figures.}
\label{fig:lif_correlations}
\end{figure}
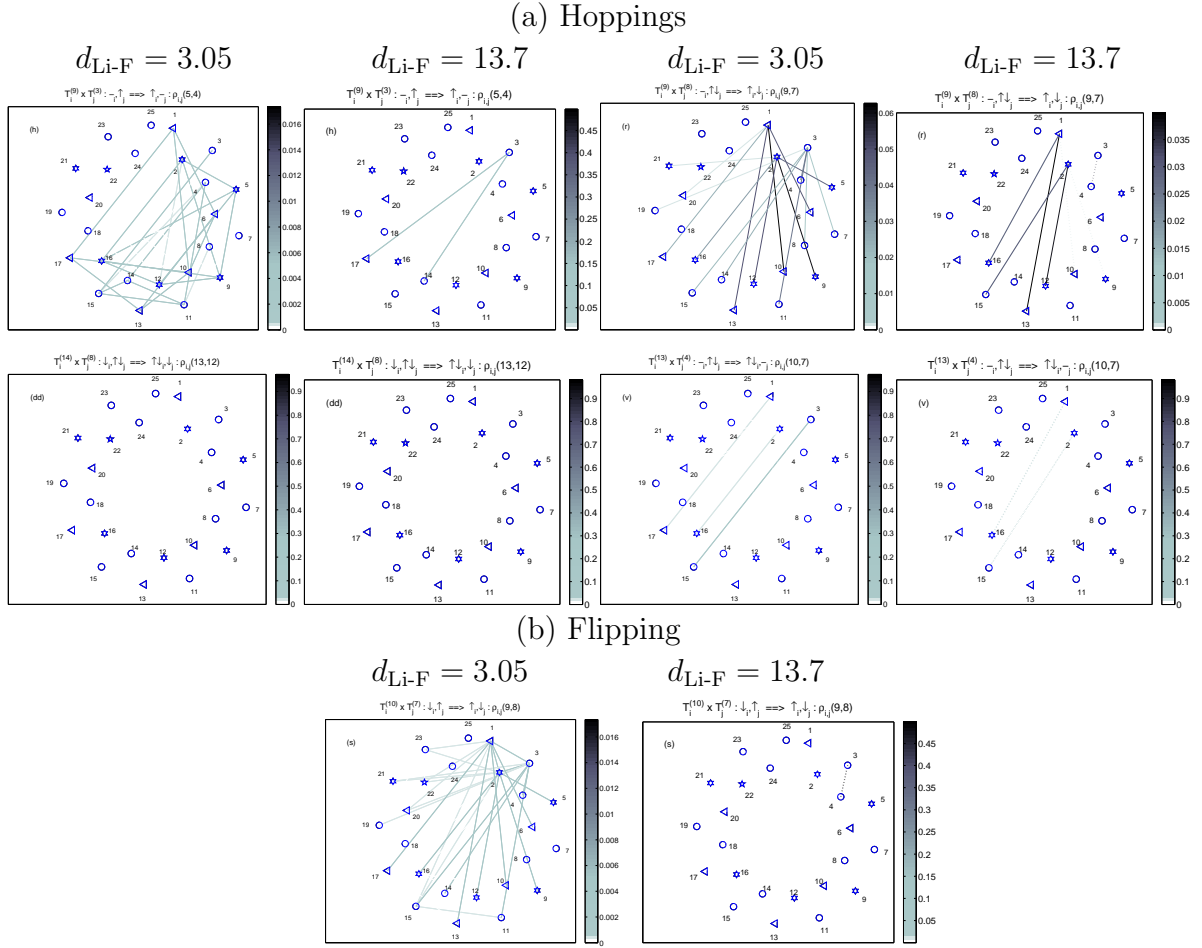

\begin{table}[t]
\centering
\begin{tabular}{c||c|c|c|c}
\hline
\hline
 & {$|$\zero$\rangle_i$} & {$|$\down$\rangle_i$} & {$|$\up$\rangle_i$} & {$|$\double$\rangle_i$} \\
\hline
\hline
{$|$\zero$\rangle_i$}  & ${\cal T}^{(1)}_i$ & ${\cal T}^{(2)}_i$ &  ${\cal T}^{(3)}_i$ & ${\cal T}^{(4)}_i$ \\
\hline
{$|$\down$\rangle_i$} & ${\cal T}^{(5)}_i$ & ${\cal T}^{(6)}_i$ & ${\cal T}^{(7)}_i$ & ${\cal T}^{(8)}_i$\\
\hline
{$|$\up$\rangle_i$}  & ${\cal T}^{(9)}_i$ & ${\cal T}^{(10)}_i$ &  ${\cal T}^{(11)}_i$ & ${\cal T}^{(12)}_i$ \\
\hline
{$|$\double$\rangle_i$} & ${\cal T}^{(13)}_i$ & ${\cal T}^{(14)}_i$ & ${\cal T}^{(15)}_i$ & ${\cal T}^{(16)}_i$\\
\hline
\hline
\end{tabular}
\caption{Single-orbital operators describing transitions
  between single-orbital basis states for a $S=1/2$ spin system.}
\label{tab:primitiveop-f12}
\end{table}
\begin{table}[t]
\centering
{\tiny
\renewcommand{\tabcolsep}{0.08cm}   
\begin{tabular} {C||C|CC|CC|C|CCCC|C|CC|CC|C}
\hline\hline
& \multicolumn{1}{D{0.8cm}|}{{\tiny $n$=$0$, $s_z$=$0$}} 
& \multicolumn{2}{D{1.6cm}|}{{\tiny $n$=$1$, $s_z$=-$\frac{1}{2}$}}
& \multicolumn{2}{D{1.6cm}|}{{\tiny $n$=$1$, $s_z$=$\frac{1}{2}$}} 
& \multicolumn{1}{D{0.8cm}|}{{\tiny $n$=$2$, $s_z$=-$1$}} 
& \multicolumn{4}{D{3.2cm}|}{{\tiny $n$=$2$, $s_z$=$0$}} 
& \multicolumn{1}{D{0.8cm}|}{{\tiny $n$=$2$, $s_z$=$1$}} 
& \multicolumn{2}{D{1.6cm}|}{{\tiny $n$=$3$, $s_z$=-$\frac{1}{2}$}} 
& \multicolumn{2}{D{1.6cm}|}{{\tiny$n$=$3$, $s_z$=$\frac{1}{2}$}}
& \multicolumn{1}{D{0.8cm}}{{\tiny $n$=$4$, $s_z$=$0$}}\\
\cline{2-17}
$\rho_{i,j}$ & \zero \!\zero & \zero \!\down & \down \!\zero & \zero \!\up &   \up \!\zero & \down \!\down &
 \zero \!\double & \down \!\up & \up \!\down & \double \!\zero 
 & \up \!\up &  \down \!\double  & \double \!\down  &\up \!\double &\double \!\up & \double \!\double\\
\hline
\hline
\zero \!\zero     & (1,1) & & & & & & & & & & & & & &  & \\
\hline
\zero \!\down     &  & (1,6) & (2,5) & & & & & & & & & & & & &\\
\down \!\zero     &  & (5,2) & (6,1) & & & & & & & & & & & & &\\
\hline
\zero \!\up       &  &     &  & (1,11) & (3,9) & & & & & & & & & & &\\
\up \!\zero       &  &     &  & (9,3)  & (1,11) & & & & & & & & & & &\\
\hline
\down \!\down     &  &     &  &      &   & (6,6) & & & & & & & & & &\\
\hline
\zero \!\double   &  &     &  &      &   &  & (1,16) & (2,15) & (3,14) & (4,13) & & & & & &\\
\down \!\up       &  &     &  &      &   &  & (5,12) & (6,11) & (7,10) & (8,9) & & & & & &\\
\up \!\down       &  &     &  &      &   &  & (9,8)  & (10,7) & (11,6) & (12,5) & & & & & &\\ 
\double \!\zero   &  &     &  &      &   &  & (13,4) & (14,3) & (15,2) & (16,1) & & & & & &\\
\hline
\up \!\up         &  &     &  &      &   &  &      &      &      &   & (11,11) & & & & &\\
\hline
\down \!\double   &  &     &  &      &   &  &      &      &      &   &   & (6,16) & (8,14) & & &\\
\double \!\down   &  &     &  &      &   &  &      &      &      &   &   & (14,8) & (16,6) & & &\\
\hline
\up \!\double     &  &     &  &      &   &  &      &      &      &   &   &      &   & (11,16) & (12,15) &\\
\double \!\up     &  &     &  &      &   &  &      &      &      &   &   &      &   & (15,12) & (16,11) &\\
\hline
\double \!\double &  &     &  &      &   &  &      &      &      &   &   &      &   &       &   & (16,16)\\
\hline\hline
\end{tabular}}
\caption{The two-orbital reduced density matrix $\rho_{ij}$ for SU(2) fermions
expressed in terms of single-orbital operators,
${\cal T}^{(m_i)}_i$ with $m_i=1,\dots,16$.
For better readability only the operator number indices $m$ are shown,
that is, $(m_i,m_j)$ corresponds to $\langle {\cal T}^{(m_i)}_{i}{\cal T}^{(m_j)}_{j}\rangle$.
Here $N_c$ and $S^z$ denote the particle-number and $z$ spin component quantum numbers of the two orbitals.}
\label{tab:fermion-ops-2}
\end{table}
Illustrating these, 
some generalized correlation functions are plotted for the LiF molecule in Fig.~\ref{fig:lif_correlations}.
As was mentioned in the begining of this section\cite{Barcza-2014,Fertitta-2014},
the generalized correlation functions (matrix elements for $\varrho_{ij}$)
are connected to the values of the mutual information $I_{ij}$,
which is plotted in Fig.~\ref{fig:I_opt} later.

\subsection{Methods based on block transfromation procedures}
\label{sec:num.block}

\subsubsection{Block renormalization group method (BRG)}
\label{sec:num.block.brg}

\begin{figure}
\centering
\setlength{\unitlength}{40pt}	
\begin{picture}(9,4)
\put(0,0){\includegraphics{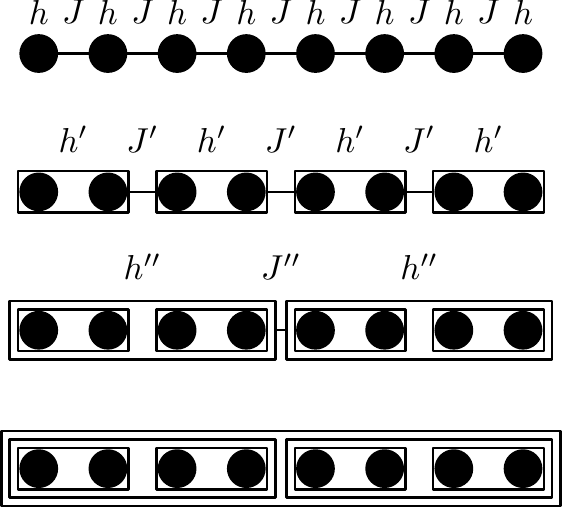}}
\put(5,0){\includegraphics{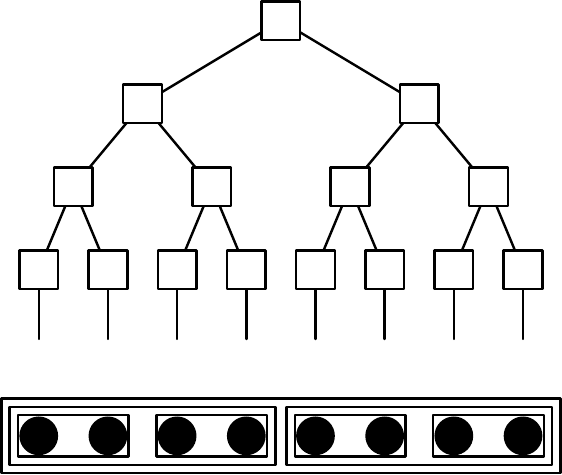}}
\put(0,4){\makebox(0,0)[r]{\strut{}(a)}}
\put(5,4){\makebox(0,0)[r]{\strut{}(b)}}
\end{picture}
\caption{Schematic plot of the Block Renormalization Group (BRG) method 
as bloc transfromation procedure where $h$ and $J$ label on-orbital and 
nearest neighbor interaction, respectively (a), and as a tree-network (b).} 
\label{fig:brg} 
\end{figure}

One of the first attempts to approximate the full configuration Hilbert 
space 
$\Lambda^{(d)}=\otimes_{i=1}^{d}\Lambda_i$ ($\dim \Lambda_i=q$)
of a $d$-orbital system  
goes back to the late 1960's when Kadanoff invented
the \emph{Block Spin Renormalization Group} method
and applied it to the two-dimensional Ising model\cite{Kadanoff-1966}.
This was later extended to quantum systems in one dimension called 
\emph{Block Renormalization Group} (BRG) method\cite{Drell-1977,Jullien-1978}.
%
%
The main idea of the method is to group $d_s$ number of orbitals into blocks.
The total Hamiltonian is then written as a sum of 
terms corresponding to the interactions within the blocks ({\em intrablock} Hamiltonian) and  
terms corresponding to the interactions between the blocks ({\em interblock} Hamiltonian). 
The unitary matrix $\mathbf{O}$ introduced in Sec.~\ref{sec:num.basic.toMPS} 
is formed from the $q$ lowest eigenstates of the intrablock Hamiltonian 
and operators are transformed to a new basis using Eq.~\ref{eq:transform}. 
Using the transformed operators the
interblock Hamiltonian can also be expressed. Truncating the Hilbert 
space of the blocks and keeping only $q$ 
states per block ensures that one can rescale the interaction strengths 
({\em flow equations}) and thus the original form of the Hamiltonian is 
retained.
In the next iteration step the $d_s$-blocks are collected.
The schematic plot of the procedure is shown in Fig~\ref{fig:brg}(a).
The procedure is repeated until 
subsequent iterations do not change the interaction strengths, i.e., 
until the so-called {\em fixed point} of the RG 
transformation is reached when measurable quantities corresponding to the 
$d\rightarrow\infty$ limit can be calculated. 
While this method gave reasonably good results for some one-dimensional 
models with local interactions, using such systematic change of basis and 
truncation led to loss of information in each iteration step and the 
accumulation of the error
hindered the application of the method for more complex problems.
In case of systems with finite number of orbitals this block transformation procedure 
can also be carried out until 
all orbitals are included in a single block 
and the approximated ground state energy can be calculated.
This corresponds to the root
for the Hierarchial Tucker format
discussed in Sec.~\ref{sec:tensor.ht}.
Due to the dramatic truncation of the states 
and non-local interactions this procedure cannot be applied efficiently in 
quantum chemistry.
However, the BRG method also serves the basis of hierarchal tensor
representation and tree tensor network state ansatz discussed in 
Sec.~\ref{sec:num.block.ttns}.
Recently, extension of the method known as 
the \emph{Multiscale Entanglement Renormalization Ansatz}\cite{Evenbly-2011} (MERA)
gave a new impetus to its application for strongly correlated systems. 

\subsubsection{Numerical renormalization group method (NRG)}
\label{sec:num.block.nrg}
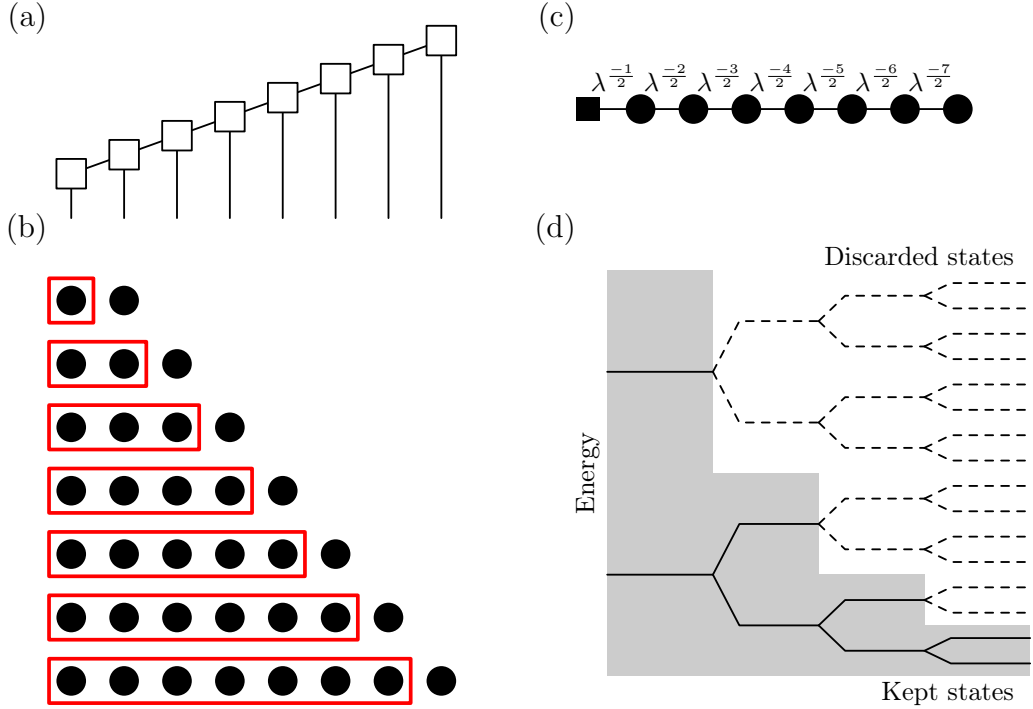
\begin{figure}[t]
\centering
\setlength{\unitlength}{40pt}   
\begin{picture}(9,6.5)
\put(0,0){\includegraphics{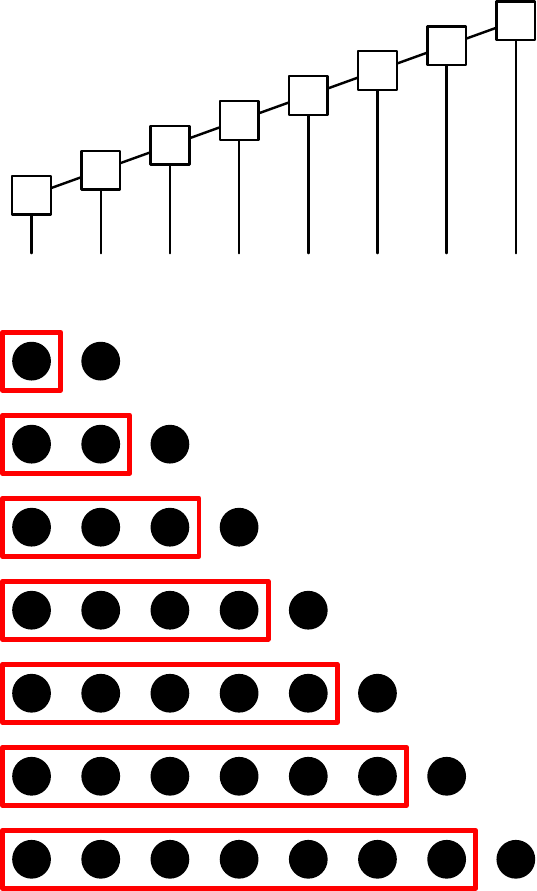}}
\put(5,0){\includegraphics{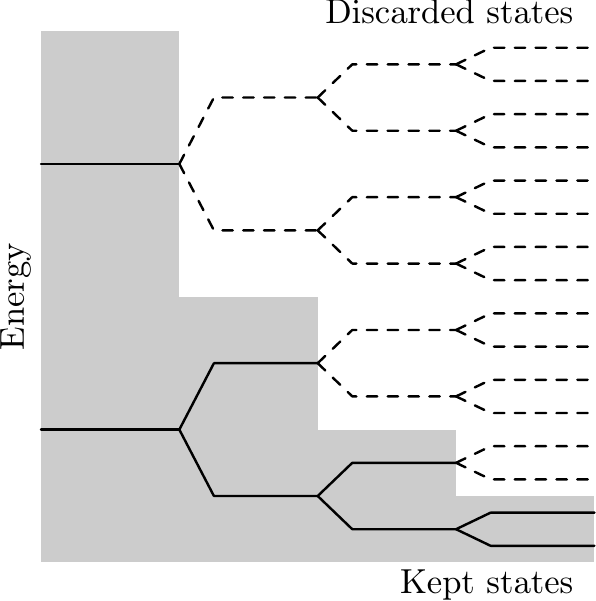}}
\put(5,5.5){\includegraphics{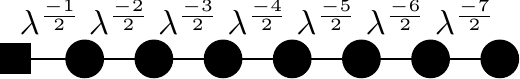}}
\put(0,6.5){\makebox(0,0)[r]{\strut{}(a)}} 
\put(0,4.5){\makebox(0,0)[r]{\strut{}(b)}} 
\put(5,6.5){\makebox(0,0)[r]{\strut{}(c)}}
\put(5,4.5){\makebox(0,0)[r]{\strut{}(d)}}
\end{picture}
\caption{Schematic plot of the Numerical Renormalization Group (NRG) method 
 a block-decimization procedure (b) leading to a tree-network (a).
Hamiltonian on the Wilson chain of length $d$: the hopping is decreasing exponentially (c).
A complete basis of a Wilson chain
represented as the exponentially increasing number of energy levels
belonging to the successive iterations.
Continuous/dashed lines represent kept, low-energy/discarded, high-energy levels, respectively. 
For the consecutive
iteration steps the distances between the levels illustrates how the
energy resolution of NRG gets exponentially refined (d).
} 
\label{fig:nrg} 
\end{figure}

Another variant of the RG method, 
known as the \emph{Numerical Renormalization Group} (NRG) method 
shown in Fig.~\ref{fig:nrg}  is due to Wilson\cite{Wilson-1975}. 
In the NRG related Hamiltonian
an impurity interacts with a local fermion. The 
dynamics of this fermion is described by a
semi-infinite one dimensional network, also know as the Wilson chain.
The impurity sits on the left 
side and electrons can move along the chain 
with an exponentially
decreasing hopping amplitude $\lambda^{-j/2}$.
Therefore, each orbital represents a different energy scale. Starting
with the very left orbital, new blocks including $l$ orbitals are formed 
by adding orbitals systematically to the block, i.e., 
$\Xi^{(\mathrm{L})}=\Xi^{(\mathrm{l})}\otimes \Lambda_{l+1}$ where
in the first step $\Xi^{(\mathrm{l})}=\Lambda_1$.
In each iteration step the block Hamiltonian is solved and the unitary transformation matrix $\mathbf{O}$ is formed
from eigenstates corresponding to the lowest $M$ eigenvalues. The block Hamiltonian is rescaled based on the 
decay rate of the hopping and the intrablock Hamiltonian is determined on the new basis.  
Another major difference compared to the BRG method
is that in NRG $q<M\ll q^d$ states are kept, thus the original form of the Hamilton is lost. Due to the appearance
of new operators during the iteration scheme flow equations described above cannot be studied.
The change in the energy spectrum, however, can be analyzed and once subsequent iterations leave the spectrum unchanged
the fix point is reached. This approach works well due to the separation of energy scales. A problem, however,
arises for lattice models when $\lambda\rightarrow 1$ and error starts
to accumulate significantly for increasing block size. 
This hindered the application of NRG to large lattice models. 
Quite recently, an extension of the method using a similar blocking structure as in DMRG 
has led to the development of the so called density matrix numerical renormalization group (DM-NRG)
which allows us to study more complex problems\cite{Hofstetter-2000,Anders-2005,Peters-2006,Toth-2008,Weichselbaum-2007}.

\subsubsection{Density matrix renormalization group method (DMRG)} 
\label{sec:num.block.dmrg}
In order to circumvent problems discussed for BRG and NRG, 
in the two-site variant of the \emph{Density Matrix Renormalization Group} (DMRG) 
method\cite{White-1992b}
$\Lambda^{(d)}$ is approximated by a tensor product space of four tensor spaces, i.e.,
$\Xi^{(d)}_{\rm DMRG}=\Xi^{(\mathrm{l})}\otimes\Lambda_{l+1}\otimes\Lambda_{l+2}\otimes\Xi^{(\mathrm{r})}$.
This is called superblock and
the basis states of the blocks are optimized by successive application of 
the singular value decomposition as discussed in Secs.~\ref{sec:tensor.tt} 
and \ref{sec:num.ent.toSVD}. 
Here we use the convenient notations that
the whole system, consisting of $d$ orbitals $1,2,\dots d$, is partitioned into blocks (subsystems),
for which we use the labels $(\mathrm{L})$, $(\mathrm{l})$, $(\mathrm{R})$ and $(\mathrm{r})$.
$(\mathrm{l})$ simply means the block composed of the first $l$ orbitals, that is, $(\mathrm{l})=\{1,2,\dots,l\}$.
An extended block composed of the first $l+1$ orbitals is denoted as $(\mathrm{L})=\{1,2,\dots,l,l+1\}$.
The other part of the system is $(\mathrm{R})=\{l+2,l+3,\dots,d\}$, while $(\mathrm{r})=\{l+3,\dots,d\}$.
The $d$-orbital wavefunction is, therefore, written as
\begin{equation}
 |\Psi_{\rm DMRG}\rangle = \sum_{m_{(\mathrm{l})} \alpha_{l+1} \alpha_{l+2} m_{(\mathrm{r})}} 
U_{\rm DMRG}(m_{(\mathrm{l})}, \alpha_{l+1}, \alpha_{l+2}, m_{(\mathrm{r})})
 |\xi^{(\mathrm{l})}_{m_{(\mathrm{l})}}\rangle \otimes 
 |\phi^{\{l+1\}}_{\alpha_{l+1}}\rangle \otimes 
 |\phi^{\{l+2\}}_{\alpha_{l+2}}\rangle \otimes  
 |\xi^{(\mathrm{r})}_{m_{(\mathrm{r})}}\rangle
\label{eq:psi-dmrg}
\end{equation}
where the tensor $U_{\rm DMRG}$ is 
determined by an iterative diagonalization of 
the corresponding so called superblock Hamiltonian.
The dimensions of the spaces of the local \emph{left block} including $l$ orbitals 
and the \emph{right block} with $r=d-l-2$ orbitals 
are denoted with $M_l={\rm \dim}\ \Lambda^{(\mathrm{l})}$ and 
$M_r={\rm \dim}\ \Lambda^{(\mathrm{r})}$, respectively.
Since ${\rm \dim}\ \Lambda_{l+1}={\rm \dim}\ \Lambda_{l+2} = q$, the 
resulting dimensionality of 
the DMRG wave function is 
${\rm \dim}\ \Xi^{(d)}_{\rm DMRG}=q^2M_lM_r\ll q^d$.
\begin{figure}
\centering
\includegraphics{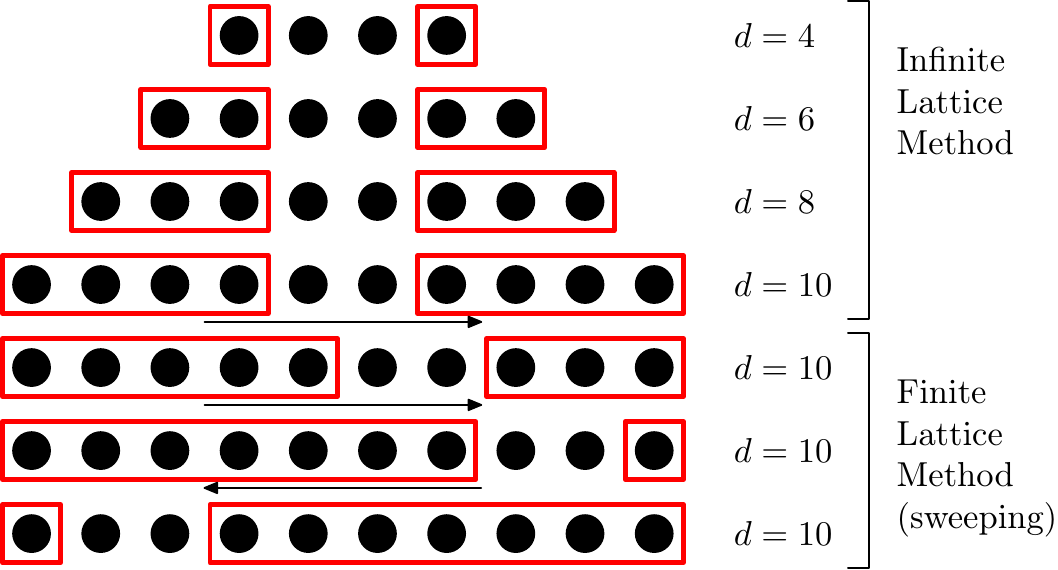}
\caption{Decomposition of the $d$-orbital Hilbert space into four subsystems 
called superblock. The $d$-orbital Hilbert space is built
iteratively from a left block including $l$ active orbitals and the right 
block from $r$ active orbitals. The size of the two blocks is 
increased in each iteration step until $l+2+r=d$. 
In the following steps the $d$-orbital system is partitioned 
asymmetrically, i.e. the size of left block is increased systematically 
while the size of the right block is decreased until $l=d-3$ and $r=1$. 
The same procedure is repeated in the opposite direction until $l=1$ and 
$r=d-3$. This procedure is called sweeping (macro-iteration step).}
\label{fig:superblocka} 
\end{figure}
\begin{figure}[t]
\centering
\includegraphics{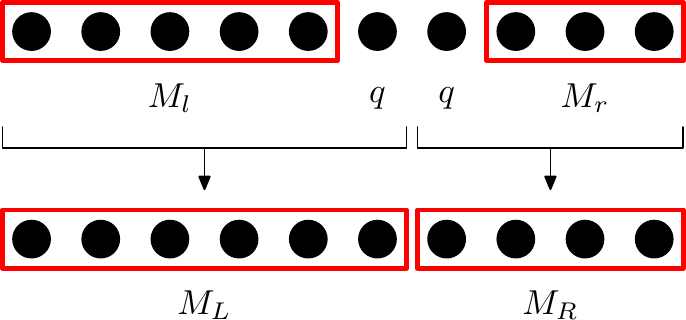}
\caption{Schematic plot of a DMRG iteration step in order
to increase block sizes and form a bipartite representation for 
the singular value decomposition.
$(\mathrm{l})$ and $(\mathrm{r})$ denote the left and right block
of length $l$ and $r$, and of dimension $M_l$ and $M_r$, 
respectively, $\bullet$ stands for the
intermediate orbitals ($\{l+1\}$ and $\{l+2\}$) with dimension $q$.
The blocks $(\mathrm{L}) = (\mathrm{l})\bullet, (\mathrm{R}) = \bullet (\mathrm{r})$ have dimension
$M_L$ and $M_R$, respectively.}
\label{fig:superblockb} 
\end{figure}

In the \emph{original version} of the DMRG, 
introduced to treat finite one-dimensional lattice models\cite{White-1992b}, 
the Hilbert space of a lattice with $d$ sites 
is built iteratively starting with four sites as shown in Fig.~\ref{fig:superblocka}.  
In each iteration step, the Hilbert space 
$\Xi^{(\mathrm{L})}$ 
of an enlarged 
block $(\mathrm{L})$ is formed from the tensor product of the Hilbert spaces  
of the block $\Xi^{(\mathrm{l})}$ and the adjacent site $\Lambda_{l+1}$ 
-- similarly  $\Xi^{(\mathrm{R})}$ from $\Lambda_{l+2}$ and $\Xi^{(\mathrm{r})}$ --
and transformed to a new {\em truncated} basis by using a unitary operation
based on singular value decomposition as discussed in section \ref{sec:num.ent.toSVD}. 
Therefore, in each iteration step the size of the effective system
is increased by two until the desired length $d$ is achieved. This procedure 
is called \emph{infinite-lattice procedure}. 
In the following steps the $d$-site system is partitioned asymmetrically, 
i.e. the size of left block is increased systematically while
the size of the right block is decreased until $l=d-3$ and $r=1$. In 
each iteration step, the approximated 
Hilbert space of the left block (called {\em system block}) is improved 
as it interacts with the right block (called {\em environment}). 
The same procedure is repeated in the opposite direction until 
$l=1$ and $r=d-3$ when the left block becomes the environment 
block and the right block the system block. 
This procedure is called \emph{sweeping} (macro-iteration)
and it is a part of the so called \emph{finite-lattice method.}
For more detailed derivations we refer to the original papers and review articles\cite{White-1992b,White-1993,Schollwock-2005}.

In analogy, in the infinite-lattice procedure one can say that the $d$-orbital Hilbert space is built 
iteratively by forming 
$l$-orbital and $r$-orbital blocks from the one-orbital Hilbert spaces  
starting with an \emph{``active space''} including only four orbitals. In each 
iteration step the number of active orbitals
is increased by two until all the $d$ orbitals become active, 
i.e., part of either the left or right block.
This procedure serves as the initialization of the MPS
network with $d$ component tensors. When the network is formed the
elements of the $\mathrm{A}_i$ matrices are random numbers. The infinte 
lattice method can be viewed as a procedure to start with four 
``active'' component 
tenors by setting the remaining $d-4$ component tensors to trivial.
This means that
the $m_{i-1}$ and $m_i$ indices of the corresponding 
$A_i(m_{i-1},\alpha_i,m_i)$ takes only the value $1$, and 
$A_i(1,\alpha_i,1)=\delta_{1,\alpha_i}$ that is, $A_i(1,1,1)=1$ and the others 
are $0$. 
In each
iteration step, the number of ``active'' component tensors is increased
by two until no component tensors are set to trivial.

\begin{figure}
\centering
\includegraphics[scale=0.9]{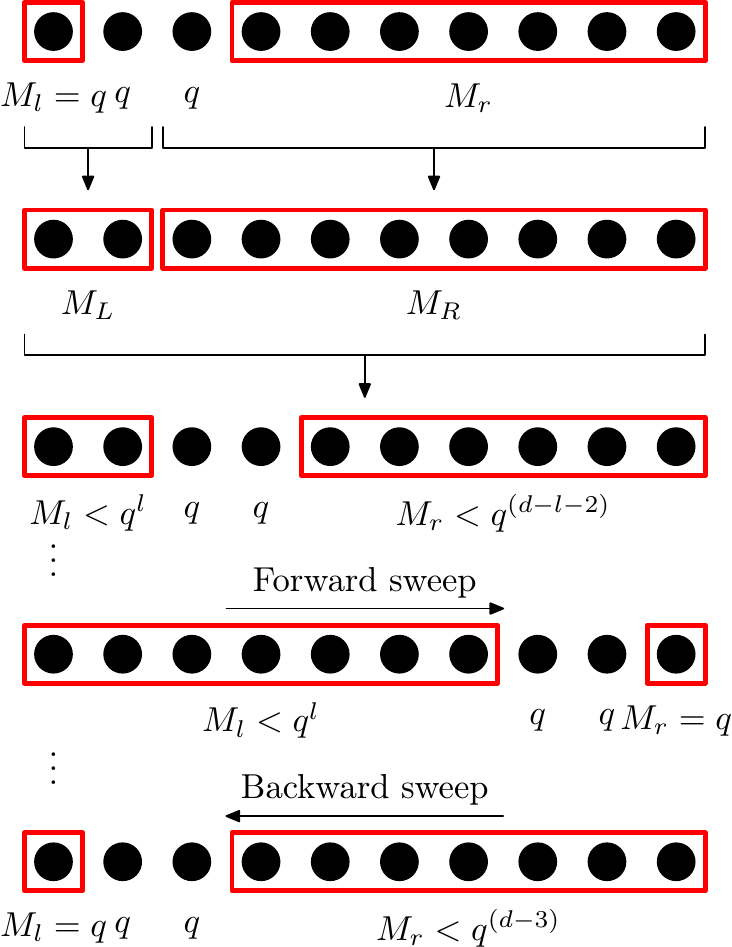}
\caption{A modified initialization of the tensor network with $d$
orbitals used in the Dynamically Extended Active Space (DEAS) procedure.
In the DEAS procedure one starts with a superblock structure
with $l=1$ and $r=d-3$ and use an approximated Hilbert space 
with diemnsion $M_r\ll q^r$.}
\label{fig:DEAS}
\end{figure}

In quantum chemistry, it is more efficient to start with 
an initial network which already corresponds to the finite system 
with $d$ orbitals as has been introduced through the \emph{Dynamically 
Extended Active Space} (DEAS) procedure\cite{Legeza-2003b}.
In the DEAS procedure one starts with a superblock structure
with $l=1$ and $r=d-3$, as is shown in Fig.~\ref{fig:DEAS}, 
and carries out the forward and backward sweeping procedure, i.e.,
the finite lattice method as described above.  
A cruical problem, however, is that during the first sweep when 
the left block is optimized the right block Hilbert space
has to be approximated with $M_r\ll q^{r}$ basis states. 
An efficient method to carry out such optimization will be discussed 
in Sec.~\ref{sec:num.optim.opt-init} based on the Configuration 
Interaction (CI) procedure.

Let us highlight the main aspect of DMRG procedure once again:
If one could represent the Hilbert spaces of the 
four subsystems used in the two-site DMRG exactly using one-orbital 
basis states then in the first step of the DEAS procedure this would mean 
$M_l=q$ and $M_r=q^{d-3}$ and
$\Lambda^{(d)}_{\rm DMRG}=\Lambda^{(\mathrm{l})}\otimes\Lambda_{l+1}\otimes\Lambda_{l+2}\otimes\Lambda^{(\mathrm{r})}$.
By traversing through the system back-and-forth the left and right block
Hilbert spaces are transformed and truncated,
and after a full sweep the approximated subspace is given as
$\Xi^{(d)}_{\rm DMRG}=\Xi^{(\mathrm{l})}\otimes\Lambda_{l+1}\otimes\Lambda_{l+2}\otimes\Xi^{(\mathrm{r})}$.
Therefore, 
the $d$-orbital wavefunction written in terms of one-orbital basis is converted 
to an approximated multi-orbital basis in 
$\Xi^{(d)}_{\rm DMRG}=\Xi^{(\mathrm{l})}\otimes\Lambda_{l+1}\otimes\Lambda_{l+2}\otimes\Xi^{(\mathrm{r})}$,
where $\dim \Xi_{\rm DMRG}^{(d)}\ll\dim\Lambda^{(d)}$ 
depending on the level of truncation.

A main difference compared to the BRG and NRG methods is how
the transformation matrix $\mathbf{O}$ is constructed. 
In a given iteration step 
(see Fig.~\ref{fig:superblockb})
the $(\mathrm{l})\bullet$ composite system is combined to one subsystem 
$(\mathrm{L})$ with $\Xi^{(\mathrm{L})}=\Xi^{(\mathrm{l})}\otimes\Lambda_{l+1}$ 
and $\bullet (\mathrm{r})$ to another one $(\mathrm{R})$ 
with $\Xi^{(\mathrm{R})}=\Lambda_{l+2}\otimes\Xi^{(\mathrm{r})}$.
This leads to $\Xi^{(\mathrm{L})}\otimes\Xi^{(\mathrm{R})}=\Xi^{(d)}_{\rm DMRG}\subseteq \Lambda^{(d)}$ and
the bipartite representation of the wavefunction is formed as 
\begin{equation}
|\Psi_{\rm DMRG} \rangle = \sum_{m_{(\mathrm{L})}m_{(\mathrm{R})}} 
U_{\rm DMRG}(m_{(\mathrm{L})},m_{(\mathrm{R})}) |\phi^{(\mathrm{L})}_{m_{(\mathrm{L})}}\rangle \otimes 
|\phi^{(\mathrm{R})}_{m_{(\mathrm{R})}}\rangle. 
\label{eq:bipartite}
\end{equation}
According to 
section \ref{sec:num.ent.toSVD},
using singular value decomposition it can be written as a single sum of 
tensor products.  
The new basis states $|\xi^{(\mathrm{L})}_{m_{(\mathrm{L})}}\rangle$ and $|\xi^{(\mathrm{R})}_{m_{(\mathrm{R})}}\rangle$ given in 
Eq.~(\ref{eq:schmidt}) are obtained
by diagonalizing the reduced subsystem density matrices $\rho^{(\mathrm{L})}$ and $\rho^{(\mathrm{R})}$, see Eq.~(\ref{eq:rhoA}).
The transformation matrix $\mathbf{O}$ introduced in section \ref{sec:num.basic.toMPS} is 
formed from eigenstates $|\xi^{(\mathrm{L})}_{m_{(\mathrm{L})}}\rangle$ 
(or $|\xi^{(\mathrm{R})}_{m_{(\mathrm{R})}}\rangle$) corresponding to the 
$M_l^{\rm kept}\leq M_lq$ (or $M_r^{\rm kept}\leq M_rq$)   
largest eigenvalues $\omega_m$. 
Due to the truncation of basis states the so-called {\em truncation error}
is defined as the sum of  
the truncated number of eigenvalues of the reduced subsystem 
density matrix deviates from unity, i.e.,
\begin{equation}
\label{eq:errtrunc}
\delta \varepsilon_{\rm TR} = 1-\sum_{m=1}^{M_{\rm kept}} \omega_m.
\end{equation}
Operators of the enlarged blocks 
are transformed to this new basis as 
$(\mathbf{X}_i\mathbf{Y}_{l+1})^{(l+1)} = \mathbf{O} (\mathbf{X}^{(\mathrm{l})}_i\otimes \mathbf{Y}^{\{l+1\}}_{l+1}) \mathbf{O}^\dagger$,
where $\mathbf{X}_i$ and $\mathbf{Y}_{l+1}$ are $M_l\times M_l$ and $q\times q$ matrices, respectively. 
The number of block states, $M_l$ and $M_r$, required to achieve sufficient convergence can be regarded 
as a function of the level of entanglement among the molecular orbitals. Hence the maximum number of 
block states $M_{\rm max} = \max{(M_l,M_r)}$\ determines the accuracy of a DMRG calculation\cite{Legeza-2003a,Schollwock-2005}
as will be investigated in the next section.

If the transformation matrix $\mathbf{O}$ in each iteration step is reindexed according to the procedure explained 
in Sec.~\ref{sec:num.basic.toMPS} and the corresponding $\mathbf{B}$ matrices are stored within a full sweep then the 
DMRG wavefunction for a given superblock partitioning can be written in MPS form\cite{Ostlund-1995,Verstraete-2004b,Verstraete-2008} as
\begin{equation}
\begin{split}
|\Psi_{\rm DMRG}\rangle = \sum_{m_{(\mathrm{l})} \alpha_{l+1} \alpha_{l+2} m_{(\mathrm{r})}}
 &U_{\rm DMRG}(m_{(\mathrm{l})},\alpha_{l+1},\alpha_{l+2},m_{(\mathrm{r})})\\ \times
&(\mathbf{B}_l(\alpha_l)\ldots \mathbf{B}_{2}(\alpha_2))_{m_{(\mathrm{l})};\alpha_1}
(\mathbf{B}_{l+3}(\alpha_{l+3})\ldots \mathbf{B}_{d-1}(\alpha_{d-1}))_{m_{(\mathrm{r})};\alpha_d}\\ \times
&|\phi^{\{1\}}_{\alpha_1}\rangle \otimes\dots\otimes 
|\phi^{\{l+1\}}_{\alpha_{l+1}}\rangle\otimes
|\phi^{\{l+2\}}_{\alpha_{l+2}}\rangle\otimes\dots\otimes
|\phi^{\{d\}}_{\alpha_d}\rangle.
\label{eq:full-tensor}
\end{split}
\end{equation}
Therefore, DMRG can be viewed as an efficient method to generate the 
optimized set of $\mathbf{A}_i$ ($\mathbf{B}_i$) matrices
used to construct the MPS representation of the $d$-orbital wavefunction. 
Since in this representation the $d$-orbital wavefunction is written as a 
linear combination of the tensor product of the 
one-orbital basis (CI coefficients), it allows one to connect the DMRG 
wavefunction to conventional quantum chemical techniques.
For example, the CI-coefficients of the most relevant terms can be 
determined\cite{Boguslawski-2011}.

Concluding this section, 
different one-dimensional representation of tensor network state algorithms, 
i.e, matrix product state methods have been developped in the various 
communities.
In this one-dimensional optimization scheme the network is built from matrices. The TT and MPS approaches
are ``wavefunction'' oriented description of the problem while DMRG is 
more like an ``operator'' representation of the problem. 
In the TT and MPS the physical indices 
are for local $q$ dimensional tensor spaces 
thus operators are $q\times q$ matrices 
but the $\mathbf{A}(\alpha)$ matrices must be stored.
The norm is calculated by simply connecting the physical indices vertically. 
In contrast to this, in the DMRG description
when the network is separated to a left and a right part,
the operators of the left and right part are represented on a multi-orbital tensor space 
of dimension $M_l$ and $M_r$, respectively, 
where both are much larger than $q$.
Therefore the corresponding matrices of dimensions $M_l\times M_l$ and 
$M_r\times M_r$ must be stored during the iterative minimization procedure. 
In the quantum chemistry framework long range Coulomb 
interactions are given by the 4-th order
tensor $V_{ijkl}$ of equation (\ref{eq:doubleint}) thus the number of renormalized operators scales as $\mathcal{O}(d^4)$. 
Using, however, an efficient factorization of 
the interaction terms distributed among the various subsystems\cite{Xiang-1996},
this scaling can be reduced to $\mathcal{O}(d^2)$, see in section \ref{sec:num.block.factor}.
Therefore, the required memory to store operators in a given QC-DMRG iteration step assuming $M_l=M_r\equiv M$ is $\mathcal{O}(M^2d^2)$. 
The computational cost of a given QC-DMRG step scales as $\mathcal{O}(M^3d^2)$ and for a full sweep $\mathcal{O}(M^3d^3)$. 
A main advantage of the DMRG method is, however, that in each iteration step the core tensor is 
optimized so orthogonalization of the left and
right block states are guaranteed\cite{Schollwock-2005}.

\subsubsection{Higher dimensional network: Tree tensor network state (TTNS)}
\label{sec:num.block.ttns}

\begin{figure}
\centering
\setlength{\unitlength}{40pt}   
\begin{picture}(10,3.3)
\put(0,0){\includegraphics{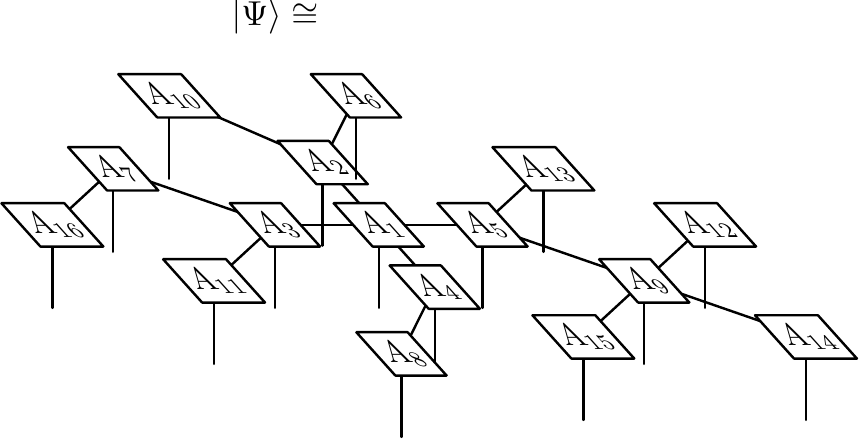}}
\put(7,1){\includegraphics{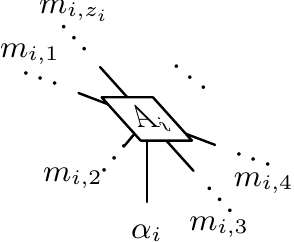}}
\put(0,3.3){\makebox(0,0)[r]{\strut{}(a)}}
\put(7,3.3){\makebox(0,0)[r]{\strut{}(b)}}
\end{picture}
\caption{Schematic plot of a higher dimensional network, for example, 
the tree tensor network state (TTNS).
Each node is represented by a tensor $\mathbf{A}_i$ of order $z_i+1$, with
$z_i$ is an orbital dependent coordination number. The network supposed
to reflect the entanglement structure of the molecule as much as possible.
The vertical  
lines are the physical indices $\alpha_i$, $i\in \{1,d\}$,
while the others that connect the orbitals are virtual ones.}
\label{fig:ttns}
\end{figure}

A natural extension of the MPS approach is to 
form an ansatz state by contracting a network of
higher order tensors\cite{Verstraete-2004a,Murg-2007,Murg-2009,Verstraete-2008,Vidal-2008,Changlani-2009,Marti-2010d,Marti-2011,Legeza-2013,Legeza-2014,Hackbusch-2010,Orus-2014},
as discussed in Sec.~\ref{sec:tensor}.
A special class of such ansatz states are the 
\emph{Tree Tensor Network States} (TTNS)\cite{Corboz-2009,Murg-2010a,Nakatani-2013,Murg-2014}
which are formed by contracting tensors according to a tree network,
as shown in Fig.~\ref{fig:ttns}.
The structure of the
tree network can be arbitrary and the coordination number can vary from
site to site.
Each tensor in the network represents a physical orbital
and is of order $z_i+1$, were $z_i$ describing the coordination number
of site $i$:
\begin{equation}
A_i(\alpha_i,m_{i,1},\ldots,m_{i,z_i}).
\end{equation}
The $z_i$ virtual indices $m_{i,1},\ldots,m_{i,z_i}$ are of dimension~$M$ and 
are contracted as the TTNS is formed.
The physical index~$\alpha_i$ is of dimension~$q$ and describes the
physical state of the orbital, e.g. the
number of up- and down-electrons on that orbital.

The TTNS is especially suitable  
to treat models in which orbitals have varying
degrees of entanglement (see Figs.~\ref{fig:s_i} and \ref{fig:I_free}):
since entanglement is transferred via the virtual bonds that
connect the sites,
sites with a larger coordination number
are better suited to represent higher entanglement.
In this way, the coordination number can be adapted according to
the entanglement of the orbitals,
and the orbitals can be arranged on the tree such that
highly entangled orbitals are close together (see later in section~\ref{sec:num.optim.opt-topology}).

An additional motivation for using a tree structure
is to take advantage of the property of the
tree tensor network ansatz that the long-range correlations differ
from the mean-field value polynomially with distance rather than
exponentially as for MPS\cite{Murg-2010a}. 
This is due to the fact that
the number of virtual bonds required to connect two arbitrary orbitals
scales logarithmically with the number of orbitals~$d$ for $z>2$,
as can be seen by
considering a Cayley-tree 
of depth~$\Delta$:
the number of sites in the tree is
\begin{equation}
d = 1+z \sum_{j=1}^{\Delta} (z-1)^{j-1} = \frac{z(z-1)^{\Delta}-2}{z-2}
\end{equation}
and thus, the maximal distance between two orbitals, $2 \Delta$, scales logarithmically with~$d$ for $z>2$.
On the other hand, for $z=2$ the number of virtual bonds required to connect two
arbitrary orbitals scales linearly in~$d$.

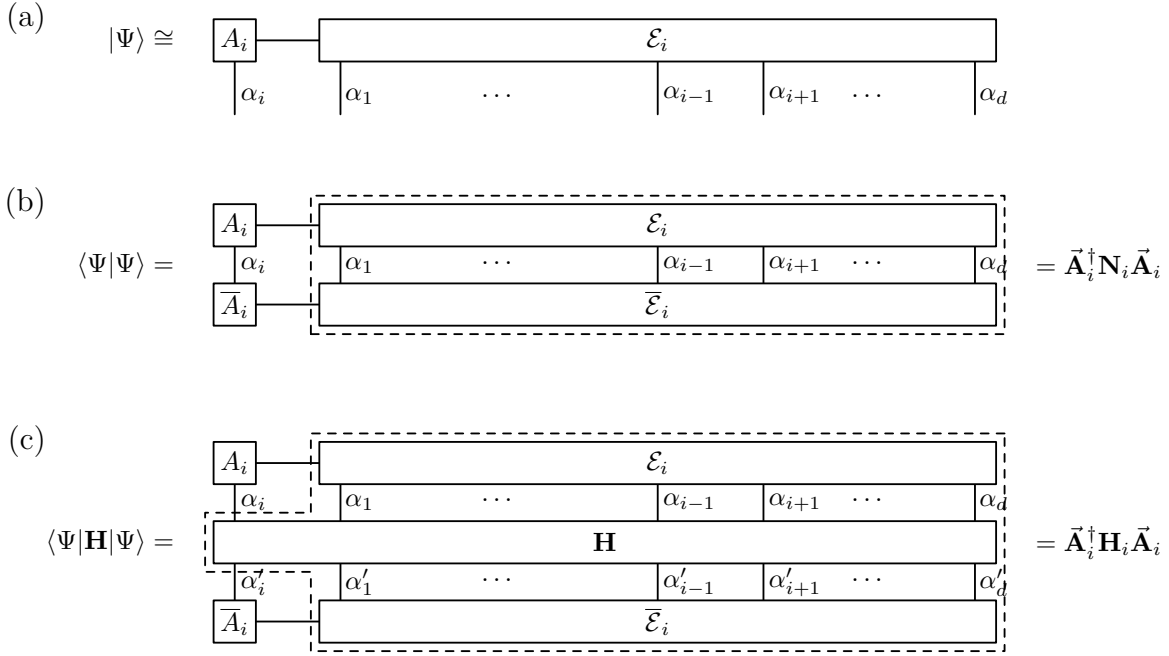
\begin{figure}[t]
\centering
\setlength{\unitlength}{40pt}   
\begin{picture}(10.5,6)
\put(0,0){\includegraphics{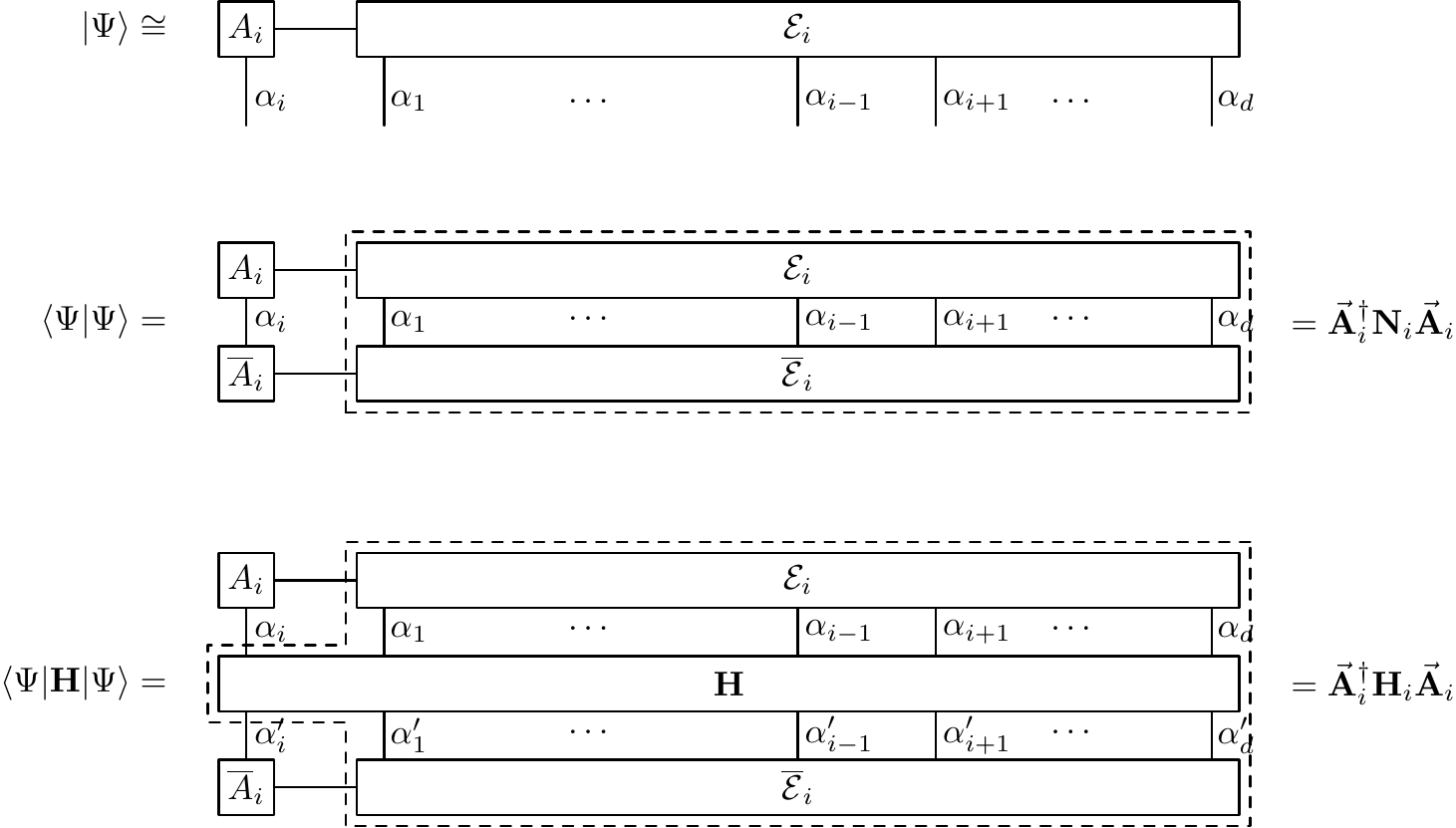}}
\put(0,6){\makebox(0,0)[r]{\strut{}(a)}}
\put(0,4.25){\makebox(0,0)[r]{\strut{}(b)}}
\put(0,2){\makebox(0,0)[r]{\strut{}(c)}}
\end{picture}
\caption{Concept of the variational optimization of tensor network states:
(a) tensor network state $|\Psi\rangle$ separated into two parts: the tensor $\mathbf{A}_i$ that is supposed to
be optimized and an environment tensor $\Varepsilon_i$ that
is formed by contracting all tensors except $\mathbf{A}_i$.
(b) norm $\langle\Psi|\Psi\rangle$ of the tensor network state defined in (a);
the norm equals to $\vec{\mathbf{A}}_i^{\dagger} \mathbf{N}_i \vec{\mathbf{A}}_i$ with
$\vec{\mathbf{A}}_i$ corresponding to the $q M^{z_i}$-dimensional vector
obtained by joining all indices of tensor $\mathbf{A}_i$,
and $\mathbf{N}_i$ represents the effective environment, drawn with dashed lines.
(c) expectation value $\langle\Psi| H |\Psi\rangle$ of $H$ with respect to the
tensor network state defined in (a);
the expectation value equals to $\vec{\mathbf{A}}_i^{\dagger} \mathbf{H}_i \vec{\mathbf{A}}_i$
with $\mathbf{H}_i$ representing the effective Hamiltonian, drawn with dashed lines.}
\label{fig:ttns_fig1}
\end{figure}

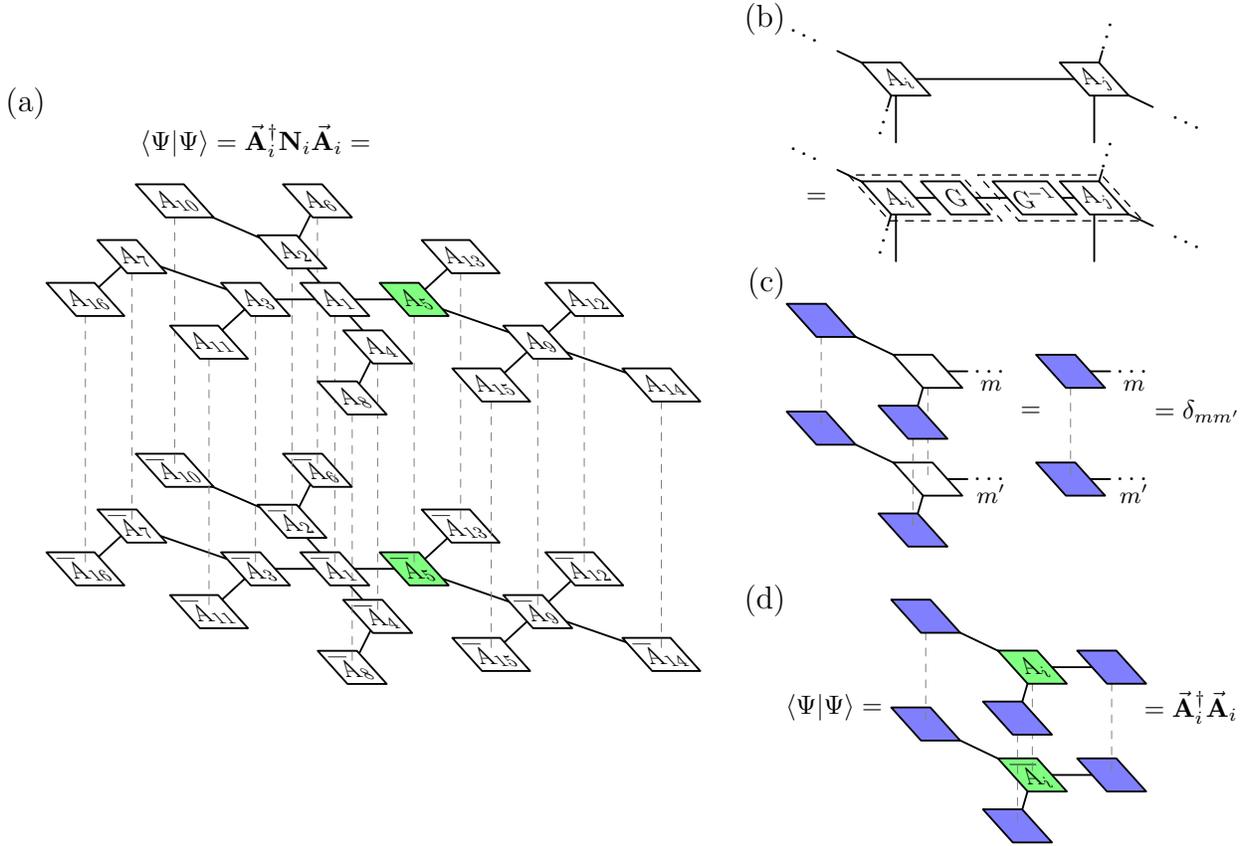
\begin{figure}[t]
\centering
\setlength{\unitlength}{40pt}   
\begin{picture}(11,8)
\put(0,1.5){\includegraphics{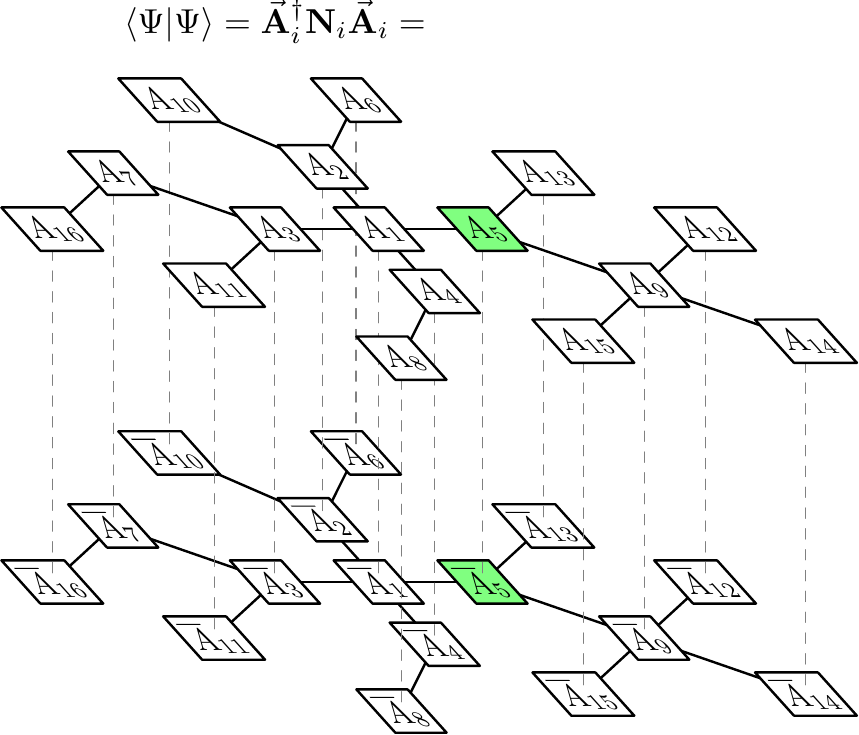}}
\put(7,5.5){\includegraphics{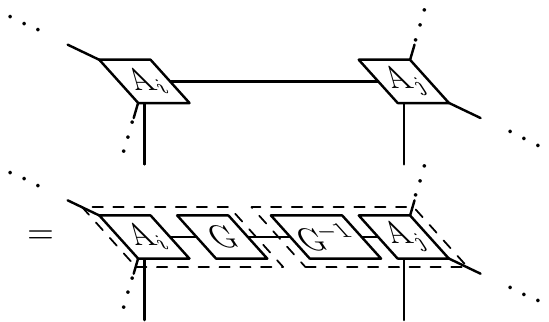}}
\put(7,2.8){\includegraphics{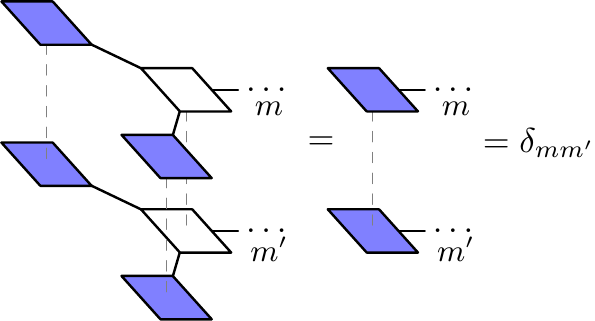}}
\put(7,0){\includegraphics{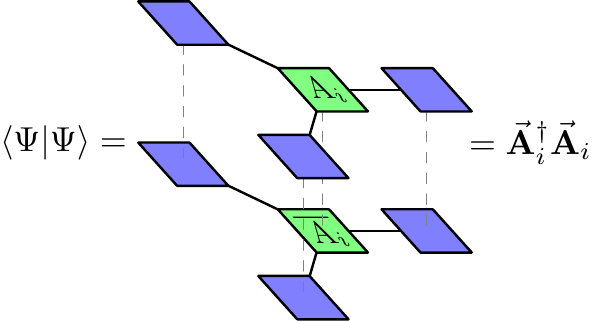}}
\put(0,7){\makebox(0,0)[r]{\strut{}(a)}}
\put(7,7.8){\makebox(0,0)[r]{\strut{}(b)}}
\put(7,5.3){\makebox(0,0)[r]{\strut{}(c)}}
\put(7,2.3){\makebox(0,0)[r]{\strut{}(d)}}
\end{picture}
\caption{ 
(a) norm $\langle\Psi|\Psi\rangle$ of the TTNS defined in Fig.~\ref{fig:ttns};
the tensor network picture of the norm corresponds to a two-layer structure,
with the ket $|\Psi\rangle$ being on top and the bra $\langle\Psi|$ on bottom.
For better readability, the contracted physical indices are drawn with dashed lines.
(b) gauge transformation in a tensor network state: the state remains
invariant if matrices $G$ and $G^{-1}$ are inserted at one bond and
merged with the adjacent tensors.
(c) orthonormalization condition imposed on all tensors (blue) except $\mathbf{A}_i$ 
(green).
(d) norm of the TTNS with all tensors except $\mathbf{A}_i$ fulfilling
the orthonormalization condition.
}
\label{fig:ttns_fig2}
\end{figure}

In the algorithmic approach to optimize the TTNS, one can use tools 
known in literature\cite{Shi-2006,Tagliacozzo-2007,Corboz-2009,Corboz-2010}
and optimize the network site-by-site as in the DMRG.
The fact that the tree tensor network does not contain any loops
allows an exact mathematical treatment\cite{Hackbusch-2010,Legeza-2014}
(see in section \ref{sec:tensor.networks}).
For $z=2$, the DMRG algorithm is recovered.
The TTNS algorithm is similar to a
DMRG calculation with~$z$ blocks instead of two,
where a block consists of all of the sites within one of the branches
emerging from site~$i$ (see Fig.~\ref{fig:ttns_fig2}(a)).

As in DMRG, the TTNS algorithm consists in the variational optimization of the tensors $\mathbf{A}_i$
in such a way that the energy is minimized (with the constraint that the norm of the state remains constant).
This is equivalent to
minimizing the functional
\begin{equation}
F = \langle\Psi| \mathbf{H} |\Psi\rangle - E \left( \langle\Psi|\Psi\rangle - 1 \right),
\end{equation}
with $|\Psi\rangle=|\Psi(\mathbf{A}_1,\ldots, \mathbf{A}_d)\rangle$.
This functional is non-convex with respect to all parameters
$\{\mathbf{A}_1,\ldots,\mathbf{A}_d\}$.
However, by fixing all tensors $\mathbf{A}_k$ except $\mathbf{A}_i$,
due to the tensor network structure of
the ansatz, it is quadratic in the parameters $\mathbf{A}_i$ associated with
one lattice site~$i$.

As depicted in Fig.~\ref{fig:ttns_fig1}(a), the tensor network state 
can be separated in two parts: the tensor $\mathbf{A}_i$ that is supposed to
be optimized and an environment tensor $\Varepsilon_i$ that
is formed by contracting all tensors except $\mathbf{A}_i$.
$\mathbf{A}_i$ is connected to the environment tensor $\Varepsilon_i$ by
$z_i$ virtual bonds, with $z_i$ being the coordination number of site~$i$.
Using this separation, it is evident that
$\langle\Psi|\Psi\rangle = \vec{\mathbf{A}}_i^{\dagger} \mathbf{N}_i \vec{\mathbf{A}}_i$
and
$\langle\Psi|\mathbf{H} |\Psi\rangle = \vec{\mathbf{A}}_i^{\dagger} \mathbf{H}_i \vec{\mathbf{A}}_i$,
as shown in Fig.~\ref{fig:ttns_fig1}(b) and (c).
$\vec{\mathbf{A}}_i$ is thereby the
reshaped $q \times M \times \cdots \times M$-tensor $\mathbf{A}_i$
into a $q M^{z_i}$-dimensional vector.
The inhomogenity $\mathbf{N}_i$ and the effective Hamiltonian $\mathbf{H}_i$
with respect to site $i$ are matrices of size $q M^{z_i} \times q M^{z_i}$
that are obtained by contracting all tensors except $\mathbf{A}_i$ 
in the tensor expressions for
$\langle\Psi|\Psi\rangle$ and $\langle\Psi|\mathbf{H} |\Psi\rangle$, respectively
(see Fig.~\ref{fig:ttns_fig1}(b) and (c)).

The optimal parameters $\mathbf{A}_i$ can be found by minimizing the quadratic function
\begin{equation}
F(\vec{\mathbf{A}}_i) = \vec{\mathbf{A}}_i^{\dagger} \mathbf{H}_i \vec{\mathbf{A}}_i - E \left( \vec{\mathbf{A}}_i^{\dagger} \mathbf{N}_i \vec{\mathbf{A}}_i - 1 \right),
\end{equation}
which is equivalent to
solving the generalized eigenvalue problem
\begin{equation}
\mathbf{H}_i \vec{\mathbf{A}}_i = E \mathbf{N}_i \vec{\mathbf{A}}_i.
\end{equation}

For a network without loops, it is always possible
to set $\mathbf{N}_i$ equal to the identity, which
accounts for numerical stability
because the generalized eigenvalue problem reduces to an ordinary one.
The reason for this possiblity is the gauge degree of freedom
that exists in tensor networks:
without changing the state, matrices $G$ and $G^{-1}$
can always be inserted at a bond and merged with the adjacent tensors,
as depicted in Fig.~\ref{fig:ttns_fig2}(b).
Because of this gauge degree of freedom,
each tensor $\mathbf{A}_j$ for $j \neq i$ can be enforced to fulfill the 
orthonormalization condition
\begin{equation}
\sum_{\vec{m}_\mathrm{in}}
A_j(\alpha_j,\vec{m}_\mathrm{in},m_\mathrm{out}) A_j(\alpha_j,\vec{m}_\mathrm{in},m_\mathrm{out}')
=\delta_{m_\mathrm{out} m_\mathrm{out}'}.
\end{equation}
Here, the ``out''-index $m_\mathrm{out}$ is the index pointing towards site~$i$,
the remaining indices are denoted ``in''-indices $\vec{m}_\mathrm{in}$.
In pictorial form, this condition is illustrated in Fig.~\ref{fig:ttns_fig2}(c).
The mathematical operation that endows tensors with the orthonormalization
condition is the $QR$-decomposition which is numerically stable\cite{Murg-2010a}.
Due to the orthonormalization condition, the
tensor network for the norm of the TTNS, as shown in Fig.~\ref{fig:ttns_fig2}(a),
can be ``cropped'' from the leaves towards site~$i$,
until only the tensors $\mathbf{A}_i$ and $\overline{\mathbf{A}_i}$ at site $i$ remain.
The norm of the TTNS then simplifies to $\langle\Psi|\Psi\rangle = \vec{\mathbf{A}}_i^{\dagger} \vec{\mathbf{A}}_i$,
which makes $\mathbf{N}_i=1$ (see Fig.~\ref{fig:ttns_fig2}(d)).

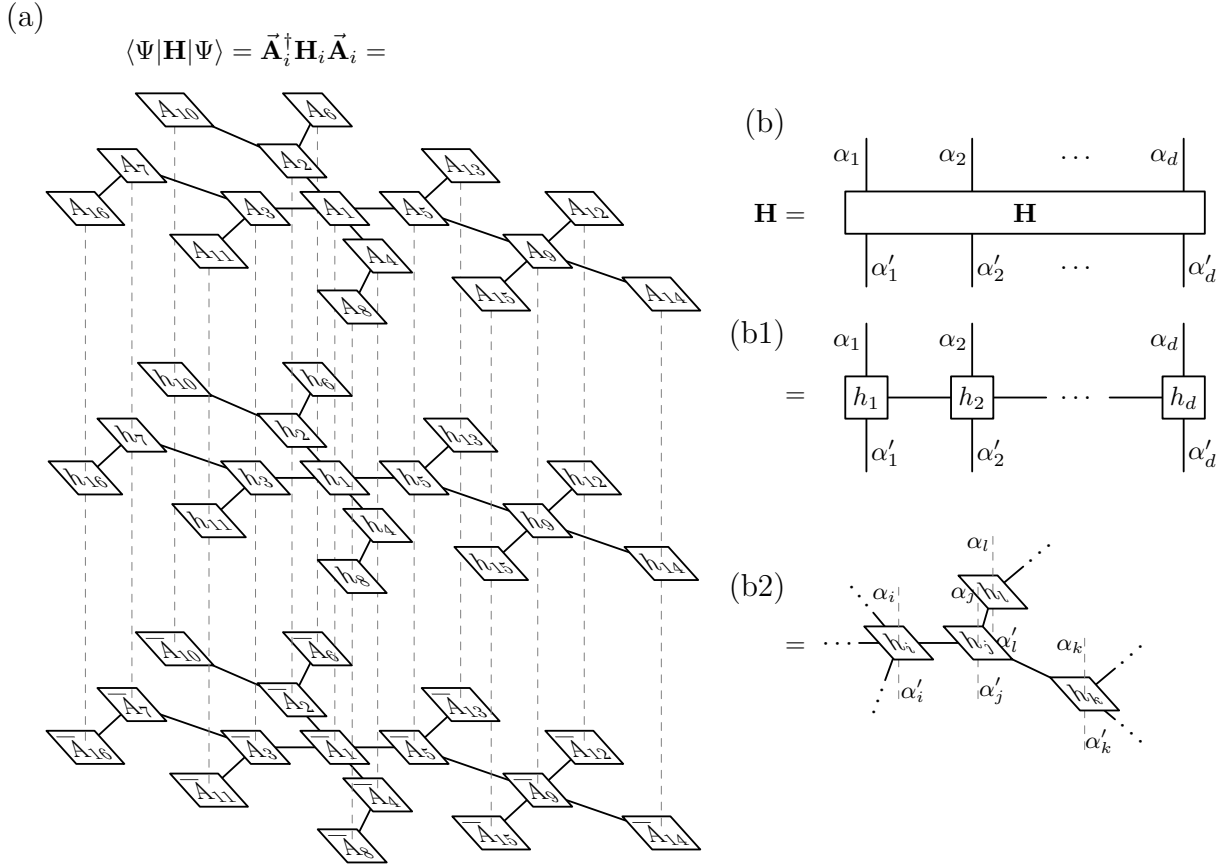
\begin{figure}[t]
\centering
\setlength{\unitlength}{40pt}   
\begin{picture}(11,8)
\put(0,0){\includegraphics{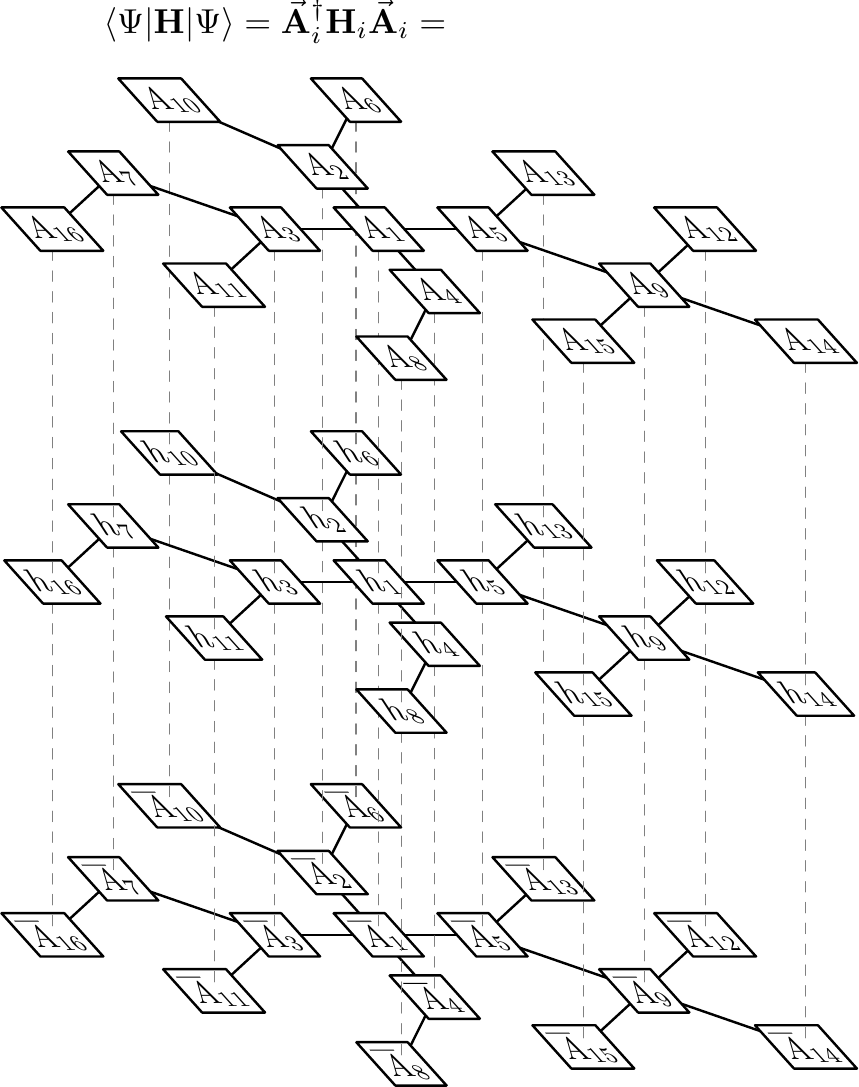}}
\put(6.7,3.7){\includegraphics{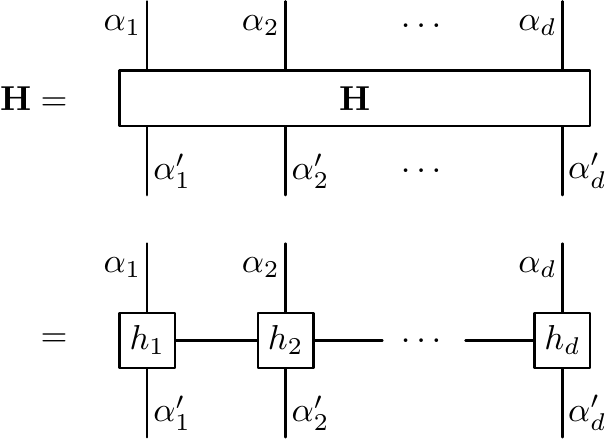}}
\put(6.7,1){\includegraphics{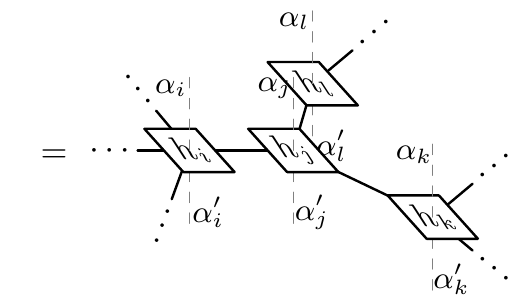}}
\put(0,8){\makebox(0,0)[r]{\strut{}(a)}}
\put(7,7){\makebox(0,0)[r]{\strut{}(b)}}
\put(7,5){\makebox(0,0)[r]{\strut{}(b1)}}
\put(7,2.6){\makebox(0,0)[r]{\strut{}(b2)}}
\end{picture}
\caption{ 
(a) expectation value $\langle\Psi|\mathbf{H} |\Psi\rangle$ with respect to 
the TTNS defined in Fig.~\ref{fig:ttns};
the tensor network picture of the expectation value corresponds to 
a three-layer structure, with 
the ket $|\Psi\rangle$ consisting of component tensors $\mathbf{A}_i$ being on top, 
the bra $\langle\Psi|$ consisting of component tensors $\overline{\mathbf{A}_i}$ on bottom,
and the Hamiltonian $\mathbf{H}$, represented as TTNO of component tensors $\mathbf{h}_i$ in the middle.
For better readability, the contracted physical indices are drawn with dashed lines.
(b) decomposition of the Hamiltonian as MPO (b1) and TTNO (b2).
}
\label{fig:ttns_fig3}
\end{figure}

The challenge that remains is to calculate the effective Hamiltonian
$\mathbf{H}_i$ of the eigenvalue problem. 
As mentioned before, it is obtained by
contracting all tensors except $\mathbf{A}_i$ and $\overline{\mathbf{A}_i}$
in the tensor network of
$\langle\Psi|\mathbf{H} |\Psi\rangle$.
In case of a TTNS, the contraction is efficient if the
Hamiltonian $\mathbf{H}$ is present in the form of a tree tensor network, as well.
The tree network of the Hamiltonian shall have the same structure as
the tree network of the state.
In analogy to the definition of the TTNS,
a tensor
\begin{equation}
h_i(\alpha_i',\alpha_i,m_{i,1},\ldots,m_{i,z_i})
\end{equation}
is associated 
to each site~$i$
with physical indices $\alpha_i'$ and $\alpha_i$ and
virtual indices $m_{i,1},\ldots,m_{i,z_i}$.
The coefficients 
$\mathbf{H}(\alpha_1',\ldots,\alpha_d',\alpha_1,\ldots,\alpha_d)$
are then obtained by contracting the virtual indices of the tensors $\mathbf{h}_i$
according to the tree network.
For $z=2$, this corresponds to the representation of the
Hamiltonian as a \emph{Matrix Product Operator} (MPO), as depicted in Fig.~\ref{fig:ttns_fig3}(b1).
For $z>2$, this concept is generalized to a \emph{Tree Tensor Network Operator} (TTNO), which is
illustrated in Fig.~\ref{fig:ttns_fig3}(b2).
In fact, for local Hamiltonians it is always possible to find a representation
as an MPO or TTNO with \emph{constant} dimension of the virtual bonds\cite{Pivru-2010}.
For non-local Hamiltonians of the form~(\ref{eq:HamiltonHd}), as arising in quantum chemistry,
it is always possible to find an MPO- or TTNO-form with bond-dimension $\mathcal{O}(d^2)$.

Once the Hamiltonian $\mathbf{H}$ is represented as TTNO 
with the same network structure as the TTNS $|\Psi\rangle$,
the tensor network form of the expectation value corresponds to a three-layer object,
as depicted in Fig.~\ref{fig:ttns_fig3}(a), with
the ket $|\Psi\rangle$ consisting of component tensors $\mathbf{A}_i$ being on top,
the bra $\langle\Psi|$ consisting of component tensors $\overline{\mathbf{A}_i}$ on bottom,
and the Hamiltonian $\mathbf{H}$, represented as TTNO of component tensors $\mathbf{h}_i$ in the middle.
By starting from the leaves and proceeding inwards towards site~$i$,
this network can be contracted efficiently (i.e. polynomially in $d$ and $M$),
yielding the expectation value $\langle\Psi|\mathbf{H} |\Psi\rangle$ and,
if $\mathbf{A}_i$ and $\overline{\mathbf{A}_i}$ are omitted, the effective Hamiltonian $\mathbf{H}_i$.
In order to reduce computational costs 
related to the diagonalization of the effective Hamiltonian
a half-renormalization scheme has also been introduced\cite{Nakatani-2013}.

For more detailed derivations we refer to the original papers\cite{Murg-2010a,Nakatani-2013,Murg-2014}.

\subsubsection{Efficient factorization of the interaction terms}
\label{sec:num.block.factor}
When the $d$-orbital system is partitioned into several subsystems the 
Hamiltonian is built from terms acting \emph{within} the subsystems and from 
terms \emph{among} the subsystems. 
During the course of the iterative diagonalization of the 
effective Hamiltonian acting on the $M_LM_R$ dimensional subspace,
the matrix vector multiplication $\mathbf H|\Psi\rangle$ is performed several times\cite{Lanczos-1950,Davidson-1975}. 
For a bipartite split using the matricization of $U$ discussed in Sec.~\ref{sec:tensor.matricization},
 $|\Psi\rangle$ is converted to a matrix with size $M_L\times M_R$
and the matrix vector multiplication is formed as two matrix-matrix multiplication of operator pairs as 
$\mathbf{X}_i^{(\mathrm{L})}\mathbf{U}(\mathbf{Y}_j^{(\mathrm{R})})^{\rm T}$ 
where $\mathbf{X}_i^{(\mathrm{L})}$ and $\mathbf{Y}_j^{(\mathrm{R})}$ are operators acting on the left and 
right subsystem, respectively. 

In order to treat long-range interactions efficiently, the
interaction terms must be factorized, thus the matrix and tensor 
algebra during the diagonalization procedure is simplified. 
This is called partial summation\cite{Xiang-1996}. 
For example, considering a two-orbital interaction in general, like the two-operator term in (\ref{eq:HamiltonHd}),
for a bipartite split of the system  
\begin{equation}
\label{eq:genHkin}
\mathbf{H}_\text{two} = 
\sum_{i,j\in (\mathrm{L})} T_{ij} {\mathbf{X}}_i^{(\mathrm{L})} {\mathbf{Y}}_j^{(\mathrm{L})} + 
\sum_{\substack{i\in (\mathrm{L})\\j\in (\mathrm{R})}} T_{ij} {\mathbf{X}}_i^{(\mathrm{L})} {\mathbf{Y}}_j^{(\mathrm{R})} + 
\sum_{\substack{i\in (\mathrm{R})\\j\in (\mathrm{L})}} T_{ij} {\mathbf{X}}_i^{(\mathrm{R})} {\mathbf{Y}}_j^{(\mathrm{L})} + 
\sum_{i,j\in (\mathrm{R})} T_{ij} {\mathbf{X}}_i^{(\mathrm{R})} {\mathbf{Y}}_j^{(\mathrm{R})}, 
\end{equation}
one of the coupling between the two subsystems (the second term above) can be simplifed as
\begin{equation}
\label{eq:genHkinfactor}
\mathbf{H}^{(\mathrm{L})(\mathrm{R})} = 
\sum_{j\in (\mathrm{R})} \left( \sum_{i\in (\mathrm{L})} T_{ij} {\mathbf{X}}_i^{(\mathrm{L})}\right) {\mathbf{Y}}_j^{(\mathrm{R})} = 
\sum_{j\in (\mathrm{R})} {\cal A}^{(\mathrm{L})(\cdot)}_j {\mathbf{Y}}_j^{(\mathrm{R})}.  
\end{equation}
Here ${\cal A}^{(\mathrm{L})(\cdot)}_j=\sum_{i\in L} T_{ij} {\mathbf{X}}_i^{(\mathrm{L})}$
is called \emph{one-orbital auxiliary operator.}

Therefore, the number of operator multiplications reduces from $d^2$ to $d$. 
Symbolically this can be written in a compact form:
we assign a label to each subsystem and form the total system by adding together the subsystems. This
sum is raised to the power given by the number of operators corresponding to the given interaction. 
For example, for the four operator term, coming from the Coulomb intercation,
$\mathbf{H}_\text{four} = \sum_{ijkl} V_{ijkl} {\mathbf{X}}_i {\mathbf{Y}}_j{\mathbf{Z}}_k {\mathbf{W}}_l$ 
 in (\ref{eq:HamiltonHd}) 
and for the bipartite
split (subsystems $(\mathrm{L})$ and $(\mathrm{R})$),
 this can be factorized as
$((\mathrm{L})+(\mathrm{R}))^4=(\mathrm{L})^4 + 4(\mathrm{L})^3(\mathrm{R}) + 6(\mathrm{L})^2(\mathrm{R})^2 + 4(\mathrm{L})(\mathrm{R})^3 + (\mathrm{R})^4$.
Constant factors comes from the permutation of indices and 
exponents show the number of operators acting within the
corresponding subsystem.
Therefore, when the first three operators act on the $(\mathrm{L})$ subsystem and 
the last operator on the $(\mathrm{R})$ subsystem then
\begin{equation} 
\mathbf{H}^{(\mathrm{L})(\mathrm{L})(\mathrm{L})(\mathrm{R})} = 
\sum_{l\in (\mathrm{R})} \left( \sum_{i,j,k\in (\mathrm{L})} V_{ijkl} {\mathbf{X}}_i^{(\mathrm{L})} 
{\mathbf{Y}}_j^{(\mathrm{L})} {\mathbf{Z}}_k^{(\mathrm{L})}\right) {\mathbf{W}}_l^{(\mathrm{R})}= 
\sum_{l\in {(\mathrm{R})}} {\cal A}^{(\mathrm{L})(\mathrm{L})(\mathrm{L})(\cdot)}_l \mathbf{W}_l^{(\mathrm{R})}\,,  
\end{equation}
thus the number of operator multiplications reduces from $d^4$ to $d$ 
by forming a three-orbital auxiliary operator ${\cal A}^{(\mathrm{L})(\mathrm{L})(\mathrm{L})(\cdot)}_l$.
Similarly, when the first two operators act on the $(\mathrm{L})$ subsystem and 
the last two operators on the $(\mathrm{R})$ subsystem then
\begin{equation}
\mathbf{H}^{(\mathrm{L})(\mathrm{L})(\mathrm{R})(\mathrm{R})}  = 
\sum_{k,l\in (\mathrm{R})} \left( \sum_{i,j\in (\mathrm{L})} V_{ijkl} 
{\mathbf{X}}_i^{(\mathrm{L})} {\mathbf{Y}}_j^{(\mathrm{L})} \right) {\mathbf{Z}}_k^{(\mathrm{R})} {\mathbf{W}}_l^{(\mathrm{R})}= 
\sum_{k,l\in (\mathrm{R})} {\cal A}^{(\mathrm{L})(\mathrm{L})(\mathrm{\cdot})(\cdot)}_{kl} \mathbf{Z}^{(\mathrm{R})}_k \mathbf{W}_l^{(\mathrm{R})}
\end{equation}
thus the number of operator multiplications reduces from $d^4$ to $d^2$ 
by froming two-orbital auxiliary operators ${\cal A}^{(\mathrm{L})(\mathrm{L})(\cdot)(\cdot)}_{kl}$.
Extensions for more subsystems used in QC-DMRG and QC-TTNS is straighforward.
For example, for subsystems $(\mathrm{l})$, $(l+1)$, $(l+2)$, $(\mathrm{r})$, the two-operator 
term is composed from the following terms as 
$((\mathrm{l})+(l+1)+(l+2)+(\mathrm{r}))^2 =
(\mathrm{l})^2 + 2(\mathrm{l})(l+1) + 2(\mathrm{l})(l+2)+2(\mathrm{l})(\mathrm{r})+
(l+1)^2 + 2(l+1)(l+2) + 2(l+1)(\mathrm{r})
+ (l+2)^2 +  2(l+2)(\mathrm{r}) +
(\mathrm{r})^2$
and the four-operator term factorizes as 
$((\mathrm{l})+(l+1)+(l+2)+(\mathrm{r}))^4=
(\mathrm{l})^4 + 4(\mathrm{l})^3(l+1) + 4(\mathrm{l})^3(l+2) + 4(\mathrm{l})^3(\mathrm{r}) + 6(\mathrm{l})^2(l+1)^2 + 12(\mathrm{l})^2(l+1)(l+2)+\ldots$
It is worth to note that symmetries of $V_{ijkl}$ can be used to reduce the number of independent terms.

Renormalization of multi-orbital operators, i.e, 
when more than one operator act in the same subsystem,
requires specal care since they cannot be calculated accurately
as a product of the renormalized operators.
For example,  
if $i,j$ belong to the same DMRG block due to the truncation of the 
Hilbert-space, $\mathbf{O}\mathbf{O}^\dagger\neq\mathbb{I}$.
\begin{equation}
\mathbf{O} (\mathbf{X}^{(\mathrm{L})}_i \mathbf{Y}^{(\mathrm{L})}_j) \mathbf{O}^\dagger \neq    
\mathbf{O}\mathbf{X}^{(\mathrm{L})}_i\mathbf{O}^\dagger  \mathbf{O}\mathbf{Y}^{(\mathrm{L})}_j\mathbf{O}^\dagger  
\end{equation}
Therefore, multi-orbital operators must be renormalized independently 
and stored.
As an example, the renormalization of a four-orbital operator
acting on the  $(\mathrm{L})=(\mathrm{l})\bullet$ composite system is 
$\mathbf{O}{\cal A}^{(\mathrm{L})(\mathrm{L})(\mathrm{L})(\mathrm{L})}\mathbf{O}^\dagger=
\mathbf{O}(\sum_{ijkl\in(\mathrm{L})}V_{ijkl}\mathbf{X}^{(\mathrm{L})}_i
                                             \mathbf{Y}^{(\mathrm{L})}_j
                                             \mathbf{Z}^{(\mathrm{L})}_k
                                             \mathbf{W}^{(\mathrm{L})}_l)\mathbf{O}^\dagger$,
where the auxuiliary operator ${\cal A}^{(\mathrm{L})(\mathrm{L})(\mathrm{L})(\mathrm{L})}$
is decomposed into further auxiliary operators as follows
\begin{eqnarray}
{\cal A}^{(\mathrm{L})(\mathrm{L})(\mathrm{L})(\mathrm{L})}&=& 
    {\cal A}^{(\mathrm{l})(\mathrm{l})(\mathrm{l})(\mathrm{l})}\otimes \mathbb I_{l+1} 
  + {\cal A}^{(\mathrm{l})(\mathrm{l})(\mathrm{l})(\cdot)}_{l+1}\otimes \mathbf{W}_{l+1} 
  + {\cal A}^{(\mathrm{l})(\mathrm{l})(\cdot)(\cdot)}_{l+1,l+1}\otimes \mathbf{Z}_{l+1} \mathbf{W}_{l+1} \nonumber\\ 
&+& 
    {\cal A}^{(\mathrm{l})(\cdot)(\cdot)(\cdot)}_{l+1}\otimes \mathbf{Y}_{l+1}\mathbf{Z}_{l+1}\mathbf{W}_{l+1}
+ \mathbb{I}^{(\mathrm{l})}\otimes V_{l+1,l+1,l+1,l+1} \mathbf{X}_{l+1}\mathbf{Y}_{l+1}\mathbf{Z}_{l+1}\mathbf{W}_{l+1}.
\end{eqnarray}

In summary, the numerical effort of the QC-DMRG and QC-TTNS algorithms has two 
major contributions.
On the one hand, the number of block states is crucial:
The numerical effort for calculating one term of the effective Hamiltonian by tensor contraction
scales as $M^{z+1}$ for trees of arbitrary coordination number $z$.
On the other hand, this calculation has to be performed for each
term in the Hamiltonian, 
and using the summation tricks as described above 
the scaling is $d^2 M^{z+1}$.
Since $\mathcal{O}(d)$ iteration steps are required for convergence,
the overall time of the algorithms scale as $d^3 M^{z+1}$.

\subsection{Optimization of convergence properties}
\label{sec:num.optim}

In order to use QC-DMRG and QC-TTNS as black box methods, it is mandatory 
to utilize various 
concepts inherited from quantum information theory 
\cite{Legeza-2003b,Legeza-2004b,Rissler-2006,Barcza-2011,Murg-2010a,Murg-2014,Fertitta-2014}. 
In this section we briefly discuss some entanglement based 
optimization procedures which are used to minimize the 
\emph{overall entanglement}, expressed as a cost function\cite{Rissler-2006,Barcza-2011}, 
\begin{equation}
I_{\text{overall}}^{}=\sum_{i,j} I_{ij} d_{ij}^\eta. 
\label{eq:cost} 
\end{equation}
Here $d_{ij}$ is the \emph{distance function} between orbital $i$ and $j$,
in the graph-theoretical sense,
$I_{ij}$ is the \emph{two-orbital mutual information} given in (\ref{eq:mut})
and $\eta$ is some exponent.
Therefore, the correlations between the pairs of orbitals is weighted by the distance $d_{ij}$.
The distance $d_{ij}$ depends on the tensor topology,
and it is defined as the length of the shortest path connencting $i$ and $j$ in the tensor network.
In the special case of MPS, the distance is simply $d_{ij}=|i-j|$.

The physical motivation behind the quantity $I_{\text{overall}}^{}$
is that in a given iteration step
the Schmidt rank is related to the number and strength of the entanglement bonds
between the left and right blocks,
thus if two highly correlated orbitals are located far from each other
then they give a large contribution until they fall into the same block.
Since the overall cost is related to the sum of the Schmidt ranks,
the major aim is to reduce the ranks for each iteration steps.
The optimization methods surveyed in this section serve 
for the manipulation of this cost function $I_{\text{overall}}^{}$
in three different ways:
by changing $d_{ij}$ by reordering the component tensors for a given tensor topology
(section \ref{sec:num.optim.opt-ordering});
by changing $d_{ij}$ by altering the tensor topology itself
(section \ref{sec:num.optim.opt-topology});
or by changing $I_{ij}$ by transforming the orbital basis
(section \ref{sec:num.optim.opt-basis}).
Besides this, there are other factors
which effects the convergence rate and computational time:
using Dynamical Block State Selection (DBSS) methods
(section \ref{sec:num.optim.opt-dbss});
using entanglement based network initialization
(section \ref{sec:num.optim.opt-init});
or reducing the Hilbert space by taking symmetries into consideration
(section \ref{sec:num.optim.opt-symm}).

\subsubsection{Error sources and data sparse representation of the wavefunction}
\label{sec:num.optim.errors}

As has been discussed before, the success and numerical efficiency of the 
QC-DMRG and QC-TTNS algorithm rely on a 
subsequent application of the singular value decomposition 
\cite{Schollwock-2011,Murg-2010a,Nakatani-2013,Legeza-2014} (section \ref{sec:num.ent.toSVD})
while the 
performance depends on the level of entanglement encoded in the wave 
function\cite{Vidal-2003a,Legeza-2003b}. 
In each DMRG (or TTNS) step, the basis states of the system block are then 
transformed 
to a new {\em truncated basis} set by a unitary transformation 
based on the preceeding SVD\cite{Schollwock-2005}. This transformation depends 
therefore on how accurately the environment is 
represented\cite{Legeza-1996,Moritz-2006}\ as well as on the level of 
truncation\cite{Legeza-2003a}.
As a consequence the accuracy of the DMRG method is governed by the 
truncation error, 
$\delta \varepsilon_{\rm TR}$, as well as by the environmental error, 
$\delta \varepsilon_{\rm sweep}$\cite{Legeza-1996}. 
The latter is 
minimized in each DMRG sweep (macro-iteration) by a successive application of 
the SVD going through the system back and forth.
Since $\dim(\Xi_{\rm DMRG})\ll \dim(\Lambda_{\rm FCI})$ 
DMRG provides a data-sparse representation of the wavefunction, thus
the sparsity can be defined as 
$\dim(\Xi_{\rm DMRG})/\dim(\Lambda_{\rm FCI})$
for a given error margin.

\begin{figure}[t]
\centering
\setlength{\unitlength}{40pt}   
\begin{picture}(11,7) 
\put(  0,  0){\includegraphics[width=0.45\columnwidth]{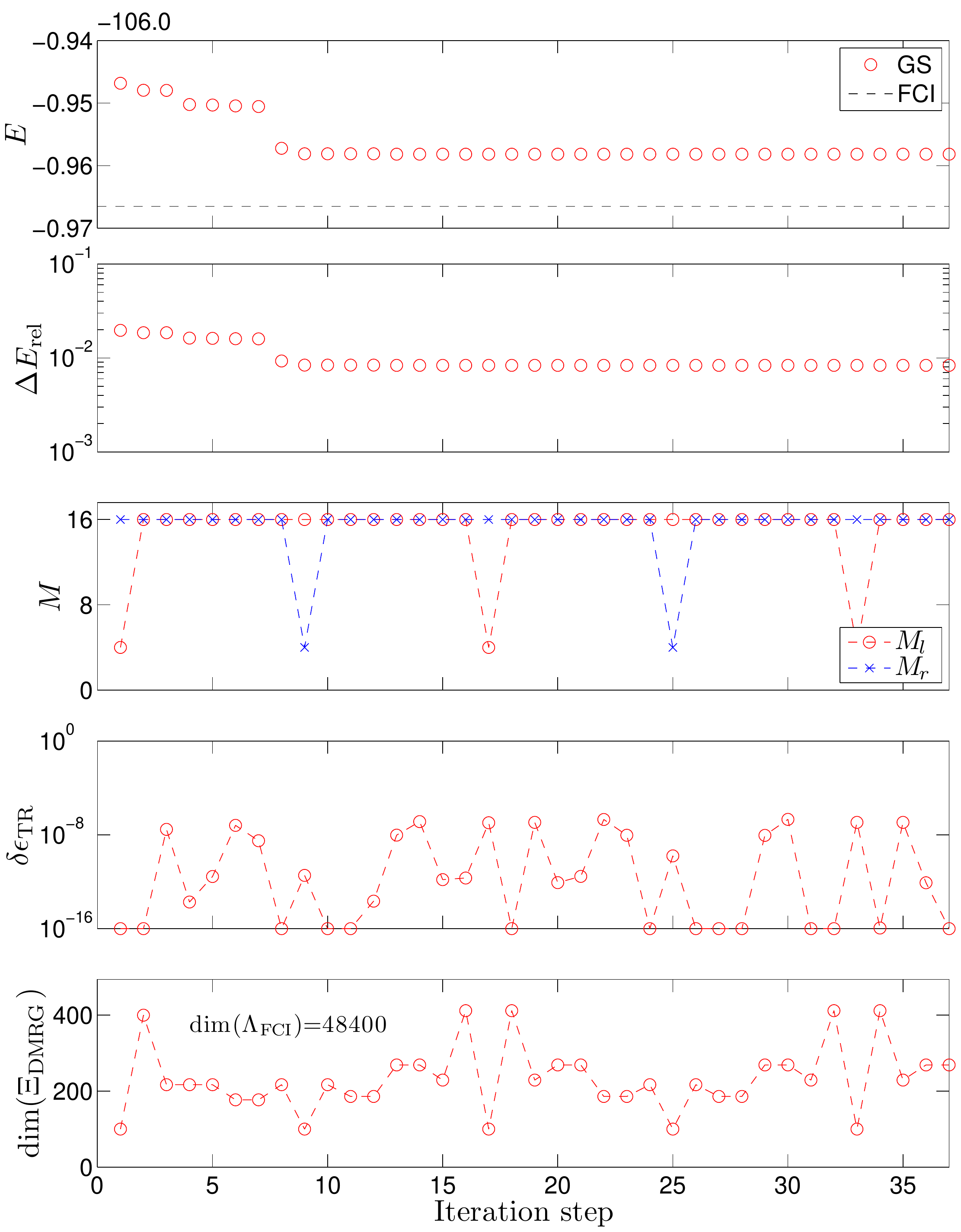}}
\put(5.5,  0){\includegraphics[width=0.45\columnwidth]{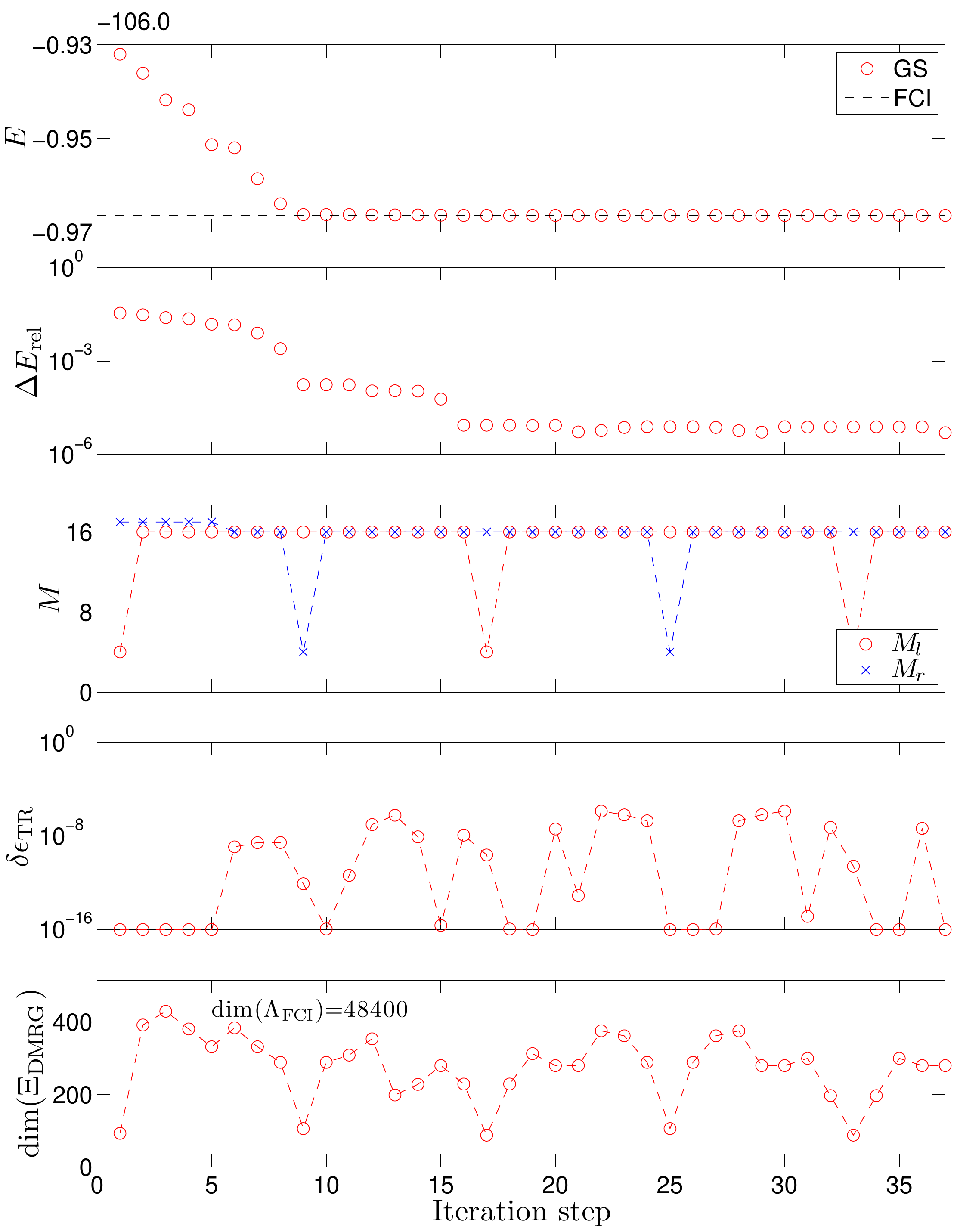}}
\put(0.5,7.0){\makebox(0,0)[r]{\strut{}(a)}} 
\put(6.0,7.0){\makebox(0,0)[r]{\strut{}(b)}}
\end{picture}
\caption{Ground state energy ($E$) in a.u., 
relative error ($\Delta E_{\rm rel}=(E_{\rm DMRG}-E_{\rm FCI})/E_{\rm FCI}$), 
number of block states ($M_l,M_r$), 
truncation error $\delta \varepsilon_{\rm TR}$, dimension of the superblock Hilbert space 
($\Xi_{\rm DMRG}$), are shown as a function of DMRG itertaion steps
for LiF at $d_{\text{Li-F}}=3.05$ a.u.~with CAS(6,12) with fixed $M_l=M_r=16$ for a non-optimized tensor hierarchy (ordering)
(a) and for an optimized tensor tensor hierarchy (ordering) (b).}
\label{fig:dmrg-run}
\end{figure}

As an example, relevant quantities as a function of DMRG iteration steps
are shown in Fig.~\ref{fig:dmrg-run} for LiF at $d_{\text{Li-F}}=3.05$ a.u.~for 
two different tensor arrangements (ordering).
Since DMRG is a variational method it converges to the full-CI energy 
from above as is apparent in the top panels of Fig~\ref{fig:dmrg-run}. Close
to the turning points when either left or right block contains a single 
orbital $M_l$ or $M_r$ drops to $4=q$. Although 
the truncation error 
fluctuates between $10^{-16}$ and $10^{-6}$ for both
tensor arrangements (ordering) and the size of the superblock
Hilbert space is at most 400, 
a much lower energy has been reached with the optimized ordering.
This clearly shows that in order to minimize 
$\delta \varepsilon_{\rm sweep}$ and avoid DMRG to converge 
to a local minima besides sweeping the tensor arrangement must also
be optimized as will be discussed below.

\begin{figure}
\centering
\setlength{\unitlength}{40pt}   
\begin{picture}(11,5) 
\put(  0,  0){\includegraphics[width=0.45\columnwidth]{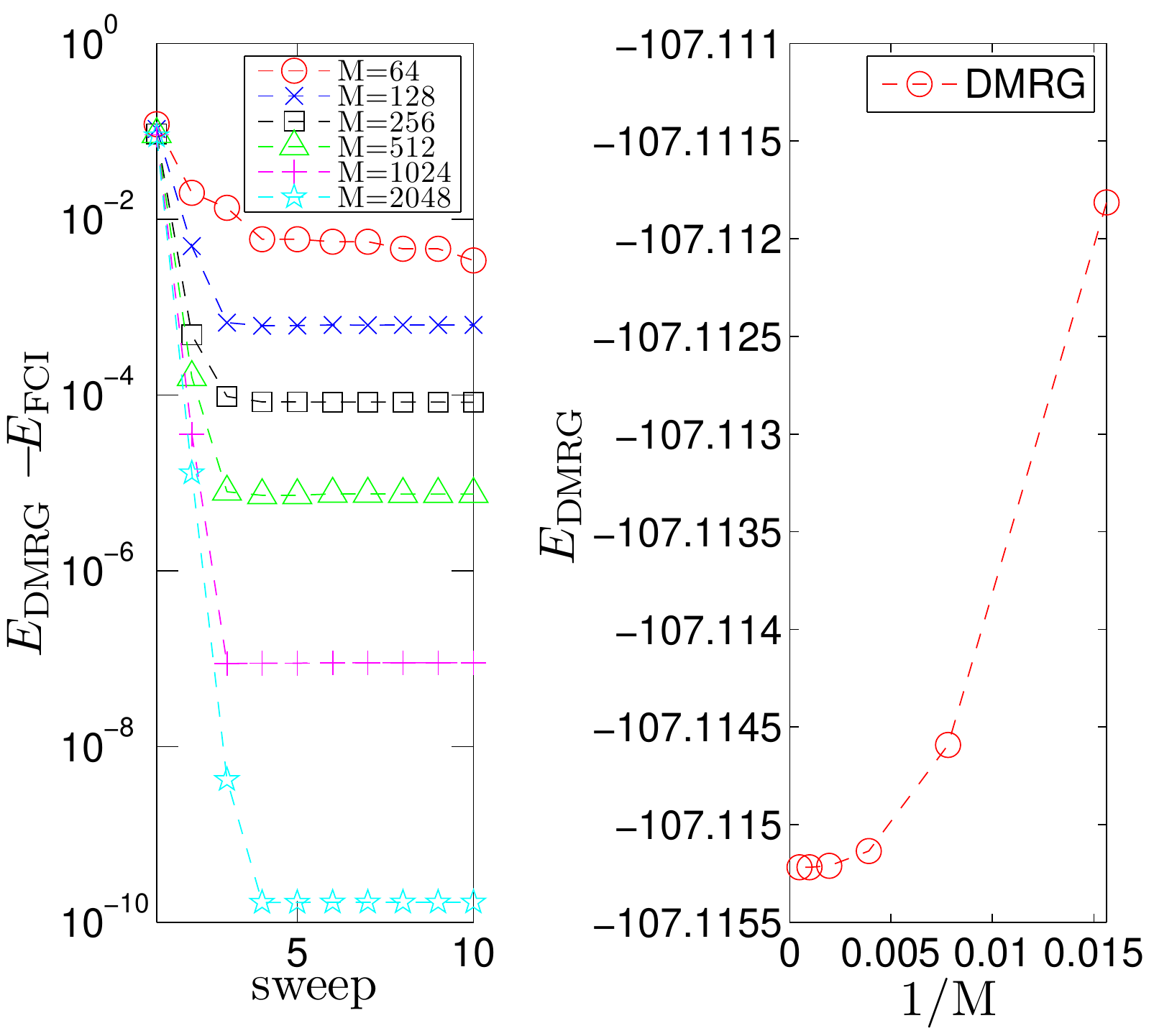}}
\put(5.5,  0){\includegraphics[width=0.45\columnwidth]{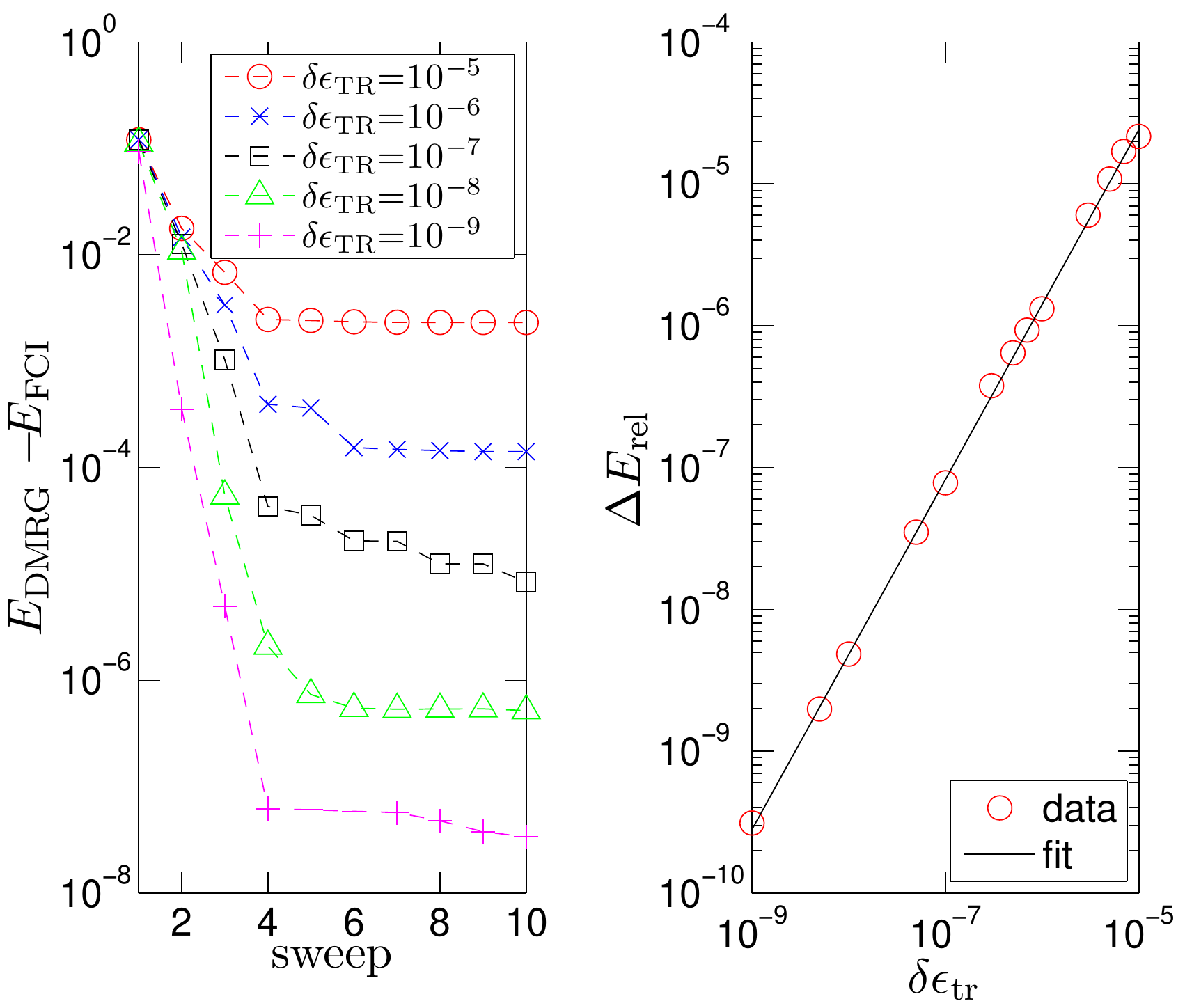}}
\put(0.2,5.0){\makebox(0,0)[r]{\strut{}(a)}}
\put(3.0,5.0){\makebox(0,0)[r]{\strut{}(b)}}
\put(5.8,5.0){\makebox(0,0)[r]{\strut{}(c)}}
\put(8.6,5.0){\makebox(0,0)[r]{\strut{}(d)}}
\end{picture}
\caption{Convergence of the ground state energy for LiF at $d_{\text{Li-F}}=3.05$ a.u. as a function of DMRG sweeping
for various fixed number of block states (a),
$E(M, \delta \varepsilon_{\rm sweep}=0)$ as a function of $1/M$(b),
ground state energy as a function of DMRG sweeping
for various fixed $\delta \varepsilon_{\rm TR}$ using DBSS procedure (c),
and
$\Delta E_{\rm rel}(\delta \varepsilon_{\rm TR}, \delta \varepsilon_{\rm sweep}=0)$ as a 
function of $\delta \varepsilon_{\rm TR}$ on a log-log scale (d). 
The solid lines are our fits.
}
\label{fig:lif-e-fixedm}
\end{figure}

Using an optimized ordering the convergence of the ground state energy for LiF 
at $d_{\text{Li-F}}=3.05$ a.u. as a function of DMRG sweepings for various fixed number of 
block states is shown in Fig~\ref{fig:lif-e-fixedm}(a).
Taking the limit\ of zero energy change between two sweeps 
$E(M, \delta \varepsilon_{\rm sweep}=0)$\ for a given $M$\ 
and assuming $M_l=M_r=M$\ 
various extraplotaion schemes as a function of $M$ 
have been introduced\cite{Chan-2002a,Legeza-2003a,Mitrushenkov-2003a,Marti-2010b,Barcza-2012} in order 
to provide 
a good estimate for the truncation-free solution.
A more rigorous extrapolation scheme is based on the truncation error\cite{Legeza-1996},
i.e., once the environmental error is eliminated, the relative error,
$\Delta E_{\rm rel}=(E_{\rm DMRG}-E_{\rm FCI})/E_{\rm FCI}$, 
is determined by $\delta \varepsilon_{\rm TR}$ as
\begin{equation}
\ln \Delta E_{\rm rel} = a \ln \delta \varepsilon_{\rm TR} + b\,.
\label{eq:extrapolation}
\end{equation}
When the number of block states are kept fixed the truncation 
error fluctuates within a full sweep (see Fig.~\ref{fig:dmrg-run})
thus the largest truncation error within a full sweep 
determines the overall accuracy.
In Fig.~\ref{fig:lif-e-fixedm}(d) the relative error of the ground state 
energy is shown as a function of the largest truncation error within 
the last full sweep on a log-log scale. The linear behavior
allows one to obtain the truncation free energy
by taking all the datapoints obtained upto a given $\delta \varepsilon_{\rm TR}$
and letting $E_\text{FCI}$ as a free parameter denoted as $E$.

\subsubsection{Targeting several states together}
\label{sec:num.optim.multi-target}

As it is possible to calculate several lowest lying eigenstates of the 
superblock Hamiltonian using the Davidson\cite{Davidson-1975} or L\'anczos\cite{Lanczos-1950} algorithm, 
more eigenstates can be targeted within a single 
QC-DMRG or QC-TTNS calculation
\cite{Legeza-2003a,Legeza-2003c,Moritz-2005b,Dorando-2007,Ghosh-2008,Liu-2013,Nakatani-2014,Wouters-2014a,Sharma-2014,Murg-2014}.

\begin{figure}[t]
\centering
\setlength{\unitlength}{40pt}   
\begin{picture}(11,7)
\put(  0,  0){\includegraphics[width=0.45\columnwidth]{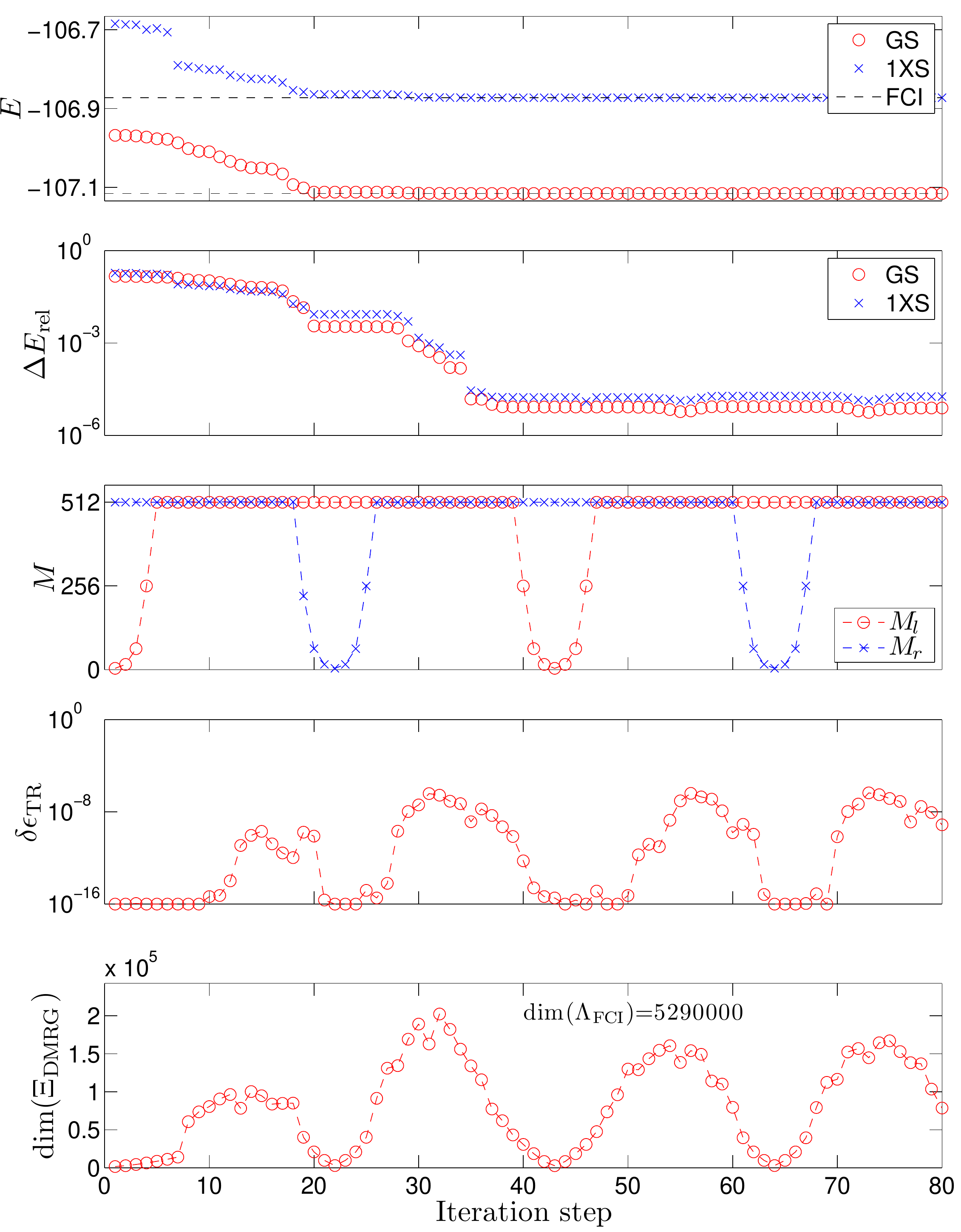}}
\put(5.5,  0){\includegraphics[width=0.45\columnwidth]{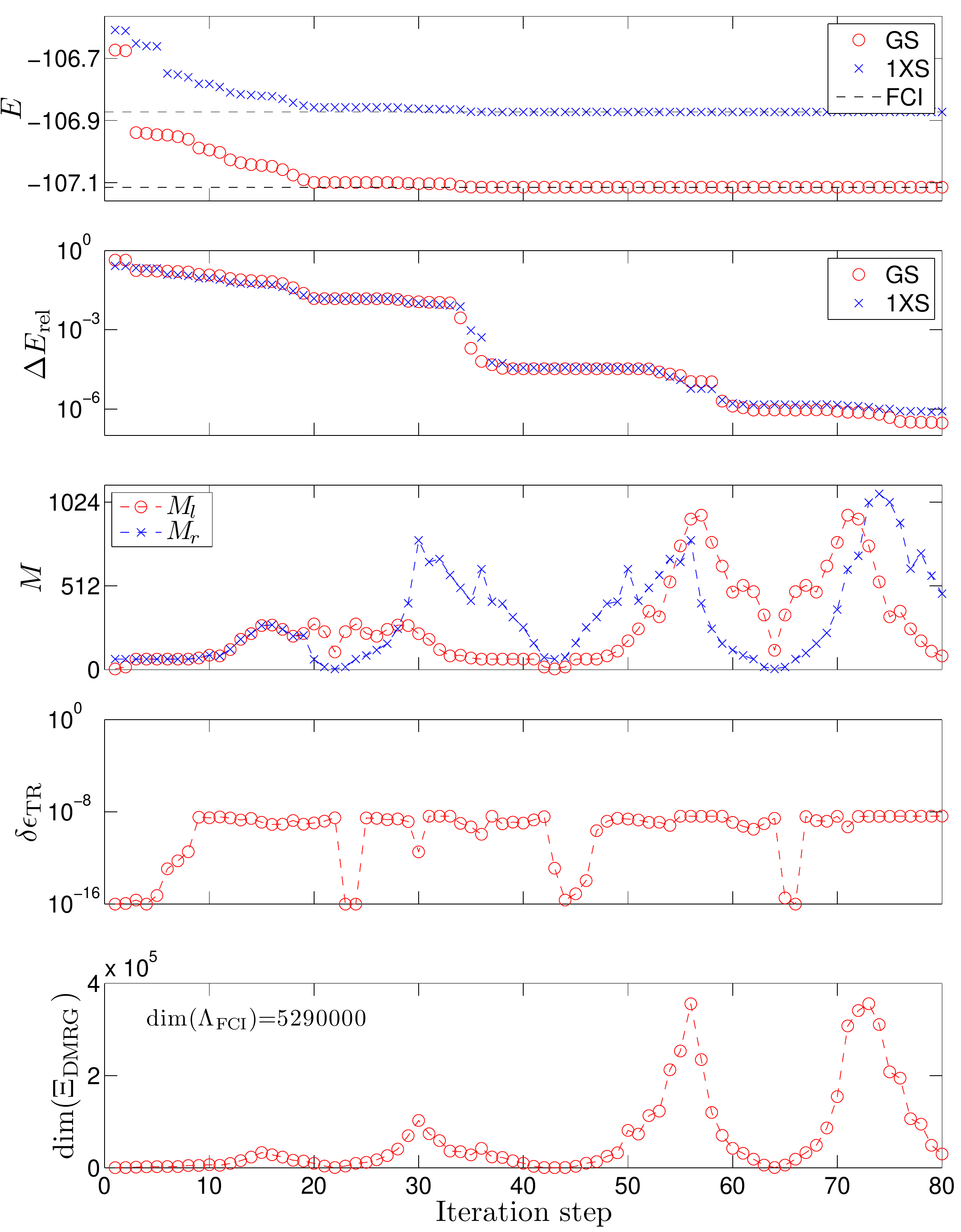}}
\put(0.5,7.0){\makebox(0,0)[r]{\strut{}(a)}}
\put(6.0,7.0){\makebox(0,0)[r]{\strut{}(b)}}
\end{picture}
\caption{
(a) Similar to Fig.~\ref{fig:dmrg-run}(b) but for LiF at $d_{\text{Li-F}}=3.05$ a.u. with CAS(6,25) and
targeting the ground state and 
excited states within a single DMRG calculation with 
$M_l=M_r=64$ block states, for optimized tensor hierarchy (ordering).
(b) Similar to (a) but using the DBSS procedure
with $M_{\rm min}=64$ and $\chi=10^{-7}$.
}
\label{fig:dmrg-run-dbss}
\end{figure}

In this case the total system is no longer treated as a pure state but 
as a mixed state with mixing weights $p_\gamma>0$ (with $\gamma=1,\ldots,n$ and $\sum_\gamma p_\gamma=1$),
the reduced subsystem density matrix 
can be formed from the reduced density matrices $\rho_\gamma$ of the lowest 
$n$ eigenstates $|\Psi_\gamma\rangle$ as
$\rho = \sum_\gamma p_\gamma \rho_\gamma$.
The optimal choice of the $p_\gamma$ distribution, however, is not established yet.
As an example, energies of the ground state and first excited state obtained
for the LiF at $d_{\text{Li-F}}=3.05$ a.u. is shown in Fig.~\ref{fig:dmrg-run-dbss}(a). 
It is worth mentioning that target states can also be formed based on the 
action of a given operator, i.e, besides the ground and excited states
one can inlcude states by applying a given operator to 
the ground state. For more details we refer to the literature\cite{Noack-2005}.

\begin{figure}
\centering
\setlength{\unitlength}{40pt}   
\begin{picture}(8.5,3)
\put(0,1.7){\includegraphics{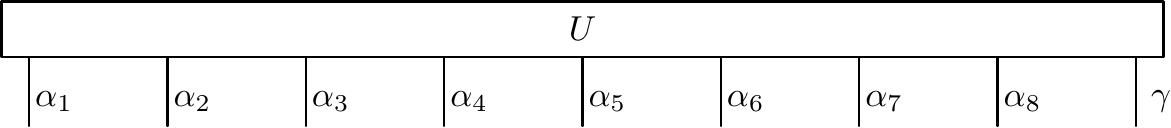}}
\put(0,0){\includegraphics{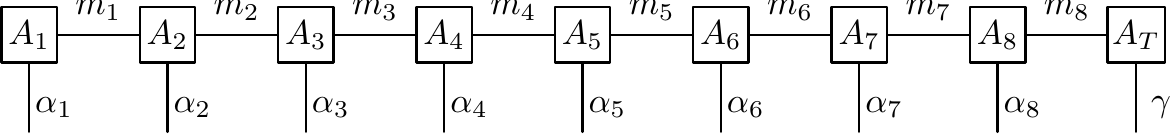}}
\put(0,3){\makebox(0,0)[r]{\strut{}(a)}}
\put(0,1.3){\makebox(0,0)[r]{\strut{}(b)}}
\end{picture}
\caption{(a) Graphical representation of the full tensor 
$U(\alpha_1,\ldots,\alpha_d,\gamma)$ when excited states
are also calculated. $\alpha$ stands for the physical indicies while 
$\gamma$ labels the excited states. (b) Tensor network representation
of the full tensor.
}
\label{fig:mps-multitarget}
\end{figure}

For multi-target states with equal weights $p_{\gamma} \equiv p=1/n$, we minimize the sum 
$ \sum_{\gamma = 1}^n \langle \Psi_{\gamma} | H | \Psi_{\gamma} \rangle$ constrained to the orthogonality condition
$\langle \Psi_\beta | \Psi_\gamma  \rangle = \delta_{\beta,\gamma} $. Clearly the minimum of this functional
is the sum of the $n$ lowest eigenvalues $E_0 + \cdots + E_{n-1} $ of the Hamiltonian $H$, and a minimizer is 
provided by the first $n $ eigenfunctions.  
In an MPS framework, the tensor $U(\alpha_1,\ldots,\alpha_d,\gamma)$
corresponding to the $\gamma^{\rm th}$ eigenstate with order $d+1$
as is shown in Fig.~\ref{fig:mps-multitarget}(a) can be expressed
as a network shown in Fig.~\ref{fig:mps-multitarget}(b). Therefore,
the network contains $d+1$ component tensors and in each optimization
step $\gamma$ is shifted through the newtwork. Although, this procedure
is commonly used in the DMRG community, it is worth mentioning that
$\gamma$ index has a different physical meaning than the $\alpha$ indices.
Quite recently alternative methods to calculate
excited states have also been itroduced\cite{Pirvu-2012,Haegeman-2012,Haegeman-2013,Wouters-2013}.

\subsubsection{Optimization of the Schmidt ranks using
dynamic block state selection (DBSS) approach and entropy sum rule}
\label{sec:num.optim.opt-dbss}

The two-orbital variant of the DMRG method has originally been employed with 
a fixed number of block states as shown above 
while the degree of entanglement between 
the DMRG blocks for a given superblock configuration is related to 
the Schmidt rank $r_{\rm Sch}$ as discussed in 
Secs.~\ref{sec:tensor} and \ref{sec:num.block.dmrg}.
Therefore, the fluctuation of the truncation error makes the
utilization of Eq.~(\ref{eq:extrapolation}) less stable.
It is more efficient
to control the truncation error $\delta \varepsilon_{\rm TR}$
at each renormalization step
and change the number of block states dynamically\cite{Legeza-2003a}.

Alternatively, one can control the truncation in terms of the quantum information
loss $\chi$, 
expressed by the von Neumann and R\'enyi entropies\cite{Legeza-2004b}. 
In a given DMRG renormalization step
denoting by $S^{(\mathrm{l})}$ the entropy of the left block of length $l$ 
and by $S_{l+1}$ the entropy of the $l+1^{\rm th}$ orbital, 
the sum of the entropies of these subsystems is reduced
by forming a larger block, 
$(\mathrm{L})\equiv(\mathrm{l})\bullet$, is given as
\begin{equation}
S^{(\mathrm{l})} + S_{l+1} - S^{(\mathrm{L})} = I^{(\mathrm{l})} \geq0,
\label{eq:information2}
\end{equation}
where the mutual information $I^{(\mathrm{l})}$ 
quantifies the correlation 
between the subsystem and the orbital
(similarly to the mutual information in Eq.~(\ref{eq:mut}), doing the same for two orbitals).
This means that 
if $I^{(\mathrm{l})} >0$ 
then we need more information 
for the description of the state of the $(\mathrm{l})$ block and the $\bullet$ separatedly
than for the description of them as a whole $(\mathrm{L})=(\mathrm{l})\bullet$,
that is, they are correlated.
A similar relation holds for the right block, $(\mathrm{R})\equiv \bullet(\mathrm{r})$, as well.
If an effective system of length $d+2$ is formed by adding
two non-interacting orbitals to the right and left ends of the chain,
all blocks containing 1 to $d$ orbitals of the original system can be 
formed by the forward and backward sweeps.
The total information gain during a half sweep can be calculated as 
$\sum_{l=1}^{d-1} I^{(\mathrm{l})}$.
In general, $I^{(\mathrm{l})}$ is also a function of subsequent sweeps.
However, once the DMRG method has converged, 
subsequent DMRG sweeps do not change $S^{(\mathrm{l})}$
and $S_l$. 
If, additionally, all $M_l=q^l$ and $M_r=q^r$ basis states of the blocks are kept at each 
iteration step, i.e., no truncation is applied, a sum rule holds, which 
relates the total information gain within a full half sweep and the sum of 
orbital entropies given as
\begin{equation}
\sum_{l=1}^{d-1} I^{(\mathrm{l})} =  \sum_{l=1}^d S_l \,,
\label{eq:i-tot}
\end{equation}
where we have used $S^{(1)}=S_1$ and $S^{(d)}=0$.

This equality, however, does not hold in practical DMRG calculations
since during the
renormalization process $S^{(\mathrm{L})}$ is reduced to 
$S_{\rm Trunc}^{(\mathrm{L})}$  due to the truncation of the basis states.
Once the DMRG method has converged,
the following 
equality should hold to a good accuracy
\begin{equation}
\sum_{l=1}^{d-1} I^{(\mathrm{l})} \simeq \sum_{l=1}^d S_l - \sum_{l=1}^{d-1} 
\left( S^{(\mathrm{L})} - S_{\rm Trunc}^{(\mathrm{L})} \right) \,.
\label{eq:i-tot-trunc}
\end{equation}
An analogous relationship holds for the backward sweep as well.
In order to control the quantum information loss,
$M_L$ (or $M_R$) is increased systematically at 
each renormalization step until the following condition holds
\begin{equation}
S^{(\mathrm{L})} - S_{\rm Trunc}^{(\mathrm{L})} < \chi\,, 
\label{eq:chi}
\end{equation}
where $\chi$ is an a priori defined error margin.
For $S^{(\mathrm{L})}$, i.e., before the truncation, 
$M_L = M_l q$ while
for $S_{\rm Trunc}^{(\mathrm{L})}$ according to
Eq.~(\ref{eq:chi})
$M_L^{\rm Trunc}\leq M_l q$ is used.
This approach guarantees that the number of
block states are adjusted according to the entanglement between the DMRG
blocks and the a priori defined accuracy can be reached.
In addition, an entropy sum rule based on Eq.~(\ref{eq:i-tot-trunc})
can be used as an alternative test of convergence\cite{Legeza-2004b}. 

In order to reduce the possibility of convergence to a local minima 
the minimum number of block states, $M_{\rm min}$ 
must also be introduced. Setting $M_{\rm min}\simeq q^3$ or $q^4$
is sufficient in most cases.
The maximum number of block states selected dynamically during 
the course of iterations denoted by $M_{\rm max}$ determines
wheter a calculation for a given accuracy can be performed
on the available computational resources.
It is worth to emphasize that this approach does not work for the one-orbital
variant of the DMRG algorithm since
the Schmidt number of a one-orbital superblock configuration 
$M_L=M_l q$ cannot be larger
than $M_r$. This prevents $M_l$ to increase above $M_r$ according 
to Eq.~(\ref{eq:schmidt}).

As an example, relevant quantities as a function of DMRG iteration steps
are shown in Fig.~\ref{fig:dmrg-run-dbss}(b) for LiF 
at $d_{\text{Li-F}}=3.05$ a.u.~for 
the optimized tensor arrangements (ordering) using the DBSS procedure
with $M_{\rm min}=64$ and 
$\chi=10^{-7}$.
In Fig.~\ref{fig:lif-e-fixedm}(c) 
the convergence of the ground state energy as a function of 
DMRG sweeping for various fixed $\delta \varepsilon_{\rm TR}$ using 
the DBSS procedure is shown.
Using Eq.~(\ref{eq:extrapolation}) 
and data points obtained 
for $\delta \varepsilon_{\rm TR}\ge10^{-6}$ after the 10$^{\rm th}$ sweep the extrapolated energy is $E=-107.11519(2)$,
for $\delta \varepsilon_{\rm TR}\ge10^{-9}$ it is $E=-107.115216925(2)$,
while $E_{\rm exact}=-107.1152169273$.

\subsubsection{Optimization of the network hierarchy (ordering) and entanglement localization}
\label{sec:num.optim.opt-ordering}

\begin{figure}
\centering
\setlength{\unitlength}{40pt}   
\begin{picture}(11,3.3)  
\put(  0,  0){\includegraphics[scale=0.31]{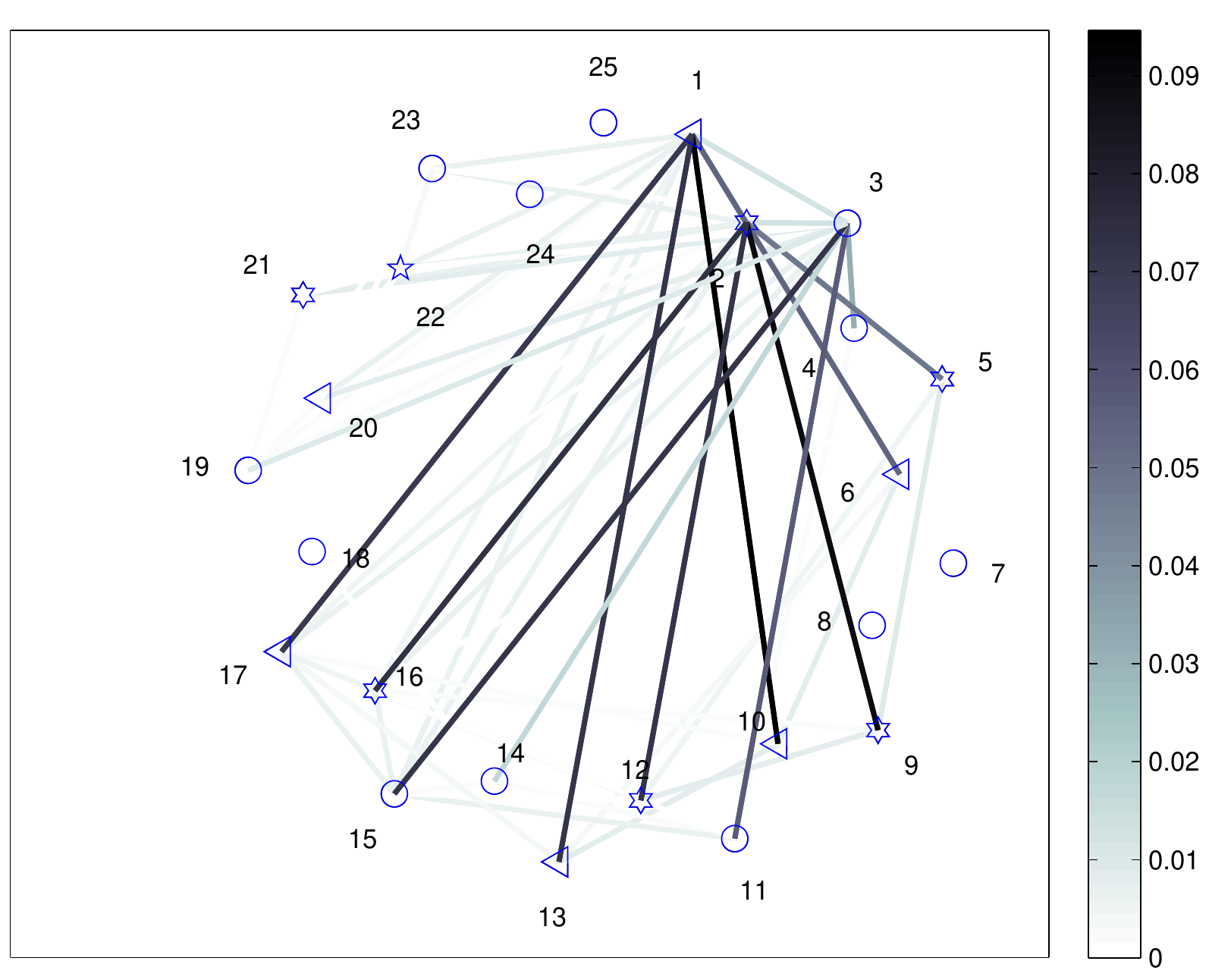}}
\put(3.8,  0){\includegraphics[scale=0.31]{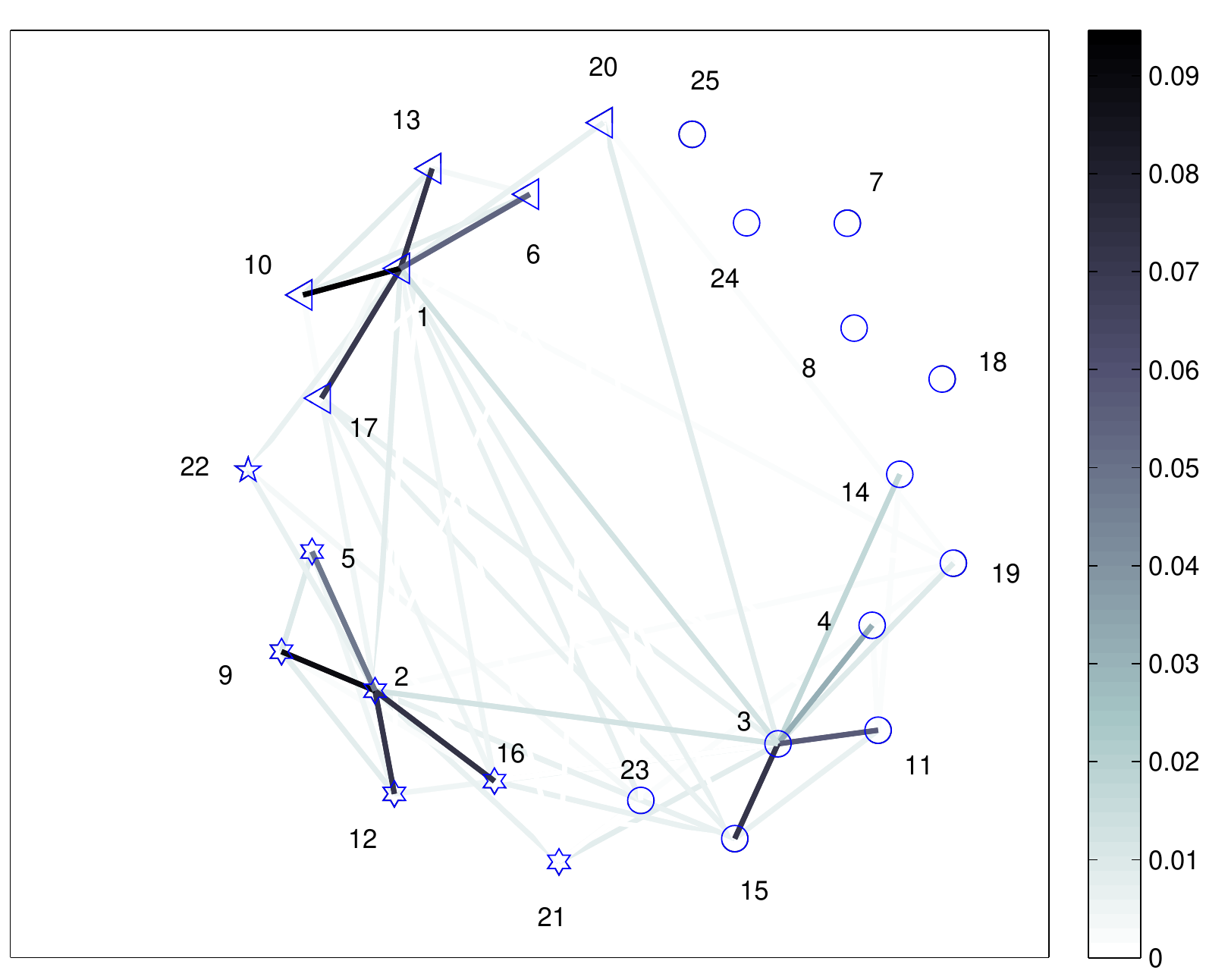}}
\put(7.6,  0){\includegraphics[scale=0.31]{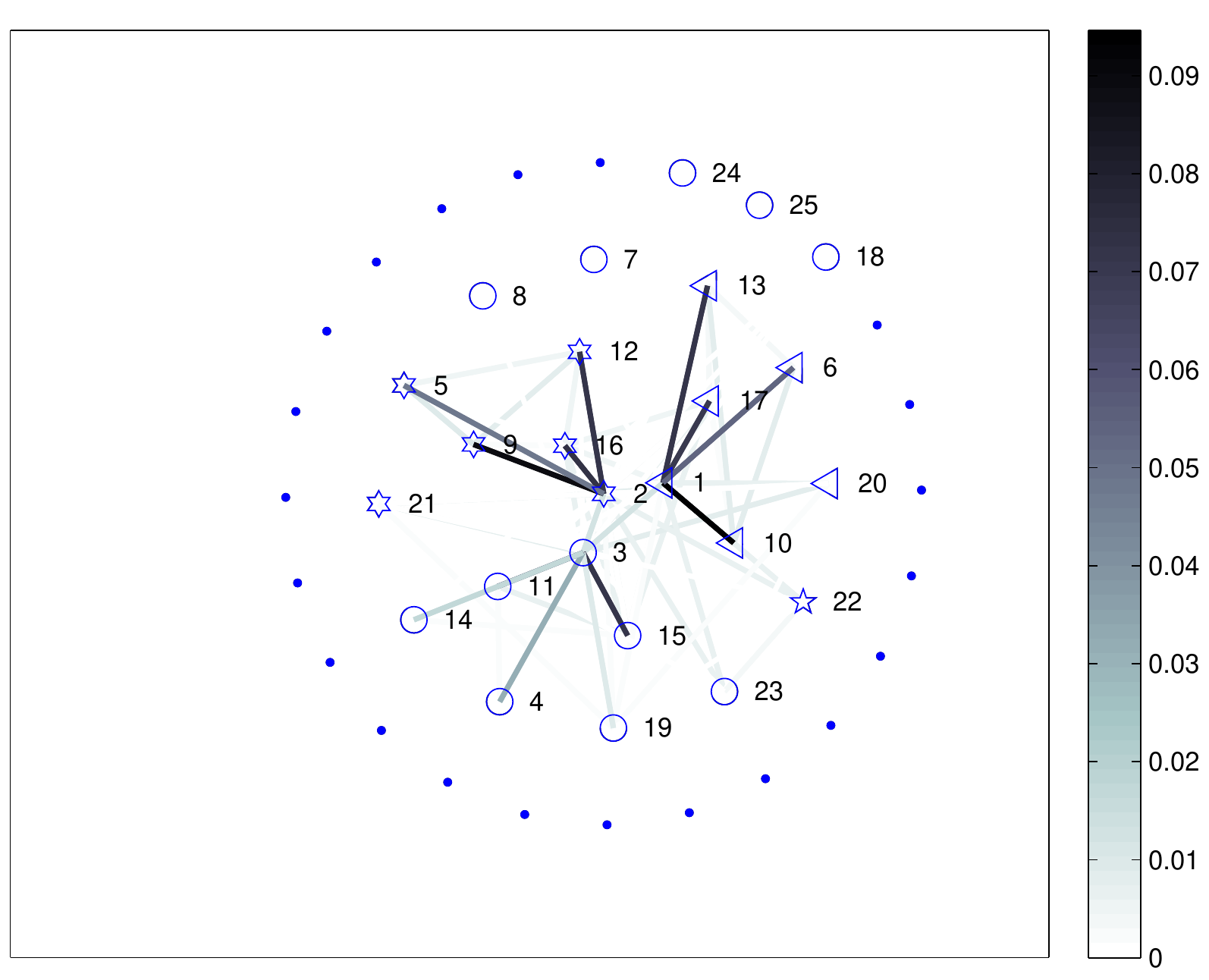}}
\put(0.5,3.3){\makebox(0,0)[r]{\strut{}(a)}} 
\put(4.3,3.3){\makebox(0,0)[r]{\strut{}(b)}} 
\put(8.1,3.3){\makebox(0,0)[r]{\strut{}(c)}}
\end{picture}
\caption{Optimization of tensor hierarchy (ordering) and topology by minimizing 
the overall entanglement $I_\text{overall}$ for the LiF at
the equilibrium bond length $r=3.05$ a.u.
(a) and (b) are for the one dimensional MPS like topology for the original 
ordering and for the optimized ordering, respectively.
(c) Shows the optimized topology on the tree (small dots indicate not 
used grid points of the tree). The total quantum information
$I_{\rm tot}$ given in (\ref{eq:Itot}) does not change but the overall entanglement 
$I_\text{overall}$ drops significantly.}
\label{fig:I_opt}
\end{figure}
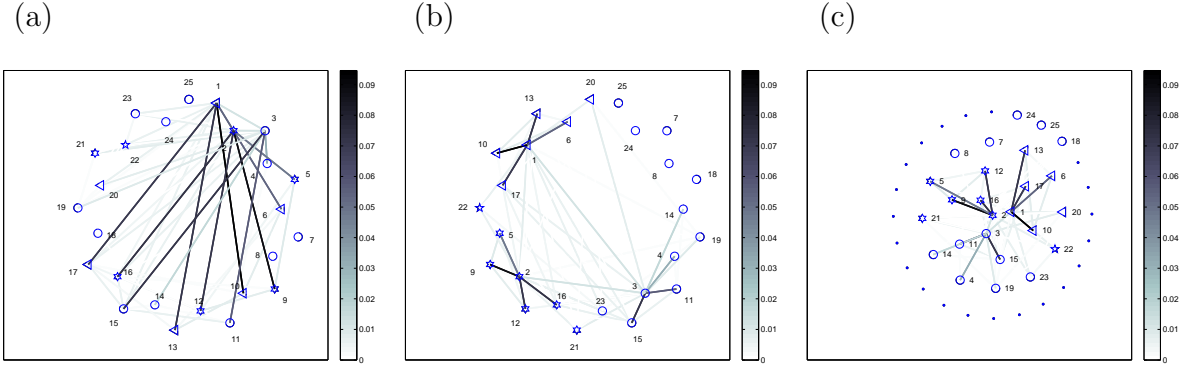

As was briefly mentioned before,
in order to use QC-DMRG as a black box method, 
first the arrangement of orbitals along a one-dimensional 
topology has to be optimized (ordering) in order to reduce the set of 
Schmidt ranks when the system is systematically partitioned into a left and 
right parts during the DMRG sweeping 
procedure\cite{Chan-2002a,Legeza-2003a,Legeza-2003b,Legeza-2003c,Mitrushenkov-2003a,Moritz-2005a,Rissler-2006,Kurashige-2009,Yanai-2010,Barcza-2011,Boguslawski-2012b,Mizukami-2013,Ma-2013,Wouters-2014a}.
This allows us to carry 
out calculations with much smaller number of block states 
using the DBSS approach\cite{Legeza-2003a,Legeza-2004b,Barcza-2011} (section \ref{sec:num.optim.opt-dbss}). 
For the one-dimensional tensor topology, i.e., for DMRG and MPS, the distance function is $d_{ij}=|i-j|$ in Eq.~(\ref{eq:cost}) and 
using $\eta=2$ has the advantage that this optimization task can be carried out 
using concepts of spectral graph theory\cite{Atkins-1998}.
It follows that the so called \emph{Fiedler vector} $x=(x_1, \dots x_d)$ 
is the solution that minimizes $F(x)=x^\dagger L x=\sum_{i,j} I_{ij} (x_i-x_j)^2$ 
subject to the constraints $\sum_i x_i=0$ and $\sum_i x_i^2=1$, 
where the graph Laplacian is $L_{ij}=D_{ij}-I_{ij}$ with the diagonal $D_{ij}=\delta_{ij}\sum_{j'} I_{ij'}$.
The second eigenvector of the Laplacian is the Fiedler vector\cite{Fiedler-1973,Fiedler-1975} which 
defines a (1-dimensional) embedding of the graph on a line that tries to respect the highest 
entries of $I_{ij}$ and the edge length of the graph. 
Ordering the entries of the Fiedler 
vector by non-increasing or non-decreasing way provides us a possible ordering.
Usually the best ordering obtained with small number of block states also provide almost 
the best ordering for calculation performed with large number of block states, thus
this task can be carried out with a limited number of block states.
As an example, non-optimal and optimized tensor orderings
for LiF at the equilibrium bond length $r=3.05$ a.u.~are shown in Figs.~\ref{fig:I_opt}(a) and (b)
for the one-dimensional network topology, respectively.
For both tensor topologies $I_{\rm tot}=1.32$, given in Eq.~(\ref{eq:Itot}), does not change 
but the overall entanglement $I_{\text{overall}}$, given in Eq.~(\ref{eq:cost}), 
drops significantly from $126.47$ to $19.63$.
As a consequence, the maximum height and the spread of the block entropy 
is reduced significantly as shown in Fig.~\ref{fig:LiF_block_entropy}(a) 
and (b). Since Schmidt ranks are related to the block entropy,
the same accurracy can be reached using much less block states and sweeps
in the optimized (ordered) case. This leads to a huge save in CPU time
and memory\cite{Legeza-2003a,Legeza-2003b,Barcza-2011,Fertitta-2014}.

\subsubsection{Optimization of the network topology}
\label{sec:num.optim.opt-topology}

Another possibility to minimize the overall entanglement $I_\text{overall}$ given by Eq.~(\ref{eq:cost}) is 
to carry out network topology optimization. 
Based on the two-dimensional entanglement graph shown in 
Fig.~\ref{fig:I_free}(a), it is clear that orbitals are entangled with each
other with different strengths. Therefore, when a tensor network is formed,
the obvious choice is to allow the coordination number $z_i$ to vary 
from orbital to orbital\cite{Murg-2010a}. 

For the tree topology, see in section \ref{sec:num.block.ttns},
$d_{ij}$ in Eq.~(\ref{eq:cost}) can be computed as the distance from the 
center to $i$, plus the distance from the center to $j$,
minus twice the distance from the center to their lowest common ancestor.
The lowest common ancestor can be obtained within a linear 
preprocessing time $\mathcal{O}(d)$ and a constant query time
using the Berkman's algorithm\cite{Berkman-1993}.

In practice, the optimal structure of the tree tensor network can be 
determined in a self-consistent way. 
First the one-orbital entropy and two-orbital
mutual information is calculated  
with $z_i=2$ and fixed small number of block states
using the ordering of orbitals for which the $T_{ij}$ and $V_{ijkl}$ 
integral files were generated  
in order to determine entropy profiles qualitatively.
Next orbitals with largest entropy values are placed close to 
the center of the network by keeping together those orbitals which are 
connected by large $I_{ij}$ bonds as is shown in Fig.~\ref{fig:I_opt}(c).
Using such an optimized tensor topology the overall entanglement 
optimized for the $z_i=2$ case can drop even further. In the present 
example for the LiF it reduces from $I_\text{overall}^{\rm MPS}=19.63$ to 
$I_\text{overall}^{\rm TTNS}=5.53$. 
As a result, the same numerical accuracy obtained with an MPS topology
could have been reached with smaller number of block states and 
using less iteration steps when the optimized 
tree topology was used\cite{Murg-2014}.   

The overall efficiency of the QC-TTNS method is determined by two major
parameters. On the one hand tensor ranks $M$ decrease by going 
from QC-DMRG to QC-TTNS, but on the order hand the orders $z$ of the tensors 
increases. Although, the computational cost of one iteration
step is proportional to $M^{z+1}$, the number of tensors 
with $z=1$ lying on the boundaries of the network increases exponentially 
when larger and larger systems are considered.
Therefore, there is an expected crossover in cpu time between
the full sweep of the QC-MPS and QC-TTNS. 
It is worth mentioning, that a two-orbital variant of the TTNS ansatz
has also been considered in which 
the $z-1$ environment blocks are mapped into one environment block
through the so-called \emph{half-renormalization} (HR) algorithm\cite{Nakatani-2013}.
At present, optimization tasks are less established 
and straightforward, thus 
further developments are mandatory in order to fully utilize the potentials 
relying behind the TTNS algorithm.

\subsubsection{Optimization of the basis using entanglement protocols}
\label{sec:num.optim.opt-basis}

\begin{figure}
\centering
\setlength{\unitlength}{40pt}   
\begin{picture}(11,4.5)
\put(  0,  0){\includegraphics[width=0.45\columnwidth]{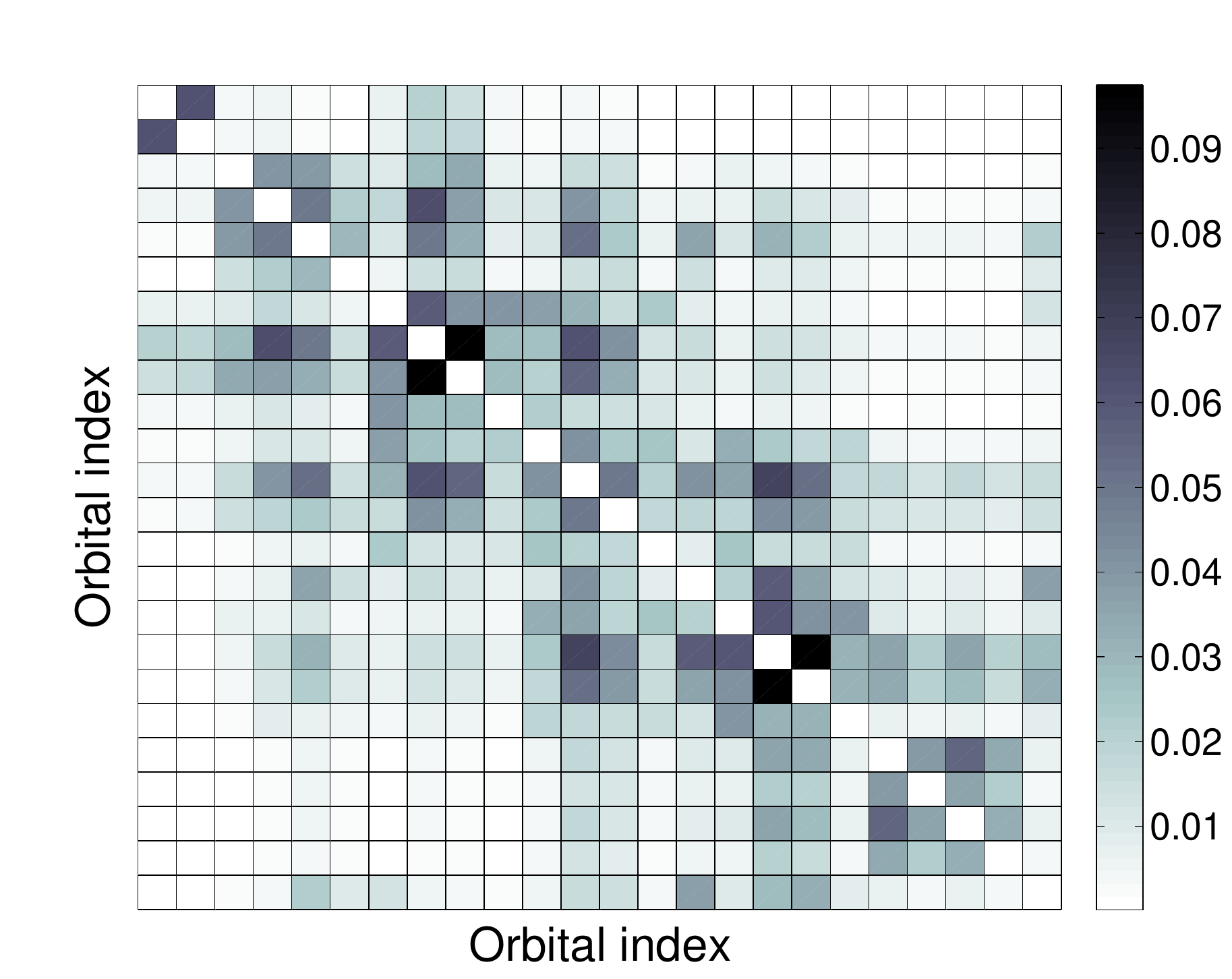}}
\put(5.5,  0){\includegraphics[width=0.45\columnwidth]{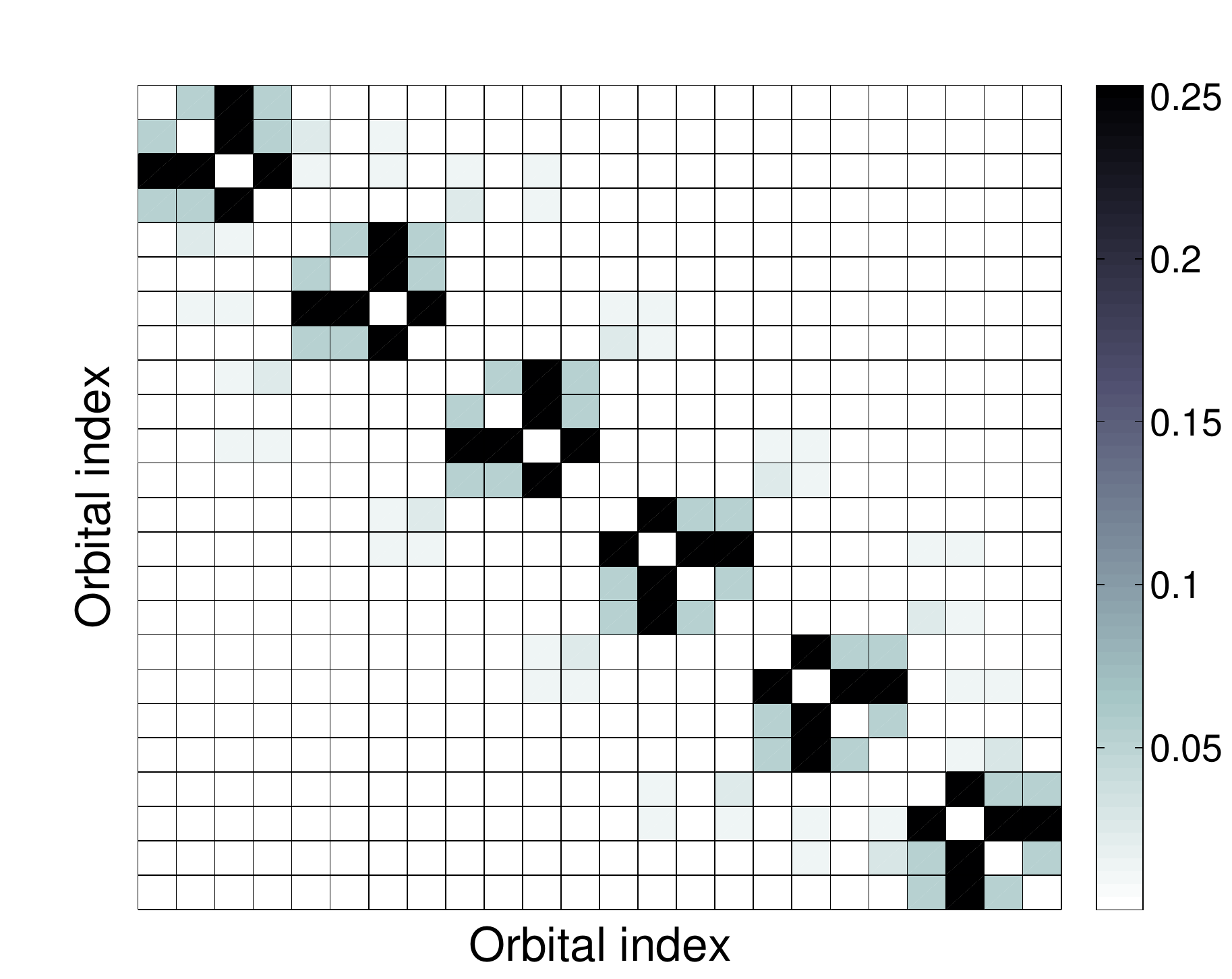}}
\put(0.5,4.5){\makebox(0,0)[r]{\strut{}(a)}} 
\put(6.0,4.5){\makebox(0,0)[r]{\strut{}(b)}}
\end{picture}
\caption{
Colorscaled plot of
two-orbital mutual information
(for optimized orbital ordering using the Fiedler vector)
for the ground state for Be$_6$ for a stretched structure, 
$d_{\rm Be-Be}=3.30$\AA, using the DMRG method with canonical (a) 
and local (b) orbitals.
 $I_{\rm tot}=7.81$, $I_\text{overall}=332.38$ with the canonical basis 
and $I_{\rm tot}=5.83$, $I_\text{overall}=58.1$ with the local basis.}
\label{fig:fertitta}
\end{figure}

In the past 15 years various orbital bases have been employed 
to study quantum chemical systems 
\cite{White-1999,Legeza-2003c,Rissler-2006,Barcza-2011,Boguslawski-2012a,Boguslawski-2012b,Ghosh-2008,Luo-2010,Mitrushchenkov-2012,Ma-2013,Wouters-2014a,Yanai-2009,Kurashige-2013}.
Although the impact of a given basis on the efficiency of the QC-DMRG
or QC-TTNS can be monitored by the convergence of the energy, 
a rigorous analysis in terms of the resulting entanglement patterns 
is mandatory in order to choose the most appropriate basis\cite{Fertitta-2014}. 
This is due to the fact, that the mutual information is 
orbital basis dependent. 
Therefore, besides orbital ordering
and optimization of tensor topology the overall entanglement 
$I_\text{overall}$ can be manipulated by changing the orbital 
basis as well. 
The performance of QC-DMRG and QC-TTNS can be optimized by using proper choice 
of the orbital basis, {\emph i.e.}, the same state can be obtained with much 
smaller number of block states\cite{Legeza-2006b,Fertitta-2014,Wouters-2014b}. 
As an example, 
entanglement patterns reported\cite{Fertitta-2014} 
for a ring of Be atoms 
using canonical HF and localized (Foster-Boys\cite{Boys-1960}) orbitals 
are shown in Fig.~\ref{fig:fertitta}.
The overall entanglement has been found to be much smaller in the 
latter case and as a consequence the same accuracy has been reached with much 
smaller number of block states. 

Therefore, a main goal is to find a basis in which entanglement is 
localized as much as possible at the orbitals of the network, what
would
guarantee that a given precision
could be attained with a smaller number of block states, 
and thus with less computational effort.
One possibility is to find the optimal basis
can be obtained by a canonical transformation of  
the fermionic modes using an $d\times d$
unitary matrix $U$, see section~\ref{sec:qchem.xmpleHFchbasis}. 
In general, there are two ways to implement the basis transformation: 
one is based on the state and the other is based on 
the Hamiltonian, i.e., 
\begin{displaymath}
E(U) \equiv 
\langle \Psi| U H U^\dagger |\Psi\rangle=
\langle \Psi(U)| H |\Psi(U)\rangle =
\langle \Psi| H(U) |\Psi\rangle.
\end{displaymath}
Since $E(U)$ is a non-convex function of the parameters~$U$, 
it is a highly non-trivial problem to find the absolute minimum. 
Gradient search has been applied\cite{Murg-2010a} to the 
function $E(U)$ expressed as
\begin{displaymath}
E(U) =
\sum_{i j} \tilde{T}(U)_{i j} \expect{a_i^{\dagger} a_j} +
\sum_{i j k l} \tilde{V}(U)_{i j k l} \expect{a_i^{\dagger} a_j^{\dagger} a_k a_l}
\end{displaymath}
with
$\tilde{T}(U)  =  U T U^{\dagger}$ and
$\tilde{V}(U)  =  (U \otimes U) V (U \otimes U)^{\dagger}$,
see equation (\ref{eq:htraf}).
In this case, the correlation functions $\expect{a_i^{\dagger} a_j}$ and
$\expect{a_i^{\dagger} a_j^{\dagger} a_k a_l}$ could be calculated with
respect to the original state since they are independet of the
parameters in $U$. The function $E(U)$ in this form and its gradient
could be calculated explicitly
and efficiently for different parameter sets~$U$, which made 
the gradient search feasible.
This assures that the energy decreases significantly in the
course of the algorithm
since the orbital optimization is performed repeatedly
during the course of the network optimization.

\subsubsection{Optimization of the network initialization based on entanglement}
\label{sec:num.optim.opt-init}

Besides ordering, network topology and basis states optimization 
the optimal performance of SVD based methods is strongly effected 
by the initial conditions,
or in other words by the initial matrix and tensor configurations.
If a poorly approximated starting 
configuration is used, the convergence can be very slow
and the DMRG can even be trapped in local minima\cite{Legeza-2003a,Legeza-2003b,Moritz-2006}. In the past decade various solutions have been introduced in order to 
optimize network initialization\cite{Mitrushenkov-2001,Chan-2002a,Legeza-2003b,Moritz-2006,Barcza-2011,Wouters-2014a}. In the following we focus on an entanglement based procedure.

Having a tensor network with a given topology and hierarchy (ordering)
the elements of the component tensors are random numbers 
in the first ietartion step. 
In QC-DMRG and QC-TTNS methods various $\Xi$ truncated Hilbert spaces 
can be formed from different subsets of the corresponding basis states
in order to approximate $\Lambda$ Hilbert space.
In other words, for a given partitioning of the system into blocks
various environment blocks can be generated for a given system block.  

\begin{figure}
\centering
\includegraphics{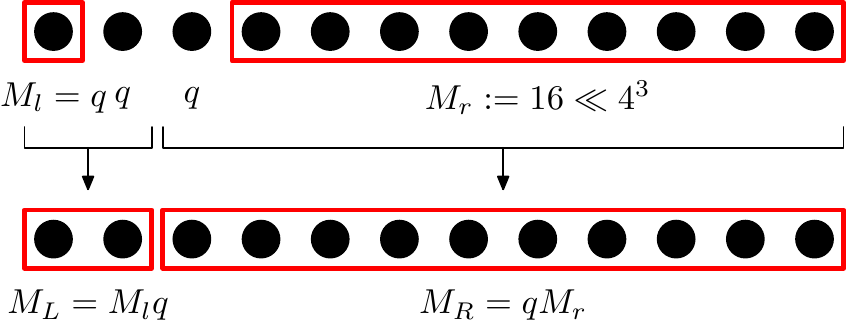}
\caption{A superblock configuration with $l=1$ and $r=9$.}
\label{fig:dmrg-deas-split}
\end{figure}

\begin{figure}
\centering
\includegraphics[width=0.40\columnwidth]{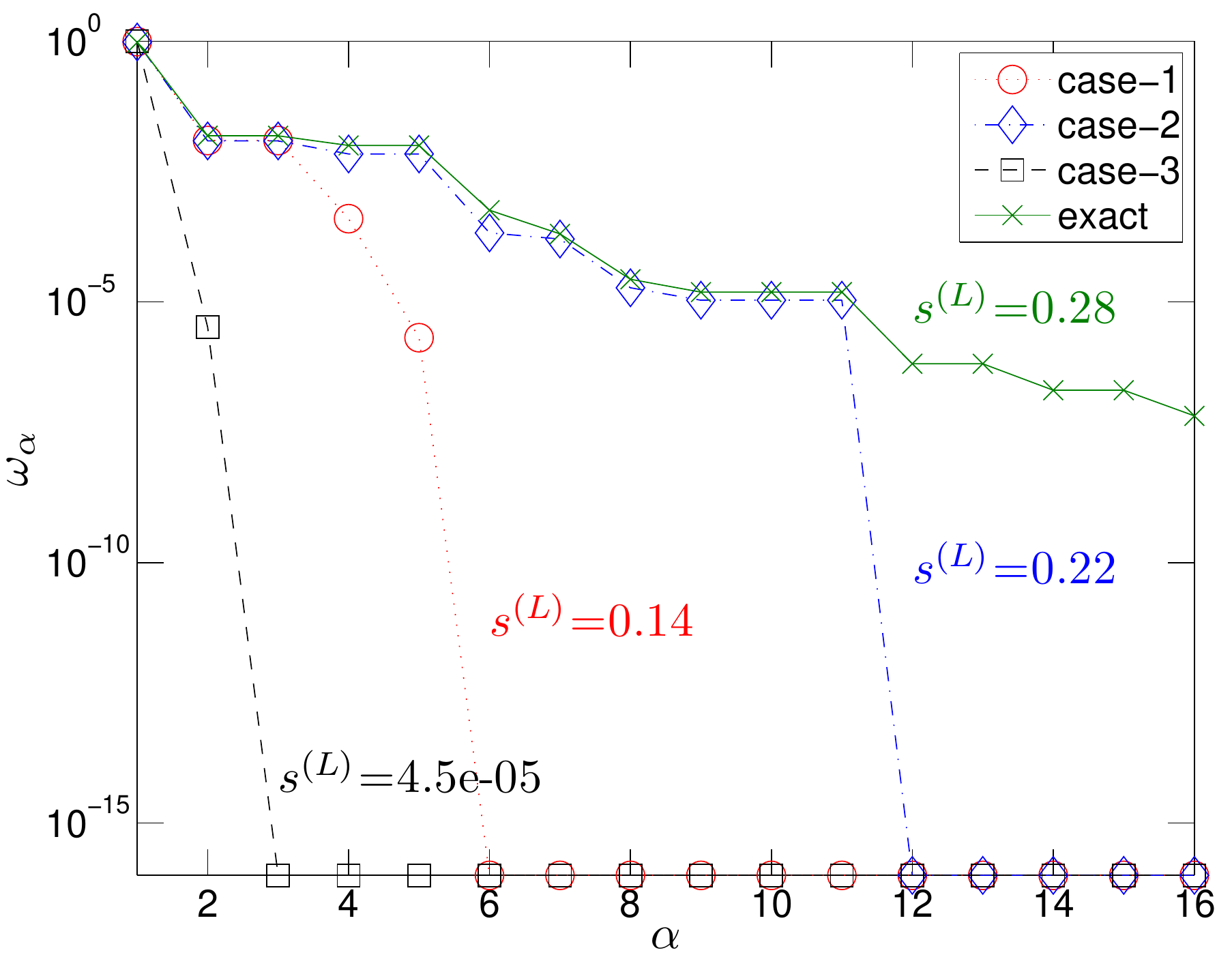}
\caption{Eigenvalue spectrum of $\rho^{(\mathrm{L})}$              
(and $\rho_{(\mathrm{R})}$) corresponding to the three subsets of 
environmnet states
shown in table~\ref{tab:dmrg-deas-split}.
}
\label{fig:lif-kullback}
\end{figure}

In case of the two-orbital QC-DMRG the optimization starts
with a superblock configuration
$l=1$ and $r=d-3$ as shown in Fig.~\ref{fig:DEAS}.
When the SVD is performed, the eigenvalue spectrum of the 
reduced density matrix of the $\mathrm (L)$ block 
depends on how the truncated basis was
formed for the right block. 
Since the exact representation
of the right block 
would require $M_r=q^{d-l-2}$ states, which is too 
large for large $d$,
only a subset of orbitals is included to form the active space.
As an example, three different environment blocks formed from
three different subsets of $M_r=16$ 
basis states 
(or $M_r=17$ due to spin reflection symmetry)
obtained for the 
superblock configuration $l=1$ and $r=9$ 
(see Fig.~\ref{fig:dmrg-deas-split}) of the 
LiF molecule for $d_{\text{Li-F}}=3.05$ a.u.~with CAS(6,12) 
are shown in table~\ref{tab:dmrg-deas-split}.
Using an ordering according to the energy, the 
first, second and third orbitals are the Hartree-Fock (HF) orbitals.
The selected $M_r=16$ environment states together with 
the $M_lq^2=64$ states of the $\mathrm (l)\bullet\bullet$ composite system
fulfill the conservation of total number of particles 
with up and down spins, i.e., 
$N_\downarrow^\mathrm{(l)}+
N_{l+1,\downarrow}+
N_{l+2,\downarrow}+
N_\downarrow^\mathrm{(r)}=3$ and
$N_\uparrow^\mathrm{(l)}+
N_{l+1,\uparrow}+
N_{l+2,\uparrow}+
N_\uparrow^\mathrm{(r)}=3$.
\begin{table}[t]
\renewcommand{\tabcolsep}{0.08cm}
{\footnotesize
\centering
  \begin{tabular}{|c|ccccccccc|c|}
  \hline
  \multicolumn{1}{ |c| }{$\alpha_r$}   & \multicolumn{9}{ |c| }{configuration 1} &  \multicolumn{1}{ |c| }{CI}\\
  \hline
  \hline
  1 & \zero &\zero &\zero & \zero & \zero & \zero & \zero & \zero & \zero & 0 \\ 
  2 & \zero &\zero &\zero &\zero &\zero &\zero &\zero &\zero &\down & 1 \\ 
  3 & \zero & \zero &\zero &\zero &\zero &\zero &\zero &\zero &\up & 1 \\ 
  4 & \zero & \zero &\zero &\zero &\zero &\zero &\zero &\zero &\double & 2 \\ 
  5 & \zero &\zero &\zero &\zero &\zero &\zero &\zero &\down &\zero & 1 \\ 
  6 & \zero &\zero &\zero &\zero &\zero &\zero &\zero &\up   &\zero  & 1 \\ 
  7 & \zero &\zero &\zero &\zero &\zero &\zero &\zero &\down &\down & 2 \\  
  8 & \zero &\zero &\zero &\zero &\zero &\zero &\zero &\up &\up & 2 \\ 
  9 & \zero &\zero &\zero &\zero &\zero &\zero &\zero &\down &\up & 2 \\ 
  10 & \zero &\zero &\zero &\zero &\zero &\zero &\zero &\up &\down & 2 \\ 
  11 & \zero &\zero &\zero &\zero &\zero &\zero &\zero &\double &\zero & 2 \\ 
  12 & \zero &\zero &\zero &\zero &\zero &\zero &\down &\zero &\zero & 1 \\ 
  13 & \zero &\zero &\zero &\zero &\zero &\zero &\up &\zero &\zero & 1 \\ 
  14 & \zero &\zero &\zero &\zero &\zero &\zero &\down &\zero &\up & 2 \\ 
  15 & \zero &\zero &\zero &\zero &\zero &\zero &\up &\zero &\down & 2 \\ 
  16 & \zero &\zero &\zero &\zero &\zero &\zero &\down &\down &\zero & 2 \\ 
  17 & \zero &\zero &\zero &\zero &\zero &\zero &\up &\up &\zero & 2 \\ 
  \hline
  \hline  
    & E & E & E & E & E & E & A & A & A & \\
  \hline
  \end{tabular}
\hspace{0.1cm}
  \begin{tabular}{|c|ccccccccc|c|}
  \hline
  \multicolumn{1}{ |c| }{$\alpha_r$}   & \multicolumn{9}{ |c| }{configuration 2} &  \multicolumn{1}{ |c| }{CI}\\
  \hline
  \hline
  1 & \zero &\zero &\zero & \zero & \zero & \zero & \zero & \zero & \zero & 0 \\ 
  2 & \zero &\zero &\zero &\zero &\zero &\down &\zero &\zero & \zero & 1 \\ 
  3 & \zero & \zero &\zero &\zero &\zero &\up  &\zero &\zero &\zero  & 1 \\ 
  4 & \zero & \zero &\zero &\zero &\zero &\double &\zero &\zero &\zero  & 2 \\ 
  5 & \zero &\zero &\zero &\zero &\zero &\zero &\down &\zero &\zero & 1 \\ 
  6 & \zero &\zero &\zero &\zero &\zero &\zero &\up   &\zero    &\zero  & 1 \\ 
  7 & \zero &\zero &\zero &\zero &\zero &\down &\down &\zero &\zero  & 2 \\  
  8 & \zero &\zero &\zero &\zero &\zero &\up &\up     &\zero &\zero  & 2 \\ 
  9 & \zero &\zero &\zero &\zero &\zero &\up &\down   &\zero &\zero  & 2 \\ 
  10 & \zero &\zero &\zero &\zero &\zero &\down &\up  &\zero &\zero  & 2 \\ 
  11 & \zero &\zero &\zero &\zero &\zero &\zero &\double &\zero &\zero & 2 \\ 
  12 & \zero &\zero &\zero &\zero &\zero &\zero &\zero &\zero   &\down & 1 \\ 
  13 & \zero &\zero &\zero &\zero &\zero &\zero &\zero &\zero   &\up & 1 \\ 
  14 & \zero &\zero &\zero &\zero &\zero &\down &\zero  &\zero &\down & 2 \\ 
  15 & \zero &\zero &\zero &\zero &\zero &\up   &\zero  &\zero &\up & 2 \\ 
  16 & \zero &\zero &\zero &\zero &\zero &\up &\zero &\zero &\down & 2 \\ 
  17 & \zero &\zero &\zero &\zero &\zero &\down &\zero &\zero &\up & 2 \\ 
  \hline
  \hline  
    & E & E & E & E & E & A & A & E& A & \\
  \hline
  \end{tabular}
  \hspace{0.1cm}
  \begin{tabular}{|c|ccccccccc|c|}
  \hline
  \multicolumn{1}{ |c| }{$\alpha_r$}   & \multicolumn{9}{ |c| }{configuration 3} &  \multicolumn{1}{ |c| }{CI}\\
  \hline
  \hline
  1 & \zero &\zero &\zero & \zero & \zero & \zero & \zero & \zero & \zero & 0 \\ 
  2 & \zero &\zero &\zero &\down &\zero &\zero &\zero &\zero &\zero  & 1 \\ 
  3 & \zero & \zero &\zero &\up &\zero &\zero &\zero &\zero &\zero  & 1 \\ 
  4 & \zero & \zero &\zero &\double &\zero &\zero &\zero &\zero &\zero & 2 \\ 
  5 & \zero &\zero &\zero &\zero &\down &\zero &\zero &\zero  &\zero & 1 \\ 
  6 & \zero &\zero &\zero &\zero &\up   &\zero &\zero &\zero &\zero  & 1 \\ 
  7 & \zero &\zero &\zero &\up &\down &\zero &\zero &\zero &\zero  & 2 \\  
  8 & \zero &\zero &\zero &\down &\up &\zero &\zero &\zero &\zero  & 2 \\ 
  9 & \zero &\zero &\zero &\zero &\double &\zero &\zero &\zero &\zero & 2 \\ 
  10 & \down &\zero &\zero & \zero & \zero & \zero & \zero & \zero & \zero & 1 \\ 
  11 & \up &\zero &\zero & \zero & \zero & \zero & \zero & \zero & \zero & 1 \\ 
  12 & \down &\zero &\zero &\up &\zero &\zero &\zero &\zero &\zero & 2 \\ 
  13 & \up   &\zero &\zero &\down &\zero &\zero &\zero &\zero &\zero & 2 \\ 
  14 & \down &\zero &\zero &\zero &\up &\zero &\zero &\zero &\zero & 2 \\ 
  15 & \up   &\zero &\zero &\zero &\down &\zero &\zero &\zero &\zero  & 2 \\ 
  16 & \double &\zero &\zero &\zero &\zero &\zero &\zero &\zero &\zero & 2 \\ 
  \hline
  \hline  
    & A & E & E & A & A & E & E & E & E & \\
  \hline
  \end{tabular} 
}
\caption{Three different subsets of states are formed from $M_r=16$ states 
(or $M_r=17$ due to spin reflection symmetry)
expressed explicitely in an one-orbital basis
which 
together with the $M_lq^2=64$ $\mathrm (l)\bullet\bullet$ subsystem states
fulfill the conservation of total number of particles with up and down spins.
Labels E and A stands for empty and active orbitals respectively.
In the very right columns the corresponding CI levels are indicated.
The CAS vectors describing the three configurations are
$( 1, 2, 3, 4, 5, 6, 7, 8, 9,10,11,12)$,
$( 2, 1, 9,10, 3,12,11, 6, 5, 4, 8, 7)$, and
$( 7, 8, 4, 5, 6,11,12, 3,10, 9, 1, 2)$, respectively.}
\label{tab:dmrg-deas-split}
\end{table}
By forming the bi-partite spliting of the system with $L=2$ and $R=10$
the eigenvalue spectrum of $\rho^{(\mathrm{L})}$ 
(and $\rho^{(\mathrm{R})}$) corresponding to the three subsets and 
the one corresponding to the exact solution obtained by $M_r=8000$
block states are shown in Fig.~\ref{fig:lif-kullback}.
It is obvious that the block entropy of the system block, $S^{(\mathrm{L})}$, 
depends on the basis states used to construct the ($\mathrm{R}$) environment 
block, as it increased from $10^{-5}$ to $0.22$. 
Therefore, $S^{(\mathrm{L})}$ should be maximized by finding the best 
representation of the environment block for a given superblock 
configuration and target state, i.e, to get as close as possible
to the exact solution, in the present case to $S^{(\mathrm{L})}=0.28$.

This can be achieved by including highly entangled orbitals
from the very beginning in the calculations.
Therefore, in order to achieve fast and stable 
convergence the active space has to be expanded iteratively using orbitals 
with largest one-orbital entropy values. The sequence by which orbitals 
are taken into account is determined by the so called CAS-vector, which is 
simply a rendered sequence of orbital indices with decreasing one-orbital
entropy value. The initial CAS vector can be determined
based on the chemical character of the molecule or in a self-consistent 
fashion based on the single-orbital entropies.
These features are incorporated in the  DEAS procedure\cite{Legeza-2003b},
see in section \ref{sec:num.block.dmrg},
starting with superblock configuration as shown in 
Fig.~\ref{fig:dmrg-deas-split}. 

This approach has also been extended by including protocols based on
the Configuration Interaction (CI) 
procedure\cite{Legeza-2004a,Workshop-CECAM-2010}. 
In standard CI techniques, the trial wave function is written
as a linear combination of determinants with expansion coefficients 
determined by requiring
that the energy should be minimized\cite{Jensen-1999}. 
The number of determinants included in the 
CI wave function expansion is increased systematically
in order to achieve a better accuracy. 
The exact wave function can be expressed as
\begin{equation}
\label{eq:CIHFSDT}
\Psi_{\rm CI} = a_{\rm HF} \Phi_{\rm HF} + \sum_S a_S \Phi_S + 
                 \sum_D a_D \Phi_D + \sum_T a_T\Phi_T + \dots
\end{equation}
where determinants indicated by the subscripts $S$,\ $D$,\ $T$,\ $Q$
are singly, doubly, triply, quadruply, etc. excited relative
to the HF configuration.

If the HF orbitals 
are known, one can keep only those right block states which
together with the $lq^2$ states of the $\mathrm (l)\bullet\bullet$ 
composite system describe an excitation corresponding to a given CI-level.
In the first iteration step this can be determined explicitly and
the various CI excitations corresponding to basis states 
shown in Fig.~\ref{fig:dmrg-deas-split} are given in the right column.
In subsequent iteration steps, 
HF and non-HF orbitals can get mixed in renormalized multi-orbital 
basis states and they thus cannot be labeled by the CI excitation level.
Nevertheless, the maximum CI level that block states could correspond
to depends on the number of HF orbitals falling into the given block. 
Since the segment of the HF-orbitals belonging to the right(environment) block 
is known, the restricted subspace of the environment block can be formed for 
a given CI-level in the CI-DEAS procedure. Therefore, the right block 
contains states for a given CI-level while the total wave function can contain 
higher excitations as well due to the correlation between the two blocks.
This procedure allows one to control the minimum CI-level to be used
and a double optimization is carried out in each iteration step.
On the one hand,
the environment block states are constructed at each iteration step
based on the left block basis states, thus they are optimized for the 
renormalized system (left) block. On the other hand, during the SVD 
step the left block
states are optimized according to a well represented environment block,
thus the reduced density matrix is well defined and 
block states can be selected efficiently based on the entropy considerations
(DBSS, see in section \ref{sec:num.optim.opt-dbss}). 
This procedure guarantees that several highly entangled orbitals are 
correlated from the very beginning and both static and dynamic correlations 
are taken into account, which helps to 
avoid convergence to local minima.
Since a significant part 
of the correlation energy can be obtained in this way, usually at the 
end of the initialization procedure, i.e., after one-half sweep, 
chemical accuracy is reached.
The starting value of $M_r$ ($M_{\rm start}$) is set
prior to the calculation, but
during the iteration procedure $M_r$ is adjusted as 
$M_r = \max(M_l,M_{\rm start})$ in order to
construct at least as many environment states as the left block has
(to avoid zero Schmidt values).

The CI-DEAS procedure also has an important technical aspect.
Based on the selected $M_r$ basis states  
orbitals of the right block can be identified as doubly filled (D), 
empty (E) or active (A). If only the empty states appear 
in a given column of the configurational space as
shown in table \ref{tab:dmrg-deas-split}
the orbital is 
considered as empty, while if only the doubly filled state appears
it is considered as doubly filled. 
Otherwise, the orbital is active. 
This is indicated explicitely in the last rows of 
table \ref{tab:dmrg-deas-split}.
It has been shown that empty orbitals can be neglected, while a partial 
summation over the doubly filled orbitals gives some
corrections to the terms obtained by the partial summation over
the active orbitals. 
Therefore, the effective size of the environment block can be
reduced to the number of active orbitals\cite{Legeza-2003b,Legeza-2004a}.
Usually the number of active orbitals in the environment block range from 
5 to 10  which allows one to use larger $M_{\rm start}$ without a 
significant increase in computational time.

\begin{figure}
\centering
\includegraphics[width=0.43\columnwidth]{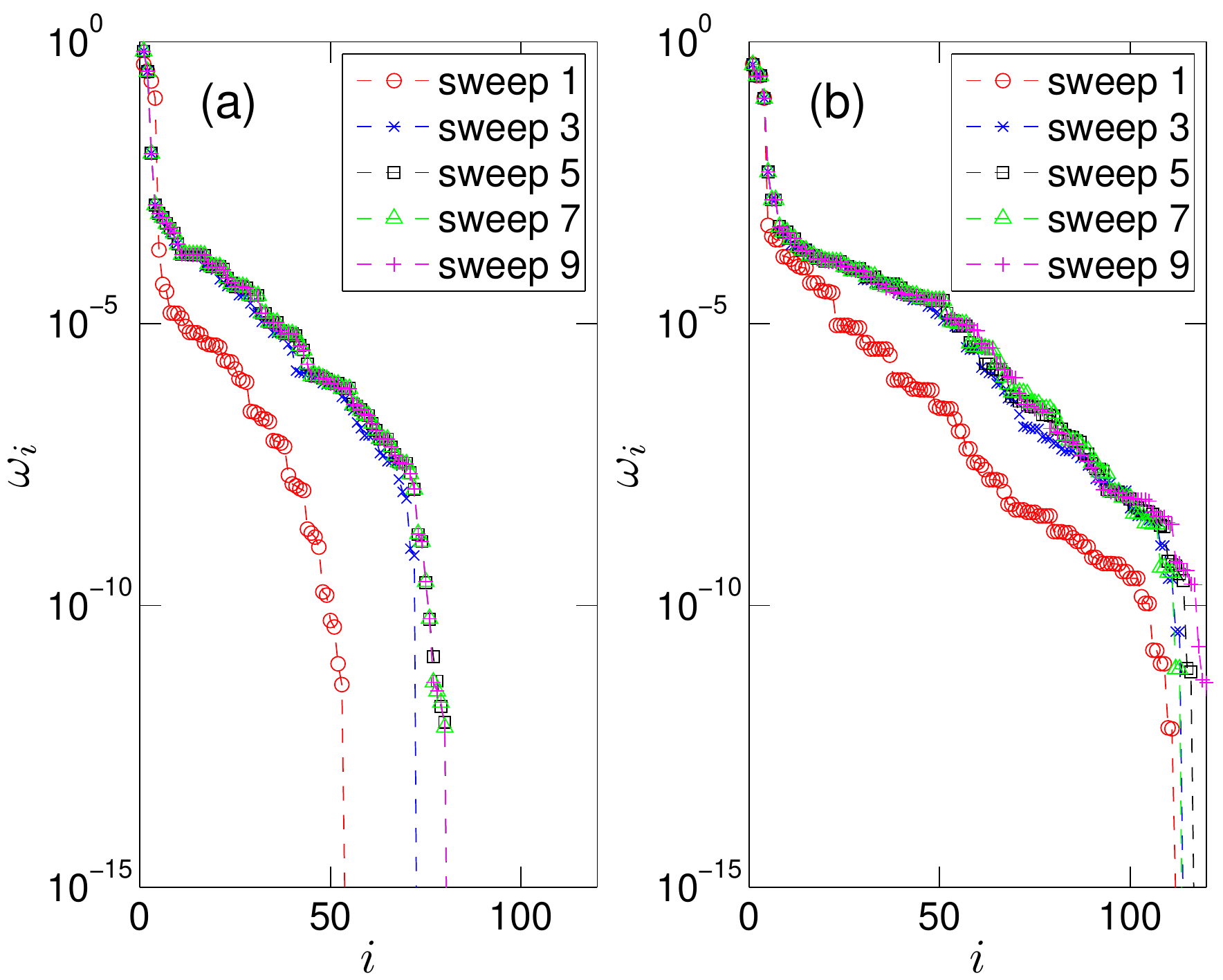}
\caption{Eigenvalue spectrum of the reduced subsytem density matrix 
for a block of $l=12$ contiguous orbitals as a function of DMRG sweeping
for the LiF at $d_{\text{Li-F}}=3.05$ a.u.~using a 
non-optimized initialization procedure (a) and the CI-DEAS 
procedure (b). 
}
\label{fig:rho-spectrum}
\end{figure}

As an example, the eigenvalue spectrum of the reduced subsytem density matrix 
for a block of $l=12$ contiguous orbitals as a function of DMRG sweeping
for the LiF CAS(6,25) at $d_{\text{Li-F}}=3.05$ a.u.~is shown in Fig.~\ref{fig:rho-spectrum}
using a non-optimized initialization procedure (a) and the CI-DEAS 
procedure (b). 

Inclusion of the CI-DEAS procedure into the QC-TTNS method is
straightforward. The only difference is that in a given iteration
step of the wramup sweep instead of two, $z_i-1$ environmnet blocks
has to be formed.

\subsubsection{Optimization of the sparsity using symmetries}
\label{sec:num.optim.opt-symm}

As has been introduced in Sec.~\ref{sec:num.basic.symm}, 
symmetry operators (with eigenvalues $Q$, called quantum numbers) can be used  
to decompose the Hilbert space into subspaces (sectors)\cite{Cornwell-1997}.
Therefore, the efficiency of the QC-DMRG and QC-TTNS methods 
can be increased significantly by applying quantum numbers. 
These include Abelian symmetries as particle number,
spin projection\cite{White-1999}, spin reflection\cite{Legeza-1997} and Abelian 
point group symmetries\cite{Legeza-2003b,Legeza-2003c,Chan-2004b}
and even non-Abelian symmetries\cite{Sierra-1997,McCulloch-2000,McCulloch-2001,McCulloch-2002a,McCulloch-2007,Pittel-2006,Rotureau-2006,Toth-2008,Legeza-2008b,Thakur-2008,Zgid-2008a,Singh-2010a,Singh-2010b,Singh-2012,Sharma-2012a,Weichselbaum-2012,Wouters-2012,Wouters-2014a,Weichselbaum-2012}.
In the latter case, the situation, however, becomes more complicated.
   
If symmetry generators commute with each other, the
eigenstates of the Hamiltonian form degenerate multiplets, 
$|\phi_{\alpha,{\underline Q}_\alpha, {\underline Q}_\alpha^z}\rangle$, 
that are classified by their label $\alpha$, 
the quantum numbers ${\underline Q}_\alpha$,
and the internal quantum number ${\underline Q}_\alpha^z$.
The dimension of a subspace (sector) $\alpha$ depends uniquely
on its quantum numbers $\underline{Q}_\alpha$, i.e.~$\dim(\alpha)=\dim(\underline Q_\alpha)$.
In the following, we use the shorthand notation introduced 
in Sec.~\ref{sec:num.basic} and write
$|\phi_{\alpha,{\underline Q}_\alpha, {\underline Q}_\alpha^z}\rangle$, 
as $|{\alpha,{\underline Q}_\alpha, {\underline Q}_\alpha^z}\rangle$.

In general, the symmetry operators of the Hamiltonian are
the representations $\mathcal{U}$
of the symmetry group $\mathcal{G}$ on the Hilbert space,
acting as
\begin{equation}
\mathcal{U}(g)\;\mathbf{H}\; \mathcal{U}^{-1}(g)= \mathbf{H},
\end{equation}
where $\mathcal{U}(g)$ is the unitary representation for the symmetry $g\in \mathcal{G}$.
Specially, if the symmetries are \emph{local} 
in the sense that they decompose into unitary
operators which commute with each other and act independently 
at different orbitals, then
not only the whole Hamiltonian 
but also every local and interaction Hamiltonian are invariant under the group $\mathcal{G}$.
Furthermore, $\mathcal{G}$ and correspondingly
$\mathcal{U}$ can be decomposed
into a direct product of $\Gamma$ subgroups
 $\mathcal{ G}_{\gamma}$ $(\gamma = 1,\dots,\Gamma)$, each acting 
independently on every orbital,
\begin{eqnarray}
\mathcal{G} &=&  \mathcal{G}_{1}\times\mathcal{G}_{2}\times\dots\times\mathcal{G}_{\Gamma}, \\
\mathcal{U}(g) &=& \prod_{\gamma=1}^{\Gamma} \mathcal{U}_\gamma (g_{\gamma}) =
\prod_{\gamma=1}^{\Gamma}\prod_i \mathcal{U}_{\gamma,i} (g_{\gamma}).
\end{eqnarray}
Once a specific decomposition of the symmetry is obtained,
$\Gamma$ number of quantum numbers classify the
irreducible subspaces (multiplets) of the subsystem Hamiltonians 
$\underline{Q} = \left\{Q^1, Q^2,\dots,Q^{\Gamma}\right\}$
and states within the multiplet are then labeled by the {\em internal} quantum 
numbers
$\underline {Q}^z =\left\{ Q^{1,z}, Q^{2,z},\dots,Q^{\Gamma,z} \right\}$.
The dimension of a subspace $\alpha$ depends uniquely
on its quantum numbers $\underline{Q}_\alpha$, 
i.e.~${\rm dim}(\alpha)={\rm dim}(\underline Q_\alpha)=\prod_{\gamma=1}^\Gamma\dim(Q_\alpha^\gamma)$.

Operators can also be arranged into irreducible tensor operators,
and an irreducible tensor operator multiplet $A$ is correspondingly
described  by quantum numbers $\underline a$, while members of the multiplet are labeled
by $\underline a^z$ with $\underline a$ and $\underline a^z$ being
$\Gamma$-component vectors. 
The Wigner--Eckart theorem\cite{Weyl-1927,Wigner-1939,Wigner-1931,Wigner-1959}
tells us that, apart from trivial group
theoretical factors (Clebsch-Gordan coefficients), the matrix elements 
of the members of a given operator multiplet and states within 
two multiplets, $|{\alpha,Q_\alpha,Q^z_\alpha}\rangle$ and 
two multiplets, $|{\alpha^\prime,Q_{\alpha^\prime},Q^z_{\alpha^\prime}}\rangle$  
are simply related by  
\begin{equation}
\left < {\alpha,{\underline Q}_{\alpha} \underline Q^z_{\alpha}} \right|  
A_{\underline a, \underline a_z}
 \left | {\alpha^\prime , \underline Q_{\alpha^\prime} \underline {Q}^z_{\alpha^\prime}}\right > = 
\left < \alpha \right . \parallel A\parallel \left . \alpha^\prime  \right > 
\left < \underline Q_{\alpha}  \underline Q^z_{\alpha^\prime} \right | \underline a, \underline a_z; 
\left .  \underline Q _{\alpha} \underline {Q}^z_{\alpha^\prime}\right >\;
\label{eq:Wigner-Eckart}
\end{equation}
where $\left < \alpha\right . \parallel A\parallel \left . \alpha^\prime  \right > $ 
denotes the reduced (invariant) matrix element of $A$, and the 
generalized Clebsch--Gordan coefficients are simply defined as 
\begin{equation}
\left<\underline{Q}^{}_\alpha\;\underline{Q}^z_\alpha\;\left|\underline{a}\;\underline{a}^{z};\underline{Q}^{}_{\alpha^\prime}\;
\underline{Q}^z_{\alpha^\prime}\right>\right.
\equiv \prod_{\gamma=1}^{\Gamma}\left<Q^{\gamma}_\alpha\;Q^{\gamma,z}_{\alpha}\;
\left|a^{\gamma}\;a^{\gamma,z}\;Q^{\gamma}_{\alpha^\prime};Q^{\gamma,z}_{\alpha^\prime}\right>\right..
\label{eq:general-Clebsch}
\end{equation}

In the presence of symmetries, 
one has to use the Clebsch-Gordan coefficients
to build $(\mathrm{L})$ block states from the block $(\mathrm{l})$ and orbital $\bullet$
states that transform as irreducible
multiplets under the symmetry transformations, $\mathcal{U}(g)$,
\begin{equation}
\begin{split}
\left | {m_{(\mathrm{L})},  \underline{Q}_{m_{(\mathrm{L})}}, \underline {Q}_{m_{(\mathrm{L})}}^z} \right > 
&\equiv
\sum_{\underline {Q}_{m_{(\mathrm{l})}}^z, \underline {Q}_{\alpha_{l+1}}^z}
\left < \underline{Q}_{m_{(\mathrm{L})}} \underline{Q}_{m_{(\mathrm{L})}}^z \right |
\underline{Q}_{\alpha_{l+1}} \underline{Q}_{\alpha_{l+1}}^z; \left . \underline{Q}_{m_{(\mathrm{l})}} \underline {Q}_{m_{(\mathrm{l})}}^z  \right >^*\\ 
&
\left |{\alpha_{l+1}, \underline{Q}_{\alpha_{l+1}}, \underline{Q}_{\alpha_{l+1}}^z} \right >\otimes 
\left | {m_{(\mathrm{l})}, \underline{Q}_{m_{(\mathrm{l})}}, \underline{Q}_{m_{(\mathrm{l})}}^z}\right > , 
\end{split}
\label{eq:new-states}
\end{equation}
Therefore, subsystem Hamiltonians have a block-diagonal structure and subsystem
reduced density matrices are also scalar under symmetry operators.
This decomposition property is crucial for using symmetries in the  QC-DMRG
and QC-TTNS calculations in order to boost their performance.

To give a simple example, 
let us take into account the spin and charge symmetries, i.e., 
$\mathcal{G}=\mathcal{G}_\text{spin}\times\mathcal{G}_\text{charge}$.
If we use only 
$\mathcal{G}_\text{spin}=U(1)$ and $\mathcal{G}_\text{charge}=U(1)$ 
symmetries then one has two hopping operators,  
$\mathbf{c}^\dagger_{i, \uparrow}$ and $\mathbf{c}^\dagger_{i, \downarrow}$ as
defined in Sec.~\ref{sec:num.basic}. 
In contrast to this, if we use $\mathcal{G}_\text{spin}=SU(2)$ spin symmetry, (while $\mathcal{G}_\text{charge}=U(1)$ remains the same as before)
$\mathcal{U} = \mathcal{U}_{\rm spin} \mathcal{U}_{\rm charge}$,
with $\mathcal{U}_{\rm spin} = \prod_i  \mathcal{U}_{{\rm spin},i}$,
then only one hopping operator remains
since matrix elements of $\mathbf{c}^\dagger_{i,\uparrow}$ and $\mathbf{c}^\dagger_{i,\downarrow}$
are related with each other by symmetry and they form a single operator
multiplet $\mathbf{c}^\dagger_i=\{\mathbf{c}^\dagger_{i,\downarrow},\mathbf{c}^\dagger_{i,\uparrow}\}$ of spin 1/2. 
The matrix elements of such multiplet are
determined using the Wigner-Eckart theorem
\begin{equation} 
\langle\mu\parallel \mathbf{c}^\dagger_i\parallel\nu\rangle=
\begin{pmatrix}
 0 & 0 & 0         \\
 1 & 0 & 0         \\
 0 & -\sqrt{2} & 0
\end{pmatrix},
\end{equation} 
where the original $\mathbb{C}^4$ space is reduced to $\mathbb{C}^3$ 
since only three basis states remain $\mu,\nu\in\{|-\rangle$, $|\uparrow\rangle$, $|\uparrow\downarrow\rangle\}$.
When the system is half-filled, utilization of 
$\mathcal{G}_\text{charge}=SU(2)$ symmetriy besides the $\mathcal{G}_\text{spin}=SU(2)$ spin symmetry
is straightforward.
In this case the hopping operator becomes a $2\times2$ matrix
\begin{equation} 
\begin{pmatrix}
 0 & \sqrt{2}     \\
 -\sqrt{2} & 0
\end{pmatrix},
\end{equation} 
where the original $\mathbb{C}^4$ space further reduces to $\mathbb{C}^2$ 
since only two basis states remain, $|\phi_\mu\rangle,|\phi_\nu\rangle\in\{|\uparrow\rangle,|\uparrow\downarrow\rangle\}$.
A detailed derivation of reduced operators and construction of the 
block states prepared as a pedagogical introdcution to the field
 can be found in the literature\cite{Legeza-2008b},
and the related free \texttt{C++} sourcecode can be downloaded from 
\texttt{http://www.phy.bme.hu/\textasciitilde{}dmnrg/}\cite{Legeza-2008c}.
Other free source codes with SU(2) spin symmetries 
are also available\cite{Sharma-2012b,Wouters-2014d}. 
Utilization of symmetries allows one to target states with given symmetries 
and to keep $M$ number of multiplets which correspons to significantly more 
$U(1)$ states what is crucial 
in order to achieve good numerical accuracy.

\subsubsection{Stability of the wavefunction}
\label{sec:num.optim.stability}

Traditional post-HF quantum chemical methods like CI or Couple Clusters (CC)
systematically improve a refernce wavefunction (often only the HF determinant, as in (\ref{eq:CIHFSDT}))
by inclusion of single, double, and higher excitations
in the wave operator.
In case of CI the wave operator takes a linear form, while CC uses a
more sophisticated exponential ansatz.

\begin{figure}
\centering
\setlength{\unitlength}{40pt}   
\begin{picture}(11,4.5) 
\put(  0,  0){\includegraphics[width=0.43\columnwidth]{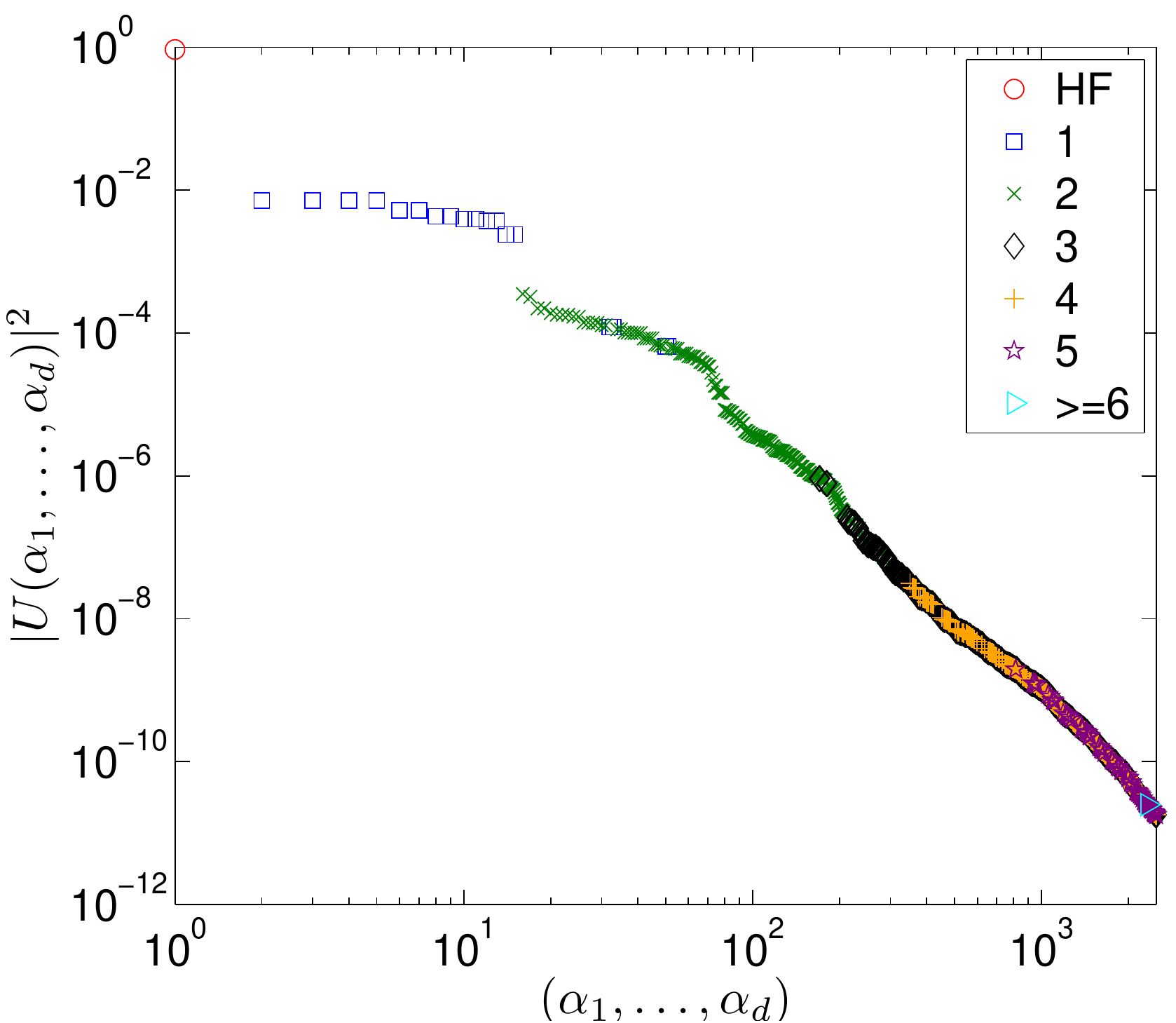}}
\put(5.5,  0){\includegraphics[width=0.43\columnwidth]{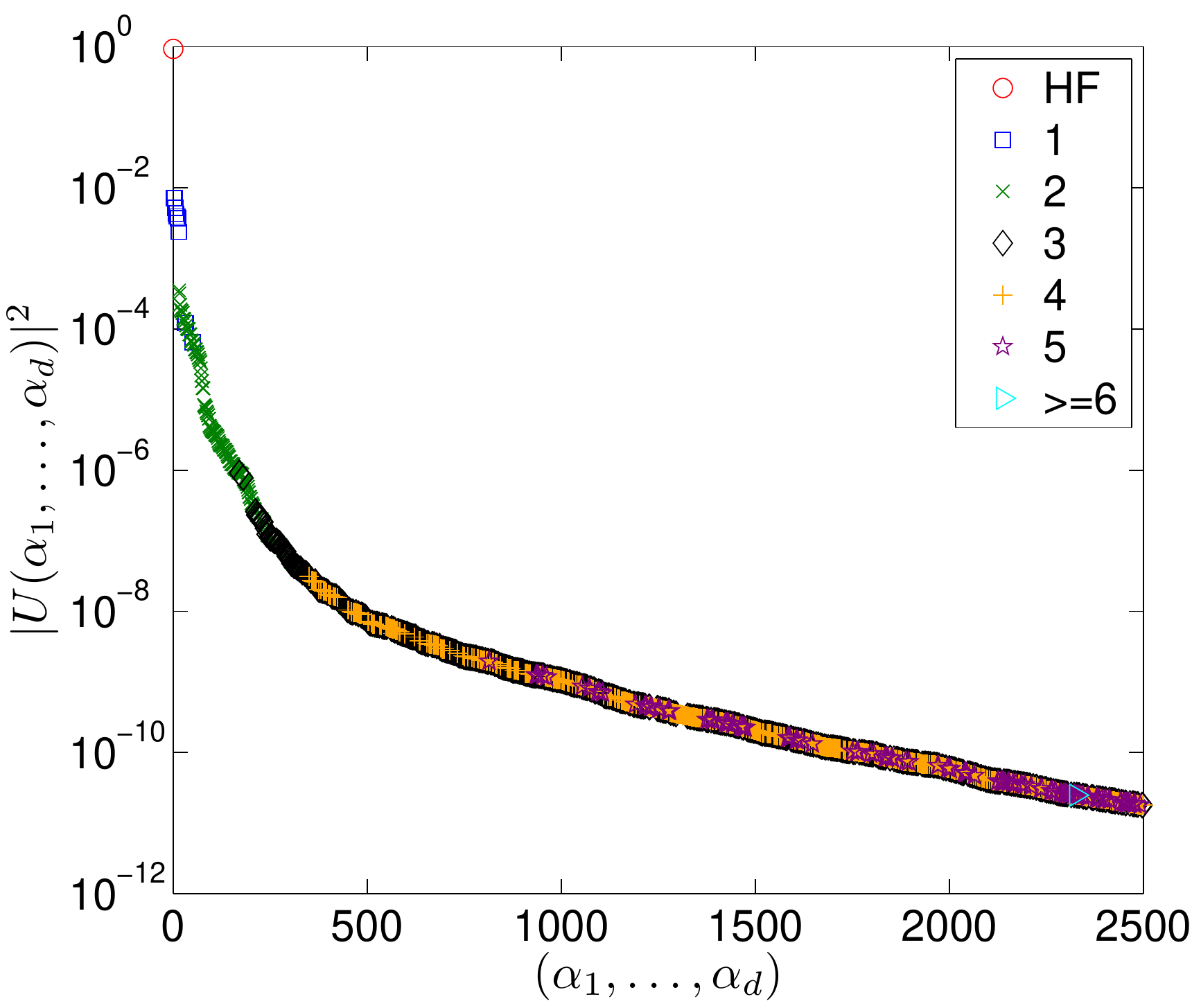}}
\put(0.5,4.5){\makebox(0,0)[r]{\strut{}(a)}}
\put(6.0,4.5){\makebox(0,0)[r]{\strut{}(b)}}
\end{picture}
\caption{
The $|U(\alpha_1,\ldots\alpha_d)|^2$ weight of the $d$-orbital basis states 
corresponding to the various CI-excitations 
are shown by different colors in a descending order in 
a log-log (a) and in a log-lin scale (b)
for the LiF for $d_{\text{Li-F}}=3.05$ with CAS(6,12).
HF state, SCI, DCI, etc are indicated by red, blue, green, 
etc. colors respectively. 
}
\label{fig:lif-full-tensor}
\end{figure}

In contrast to these, QC-DMRG and QC-TTNS take into account all the various 
excitations 
picking up the most important ones by minimizing the energy. 
As an example, the
$|U(\alpha_1,\ldots\alpha_d)|^2$ weights
casted according to an excitation level with respect to the
HF refernce wavefunction
are shown in Fig.~\ref{fig:lif-full-tensor}.
It demonstrates that higher excitation levels can be important
to provide a qualitatively correct description of the wavefunction.
(The elements of the full tensor $U(\alpha_1,\ldots\alpha_d)$ can be extracted,
according  to Eq.~(\ref{eq:full-tensor}).
But note that recovering all components of $U(\alpha_1,\ldots\alpha_d)$ 
cannot be done efficiently as its size scales exponentially.
However as a good approximation of the full CI wavefunction,
the Monte Carlo algorithm was used to recover the most important tensor 
components\cite{Boguslawski-2011}. )

\begin{figure}
\centering
\setlength{\unitlength}{40pt}   
\begin{picture}(11,4.5)  
\put(  0,  0){\includegraphics[width=0.45\columnwidth]{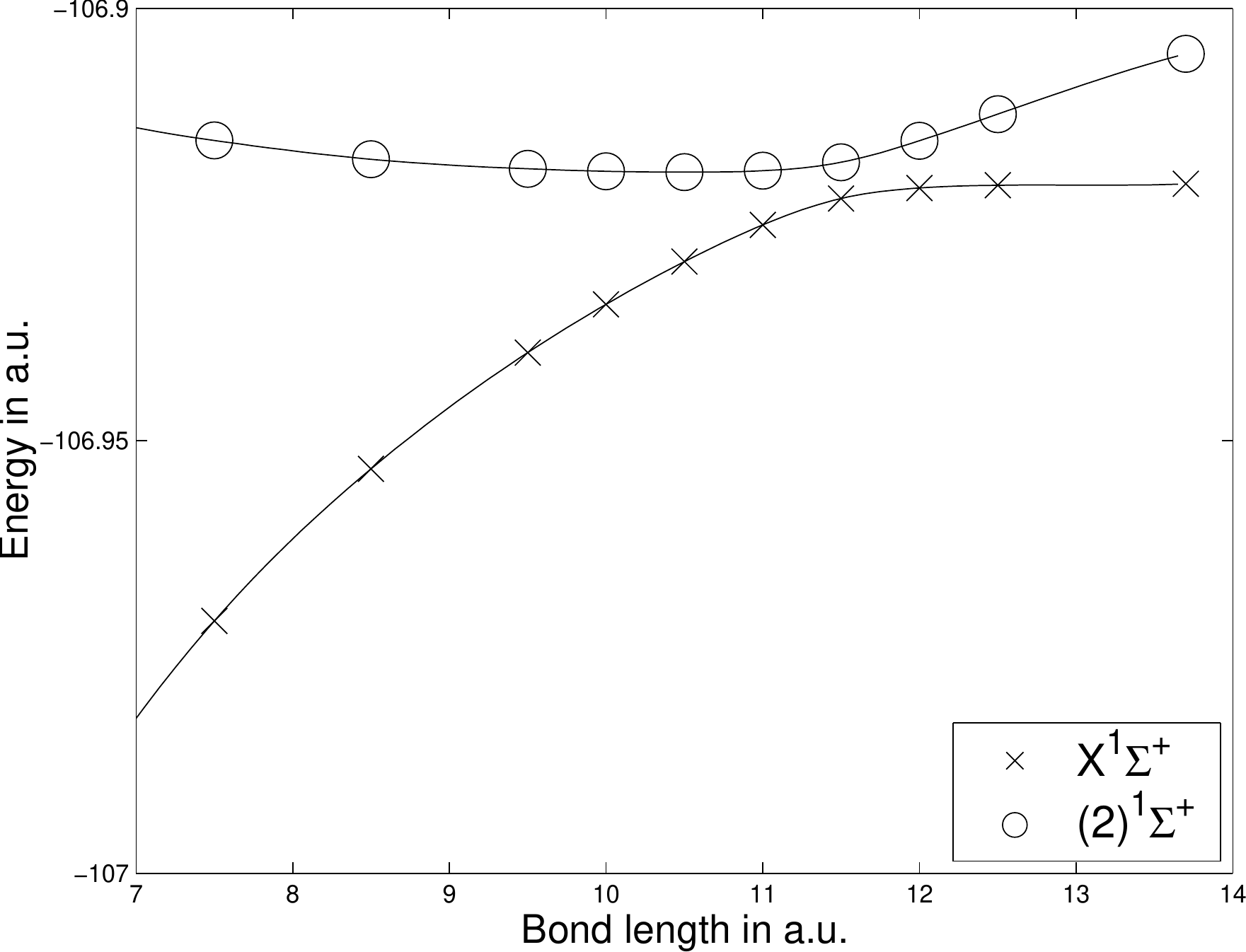}}
\put(5.5,  0){\includegraphics[width=0.45\columnwidth]{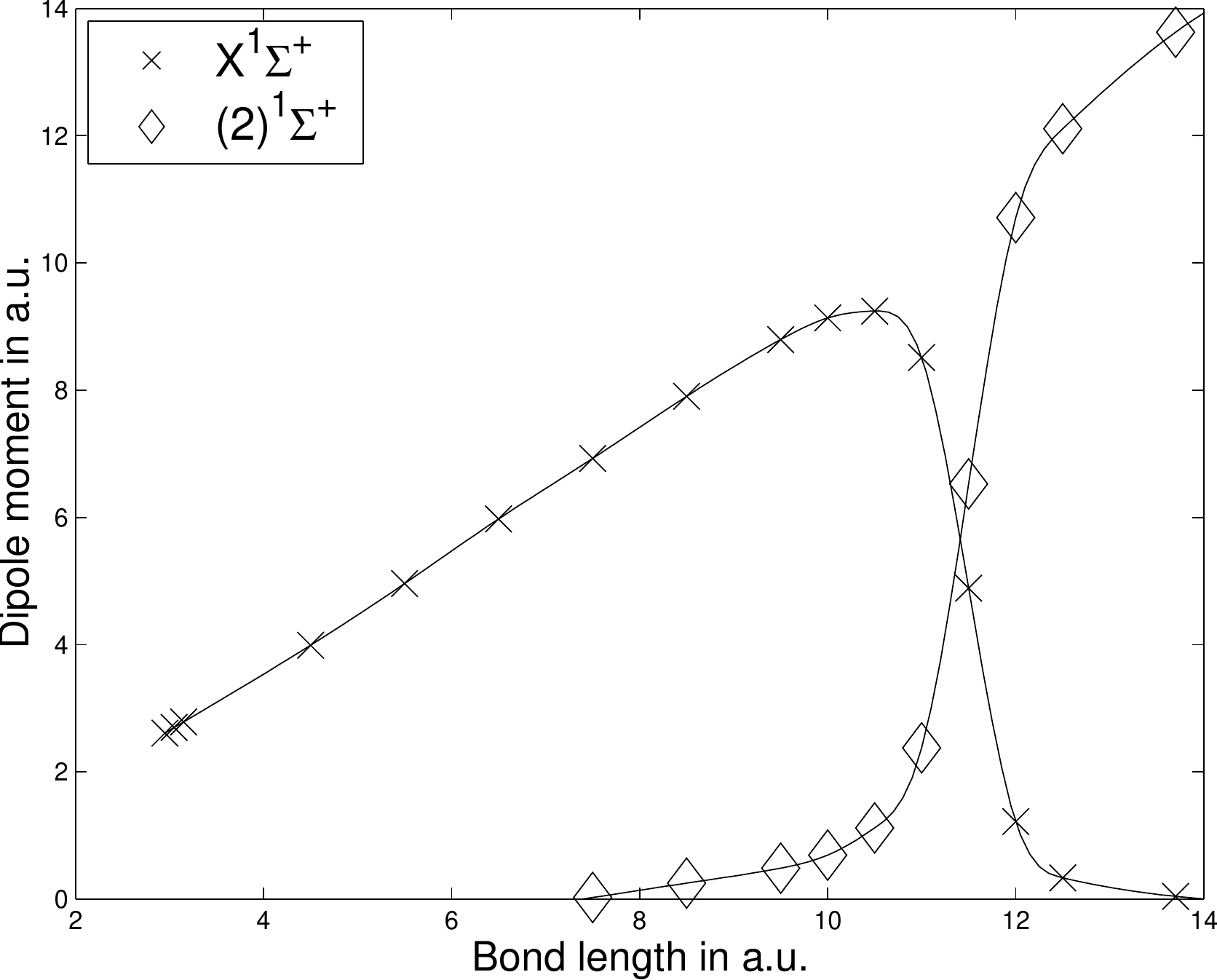}}
\put(0.5,4.5){\makebox(0,0)[r]{\strut{}(a)}} 
\put(6.0,4.5){\makebox(0,0)[r]{\strut{}(b)}}
\end{picture}
\caption{(a) Energy of the two lowest $^1\Sigma^+$ states 
as a function of the bond length obtained 
with $\delta\varepsilon_{\rm TR}=10^{-6}$ and $M_{\rm min}=16$ using 
the DBSS procedure. (b) The corresponding dipole moment functions. 
}
\label{fig:lif_energy}
\end{figure}

As a consequence, if the accuracy threshold of the 
calculation is lowered, the structure of the wave function is retained 
in essence. Since the DBSS procedure takes care of the change in the 
entanglement as the system parameters are adjusted, for example, 
when the bond length in LiF is changed,
the various calculated quantities
are continuous functions for a given  
$\delta \varepsilon_{\rm TR}$. 
For the ionic--neutral curve crossing in LiF\cite{Legeza-2003c} 
this has been demonstrated for 
the two lowest $^1\Sigma^+$ states and the dipole moment function 
as illustrated in Fig.~\ref{fig:lif_energy}(a) and (b)
for $\delta\varepsilon_{\rm TR}=10^{-6}$ and $M_{\rm min}=64$.
In addition, when
parameters were cutted drastically and very small value of
$M_{\rm min}$ and large $\delta\varepsilon_{\rm TR}$ were used the dipole 
moment deviated more significantly from the full-CI results but
they remained continuous even close to the avoided crossing. 
Therefore, the most important components of the wave function are
included by the SVD procedure which provides a stable representation of
the wavefunction.  
Similar results have been reported for the QC-TTNS method\cite{Murg-2014}.

\subsubsection{Possible black-box QC-DMRG and QC-TTNS}
\label{sec:num.optim.blackbox}

A possible black-box QC-DMRG and QC-TTNS can be composed of two 
phases: the {\em preprocessing phase} in which the ordering, network 
topology and CAS-vector are optimized using fixed small number of block states 
and the {\em production phase} in which an accurate calculation is performed 
using the DBSS procedure in order to reach an a priory set error margin. 
In the preprocessing phase, one can use the ordering for which the integral 
files were generated and a random CAS vector using limited number of block 
states. After a full sweep the one-orbital entropy can 
be calculated
from which the CAS vector can be determined. In a similar way 
the two-orbital mutual information and the optimal ordering 
can be calculated using the Fiedler vector. Next a DMRG calculation can 
be carried out with the optimized ordering and CAS-vector and the whole 
cycle is repeated until we obtain lower total energy. 
In the next step this procedure is repeated, but with larger number 
of block states. 
The preprocessing phase takes only a small fraction of the total 
computational time.

\subsection{Miscellaneous}
\label{sec:num.misc}

\subsubsection{Simulation of real materials, geometrical optimization and excited states}
\label{sec:num.misc.PDA}

As an example, we demonstrate on \emph{poly-diacetylene} (PDA) chains 
that MPS based methods can be used
very efficiently to simulate 
strongly anisotropic materials in terms of effective Hamiltonians.

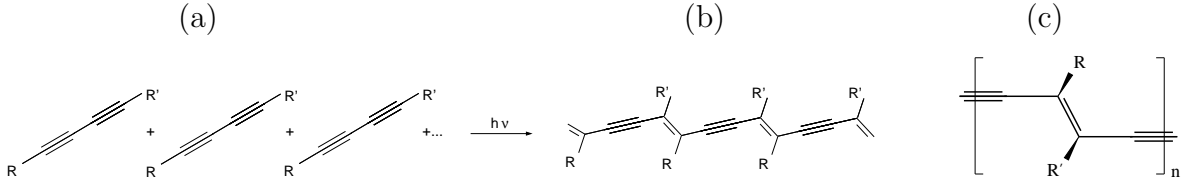
\begin{figure}
\centering
\setlength{\unitlength}{40pt}
\begin{picture}(11,1.5)
\put(  0,  0){\includegraphics[width=0.7\columnwidth]{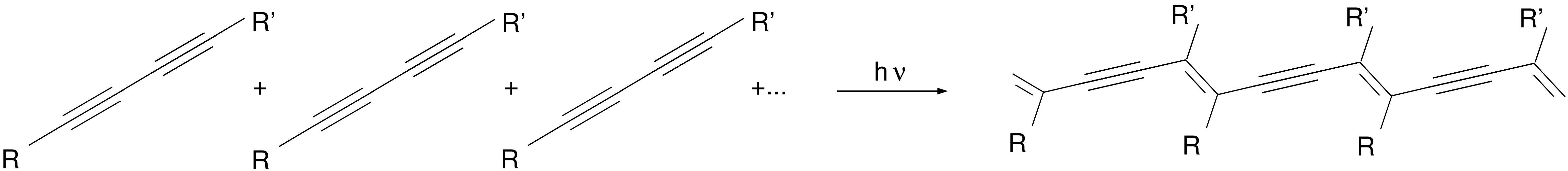}}
\put(9.0,  0){\includegraphics[width=0.18\columnwidth]{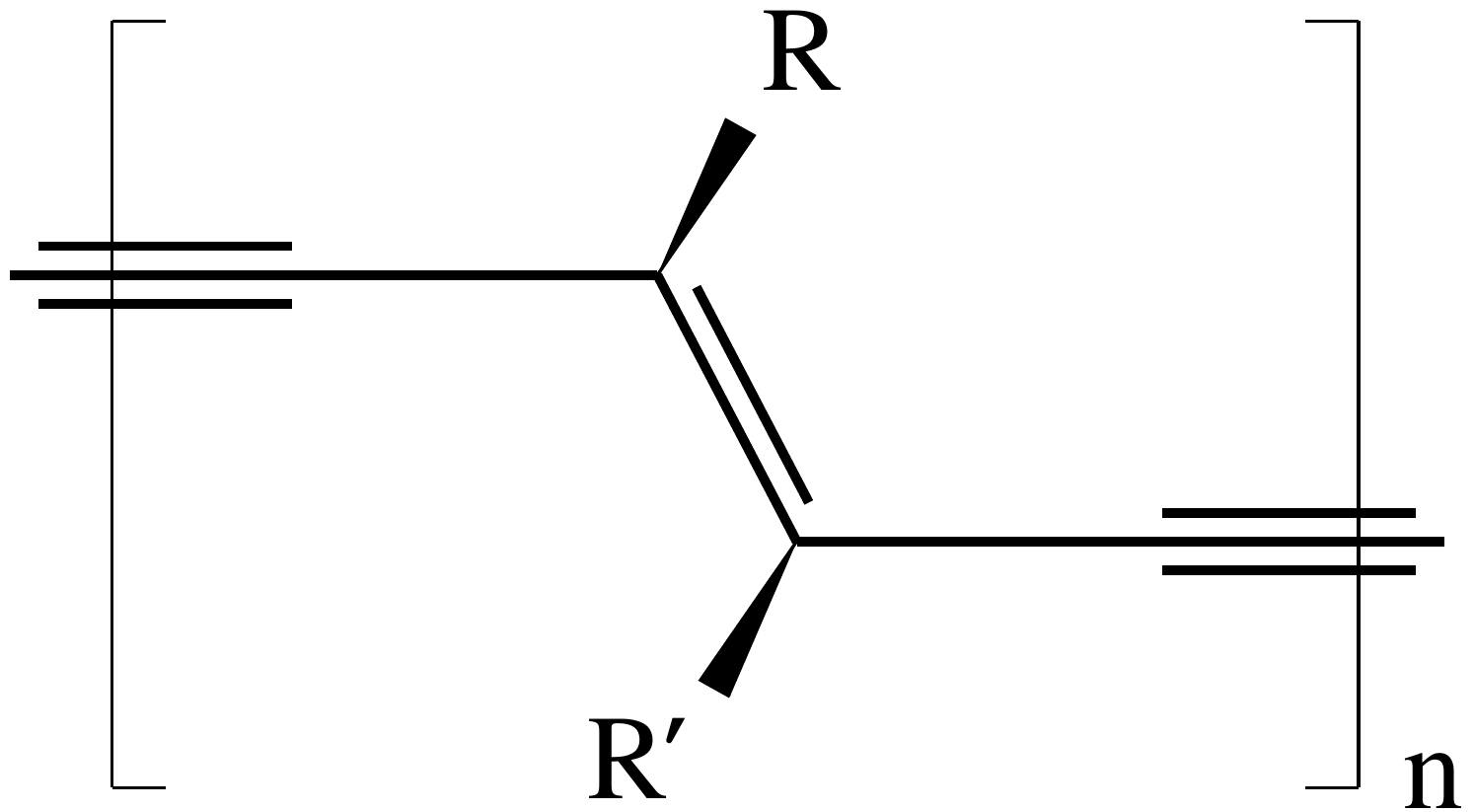}}
\put(2.0,1.5){\makebox(0,0)[r]{\strut{}(a)}}
\put(6.8,1.5){\makebox(0,0)[r]{\strut{}(b)}}
\put(10 ,1.5){\makebox(0,0)[r]{\strut{}(c)}}
\end{picture}
\caption{(a) Single crystals of diacetylene monomers prepared experimentally where
the monomer unit has the structure R-C$\equiv $C-C$\equiv$C-R. In the
$n$BCMU family R = (CH$_2$)$_n$-OCONH-CH$_2$-COO-(CH$_2$)$_3$CH$_3$.
(b) Polymerization induced by temperature or ultraviolet light leading to
polydiacetylenes (C$_4$R$_2$)$_x$.
(c) Lewis structure of a poly-diacetylene unit cell with 
single, double and triple bond lengths
$r_t=1.20$\AA, $r_d=1.36$\AA$\,$ and $r_s=1.43$\AA, respectively.}  
\label{fig:pda_pda_polimerization}
\end{figure}

PDA chains dispersed with low concentration
in their monomer single-crystal matrix are prototypical quasi one-dimensional
materials\cite{Bloor-1985,Cantov-1984,Sariciftci-1998,Schott-2006}.
The structural disorder in the chains
and their surrounding matrix is tiny,
thus these materials form the perfect testing-ground for
theoretical model studies describing interacting electrons
on perfectly ordered chains.
In addition, the electronic excitation energies of the diacetylene monomers
are much higher than those of the polymer, and the
electronic excitations of the chain in the energy range of visible light
can be measured with a very high accuracy\cite{Spagnoli-1994}.
Polymerization induced by temperature or ultraviolet light is shown 
in Fig.~\ref{fig:pda_pda_polimerization}.

The opto-electronic properties of the PDAs
are determined by two main correlation effects: 
the mutual interaction of the electrons and their interaction 
with the lattice potential\cite{Dubin-2006,Schott-2006}.
\begin{figure}
\begin{minipage}{0.45\linewidth}
\includegraphics[width=1.0\columnwidth]{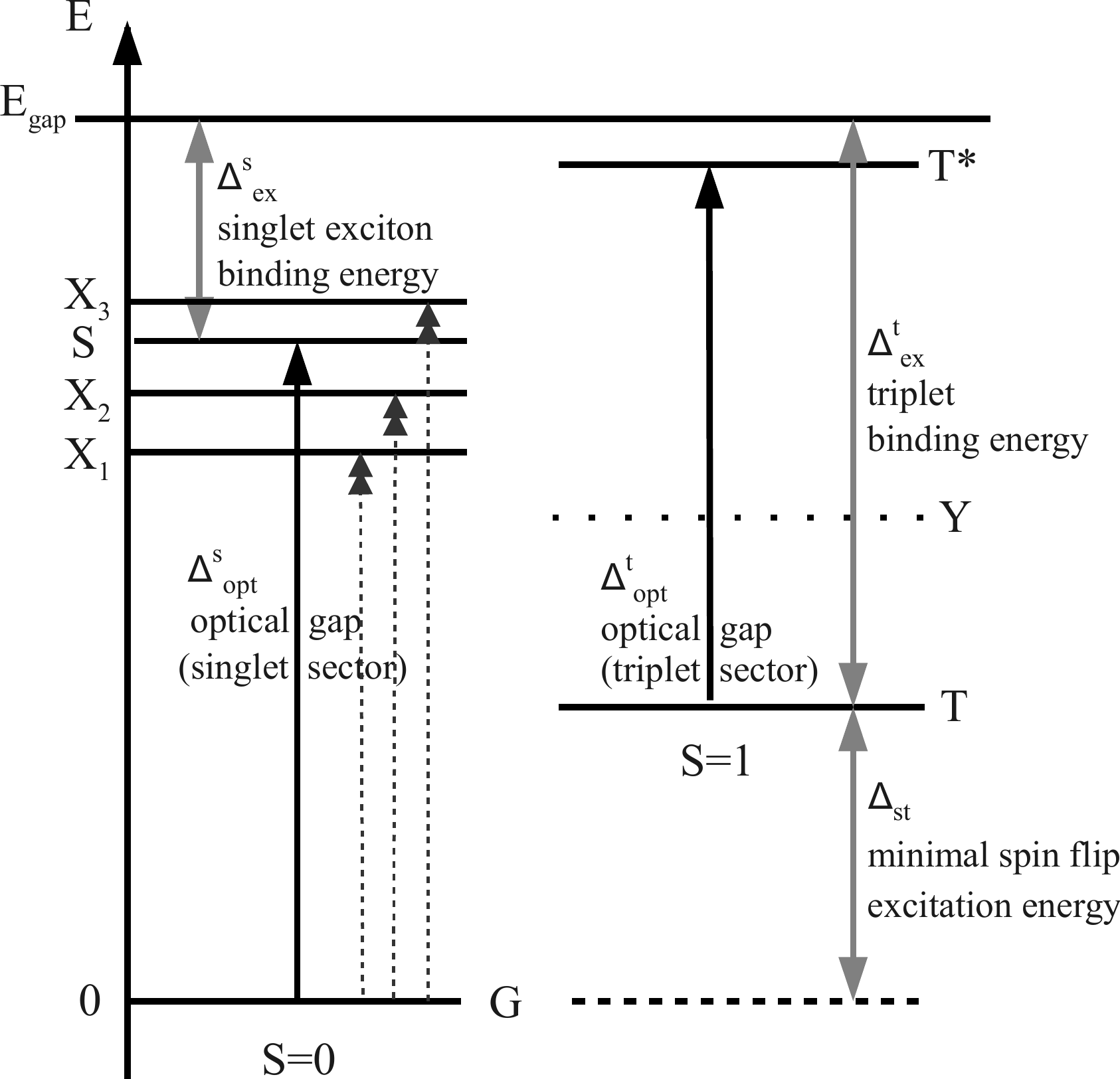}
\end{minipage}
\begin{minipage}{0.25\linewidth}
{\small
\begin{tabular}{|l|l|l|}
\hline
Energy/eV & 3BCMU & DMRG\\
\hline\hline
$E_{\rm X_1}$ &  {\sl 1.5} &  1.74 [1.94] \\
$E_{\rm X_2}$ &  {\sl 1.7} &  1.85$^{\rm a}$ [2.0]$^{\rm a}$ \\
$E_{\rm S}=\Delta_{\rm opt}^{\rm s}$ &  {\bf 1.896} &  2.00 [2.05] \\
$E_{\rm X_3}$ &  {\sl 2.0} &  \\
$E_{\rm gap}$ &  {\bf 2.482} & 2.45 [2.47] \\
$\Delta_{\rm ex}^{\rm s}=E_{\rm gap}-E_{\rm S}$
&{\bf 0.586} & 0.45 [0.42]  \\
$E_{\rm T}=\Delta_{\rm st}$
&  {\sl 1.0\hphantom{00}} $\pm$ {\sl 0.05}  & 1.00 [1.06]\\
$E_{{\rm T}^*}=\Delta_{\rm st}+\Delta_{\rm opt}^{\rm t} $
& {\sl 2.36\hphantom{0}} $\pm$ {\sl 0.05}   & 2.25\\
$\Delta_{\rm opt}^{\rm t}$ &  {\bf 1.360}   & 1.25 [1.28]\\
$\Delta_{\rm ex}^{\rm t} =E_{\rm gap}-E_{\rm T}$
&  {\sl 1.5\hphantom{00}} $\pm$ {\sl 0.05}  &  1.45 [1.40]\\
\hline
\end{tabular}
}
\end{minipage}
\caption{(left) Energy levels of in-gap states in the spin-singlet
and spin-triplet sectors.
Single-tip arrows: optical
absorption spectroscopy;
double-tip arrows: two-photon absorption spectroscopy.
Double arrows: binding energies (gaps).
G:~singlet ground state ($1{}^1A_g$); S: singlet exciton ($1{}^1B_u$);
X$_1$, X$_2$, X$_3$: singlet dark states ($m{}^1A_g$);
T: triplet ground state ($1{}^3B_u$); T$^*$: optical
excitation of the triplet ground state ($1{}^3A_g$);
Y: dark triplet state ($m{}^3B_u$).
(right) 
Excitation energies in 3BCMU at low temperatures.
All energies are measured in eV relative to the energy
of the ground state, $E_{\rm G}=0$.
Bold number: directly measured; italic number: estimate.
For DMRG results
the numbers in square brackets give the excitation energy
for the rigid-lattice transition from~G
($E_{\rm gap}$, $E_{\rm S}$, $E_{{\rm X}_{1,2}}$, $E_{\rm T}$)
and from~T ($\Delta_{\rm opt}^{\rm t}$).
}
\label{fig:pda_spectrum}
\end{figure}
In contrast to inorganic semiconductors,
the exciton binding energy in PDAs amounts to about
20\% of the single-particle gap, 
thus Coulomb interaction is substantial and effective 
and the electron-electron interaction must be treated very 
accurately. Due to such high computational demand
earlier attempts based on density-functional theory calculations
of the bare band structure in local-density approximation (LDA) failed 
to reproduce the experimentally measured excitation spectrum
\cite{Rohlfing-1999,Horst-1999}.

In contrast to this, using the DMRG method and by correlating 
some 100 electrons on 100 orbitals together with a 
geometrical optimization based on the Hellmann--Feynman theorem, i.e.,
by minimizing the force-field induced by the electron 
distribution\cite{Race-2003}, very accurate energy spectrum
can be obtained\cite{Barcza-2013}.
This is shown in Fig.~\ref{fig:pda_spectrum} and the experimentally
measured and DMRG calculated results are also summarized in 
the corresponding table.
In addition, the calculated geometrical structure agrees perfectly
with the experimental data, i.e., the single, double and triple bonds
are estimated as $r_t=1.22$, $r_d=1.37$ and $r_s=1.43$.

\subsubsection{Four-component density matrix renormalization group}
\label{sec:num.misc.4cDMRG}

\begin{figure}
\centering
\setlength{\unitlength}{40pt}   
\begin{picture}(6.5,5)
\put(  0,  0){\includegraphics[scale=0.58]{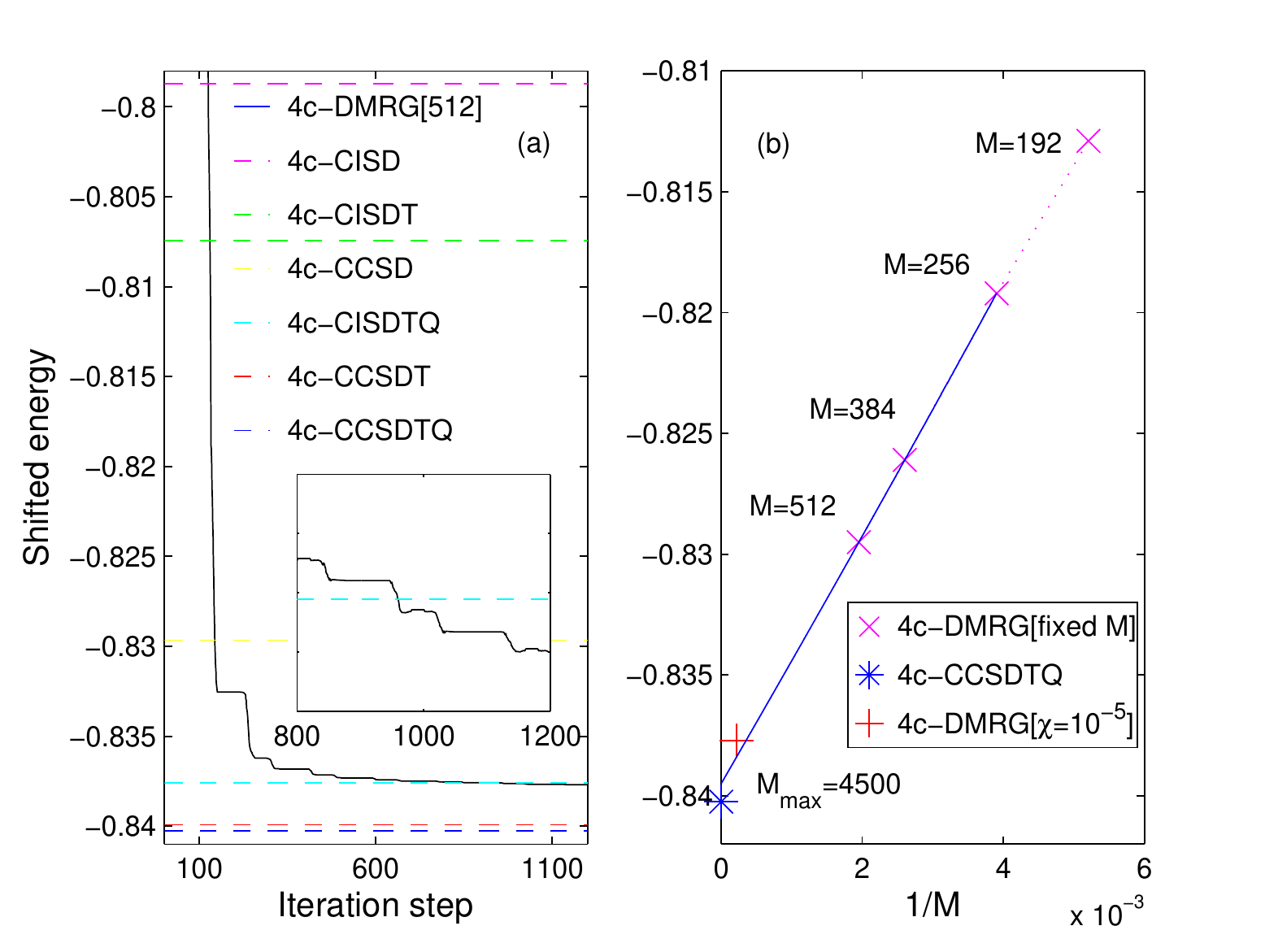}}
\put(0.7,5){\makebox(0,0)[r]{\strut{}(a)}}
\put(3.6,5){\makebox(0,0)[r]{\strut{}(b)}}
\end{picture}
\caption{(a) Convergence of the ground state energy (shifted by 20275 
$E_{\rm h}$) as a function of iteration steps of the 4c-DMRG(14,94)
($M_{\rm max}=4500$, $M_{\rm min}=1024$, $M_{\rm min}^{\rm DEAS}=2048$, $\chi=10^{-5}$) approach
at $r_e^{\rm exp}=1.872$ \AA.
Reference energies calculated by various CI and CC wave function models are also given as horizontal lines.
The inset shows that the 4c-DMRG energy drops below the 4c-CI-SDTQ energy.
(b) Extrapolation of DMRG energies E($M$)-20275 $E_{\rm h}$ 
for fixed $M$ values towards the limit $E(M\rightarrow\infty)$-20275 $E_{\rm h}$.
Figure is taken from arxiv:1312.0970.}
\label{fig:tlh-energy}

\end{figure}

Quite recently, the first implementation of the \emph{relativistic} quantum 
chemical two- and four-component density matrix renormalization group 
algorithm (2c- and 4c-DMRG) has also been presented\cite{Knecht-2014}.
This method includes a variational description of scalar-relativistic effects 
and spin--orbit coupling. By correlating 14 electrons on 94 spinors
and employing the Dirac--Coulomb Hamiltonian
with triple-$\zeta$\ quality basis,
the \emph{Potential Energy Surface} (PES) and spectroscopic 
constants have been obtained for the \emph{thallium hydride molecule.}
Utilizing the various entanglement based optimization techniques discussed in 
Sec.~\ref{sec:num.optim}, the 
CCSD reference energy has been reproduced even after the first DMRG sweep as
is shown in Fig.~\ref{fig:tlh-energy}. 
Although the 4c-CCSDTQ reference energy could not be reached with a 
maximum of $M=4500$ block states, 
the resulting 4c-DMRG potential energy curve did not only 
effectively reproduced the shape of the 
4c-CCSDTQ potential energy curve but also yielded accurate spectroscopic 
constants as extracted from a fourth-order polynomial fit.
Since QC-DMRG picks up all excitations 
required to describe the wave function to a given accuracy
the general structure of the wave function 
is preserved and could have been determined even with smaller $M$ values. 
By making the best of entanglement optimization the new 2c- and 4c-DMRG 
method is expected to become an efficient approach 
for heavy-element molecules that exhibit 
rather strong multi-configurational character in their ground- and 
excited states. Development of a 2c- and 4c-TTNS method is straightforward.

\subsubsection{Possible technical developments: hybrid CPU/GPU parallelization}
\label{sec:num.misc.GPU}

The original DMRG algorithm, introduced by S.~R.~White,
was formulated as a single threaded algorithm\cite{White-1992b}.
In the past various works have been 
carried out to accelerate the DMRG algorithm on shared\cite{Hager-2004,Alvarez-2012} 
and distributed memory\cite{Chan-2004,Kurashige-2009,Yamada-2011,Rincon-2010} 
architectures.
One of the first parallelizations was converting the projection 
operation to matrix-matrix multiplications and accelerating them via OpenMP interface\cite{Hager-2004}.
A similar approach has been presented for distributed memory environment 
(up-to 1024 cores) optimizing the communication between the cores\cite{Yamada-2011}, 
while the acceleration of the computation of correlation function has
 also been investigated\cite{Rincon-2010}. 
A novel direction for 
parallelization via a modification of the original serial DMRG algorithm
have also been introduced\cite{White-2013}. 

\emph{Graphical Processing Unit} (GPU) has been successfully employed in 
neighboring research 
areas to accelerate matrix operations. GPU is used to accelerate 
tensor contractions in \emph{Plaquette Renormalization States} (PRS)\cite{Yu-2011},
which can be regarded as 
an alternative technique to tensor network states (TNS) or the DMRG algorithm.
The \emph{second-order Spectral Projection} (SP2) algorithm has been 
accelerated, which is an alternative technique to calculate the 
density matrix via a recursive series of generalized matrix-matrix 
multiplications\cite{Cawkwell-2012}

\begin{figure}[htb]
  \centerline{
  \includegraphics[scale=0.8]{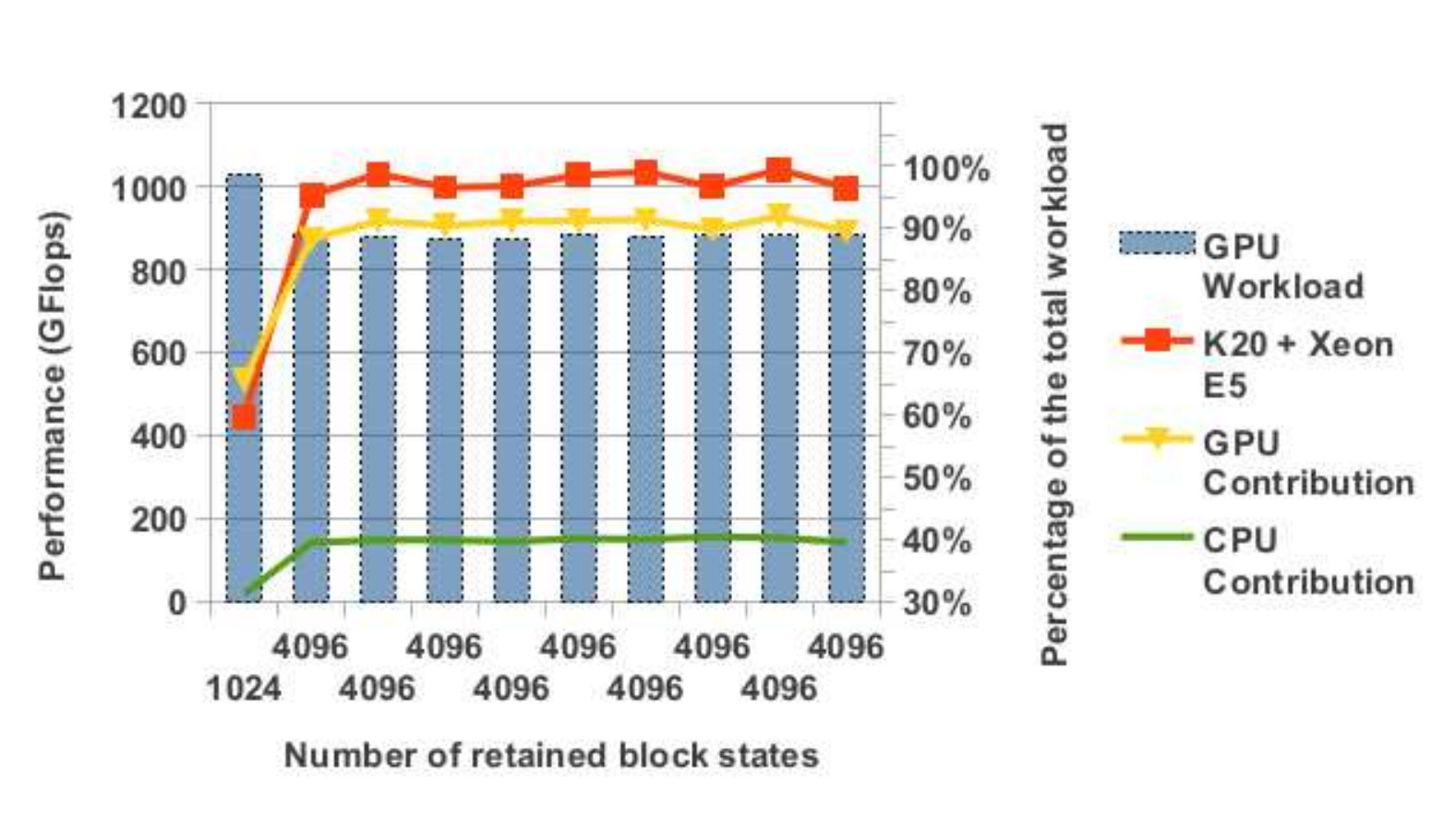}
  }
  \caption{Performance results of the hybrid CPU-GPU acceleration of 
the projection operation
for the Hubbard model on Intel Xeon E5-2640 2.5GH CPU + NVidia K20 GPU:
1071 GFlops and $\times$3.5 speedup is reached. 
(Theoretical maximum is 1.17 TFlops)
Blue bars associated to the secondary vertical axis indicate the ratio of the current GPU workload.
}
  \label{fig:gpu}
\end{figure}

Quite recently, it has been investigated
how the DMRG method can utilize the enormous computing capabilities of novel kilo-processor architectures: 
\emph{Graphical Processing Unit} (GPU) and \emph{Field-Programmable Gate Array} (FPGA)\cite{Nemes-2014}.
In case of GPU a smart hybrid CPU-GPU acceleration has been presented, which tolerates problems exceeding 
the GPU memory size, consequently, supporting wide range of problems and GPU configurations.
Contrary to earlier acceleration attempts not only the projection operation 
was accelerated, 
but further parts of the diagonalization were also computed on the GPU.
Reported results on the one-dimensional Hubbard model for a mid-range 
(Intel Core-i7 2600 3.4 GHz CPU + NVidia GTX 570 GPU)  and on a high-end 
configuration (Intel Xeon E5-2640 2.5 GHz  CPU + NVidia K20 GPU) showed
that if the workload is properly distributed 
(see Fig.~\ref{fig:gpu})
the mid-range configuration with GPU can be approximately 
$2.3$-$2.4$ times faster than without GPU, 
while the high-end configuration can be accelerated by $3.4$-$3.5$ times 
using the GPU.

The GPU architecture has been found to be a promising accelerator, 
as the most time-dominant step of the algorithm, the projection operation, 
can be formulated as 
independent dense matrix multiplications, which are ideal workload for GPUs.
Moreover, in case of high-end GPUs the acceleration of the projection 
is so remarkable, 
that it is worth to consider the acceleration of the rest of the algorithm 
to obtain a decent overall speed-up. 
Therefore, extensions to treat ab-initio quantum chemical 
applications
and a straightforward generalization 
of the algorithm to accelerate tensor network state  
algorithms\cite{Orus-2014} 
are promising research directions.



\section{Summary and outlook}


In the past decade, we have witnessed a breakthrough in electronic structure 
calculations due to the \emph{Density Matrix Renormalization Group} (DMRG) method 
which has become
a viable alternative to conventional 
multiconfiguration wave function approaches.
Inclusion of the concepts of entanglement from quantum information theory has
paved the road for identifying highly correlated molecular orbitals leading to 
an efficient construction of active spaces and for characterizing the various 
types of correlation effects relevant for chemical bonding. Quite recently, 
a reformulation of DMRG in terms of \emph{Matrix Product States} (MPS) 
has shown that it is only one special case in a much 
more general set of methods, the \emph{Tensor Network States} (TNS), 
which is expected to even outperform DMRG/MPS in the near future.

A special class of such ansatz states are the
\emph{Tree Tensor Network States} (TTNS). The mathematically rigorous analysis 
of these tensor trees has been completed only partially and many open questions remain,
concerning for example numerical procedures, but also more theoretical concepts
of differential and algebraic geometry.

In the \emph{quantum chemsitry version} of the method (QC-TTNS),
the wave function with variable tensor order
is formulated as products of tensors in a multiparticle basis spanning a
truncated Hilbert space of the original CAS-CI problem.
The tree-structure is advantageous since the distance between two 
arbitrary orbitals in the tree scales only logarithmically with the number 
of orbitals, whereas the scaling is linear in the MPS array. Therefore,
the TTNS ansatz is better suited for multireference
problems with numerous highly correlated orbitals.

The underlying benefits of QC-TTNS is, however, far from fully exploited 
and the optimization of the method is far more complicated. Due to the more 
advanced topology, several optimization tasks and problems arise which do not 
have counterparts in the MPS formulation.  
Therefore, there is a tedious work still ahead of us.

\section{Acknowledgments}

This research was supported by the
European Research Area Chemistry (ERA-Chemistry) in part by
the Hungarian Research Fund (OTKA) under Grant No. NN110360 and K100908,
the DFG SCHN 530/9-1 project under Grant No. 10041620
and FWF-E1243-N19. V.M. and F.V. acknowledge support from the
SFB-ViCoM project.
\"O.L. acknowledges support from the Alexander von Humboldt foundation
and from ETH Z\"urich during his time as a visiting professor.
The authors are also grateful for Libor Veis, 
Jen{\H o} S\'olyom, P\'eter Nagy, and Tibor Szilv\'asi 
for helpful discussions and comments. 


\bibliographystyle{plain}
\bibliography{mps_review-bibliography}{}

\end{document}